# Higher-dimensional field theories from type II supergravity

Marco Fazzi

Université Libre de Bruxelles
Faculté des Sciences
Physique Théorique et Mathématique

# Higher-dimensional field theories from type II supergravity

**Thèse présentée en vue de l'obtention du titre de Docteur en Sciences**

Marco Fazzi

Année académique 2015–2016

Directeur de thèse : Andrés Collinucci

UNIVERSITÉ LIBRE DE BRUXELLES

fnrs
LA LIBERTÉ DE CHERCHER

Electronic version.

## Acknowledgments

The work presented in this thesis was carried out at the *Université Libre de Bruxelles*, in the *Physique Théorique et Mathématique* group of the Department of Physics. It was supported by a grant of the Belgian F.R.S–FNRS of which the author has been a Research Fellow (*Aspirant*), and partly by the ERC advanced grant "SyDuGraM", by IISN (grant 4.4514.08) and by the Communauté Française de Bélgique through the ARC program. Part of the work was carried out at UCSB. The author wishes to thank the theoretical physics group at UCSB and the KITP for hospitality.

We wish to thank R. Argurio, A. Collinucci, D. R. Morrison, and A. Tomasiello for constant support, guidance, and collaboration at various stages of this thesis. Thanks extend to our colleagues and friends F. Apruzzi, E. Conde Pena, S. Giacomelli, A. Marzolla, M. Moskovic, A. Passias, D. Redigolo, D. Rosa, and L. Tizzano, for making these four years joyful and interesting.

## Original contributions

The original contributions contained in this thesis are based on the following published papers:

[I] F. Apruzzi, M. Fazzi, D. Rosa, and A. Tomasiello. "All $AdS_7$ solutions of type II supergravity". *JHEP* 04 (2014), [064]. arXiv: 1309.2949 [hep-th] (cit. on pp. 5–8, 81, 84, 131, 153, 164, 173).

[II] F. Apruzzi, M. Fazzi, A. Passias, D. Rosa, and A. Tomasiello. "$AdS_6$ solutions of type II supergravity". *JHEP* 11 (2014). [Erratum: *JHEP* 05 (2015), [012]], [099]. arXiv: 1406.0852 [hep-th] (cit. on pp. 7, 8, 177, 183).

[III] F. Apruzzi, M. Fazzi, A. Passias, and A. Tomasiello. "Supersymmetric $AdS_5$ solutions of massive IIA supergravity". *JHEP* 06 (2015), [195]. arXiv: 1502.06620 [hep-th] (cit. on pp. 7, 8, 69, 127).

[IV] F. Apruzzi, M. Fazzi, A. Passias, A. Rota, and A. Tomasiello. "Six-Dimensional Superconformal Theories and their Compactifications from Type IIA Supergravity". *Phys. Rev. Lett.* 115.6 (2015), [061601]. arXiv: 1502.06616 [hep-th] (cit. on pp. 69, 81, 127).

# Contents

# List of Figures

# List of Tables

# Chapter 1

# Introduction

One of the most outstanding predictions of string theory is the existence of interacting strongly-coupled fixed points in dimensions higher than four, namely five and six.

### Field theory

On general grounds, defining such a field theory as a gauge theory poses some obvious difficulties: Since a gauge field $A$ has mass dimension $\left[M^{(d-2)/2}\right]$ in $d$ spacetime dimensions, and the gauge coupling $\left[g_{\rm YM}^{-2}\right] = \left[M^{d-4}\right]$, the latter must be dimensionful in dimensions other than four. In particular, for $d > 4$ it grows at high energy signaling nonrenormalizability of the Yang–Mills interaction $\operatorname{Tr} F^2$. On the contrary, as we decrease the energy scale the gauge kinetic term becomes irrelevant $(\left[F^2\right] = \left[M^d\right])$ and we eventually hit a *free fixed point* of the RG flow: The gauge coupling vanishes and the theory is free of any interaction. Reabsorbing the gauge coupling into the gauge field, $A \to A' \equiv gA$, we now have $[A'] = \left[M^1\right]$ and $\left[F'^2\right] = \left[M^4\right]$ in any dimension $d$. The noncanonically normalized interaction then reads

$$\frac{1}{g_{\rm YM}^2} \operatorname{Tr} F'^2 \ , \tag{1.1}$$

and

$$\frac{1}{g_{\rm YM}^2} \to 0 \quad \text{as} \quad M \to 0 \quad \text{for} \quad d > 4 \ . \tag{1.2}$$

Obviously this discussion assumes that the gauge theory description, i.e. the Yang–Mills Lagrangian (1.1), be valid at some energy scale, and then applies the naive power counting argument. Something has to happen for the theory to violate such assumptions. For instance we could imagine that, at some very high energy scale (in the so-called $UV$), the Yang–Mills coupling $g_{\rm YM}^2$ blows up, rendering the gauge theory description inadequate. The coupling does not change anymore with energy, the theory is at a fixed point, but the dynamics of the theory is strongly-coupled, $g^2_{\text{YM}} \to \infty$. We reached a UV *interacting fixed point* theory.

It is believed that an interacting fixed point enjoys the full conformal symmetry, not only scale invariance (being a fixed point of the RG flow its $\beta$ function vanishes by definition).[1] Moreover, since it is connected by RG flow evolution to a free field theory, this should be viewed as an indication that it too is a local quantum field theory [2]. However, for six-dimensional theories with supersymmetry such a local description is far from being obvious, and in fact is not known at present.

In supersymmetric theories in six dimensions which are also conformally invariant the lack of a lagrangian description is mainly due to the presence of infamous *tensor multiplets* of the $\mathcal{N} = (2,0)$ and $\mathcal{N} = (1,0)$ superconformal algebras [3]. These multiplets contain a two-form which has a self-dual three-form as field strength. Writing down a kinetic term for such a field is a nontrivial challenge. Partial results have been obtained in the form of classical equations of motion for such (abelian and nonabelian) tensor multiplets [4], "pseudo-Lagrangians" for $(2,0)$ tensor multiplets coupled to $(1,0)$ vector multiplets to be accompanied by the self-duality constraint on the three-form [5], approaches à la Pasti–Sorokin–Tonin [6] whereby the self-duality constraint is built-in in the Lagrangian [7]. It is not clear however whether such (pseudo)lagrangian descriptions would go through to the quantum regime, nor is it known how to generalize these tentative constructions to the full nonabelian case of many tensor multiplets interacting with more ordinary matter (e.g. hypermultiplets).

In any case, assuming the local field theory philosphy is correct, one can set out to study [2, 8, 9] which conditions theories with the minimal amount of supersymmetry in five and six dimensions (i.e. eight Poincaré $Q$ supercharges) would have to satisfy to be sensible at the quantum level. In six dimensions such conditions reduce to the requirement of gauge anomaly cancellation, given that the superalgebra is chiral.

## (Super)conformal field theories

The hypothesis that the gauge theory description may not always be valid, and that we should trade the latter for a conformal theory whenever the validity is not warranted, has some important consequences at the quantum level. First of all, the interpretation of particles as states in a Hilbert space classified by quantum numbers (mass, spin, etc.) is impossible. Since the theory is a fixed point of the RG flow, the $\beta(g) = M\,\partial_M g$ function of the gauge coupling $g$ vanishes. The coupling is said not to "run": It does not depend on the energy scale $M$ at which we probe the theory. In other words, there is just no characteristic mass scale in the theory. This reflects the fact that the mass operator $P_\mu P^\mu$ (the square of the generator of translations, the momentum operator $P^\mu$) does not commute with the generator of dilatation $D$ (in the conformal algebra): Mass, as well as energy, can be freely rescaled to any value. The particle spectrum is not discrete; rather it is a continuum of states. The physical observables will then be associated with (gauge invariant) operators with "good" conformal transformation properties (they will be classified according to their *conformal dimensions*), and will have to satisfy stringent constraints coming from the conformal algebra.

[1] For an updated discussion on scale vs. conformal invariance see the review [1].

All information in the theory is contained in correlation functions of such operators. This is where holography enters the game, as will be explained later on.

Moreover, if we are interested in supersymmetric field theories, we should make sure that the conformal algebra can be extended to the full *superconformal* one. Six is the highest spacetime dimension in which the supersymmetry algebra can be enhanced to a superconformal one [3]. This fact makes a strong case for studying six-dimensional theories. It is believed that a better understanding of the higher-dimensional "mothers" of four-, three-, etc. dimensional theories will ultimately help us understand nonperturbative aspects of their dynamics. Regardless of the field theory applications, understanding these higher-dimensional theories turns out to be important to understand the nonperturbative dynamics of *string theory* itself.

### String theory

The striking observation is that string theory provides us with many beautiful ways to engineer field theories in higher dimensions displaying such exotic properties: Conformal field theories with the right amount of supersymmetry, defying the usual gauge theory description and describing the propagation of massless (extended) objects which cannot be given the interpretation of particles sweeping out a worldline.

At the quantum level string theory is perturbatively well-defined. As we will see, at small string length $l_s$ the theory supported on the worldvolume of a brane, an extended object with charge and tension, reduces to a field theory with the usual properties. This result rests upon our ability to identify the spectrum of massless excitations of an open string ending on such extended object. If we consider an arbitrarily complicated configuration of branes, of different dimensions and kinds, then the field theory describing the massless spectrum and the interactions among its constituents will reflect the complication. If we believe that a proper quantum definition of string theory exists, then we are led to accept the existence of interacting quantum field theories arising from complicated *brane configurations* although we might not be able to satisfactorily describe them with field theory methods available at present.

The quest is then finding a suitable *decoupling limit* whereby we can neglect the gravitational interaction present throughout the spacetime string theory is defined on, and be left with an interacting theory of quantum fields, but no gravity. This limit typically involves taking the Planck mass to infinity and the string length $l_{\mathrm{s}}$ to zero, while keeping the string coupling $g_{\mathrm{s}}$ fixed.

Moreover, certain movements of the branes present in the configuration will be identified with irrelevant deformations in field theory. For instance, in configurations of M5-branes in M-theory or NS5-branes in type IIA string theory (which will be discussed at length in the thesis), we can take all the branes to lie on top of each other. This is usually the point where our field theory understanding of the brane configuration breaks down: An infinite tower of massless states appears in the spectrum, and we are led to conjecture that the field theory is actually (super)conformal, and that it might be a strongly-coupled fixed point of the RG flow in a certain number of dimensions (six, for M5's or NS5's).

The possibility of *embedding* a field theory inside string theory in such a way is usually assumed to be a proof of existence of the former, and we shall pursue this idea throughout the thesis.

The first construction of a six-dimensional interacting fixed point in string theory was proposed by Witten [10], who introduced the famous $\mathcal{N} = (2,0)$ theories. They admit a simple classification in terms of the $A$, $D$, $E$ discrete subgroups of SU(2). To engineer one of such theories, consider type IIB string theory compactified on a so-called *K3 surface*. At certain loci of its moduli space, this four-dimensional manifold can degenerate and develop a *singularity*. This singularity is locally modeled by the origin of the $\mathbb{C}^2/\Gamma$ *orbifold*, with $\Gamma < \mathrm{SU}(2)$ of type $A$, $D$ or $E$. We can resolve the singularity at the origin of this space by a sequence of blow-ups, that is by excising the singular point and replacing it by a chain of two-cycles, $\mathbb{P}^1 \cong S^2$, intersecting according to the Dynkin diagram of type $A$, $D$ or $E$. The massless objects of the theory arise from D3-branes wrapping these two-cycles and extending along two directions in the external flat space (which is six-dimensional). When the cycles are blown down the tension of the D3's vanishes, and we are left with *tensionless strings* propagating in flat space.

Later, Strominger [11] gave an alternative description of the $(2,0)$ theories of type $A$. In this picture, the field theory is the worldvolume theory of coincident M5-branes probing flat space in M-theory. If we now consider M2-branes suspended between the M5's, we have a good candidate for tensionless extended objects. In fact, an open M2-brane ending on an M5 acts like a $(1+1)$-dimensional object (a string) from the perspective of the worldvolume of the latter; its tension goes like the distance between the M5's. When the latter coincide, the M2 becomes tensionless.

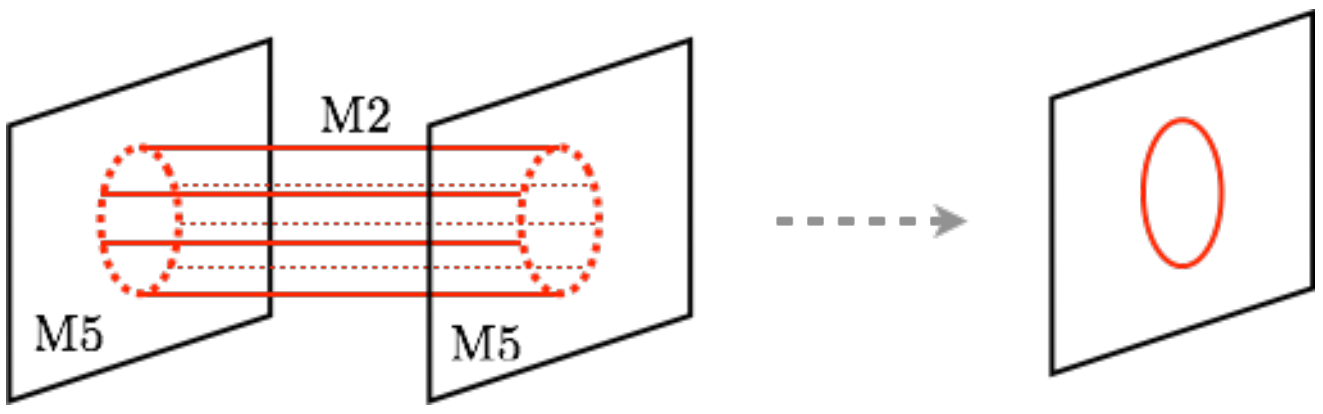

On the left side an open M2-brane suspended between two consecutive M5-branes. When the latter are brought on top of each other, the M2 becomes tensionless (its tension being proportional to the distance between the M5's), and it acts like a $(1+1)$-dimensional object on their worldvolume (i.e. a string).

Until very recently, little was known about string theory embeddings of their less supersymmetric cousins, namely six-dimensional $(1,0)$ theories. They enjoy half the amount of supersymmetry of the $(2,0)$ ones (i.e. only eight $Q$ supercharges). This suggests that we need some new ingredient to be added to the original M-theory and type IIB constructions proposed by Witten and Strominger. A possible way to achieve this in M-theory is to place the M5-branes (or the type IIB NS5-branes) near a spacetime singularity, like the $\mathbb{C}^2/\Gamma$ orbifold [12, 13, 14]. The latter acts like a half-BPS object, breaking in half the supersymmetry of the M5's probing flat space.

Other, exotic, examples include the theory of small $E_8$ instantons in the $E_8 \times E_8$ heterotic string theory [15, 16, 17], or the collision of two $E_8$ singularities in F-theory [18]. A more promising class of examples was introduced in [19, 20]. It makes use of a so-called Hanany–Witten brane setup to engineer the field theory. The ingredients of these setups are NS5, D6 and, crucially, D8-branes in *massive* type IIA string theory.

Clearly, it would be desirable to have a full classification in which the aforementioned examples naturally fit. The problem was finally attacked in [21, 22, 23] from the F-theory perspective, and simultaneously in [I, 24] from the *holographic* one, to which we now turn.

## Supergravity and holography

The idea proposed in [I] is to use holography, also known as the *AdS/CFT correspondence*, to classify the type IIA supergravity backgrounds dual to the six-dimensional $(1,0)$ theories engineered by the NS5-D6-D8 Hanany–Witten setups of [19, 20]. It is important to keep in mind that this approach is limited to those six-dimensional theories admitting a *perturbative* type IIA embedding. Therefore the ensuing classification will not include *all* $(1,0)$ theories we know of. This is to be contrasted with the point of view championed in [23], according to which F-theory (the nonperturbative completion of type IIB string theory) is actually capable of engineering all $(1,0)$ and $(2,0)$ theories discovered so far, plus many more new examples.

In this thesis we will only be concerned with those $(1,0)$ theories that can be engineered via NS5-D6-D8 brane systems. The classification strategy is the following.

In [25] Maldacena discovered that the abstract concept of *holography* [26], believed to be a general property of quantum gravity [27], has a very powerful and concrete incarnation in string theory: The *AdS/CFT correspondence*. In its original formulation, the latter states that type IIB string theory compactified on the space $\mathrm{AdS}_5 \times S^5$ with $N$ quanta of $F_5$ flux is equivalent (even at the quantum level) to four-dimensional $\mathcal{N} = 4$ super-Yang–Mills with $\mathrm{SU}(N)$ gauge group. Both sides of the correspondence arise as particular limits of the worldvolume theory of $N$ D3-branes probing ten-dimensional flat space: The uniqueness of the "parent theory" should then justify the conjectured correspondence, or *duality*, between its two "sons".

Even assuming that it hold, the regime in which we can actually exploit the power of this duality is when one of the two sides is weakly-coupled, and the other is not. Performing reliable computations in the former should thus enable us to extract information about the latter. In type II string theory, such weakly-coupled regime is type II supergravity, in which all $g_s$ and $\alpha'$ corrections are usually neglected. It is a classical approximation of the full-fledged string theory in which only the massless part of the spectrum is retained, and the massive one decoupled.

In particular, the type IIB *supergravity vacuum* $\mathrm{AdS}_5 \times S^5$, obtained from the closed string spectrum the D3-branes give rise to, is perfectly under control.

It is then natural to hope that, by studying $\mathrm{AdS}_{d+1}$ vacua of type IIA/B supergravity with more general *internal spaces* (not just $S^5$, or $S^{10-(d+1)}$), we can access the strongly-coupled regime of the *dual* $d$-dimensional field theories. This expectation has been largely borne out, as a very vast literature proves. (For a nice review on the subject see [28].)

With this philosophy in mind, in [I] we set out, and managed, to classify all $\mathrm{AdS}_7$ vacua of (massive) type IIA supergravity. The dual statement is that we classified all possible six-dimensional $(1,0)$ theories admitting a perturbative type IIA string theory embedding, as an NS5-D6-D8 Hanany–Witten setup. The $AdS_7$ vacua are obtained from the latter via a procedure known as *near-horizon limit*.

## Outline and results

In part I of this thesis we review Hanany–Witten brane constructions in diverse dimensions. Our discussion will be based on references [19, 20, 29, 30, 24, 31, 32]. In part II we present the original contributions [I, II, III].

In chapter 3 we first present the gauge theory – brane worldvolume correspondence in string theory (section 3.1). We move on to introduce the main idea behind the Hanany–Witten brane setups (section 3.2), and explain how to obtain a four-dimensional gauge theory with flavors out of an NS5-D4(-D6) system in type IIA string theory (sections 3.2.1 and 3.2.2). In the type IIB D3-D5 system we see the appearance of Nahm pole boundary conditions for the gauge theory (section 3.2.3). We then explain how to decouple gravity in the brane system and be left with an interacting field theory (section 3.2.4). In the second part of the chapter (section 3.3) we generalize the NS5-D4(-D6) systems to the NS5-D6(-O6-D8-O8) ones introduced by Brunner–Karch and Hanany–Zaffaroni, engineering six-dimensional $(1,0)$ theories in (massive) type IIA string theory. We introduce all necessary ingredients – D6-branes (section 3.3.1) and their M-theory origin (section 3.3.2), D8-branes (section 3.3.3), O8-planes (section 3.3.4), O6-planes (section 3.3.5) – and highlight the differences with respect to their lower-dimensional counterparts. Finally (section 3.4) we introduce linear quiver gauge theories obtained from NS5-D6-D8 brane systems; we explain how to flow to interesting six-dimensional superconformal theories (section 3.4.1) and in which limit the latter admit a holographic dual in (massive) type IIA supergravity (section 3.4.2).

In chapter 4 we first introduce the AdS/CFT correspondence (section 4.1) – or holography – paying particular attention to its original incarnation, namely the case of $N$ D3-branes probing flat space (section 4.1.1). We apply a similar philosophy to NS5-D6-D8 brane systems, and propose how to find a holographic limit in these cases whereby one can use AdS vacua of type IIA supergravity to study $(1,0)$ field theories (section 4.1.2). In section 4.3 we introduce elements of type II supergravity needed to understand the results presented in part II: Its field content and supersymmetry equations (section 4.3.1), and the language needed to recast the latter into a system of equations for differential forms, rather than spinors (section 4.3.2). Finally we define the conformal anomaly of a superconformal field theory and explain how to compute it in holography (section 4.2); we give the details of the computation in the case of six-dimensional $(1,0)$ theories (section 4.2.1).

In chapter 5 we first explain the relation between the results in [I] and the constructions of chapter 3. We then present the results of [I] (section 5.1). The aim is to classify $\mathrm{AdS}_7$ vacua of type II supergravity. In type IIB, we show that there are none; in massless type IIA there is only the reduction of the Freund–Rubin $\mathrm{AdS}_7 \times S^4$ solution of eleven-dimensional supergravity; in massive type IIA we find an infinite class of new vacua (studied numerically). These feature the presence of D6 and/or D8-brane sources, and the internal space is topologically a three-sphere. We explain how these vacua could arise as holographic – or near-horizon – limits of the NS5-D6-D8 brane constructions of section 3.3, engineering six-dimensional $(1,0)$ superconformal field theories.

In chapter 6 we present the results of [III]. The aim is to classify $\mathrm{AdS}_5$ vacua of type IIA supergravity. We recover many vacua already studied in the literature. We also identify a new infinite class of analytic vacua in massive type IIA, within a certain simplifying Ansatz. They are in bijective correspondence with the $\mathrm{AdS}_7$ ones of [I], presented in chapter 5. The internal space is given by a fibration of a three-dimensional manifold (a distortion of the three-sphere of the $\mathrm{AdS}_7$ vacua) over a Riemann surface. We exploit the correspondence to give a fully analytic definition of the $\mathrm{AdS}_7$ vacua. The classification of both $\mathrm{AdS}_5$ and $\mathrm{AdS}_7$ vacua of massive type IIA depends on a single function of the internal space, which has to satisfy a certain ODE. These results suggest that the $\mathrm{AdS}_5$ vacua are the holographic duals to four-dimensional $\mathcal{N}=1$ superconformal field theories obtained via (twisted) compactification of the six-dimensional $(1,0)$ ones on the Riemann surface.

In chapter 7 we first introduce five-dimensional $\mathcal{N}=1$ superconformal field theories and their known string theory realizations, both in massive type IIA and in type IIB (section 7.1). We then present the results of [II]. The aim is to classify $\mathrm{AdS}_6$ vacua of type IIB supergravity which are dual to field theories engineered by $(p,q)$-fivebrane webs. We recover the only two known solutions as part of our general analysis, and we reduce the classification problem to two coupled PDEs. A solution to these PDEs would provide us with all the bosonic fields of type IIB supergravity compactified on $\mathrm{AdS}_6$. We find the internal space is constrained to be an $S^2$ fibration over a Riemann surface. We suggest how an $\mathrm{AdS}_6$ vacuum, explicitly obtained as solution to our general PDEs, could arise as near-horizon limit of a $(p,q)$-fivebrane web in type IIB string theory.

We present our conclusions in chapter 2.

# Chapter 2

# Conclusions

In this chapter we will briefly summarize the results obtained in the original contributions. The reader can equivalently come back to this chapter after having read part II of this thesis.

We have found and classified all vacua of type II supergravity of the form $\mathrm{AdS}_7 \times M_3$. In type IIB, we found that there are none (the BPS equations are impossible). In massless IIA ($F_0 \equiv 0$) there is only one analytic solution to the BPS equations, namely the reduction of the $\mathrm{AdS}_7 \times S^4$ Freund–Rubin vacuum of eleven-dimensional supergravity; it features the presence of D6 and anti-D6-brane sources at the poles of $M_3$, which is a fibration of a round $S^2$ over a finite interval (the total space is compact). (The SU(2) isometry of the fibers should realize the R-symmetry of putative dual six-dimensional $(1,0)$ SCFTs.) The poles of $M_3$ (i.e. the endpoints of the interval, where the fiber shrinks to zero size) are not regular points for the metric (the curvature diverges there), so that $M_3$ is topologically an $S^3$ distorted by the presence of the branes. The latter are also responsible for the divergent behavior of the NSNS three-form $H$. Both divergences are cured when lifting the massless solution to eleven-dimensional supergravity. In massive IIA (i.e. the Romans mass $F_0$ is not identically vanishing throughout $M_3$, and is required to satisfy its Bianchi identity $dF_0 = n_{\mathrm{D8}} \delta_{\mathrm{D8}}$, in presence of $n_{\mathrm{D8}}$ D8-brane sources) we have reduced the BPS equations and the Bianchi identities for RR and NSNS fluxes to three ODEs (on top of requiring flux quantization). We imposed boundary conditions compatible with the presence of D6-branes or D8-D6 bound states (more precisely, magnetized D8-branes with nonzero D6 charge). The ODEs can then be studied numerically, and we constructed solutions with as many stacks of magnetized D8's as one wishes. The position of each stack along the finite interval is fixed by supersymmetry to be proportional to the number of D8-branes and their D6 charge. The $S^2$ fiber that a D8 stack wraps is at finite distance from either of the poles of $M_3$. This can be interpreted as a Myers-like effect whereby a D6-brane (formerly located at one of the poles) puffs up into a D8 when $F_0$ is switched on, thereby rendering the poles regular points for the metric. The $H$ divergent behavior is cured as well. This construction provides an infinite class of fully regular and globally-defined new solutions of massive IIA supergravity. We provided an interpretation of these solutions as near-horizon limits of Hanany–Zaffaroni NS5-D6-D8 brane setups: $N$ NS5-branes turn into the quantized $H$ flux, $\int_{M_3} H \sim N$, the summand $\mu_i$ (in a partition $\mu$ of the number of D6-branes ending on the D8-branes) turns into the supergravity D6 charge smeared on the $i$-th magnetized D8-brane source, the D8-D6 fuzzy funnels ("nonabelian" bound states) specified by the partition $\mu$ turn into D8-brane sources wrapping $S^2$ fibers inside the internal space $M_3$. Thus our infinite class of $\mathrm{AdS}_7$ vacua should be seen as providing holographic duals to those six-dimensional $(1,0)$ theories admitting one in (massive) IIA supergravity.

We have found a new infinite class of vacua of massive type IIA supergravity of the form $\mathrm{AdS}_5 \times M_5$. First, we reduced the BPS equations of type IIA supergravity and the Bianchi identities for the fluxes (in absence of sources) to six PDEs for three functions of the internal space. The local form of the internal metric has been found to be that of a fibration of a three-dimensional space over a two-dimensional surface $C$, $\tilde{M}_3 \hookrightarrow M_5 \to C$, whose geometry is controlled by the PDEs. In order to solve the latter we imposed a so-called compactification Ansatz (field theory inspired). The PDEs then simplify dramatically, and reduce to a single ODE for a function $\beta$ of the fiber $\tilde{M}_3$. Moreover, some of the PDEs also impose that the base space should be a Riemann surface of genus greater than one, $C = \Sigma_g$, and that the fiber $\tilde{M}_3$ is itself a fibration of a round $S^2$ over a finite interval, $S^2 \hookrightarrow \tilde{M}_3 \to I$. Hence $\tilde{M}_3 \cong M_3$ (the internal space of massive $\mathrm{AdS}_7$ vacua). This is no coincidence, and we provided a field theory interpretation of this fact, as we will review momentarily. We then imposed boundary conditions on the ODE. Only a few are possible, and correspond to the presence of D6 or magnetized D8-brane sources in the internal space. Their position inside $\tilde{M}_3$ (i.e. along the finite interval) is fixed by supersymmetry, and turns out to be equivalent to the position of the corresponding sources inside $M_3$. We were also able to show that the system of three ODEs for the $\mathrm{AdS}_7$ vacua is fully equivalent to the single ODE for the studied subclass of $\mathrm{AdS}_5$ vacua; this allowed us to construct a universal one-to-one correspondence between the solutions in the two classes (provided we have a solution $\beta$ to the ODE we can map it to an *analytic* solution of the three ODEs, and vice versa). Given the simplicity of the ODE in $\mathrm{AdS}_5$, we were able to solve it analytically (when brane sources are added), and by means of the aforementioned map we also gave an analytic definition of the infinite class of $\mathrm{AdS}_7$ vacua which were previously known only numerically. The interpretation of this universal map (i.e. it does not depend on the specific solution it is applied to) is that we are compactifying the dual six-dimensional $(1,0)$ SCFTs on the very Riemann surface $\Sigma_g$, to produce an infinite class of four-dimensional $\mathcal{N}=1$ SCFTs admitting a holographic dual in type IIA supergravity. This is confirmed by the appearance of an abelian isometry of the internal space $\tilde{M}_3$ (to be contrasted with the bigger SU(2) isometry of $M_3$) which should realize the U(1) R-symmetry of the dual field theories. The compactification is twisted, and the twisting (i.e. the mixing of SU(2) with an SO(2) group of local transformations on $\Sigma_g$) only preserves a U(1) subgroup. Finally, by adapting the Henningson–Skenderis original computation for $(2,0)$ theories of type $A_N$, we computed holographically the conformal anomaly of some (sample) six-dimensional $(1,0)$ theories, and that of their four-dimensional $\mathcal{N}=1$ compactifications. As was to be expected, as a consequence of the universal map, the two are proportional, the coefficient being given by the area of $\Sigma_g$. It would clearly be desirable to have a field theory definition of the four-dimensional SCFTs (possibly along the lines of what happens for $\mathcal{N}=2$ class $\mathcal{S}$ theories), and not only their holographic construction.

We have reduced the classification problem of vacua of type IIB supergravity of the form $\mathrm{AdS}_6 \times M_4$ to two coupled, first-order PDEs. The BPS equations determine the local form of the metric to be that of a fibration of a round $S^2$ over a two-dimensional surface $\Sigma$. Its geometry is determined by the BPS equations. The SU(2) isometry of the fibers should realize the R-symmetry of putative dual five-dimensional $\mathcal{N}=1$ SCFTs. The RR and NSNS fluxes are determined in terms of solutions to the PDEs, and their Bianchi identities (in absence of sources) are automatically satisfied "on-shell" (i.e. upon using the PDEs). Given the high degree of nonlinearity of the two coupled equations, we were not able to solve them in full generality. We did recover from our analysis the only two known solutions in the literature, namely the abelian and nonabelian T-dual of the unique $\mathrm{AdS}_6 \times S^4$ vacuum of massive IIA. (The latter was found by Brandhuber–Oz via near-horizon limit of the original D4-D8-O8 configuration in type I' proposed by Seiberg.) Nevertheless, using the general theory of exterior differential systems, we were able to prove that the PDEs are a "well-formed" system, and that the generic solution will depend on two functions of one variable (that we can think of as being the value of the warping function and the dilaton at the boundary of $\Sigma$). We provided an interpretation of the generic solution as being the near-horizon limit of a $(p,q)$-fivebrane web in perturbative type IIB string theory. The plane in which the web extends should give rise to the surface $\Sigma$, with the $S^2$ fiber collapsing at its boundary (so that the internal space $M_4$ is compact). Fivebrane sources would then be localized at points along the boundary (sourcing the appropriate $F_3 + \tau H$ flux via a term $\int_{S^2\times I} F_3 + \tau H$ – with $\tau = i$ in perturbative IIB and $I$ a path inside $\Sigma$ encircling the source and ending on the boundary of the former). A full (or partial – imposing some simplifying Ansatz) classification of $\mathrm{AdS}_6$ vacua of type IIB supergravity would provide us with a list of five-dimensional $\mathcal{N} = 1$ SCFTs admitting a holographic dual, possibly enabling us to study aspects of the field theories.

# Part I

## Type IIA brane engineering and supergravity solutions

# Chapter 3

# Hanany–Witten-like setups in six dimensions

## 3.1. The gauge theory – brane worldvolume correspondence

From the worldsheet perspective one can study the open sector of superstring theory with two types of boundary conditions for the two ends of a string: Dirichlet (namely the position of each end is fixed) or Neumann (no momentum flows through either end of the string). Clearly, the first choice introduces fixed loci in spacetime, breaking ten-dimensional Lorentz invariance. These fixed loci can be identified with the stringy versions of the solitonic $p$-brane solutions of ten-dimensional (type II) supergravity. A one-loop computation [33, 34] reveals that they have a tension which depends on their dimension,[1]

$$T_{\mathrm{D}p} = \frac{1}{l_{\mathrm{s}}^{p+1} g_{\mathrm{s}}} \ , \tag{3.1}$$

and that they carry appropriate RR charges. $l_{\mathrm{s}}$ is the string length (related to the famous $\alpha'$ parameter by $\alpha' = l_{\mathrm{s}}^2$), which sets a characteristic length scale in the theory; $g_{\mathrm{s}}$ is the string coupling constant. We will call them D$p$-branes. Their *worldvolume* spans one timelike direction and $p$ spacelike ones, hence it has dimension $p+1$.

Quantizing the oscillator modes of the string with such boundary conditions yields a massless spectrum whose content is that of the SYM multiplet (gauge boson, scalars and gaugini): E.g. for the D9-brane we have a single $\mathcal{N}=1$ vector multiplet in ten dimensions (which does not contain any scalar). This means that the worldvolume theory of the brane is described by the massless excitations of an open string ending on the brane itself. The other D$p$ worldvolume theories for $p<9$ are obtained by dimensional reduction of the ten-dimensional $\mathcal{N}=1$ vector multiplet. This yields SYM vector multiplets in lower dimensions with the right amount of supersymmetry (namely sixteen supercharges); see tables 3.1 and 3.2.

[1] Throughout this thesis we will be suppressing factors of $2\pi$ (or multiples thereof) in most of the formulae. This does not affect the discussion in any way; it simply modifies our normalization conventions with respect to others present in the literature, e.g. the standard ref. [35].

| | 10 dimensions | $10-(p+1)$ dimensions |
|---|---|---|
| field content | $A^M$, $M=0,\dots,9$ | $A^\mu$, $\mu=0,\dots,p$ |
| | $\Psi$ s.t. | $A^m \equiv \phi^m$, $m=p+1,\dots,9$ |
| | $\Gamma_{(10)}\Psi=+\Psi$ and $\Psi^*=\Psi$ | $\Psi=\psi\otimes\eta$ |

**Table 3.1.:** The SYM vector multiplet in ten dimensions (sixteen $Q$ supercharges) includes a gauge field $A^M$ and a Majorana–Weyl spinor $\Psi$ of positive chirality. The former decomposes into a gauge field $A^\mu$ in $(p+1)$ dimensions, living on the worldvolume of a $p$-brane, plus a bunch of $9-p$ scalars $\phi^m$, parameterizing the directions inside $\text{Mink}_{10}$ transverse to the brane worldvolume. By $\Psi=\psi\otimes\eta$ we mean a (schematic) decomposition of the ten-dimensional spinor following from a decomposition of the ten-dimensional gamma matrices, $\Gamma^\mu=\gamma^\mu_{(p+1)}\otimes 1_{9-p}$ and $\Gamma^m=\gamma_{(p+1)}\otimes\gamma^m_{(9-p)}$. ($\Gamma_{(10)}$ is the chiral gamma in ten dimensions, while $\gamma_{(p+1)}$ that in $p+1$ dimensions, when this number is even.) The chirality and reality properties of $\psi$ and $\eta$ depend on $p$ (see section 4.3.1).

We should now demand conformal invariance of the worldsheet theory.[2] This gives equations of motion for the string spacetime-embedding fields, which can in turn be obtained by varying an effective action for the D$p$-brane known as *Dirac–Born–Infeld* plus *Wess–Zumino action*:

$$\begin{aligned} S_{\text{D}p} &= S_{\text{DBI}} + S_{\text{WZ}} \\ &= -T_{\text{D}p}\int d^{p+1}\xi e^{-\phi}\sqrt{-\det(g+B+l_s^2\mathcal{F})} + \mu_p \int \left(C\wedge e^{B+l_s^2\mathcal{F}}\right)_{p+1} . \end{aligned} \tag{3.2}$$

The integration should be performed on the $p+1$-dimensional worldvolume of the D$p$-brane. The subscript in the WZ integrand selects the $p+1$-form part. $(g, \phi, B)$ is the NSNS sector of the closed massless spectrum of type II string theory; it contains the graviton, the dilaton and the $B$ field (whose field strength is the NSNS flux $H$). $C$ is the sum of RR potentials defined in (4.16), and $F$ its flux. $\mu_p$ will be defined in (3.28), and we also have $T^2_{\mathrm{D}p} = \mu_p^2$ as a consequence of supersymmetry [33].

[2] A quantum field theory enjoying scale invariance in two dimensions enjoys the full conformal invariance under suitable assumptions [36, 37]. This is however not always the case for worldsheet theories (i.e. two-dimensional NLSM's) which violate some of the assumptions (typically unitarity and discreteness of the spectrum). For the latter one has to demand the full conformal invariance, which implies scale invariance (i.e. its $\beta$ function vanishes).

| 16 $Q$ supercharges | | 8 $Q$ supercharges | | 4 $Q$ supercharges | |
|---|---|---|---|---|---|
| TM | VM | VM | HM | VM | HM |
| | $d = 10\ \mathcal{N} = 1$ | | | | |
| $d = 6$ $\mathcal{N} = (2,0)$ | $d = 6$ $\mathcal{N} = (1,1)$ | $d = 6$ $\mathcal{N} = (1,0)$ (TM) | | | |
| | $d = 4\ \mathcal{N} = 4$ <br> $d = 3\ \mathcal{N} = 8$ | $d = 4\ \mathcal{N} = 2$ <br> $d = 3\ \mathcal{N} = 4$ | | $d = 4\ \mathcal{N} = 1$ <br> $d = 3\ \mathcal{N} = 2$ | |

**Table 3.2.:** Supersymmetric theories with 16, 8, and 4 $Q$ supercharges in diverse dimensions. All algebras in $d = 3, 4$ can be made superconformal with the addition of the $S$ supercharges (and new bosonic generators). VM (vector), HM (hyper), and TM (tensor) indicate the multiplets of the supersymmetry algebra (irreducible, massless representations). The six-dimensional $\mathcal{N} = (1,1)$ SYM is nonchiral and cannot be made conformal, not even at the quantum level. The other two six-dimensional theories are chiral and superconformal and contain tensor multiplets. That of the $\mathcal{N} = (2,0)$ superalgebra contains five real scalars, two left Weyl spinors, one two-form with self-dual three-form field strength, while that of the $\mathcal{N} = (1,0)$ superalgebra contains one real scalar, one left Weyl Majorana spinor, one two-form with self-dual three-form field strength.

$F$ – the *bulk* RR flux – should not be confused with the *worldvolume* flux $\mathcal{F}$ – a two-form – appearing in (3.2). The latter is the field strength of the SYM gauge field $A$, and its presence is needed in order for the Wess–Zumino term to be gauge invariant under the $B$ field gauge transformation.

Expanding the DBI action in (3.2) to lowest order in $l_s^2$ yields the SYM action in $p+1$ dimensions,[3] with coupling

$$g_{\mathrm{YM}}^2 = l_{\mathrm{s}}^{p-3} g_{\mathrm{s}} \ . \tag{3.3}$$

We will call this set of simple but powerful observations *gauge theory – brane worldvolume correspondence*.

One also has

$$g_{\mathrm{s}} \frac{1}{l_{\mathrm{s}}^4} = g_{\mathrm{s}} M_{\mathrm{s}}^4 = M_{\mathrm{Pl}}^4 \ , \tag{3.4}$$

which gives relations:

$$g_{\mathrm{YM}}^{-2} = \frac{M_{\mathrm{s}}^{p-3}}{g_{\mathrm{s}}} \ , \quad g_{\mathrm{YM}}^2 = \frac{M_{\mathrm{s}}^{7-p}}{M_{\mathrm{Pl}}^4} \ . \tag{3.5}$$

Recalling that the Planck mass $M_{\mathrm{Pl}} = \sqrt{\frac{\hbar c}{G}}$, decoupling gravity (i.e. Newton's constant $G_{\mathrm{N}} \to 0$) requires taking $M_{\mathrm{Pl}} \to \infty$; on the other hand, to decouple the massive string excitations we must take $M_{\mathrm{s}} \to \infty$ (i.e. $l_{\mathrm{s}} \to 0$). We can take these two limits in such a way as to keep $g_{\mathrm{s}}$ finite; see (3.4). The same is true for $g_{\mathrm{YM}}$ in light of the second in (3.5), yielding an interacting field theory on the brane worldvolume. (Alternatively, the $l_{\mathrm{s}} \to 0$ limit can be argued for by requiring that the expansion of the Dirac–Born–Infeld plus Wess–Zumino brane action yield pure SYM in $p+1$ dimensions with no correction terms weighted by powers of $l_{\mathrm{s}}$.)

## 3.2. The main idea

It is now possible to exploit the gauge theory – brane worldvolume correspondence to engineer interesting field theories by letting flat branes of different dimensions intersect according to specific patterns.

The original setup introduced by Hanany and Witten [39] was aimed at constructing three-dimensional field theories. The same idea can then be generalized and used to construct higher-dimensional versions [30, 29]. In this thesis we will be mostly concerned with the construction of six-dimensional field theories with $\mathcal{N} = (1,0)$ supersymmetry.

[3] This should be understood as a derivative expansion whereby we only keep the two-derivative leading term [38]. One has to expand the dilaton prefactor $e^{-\phi}$ times the (square root of the) determinant of the pull-backs of the various fields, and then keep only the leading term in $l_{\mathrm{s}}^2$ when $l_{\mathrm{s}} \to 0$. The result of this procedure contains two terms: One is the gauge kinetic term for $\mathcal{F}$ and the other is proportional to the brane worldvolume, and gives its tension. Keeping higher terms in $l_{\mathrm{s}}^2$ means keeping higher-derivative interactions. Moreover, the second term comes from the expansion of the dilaton prefactor $e^{-\phi}$, related to the string coupling by $g_{\mathrm{s}} = e^{\langle \phi \rangle}$. *Stringy corrections* can thus arise beyond tree-level in $g_{\mathrm{s}}$.

### 3.2.1. NS5-D4

To outline the strategy, consider first a flat D4-brane in $\mathrm{Mink}_{10}$: This breaks in half the supersymmetry of the type IIA background, preserving sixteen out of the thirty-two original supercharges. In fact, suppose the flat D4 fills out the $0, \ldots, 3, 6$ directions inside $\mathrm{Mink}_{10}$ (see figure 3.1); then we need to impose the condition

$$\epsilon_{\mathrm{L}} = \Gamma^0\Gamma^1\Gamma^2\Gamma^3\Gamma^6\epsilon_{\mathrm{R}} \tag{3.6}$$

on the two ten-dimensional spinors $\epsilon_{\mathrm{R,L}}$ of type IIA supergravity. The preserved supercharges are given by the combination $\epsilon_{\mathrm{L}}Q^{\mathrm{L}} + \epsilon_{\mathrm{R}}Q^{\mathrm{R}}$.[4] For any additional (stack of coincident) flat D$p$-brane(s), spanning directions $0, \ldots, p$, we need to impose [29, 40]

$$\epsilon_{\mathrm{L}} = \Gamma^0 \cdots \Gamma^p \epsilon_{\mathrm{R}} \ , \tag{3.7}$$

generically breaking by a further half the number of preserved supersymmetries.

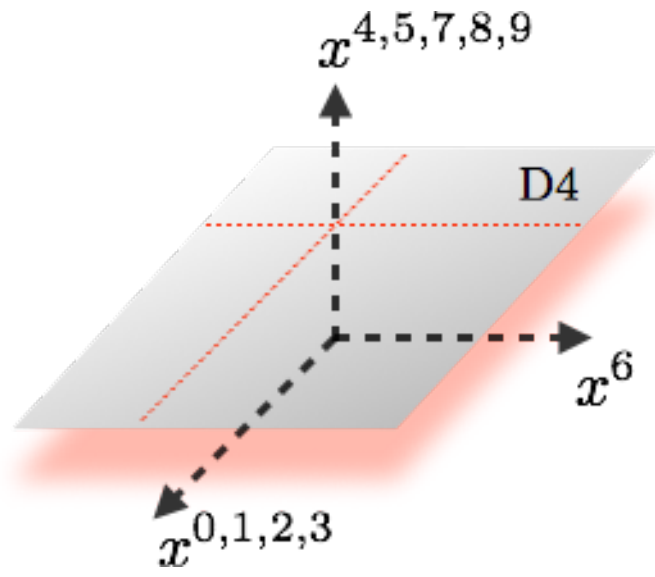

**Figure 3.1.:** An artist's impression of a single, flat D4-brane stretching across $\mathrm{Mink}_{10}$. It fills out the $0, \ldots, 3$ and 6 directions, while it is located at a point in the transverse $4, 5, 7, 8, 9$ plane.

By the gauge – brane worldvolume correspondence, we know we are engineering maximally SYM in five dimensions ($\mathcal{N} = 2$). To get some interesting physics, we wish to project out some of the degrees of freedom. This can be achieved by letting the D4-brane end on two separate NS5-branes, spanning the $0, \ldots, 5$ directions and sitting at (separate) points along the 6 direction; see figure 3.2(b).

As the NS5-branes stretch across different directions (with respect to the D4-brane), they break some more supersymmetry, according to [29, 40]

$$\epsilon_{\mathrm{L}} = \Gamma^0 \cdots \Gamma^5 \epsilon_{\mathrm{L}} \ , \quad \epsilon_{\mathrm{R}} = \Gamma^0 \cdots \Gamma^5 \epsilon_{\mathrm{R}} \ . \tag{3.8}$$

[4] If we choose a nontrivial embedding of the D4-brane inside $\mathrm{Mink}_{10}$, a lower fraction of the type IIA supercharges may be generically preserved. This depends on the existence of so-called Killing spinors in the embedding geometry.

| | $x^0$ | $x^1$ | $x^2$ | $x^3$ | $x^4$ | $x^5$ | $x^6$ | $x^7$ | $x^8$ | $x^9$ |
|---|---|---|---|---|---|---|---|---|---|---|
| D4 | $\times$ | $\times$ | $\times$ | $\times$ | $\cdot$ | $\cdot$ | $[x^6_1, x^6_2]$ | $\cdot$ | $\cdot$ | $\cdot$ |
| NS5$_1$ | $\times$ | $\times$ | $\times$ | $\times$ | $\times$ | $\times$ | $x^6_1$ | $\cdot$ | $\cdot$ | $\cdot$ |
| NS5$_2$ | $\times$ | $\times$ | $\times$ | $\times$ | $\times$ | $\times$ | $x^6_2$ | $\cdot$ | $\cdot$ | $\cdot$ |

(a)

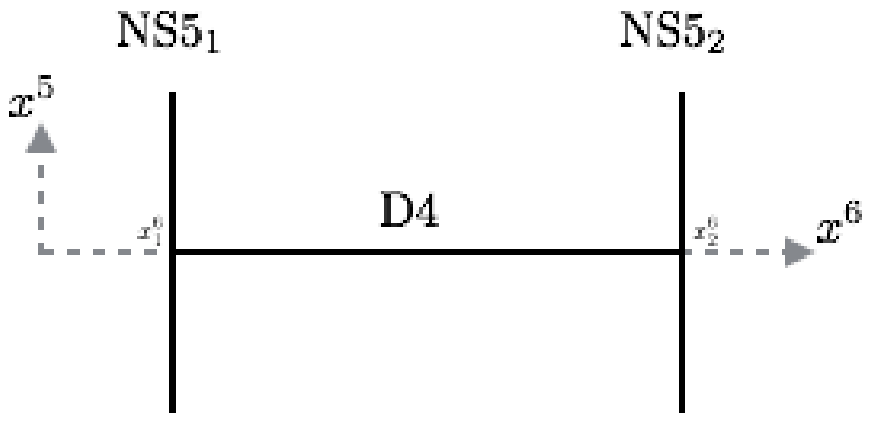

(b)

**Figure 3.2.:** In (a) we see the directions in Mink$_{10}$ filled out by the various branes: a $\times$ means that the brane fills out the corresponding direction completely; a $\cdot$ means that the brane sits at a point (which we do not specify) along the corresponding direction. The D4-brane spans a finite segment along direction 6, which justifies the Kaluza–Klein reduction (3.10). The NS5's sit at points $x^6_i$. In (b) the corresponding NS5-D4 Hanany–Witten setup.

We are now left with eight supercharges.[5] Moreover, it is known [41, 42] that a

[5]Equations (3.8) imply that the worldvolume theory of the type IIA NS5-brane should be a chiral $(2, 0)$ theory in six dimensions – both supercharges have the same chirality – despite the parent ten-dimensional theory is nonchiral. Upon reduction from M-theory, we learn that the NS5 worldvolume theory contains a tensor multiplet, as for the M5-brane, featuring a self-dual two-form sourced by open D2-branes ending on the NS5 (the reduction of open M2's ending on the M5). To perform the reduction we need to compactify M-theory on $S^1_{\mathrm{M}}$, the *M-theory circle*. Assuming the M5-brane is transverse to the circle, one of the five scalars in the tensor multiplet (which describe its position in the five-dimensional transverse space inside Mink$_{11}$) will be periodic, as it specifies its position along the circle. When we reduce along $S^1_{\mathrm{M}}$, the M5 becomes an NS5, which has four transverse directions inside Mink$_{10}$, the type IIA spacetime. For vanishing radius of $S^1_{\mathrm{M}}$, we get back to *perturbative* type IIA (i.e. $g_{\mathrm{s}} \to 0$): Four of the scalars in the tensor multiplet will now describe the position of the NS5 in the transverse space, while the latter cannot be immediately given a geometric interpretation. At strong coupling (i.e. $g_{\mathrm{s}} \gtrsim 1$), the NS5 is better seen as an M5-brane transverse to the eleventh dimension (the one parameterizing the M-theory circle), hence the fifth scalar acquires a geometric interpretation [35].

In type IIB, which is chiral, the second equation in (3.8) reads instead $\epsilon_{\mathrm{R}} = -\Gamma^0 \cdots \Gamma^5 \epsilon_{\mathrm{R}}$ [29, 40], implying that the six-dimensional worldvoulme theory exhibits nonchiral $(1, 1)$ supersymmetry. Therefore one type IIB NS5-brane will support a $d = 6$ SYM vector multiplet, with gauge coupling given by $g^2_{\mathrm{YM}} = \frac{1}{M_{\mathrm{s}}^2}$ (this is obtained by D4 ending on an NS5-brane induces a logarithmic bending of the latter. This is because the end of a brane is a charged object, and acts as source for the worldvolume fields of the brane it ends on. The induced worldvolume fields (namely RR fluxes) couple, via the DBI action in (3.2), to the scalars describing the embedding of the brane worldvolume into spacetime. This is the cause of the bending.

From the field theory perspective of SYM on the worldvolume of the D4, this can be seen as a one-loop effect. Given that the tensions of D4 (see (3.1)) and NS5 go like

$$T_{\mathrm{D4}} \sim \frac{1}{g_{\mathrm{s}}} \ , \quad T_{\mathrm{NS5}} \sim \frac{1}{g_{\mathrm{s}}^2} \ , \tag{3.9}$$

and that $g_{\mathrm{YM}}^2 \sim g_{\mathrm{s}}$ (from (3.3)), at tree-order (i.e. for $g_{\mathrm{YM}} \to 0$) the NS5 is infinitely heavy and will not be pulled by the D4. Switching on $g_{\mathrm{YM}}$ will induce loop corrections in field theory, and will make the NS5 tension finite.

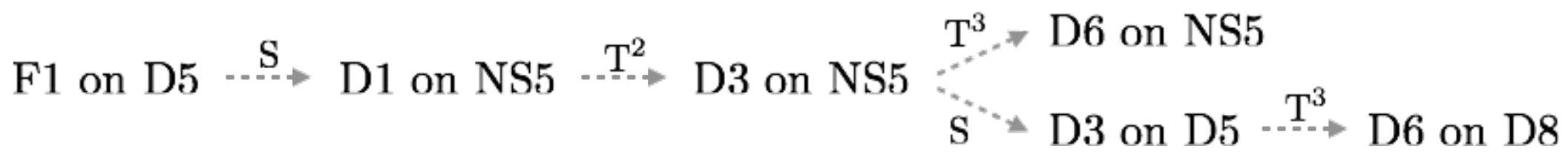

**Figure 3.3.:** A fundamental F1-string couples electrically to the type IIB NSNS $B$ field (a bulk two-form), and it can end on a D5-brane (in fact it can end on any D$p$-brane). The same $B$ field couples magnetically to the type IIB NS5-brane, on which a D1-string can end. Therefore we can use S-duality to map the F1 ending on a D5 to a D1 on an NS5. Considering the full (quantum) $\mathrm{SL}(2,\mathbb{Z})$ invariance of type IIB string theory, we can construct more general bound states of $p$ F1's and $q$ D1's (BPS for $p,q$ coprime): A $(p,q)$-string, which can end on a $(p,q)$-sevenbrane. The magnetically dual object will be a $(p,q)$-fivebrane; as a consequence, a fivebrane can also end on a sevenbrane with same $(p,q)$ "charges". The $B$ field and the RR two-form $C_2$ make up an $\mathrm{SL}(2,\mathbb{Z})$ doublet $(B_2, C_2)$, whereas the RR $C_4$ four-form forms a singlet, thus implying that a D3 is neutral under $\mathrm{SL}(2,\mathbb{Z})$ and is mapped to itself. We can then T-dualize the D1 on NS5 twice along a direction inside the NS5-brane, but transverse to the D1, yielding a D3 on NS5. An S-duality (i.e. $\begin{pmatrix} 0 & -1 \\ 1 & 0 \end{pmatrix} \in \mathrm{SL}(2,\mathbb{Z})$) maps the latter to D3 on D5, then T-dualized three times along a (common) transverse direction to D6 on D8; a different triple T-duality inside the NS5 worldvolume maps it instead to D6 on NS5. We will need the two rightmost endpoints of this chain of dualities in order to construct six-dimensional Hanany–Witten setups.

Let us go back to the gauge theory on the worldvolume of the D4. Since the latter is now stretching over a finite interval of length $L$, we are effectively applying S-duality to the D5 worldvolume theory). From table 3.1 we see that such a multiplet contains four real scalars, which will describe the position of the NS5 in the transverse space.

dealing with the Kaluza–Klein reduction of the former on the same interval: If $L$ is small, the fluctuations of the D4-brane along the 6 direction have momentum and energy proportional to $L^{-1}$. In the $L \to 0$ limit they become infinitely heavy and hence decouple, justifying a reduction à la Kaluza–Klein. This whole process leaves behind a four-dimensional field theory with eight supercharges (i.e. an $\mathcal{N}=2$ theory). Schematically:

$$\frac{1}{g^2_{\mathrm{YM}}}\int_{\mathrm{D4}} \mathrm{SYM}_{5d} \xrightarrow{\mathrm{KK}} \frac{L}{g^2_{\mathrm{YM}}}\int_{0,\dots,3} \mathrm{SYM}_{4d}\ ; \tag{3.10}$$

recalling (3.3), we can define the gauge coupling of the reduced four-dimensional theory as

$$g^2_{\mathrm{YM}4d} \equiv \frac{g^2_{\mathrm{YM}5d}}{L} = \frac{l_{\mathrm{s}} g_{\mathrm{s}}}{L}\ . \tag{3.11}$$

Hence, for a D$p$-brane suspended between two separate NS5-branes we will have:

$$g^2_{\mathrm{YM}p} = \frac{l_{\mathrm{s}}^{p-3} g_{\mathrm{s}}}{L}\ . \tag{3.12}$$

The ten-dimensional Lorentz group is broken due to the presence of the various branes, acting as fixed loci (i.e. solitonic objects) in $\mathrm{Mink}_{10}$:

$$\mathrm{SO}(1,9) \to \mathrm{SO}(1,3)_{0123} \times \mathrm{SO}(2)_{45} \times \mathrm{SO}(3)_{789} \cong \mathrm{SO}(1,3) \times \underbrace{\mathrm{U}(1) \times \mathrm{SU}(2)}_{\mathrm{U}(2)}\ . \tag{3.13}$$

The first factor is just the Lorentz group in the common directions spanned by the branes; the second and third reflect the rotational symmetry in the $4,5$ and $7,8,9$ directions respectively. In particular, we notice that the isometries of the "internal" portion of spacetime (from the D4 perspective) make up the U(2) R-symmetry of the corresponding $d=4$ $\mathcal{N}=2$ superalgebra.

To see which degrees of freedom have survived the introduction of the two NS5-branes, and which have been projected out, we need to look at the $d=5$ SYM field content and its reduction to four dimensions. The bosonic content of a vector multiplet of $d=5$ $\mathcal{N}=2$ SYM comprises a gauge field $A^\mu$ ($\mu = 0,\dots,3,6$) and five real scalars $\phi^i$ ($i=1,\dots,5$). In view of the Kaluza–Klein reduction we want to perform, we can decompose $A^\mu$ in a four-dimensional gauge field $A^m$ ($m=0,\dots,3$) plus the fifth component $A^6$; so in total we have six real scalars $(\phi^i, A^6)$. The $d=4$ $\mathcal{N}=2$ vector multiplet is just the sum of a $d=4$ $\mathcal{N}=1$ vector plus a $d=4$ $\mathcal{N}=1$ chiral multiplet. Its bosonic content is given by two real scalars, and one four-dimensional gauge field, which we identify with $A^m$. The $d=4$ $\mathcal{N}=2$ hypermultiplet is instead the sum of a $d=4$ $\mathcal{N}=1$ chiral multiplet plus an antichiral multiplet. Its bosonic content is just given by four real scalars. Thus, the two real scalars and the gauge field in the $d = 4$ $\mathcal{N} = 2$ vector multiplet make up part of the $d = 5$ $\mathcal{N} = 2$ vector multiplet; in particular, these two scalars will govern the position of the D4 in the $4, 5$ plane. The other four real scalars coming from the $d = 4$ $\mathcal{N} = 2$ hypermultiplet complete the $d = 4$ $\mathcal{N} = 2$ vector multiplet, and will govern the position of the D4 in the $6, 7, 8, 9$ plane.

Looking at picture 3.2(b), we see that the D4's position along the $7, 8, 9$ plane is fixed by the presence of the two NS5-branes it ends on; its position along the 6 direction (the finite interval it spans) fixes the fourth real scalar in the $d = 4$ $\mathcal{N} = 2$ hypermultiplet. Hence the bosonic part of the latter is projected out; by supersymmetry, the whole multiplet decouples from the spectrum (i.e. the fermions are not part of the degrees of freedom either). On the contrary, its motion along the $4, 5$ plane is left unconstrained by the NS5's, and will play the role of a free *parameter* for the field theory: We are left with (the bosonic part of) a vector multiplet, i.e. a $d = 4$ SYM theory with gauge coupling given by (3.11).[6] To get a nonabelian theory we simply need to put many D4's, say $N_{\mathrm{c}}$, on top of each other: We now have an $\mathrm{SU}(N_{\mathrm{c}})$ nonabelian vector multiplet governed by the Yang–Mills interaction.[7] We will call the D4's *color* branes.

### 3.2.2. Adding flavors

We wish to introduce additional matter in the system. This can be achieved by adding a stack of $N_{\mathrm{f}}$ semi-infinite D4-branes extending beyond the NS5-branes along the 6 direction; see figure 3.4.

The newly added D4-branes do not break any further supersymmetry and are very heavy, as they are infinitely stretched across space; hence we will discard their fluctuations. In particular the gauge bosons decouple from the spectrum. Moreover the gauge coupling of their dimensionally-reduced worldvolume theory $g^2_{\mathrm{YM}} = \frac{l_{\mathrm{s}} g_{\mathrm{s}}}{L}$ goes to zero in the $L \to \infty$ limit. We are getting an $\mathrm{SU}(N_{\mathrm{f}})$ *flavor* symmetry. In fact, an analysis of the bosonic degrees of freedom like the one above reveals that they contribute $N_{\mathrm{f}}$ $\mathcal{N} = 2$ hypermultiplets. Strings going from the infinite (finite) D4's to the finite (infinite) D4's give rise to the $\mathcal{N} = 1$ (anti)chiral multiplet in the $\mathbf{N}_{\mathrm{c}}$ ($\overline{\mathbf{N}_{\mathrm{c}}}$) (anti)fundamental representation

[6] In the nonabelian case, that is when multiple D4's are put on top of each other, the complex numbers $m_i \in \mathbb{C}_{(4,5)}$ giving the location of the $i$-th brane along the $4, 5$ plane parameterize the so-called *Coulomb branch* of the *moduli space* of the theory.

[7] As is well known, this is the standard way to engineer (S)U gauge groups in type II string theory. When all the branes are separated, strings stretching from one brane to another provide the $W$ bosons of the broken $\mathrm{U}(N_{\mathrm{c}}) \to \mathrm{U}(1)^{N_{\mathrm{c}}}$ symmetry thanks to their Chan–Paton indices; the distance between the branes is interpreted as a vev for the Higgs. In the coincident-branes limit we recover the full $\mathrm{U}(N_{\mathrm{c}})$ gauge symmetry. We can typically reduce it to $\mathrm{SU}(N_{\mathrm{c}})$ by decoupling the center-of-mass $\mathrm{U}(1)$.

| | $x^0$ | $x^1$ | $x^2$ | $x^3$ | $x^4$ | $x^5$ | $x^6$ | $x^7$ | $x^8$ | $x^9$ |
|---|---|---|---|---|---|---|---|---|---|---|
| $N_\mathrm{c}$ D4's | × | × | × | × | · | · | $[x^6_1, x^6_2]$ | · | · | · |
| $N_\mathrm{L}$ D4's | × | × | × | × | · | · | $x^6 < x^6_1$ | · | · | · |
| $N_\mathrm{R}$ D4's | × | × | × | × | · | · | $x^6 > x^6_2$ | · | · | · |
| $\mathrm{NS5}_1$ | × | × | × | × | × | × | $x^6_1$ | · | · | · |
| $\mathrm{NS5}_2$ | × | × | × | × | × | × | $x^6_2$ | · | · | · |

(a)

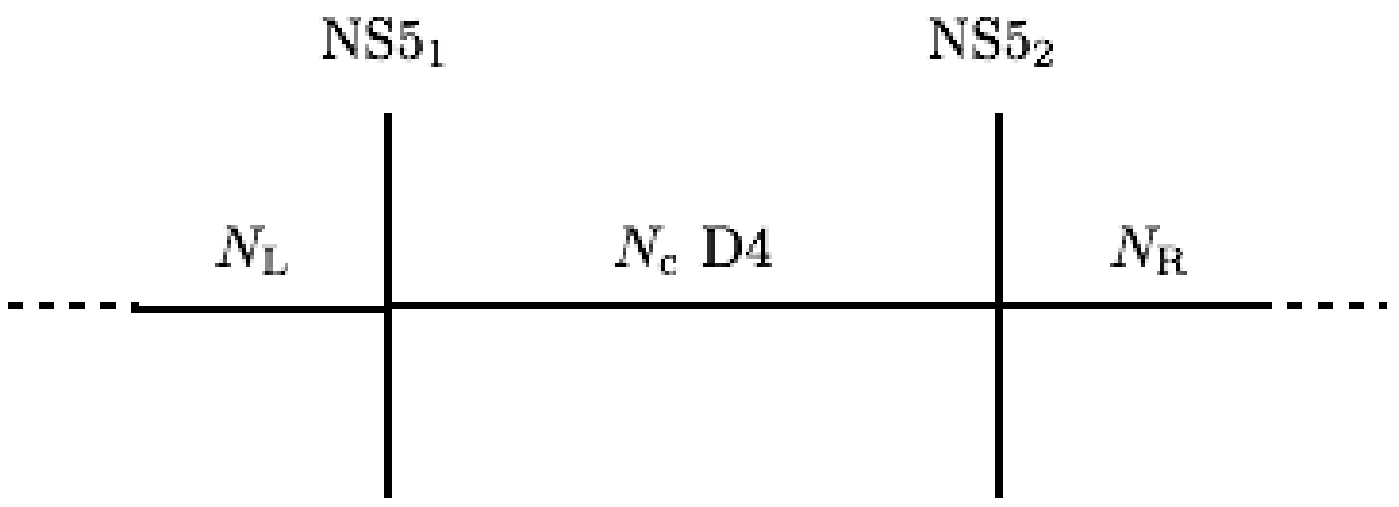

(b)

**Figure 3.4.:** Semi-infinite D4-branes extending beyond the NS5's. The s-rule fixes the numbers $N_\mathrm{c}$, $N_\mathrm{L}$, $N_\mathrm{R}$ so that they satisfy (3.14).

of the $\mathrm{SU}(N_\mathrm{c})$ gauge group on the finite D4's and $\overline{\mathbf{N}_\mathrm{f}}$ ($\mathbf{N}_\mathrm{c}$) antifundamental (fundamental) representation of the $\mathrm{SU}(N_\mathrm{f})$ flavor group.

The same reasoning allows us to introduce hypermultiplets in the bifundamental $\overline{\mathbf{N}_\mathrm{c}} \otimes \mathbf{N}'_\mathrm{c}$ representation of two consecutive gauge groups. This is achieved by introducing a third NS5-brane at a different location along the 6 direction, and having a stack of (finite) D4's stretch between the second and third NS5. Clearly, the process can be repeated with an arbitrary number $N$ of NS5-branes, producing a chain of SU gauge groups connected by bifundamental hypermultiplets. We will see that in six dimensions this number $N$ plays a crucial role in the *holographic limit* of the field theories.

Yet another modification we can do in order to add flavors to the system is to introduce D6-branes on which the semi-infinite D4-branes now end, becoming finite. See table 3.3 and figure 3.5.

The D6's just introduced do not spoil any further the supersymmetry of the system (since they are D-branes – hence only condition (3.7) applies – and they extend along directions already filled out by the D4's), and the distance between the D6 stack and the closest NS5 is assumed to be small, so that we can neglect the fluctuations of the D4's stretching between the two along the 6 direction. Once again, the strings stretching from the D6's to the D4's (and viceversa) will provide $\mathcal{N} = 2$ hypermultiplets in the $\mathbf{N}_c$ fundamental representation of the gauge group.[8] The worldvolume theory of the D6 themselves is instead neglected (it is a frozen background) as they stretch across infinite directions; they just provide fundamental hypermultiplets.

| | $x^0$ | $x^1$ | $x^2$ | $x^3$ | $x^4$ | $x^5$ | $x^6$ | $x^7$ | $x^8$ | $x^9$ |
|---|---|---|---|---|---|---|---|---|---|---|
| NS5$_i$ | × | × | × | × | × | × | $x_i^6$ | · | · | · |
| D4 | × | × | × | × | · | · | × | · | · | · |
| D6 | × | × | × | × | · | · | $x_{\text{D6}}^6$ | × | × | × |

**Table 3.3.:** Directions spanned by NS5, D4, D6-branes. We collected both color and (former) semi-infinite D4-branes under the label "D4". The D6's are localized at a point $x_{\text{D6}}^6$ along the 6 direction.

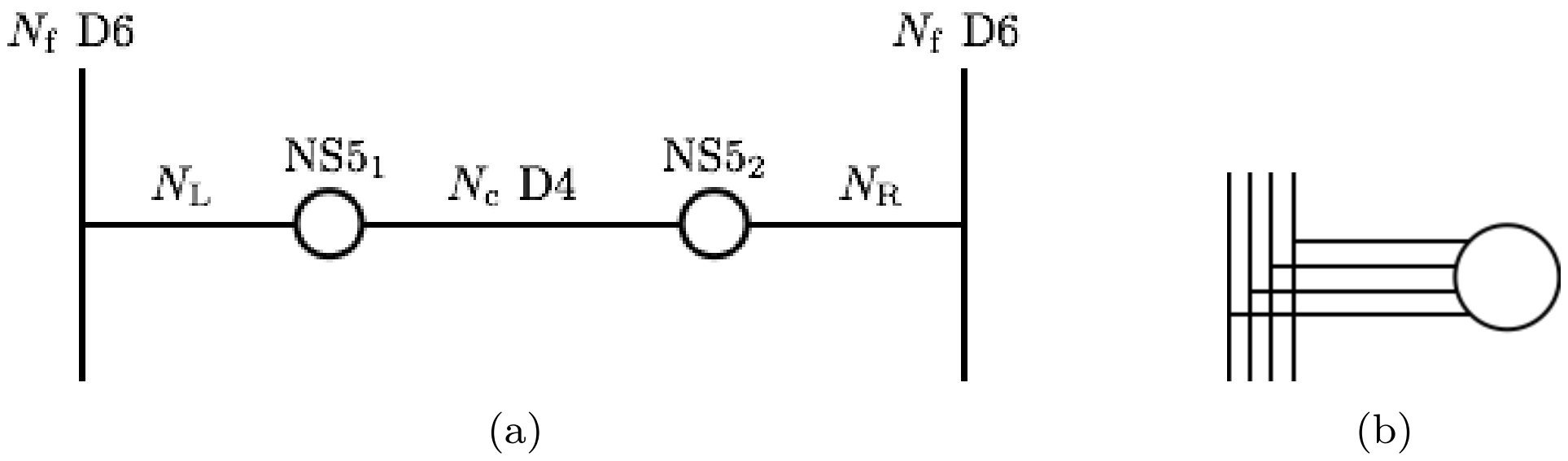

**Figure 3.5.:** In 3.5(a) we see a simple NS5-D4-D6 Hanany–Witten setup. The former semi-infinite D4's extending beyond the NS5's of figure 3.4(b) now end on stacks of $N_f$ D6-branes. The NS5-branes are here represented by circles, not to confuse them with the D6-branes, represented by vertical lines. In 3.5(b) a depiction of the s-rule: There can be at most one D4-brane suspended between an NS5 and a particular D6-brane. This clearly fixes $N_{L,R} = N_f$ in figure 3.5(a).

These two seemingly different ways of introducing flavors are actually related by so-called *Hanany–Witten moves*: when an NS5-brane passes through a flavor D6-brane, a D4 is created connecting the two [39]. Imagine having a D6 stack crossing finite D4's suspended between two consecutive NS5-branes. As we just learned, they provide hypermultiplets in the fundamental of the gauge group on the D4 worldvolume. If we now move the D6 stack to the, say, far left of the leftmost NS5, we will create semi-infinite D4's stretching from the latter to infinity. Again, these will contribute fundamental hypermultiplets, due to the strings connecting them to the finite D4's.

[8]Another modification entails rotating (some of) the NS5-branes e.g. along the $4, 5, 8, 9$ directions, or yet introducing another set of NS5's spanning $0, \ldots, 3, 8, 9$ instead of $0, \ldots, 5$ (as per table 3.2(a)). This halves the number of preserved supercharges, leaving only four ($\mathcal{N} = 1$ in $d = 4$). However, we will not be concerned with such modifications in this thesis.

One might wonder whether the number of fundamental hypermultiplets thus created (semi-infinite D4's) is the same as before (stack of D6's). Were it not, we would have an inconsistent picture. Luckily, the so-called *s-rule* [39] comes to the rescue: One and only one D4-brane is allowed to end on a certain NS5-D6 pair in a supersymmetric setup. This means that, focusing on the leftmost NS5, we should have one D6-brane for every D4 protruding out of the NS5. Bringing the D6's to infinity, we will create the same number of semi-infinite D4's. Hence:

$$N_{\mathrm{f}} = N_{\mathrm{D6}} = N_{\mathrm{D4,L}} \xrightarrow{\text{semi-infinite limit: } x^6_{\mathrm{D6,L}} \to -\infty} N_{\mathrm{f}} \ . \tag{3.14}$$

#### Higgs branch

If we considered instead a D4 stretched between two D6's, this time we would discover it is the $\mathcal{N} = 2$ vector multiplet that gets projected out, while the hypermultiplet (whose scalar part controls the D4's position along the $7, 8, 9$ plane) is preserved. This observation allows us to access the so-called *Higgs branch* of the moduli space of this $d = 4$ $\mathcal{N} = 2$ theory.

Bringing in from the far left (or far right) a D6 stack, and placing it between two consecutive NS5's (paying attention to perform the necessary Hanany–Witten moves), allows to slide the D4 segment connecting two D6's along the $7, 8, 9$ directions. This is achieved by giving a vev to the hypermultiplet scalars governing the D4 position in the $7, 8, 9$ plane. This also provides us with a new interesting way to create more complicated, interacting theories (like a CFT) from a weakly-coupled theory of vector and hypermultiplets. Bringing the D4 segments to infinity along the D6-branes will in fact initiate a *Higgs branch RG flow* that might end at an interacting fixed point [30]. We will heavily exploit this type of deformation in constructing six-dimensional theories.

### 3.2.3. D3-D5 and Nahm poles

#### Nahm's equations

Another important observation can be made by looking at the T-dual NS5-D3-D5 setup. Let us first focus on a single D3-brane on a semi-infinite interval $y > 0$. We will closely follow the discussion in [30].

The $d = 4$ $\mathcal{N} = 4$ SYM multiplet it supports can be obtained by dimensional reduction from ten dimensions, and contains one four-dimensional gauge field $A_\mu$ ($\mu = 0, \ldots, 3$), four Weyl spinors, and six real scalars, which we will group in two three-dimensional vectors $\mathbf{X}$ and $\mathbf{Y}$. From the D3 perspective, they parameterize its motion in the $4, 5, 6$ and $7, 8, 9$ planes respectively. The R-symmetry is given by the isometries of the "internal" space, that is $\mathrm{SO}(6)_{\mathrm{R}} \supset \mathrm{SO}(3)_{\mathbf{X}} \times \mathrm{SO}(3)_{\mathbf{Y}}$. The moduli space of supersymmetric vacua of the theory is parameterized by the vev's of these scalars.

| | | | | $y$ | | $\mathbf{X}$ | | | $\mathbf{Y}$ | |
|---|---|---|---|---|---|---|---|---|---|---|
| | $x^0$ | $x^1$ | $x^2$ | $x^3$ | $x^4$ | $x^5$ | $x^6$ | $x^7$ | $x^8$ | $x^9$ |
| D3 | $\times$ | $\times$ | $\times$ | $[0, +\infty)$ | $\cdot$ | $\cdot$ | $\cdot$ | $\cdot$ | $\cdot$ | $\cdot$ |
| D5 | $\times$ | $\times$ | $\times$ | 0 | $\times$ | $\times$ | $\times$ | $\cdot$ | $\cdot$ | $\cdot$ |

**Table 3.4.:** Directions spanned by D3 and D5-branes. The D5's sit at $y = 0$ along the 3 direction, while the D3's extend to infinity.

Suppose we want to find a supersymmetric vacuum of the theory engineered by a D3 spanning the interval $x^3 \equiv y \in [0, +\infty)$; see table 3.4. Clearly we have to impose suitable *boundary conditions on the SYM multiplet*, and then impose that the supersymmetry variations of the fermionic fields vanish, $\delta\Psi = 0$, where $\Psi$ collects all fermions in the SYM multiplet. The four-dimensional Lorentz invariance in the $0, \ldots, 3$ directions is broken; however it is possible to find vacua exhibiting three-dimensional invariance, by requiring $F_{\mu\nu} = F_{\mu 3} = 0$. The scalars $\mathbf{X}(y)$ and $\mathbf{Y}(y)$ are allowed to acquire a profile along the 3 direction (i.e. a vev). Furthermore, only Dirichlet boundary conditions on one of the two (but not both, nor neither) are compatible with unbroken supersymmetry [30]. Suppose $\mathbf{X}$ obeys Neumann boundary conditions, while $\mathbf{Y}$ Dirichlet. Then $\delta\Psi = 0$ is equivalent to

$$\frac{DX^i}{Dy} + \epsilon_{ijk}\left[X^j, X^k\right] \ , \quad \frac{D\mathbf{Y}}{Dy} = [\mathbf{Y}, \mathbf{Y}] = [\mathbf{X}, \mathbf{Y}] = 0 \ . \tag{3.15}$$

$D$ is the covariant derivative with respect to $A_3$.

The equation for the $X^i$ is called *Nahm's equation*, and has been studied in the context of BPS monopoles [43]. A simple solution is obtained upon prescribing a certain singular behavior for the scalars at $y = 0$ (where the D3-brane ends):

$$X^i(y) = \frac{t^i}{y} + \ldots \ . \tag{3.16}$$

The $t^i$ provide an embedding $\rho : \mathfrak{su}(2) \to \mathfrak{g}$ into the Lie algebra of the gauge group $G$ on the D3 worldvolume ($G = \mathrm{U}(N_\mathrm{c})$ for $N_\mathrm{c}$ coincident D3's), given that $t^3 = \left[t^1, t^2\right]$ (plus all cyclic permutations). At $y = 0$ we have a *simple pole* for $X^i$, while away from it the regular terms (the ellipsis in (3.16)) are leading.

These boundary conditions on the fields of the SYM multiplet can be obtained by having the D3's end on some flavor D5-branes. They are "half-BPS", i.e. they preserve half of the supersymmetry of a flat D3 that extends to $y < 0$ (instead of ending at 0 on the D5), and are automatically compatible with preserved supersymmetry (since (3.16) is a solution of (3.15), which is in turn implied by $\delta\Psi = 0$). The gauge group is broken to the commutant of (the image of) $\rho$ inside $G$.

As we said, the boundary conditions are due to the presence of D5-branes on which the D3's end. Let us consider $N_\mathrm{f}$ D5's stretching along directions $0, 1, 2, 4, 5, 6$ and sitting at $y = 0$; see table 3.4. The scalars $\mathbf{X}$ parameterize the Higgs branch of vacua of the theory and represent the motion of the color D3's along the D5's. This branch is a complex space (a hyper-Kähler manifold). As we know by now, the presence of the D5's introduces a global (flavor) $\mathrm{SU}(N_\mathrm{f})$ symmetry and new hypermultiplets $Z_i$ ($i = 1, \ldots, N_\mathrm{f}$) in the fundamental $\mathbf{N}_\mathrm{c}$. The Nahm's equation needs to be modified to account for the latter and now reads [30]:

$$\frac{\mathcal{D}\mathcal{X}}{\mathcal{D}y} + \sum_{i=1}^{N_\mathrm{f}} \delta(y - y_i)\mu_\mathbb{C}^{Z_i} = 0\ . \tag{3.17}$$

Here $\mathcal{D}$ is the covariant derivative with respect to $\mathcal{A} \equiv A_3 + iX^3$, $\mathcal{X} \equiv X^1 + iX^2$, $y_i$ is the position of the $i$-th D5 along $y$ ($y_i = 0\ \forall i$ in the case of coincident branes), and $\mu_\mathbb{C}^{Z_i}$ is the *complex moment map* for the $Z_i$ hypermultiplet. These moment maps are needed to describe the coupling of the (gauge) vector multiplet to localized "boundary" hypermultiplets [30] (as is the case for those coming from flavor D5's). The $Z_i$ can be also be understood as local complex coordinates on a hyper-Kähler manifold with $\mathrm{SU}(N_\mathrm{f})$ symmetry (the Higgs branch) in one of its complex structures.

### Fuzzy funnels

It is possible to gauge $\mathcal{A}$ away globally (i.e. $\mathcal{A} \equiv 0$), leaving

$$\partial_y\mathcal{X} + \sum_{i=1}^{N_\mathrm{f}} \delta(y - y_i)\mu_\mathbb{C}^{Z_i} = 0\ . \tag{3.18}$$

This implies that $\mathcal{X}$ is piecewise constant (equaling $\mu_\mathbb{C}^{Z_i}$) with jumps at the positions $y_i$. The interpretation of this behavior, perhaps clearer from the pole in (3.16), is that a D3 exerts a pull on the D5's distorting them into "spikes" or "funnels" where the former joins the latter. The D3 acquires two extra dimensions in the shape of a noncommutative two-sphere (topologically $S^2 \cong \mathrm{SU}(2)$), whose radius goes like $\frac{1}{y}$: At the origin $y = 0$ the sphere decompactifies and the D3 opens up into the D5.

This is in perfect agreement with the analysis in [41] of a (single) D$p$-brane ending on a D$(p+2)$-brane based on the abelian DBI action of the former.

We will see that this kind of Myers-like effect [44] is also realized in the D6-D8 system, which is one of the main ingredients of six-dimensional Hanany–Witten setups. See section 5.6.3 for a realization of this effect from the supergravity perspective.

A solution to (3.18) will once again specify an embedding of $\mathfrak{su}(2)$ into the gauge Lie algebra $\mathfrak{su}(N_\mathrm{c})$.[9]

By the Jacobson–Morozov theorem, the embedding $\rho$ defines a nilpotent element $\mu$ of the complexified Lie algebra $\mathfrak{su}(N_\mathrm{c})_\mathbb{C}$ as the image of the raising operator of $\mathfrak{su}(2)_\mathbb{C} \cong \mathfrak{sl}(2,\mathbb{C})$ (it actually defines the whole nilpotent orbit $\bar{O}_\mu \subset \mathfrak{su}(N_\mathrm{c})_\mathbb{C}$). This nilpotent element is of size $N_\mathrm{c} \times N_\mathrm{c}$ and can be put into Jordan canonical form with blocks of size $\mu_i \times \mu_i$, ordered from biggest to smallest (without loss of generality – see discussion below). This defines a certain partition of $N_\mathrm{c}$ that we also call $\mu = (\mu_i, \ldots, \mu_{N_\mathrm{c}})$, with $\mu_i \geq \mu_j$ for $i < j$, and that we can pictorially organize into a Young tableau; see figure 3.6. We define the transpose partition $\mu^t$ as that obtained by reflecting the Young tableau around an axis at $-\frac{\pi}{4}$. (It too is an ordered partition.)

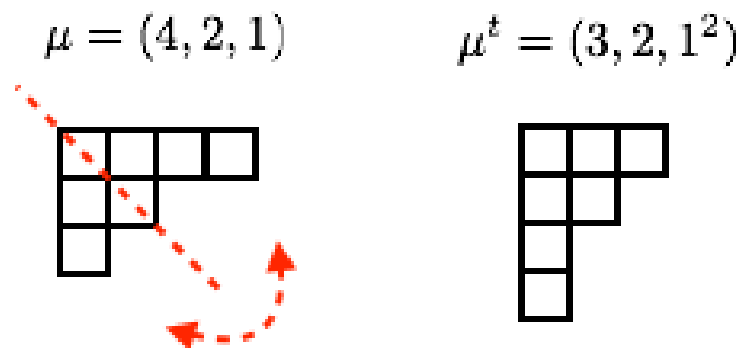

**Figure 3.6.:** A partition $\mu$ of $N_\mathrm{c} = 7$, $\mu = (4,2,1)$, and its transpose $\mu^t = (3,2,1,1) \equiv (3,2,1^2)$. The latter is obtained by reflecting the former around an axis at $-\frac{\pi}{4}$.

Clearly, all of this affects the way we group the D5's: The dimension of the $i$-th Jordan block, namely $\mu_i$, will give the number of D3's ending on the $i$-th D5 in the stack.

More precisely, given that $\mu_1 \geq \mu_2 \geq \ldots \geq \mu_{N_\mathrm{c}}$, we are requiring that the number of D3's ending on a D5 be non-increasing from right (i.e. at $y \gg 0$ – where the D3's extend to infinity) to left; see figure 3.7 and compare with figure 10 of [30]. We will call this requirement *ordering constraint.*

[9] The center-of-mass $\mathrm{U}(1) \subset \mathrm{U}(N_\mathrm{c})$ can be decoupled from the system.

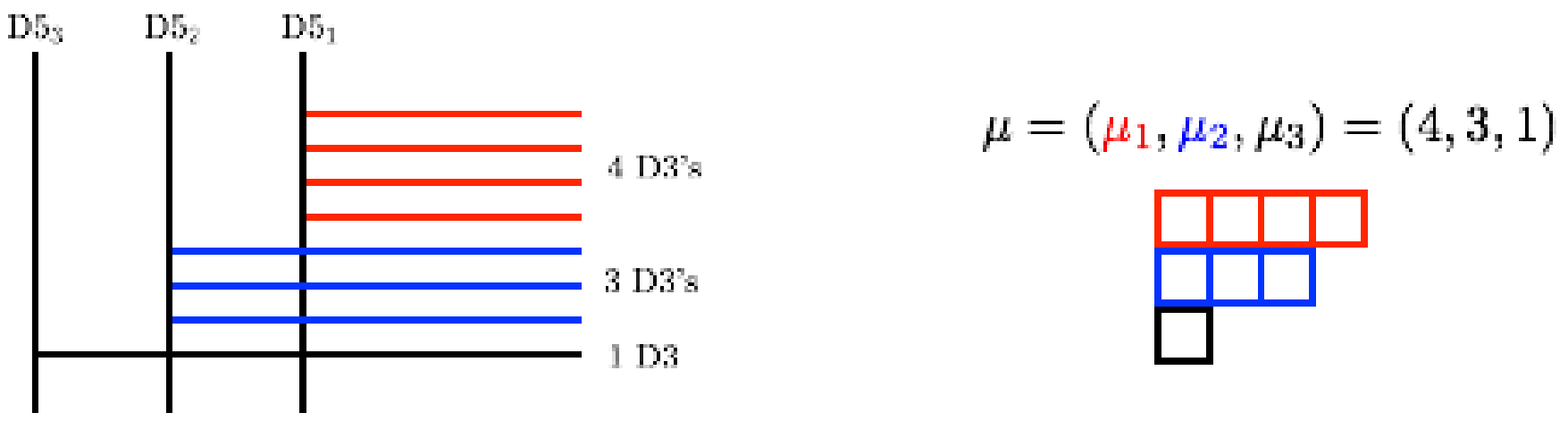

**Figure 3.7.:** D3's ending on D5-branes. Four D3's end on the first D5-brane ($\mu_1 = 4$), three end on the second ($\mu_2 = 3$), one ends on the third and last ($\mu_3 = 1$), therefore $\mu = (4, 3, 1)$ and $N_c = 8$.

The ordering constraint has as nice consequence the fact that, when all the D5's are brought on top of each other, the fuzzy funnels will not intersect each other. Rather, one created by less D3's will fit into another created by more D3's, according to the ordering constraint. See figure 3.8.

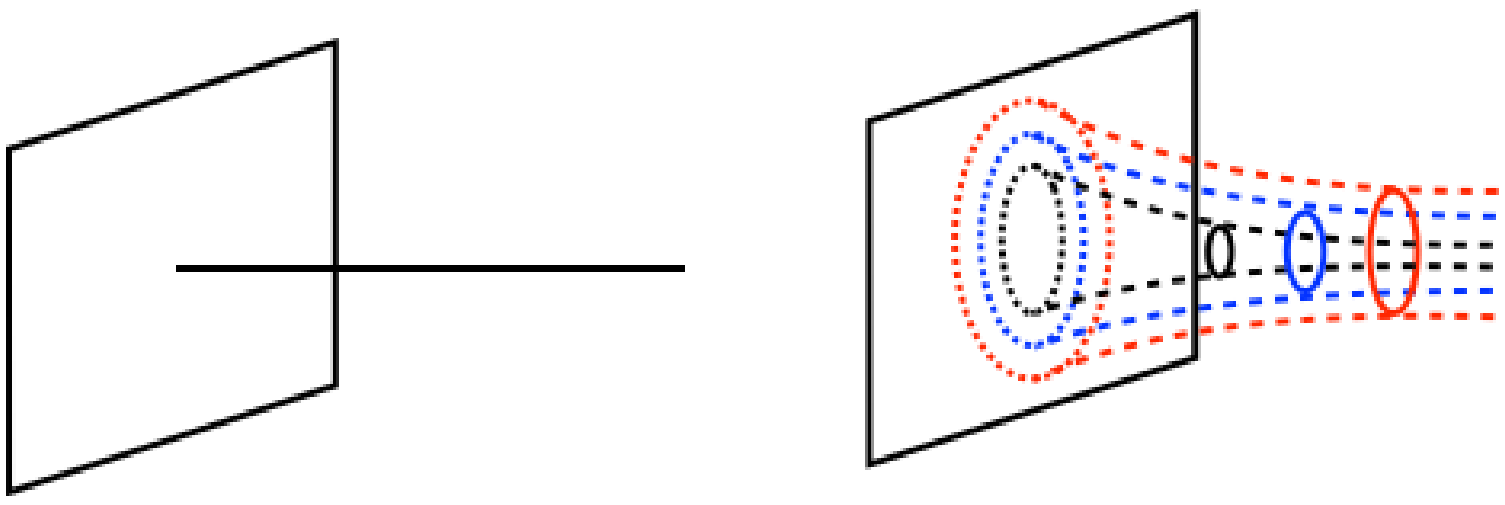

**Figure 3.8.:** On the left, the naive picture of many superposed D3's (the horizontal line) ending on D5's (the vertical plane). On the right, the correct picture: The D3's open up into the D5's at their location $y_{\mathrm{D5},i}$ along $y$ (the radius of the fuzzy sphere blows up). The ordering constraint implies that when all the D5's coincide (that is $y_{\mathrm{D5},i} = y_{\mathrm{D5},j}$ for every $i, j$) the funnels fit into each other without intersecting. Compare with figure 11 of [30].

In any case, all orderings would be in principle equivalent since we can freely move the D5's across each other, hence the biggest-to-smallest one is completely generic. Nonetheless it will prove more useful to us to assume that it holds.[10]

[10]The use of this constraint can also be justified on field-theoretic grounds using the good-bad-ugly trichotomy introduced by [45]. Roughly, "good" $\mathcal{N} = 4$ theories are those engineered by Hanany–Witten setups of the kind above that also respect the biggest-to-smallest ordering. For three-dimensional $\mathcal{N} = 4$ theories obtained by Kaluza–Klein reduction along $y$ (possible when the D3's are stretched on a finite interval), [46] has shown that a "bad" theory is just the Seiberg-like dual in three dimensions of a "good" one with a certain number of free fields. (The latter are dual to the unitary-violating monopole operators that make the theory bad in the first place.) Therefore we can always

### 3.2.4. The decoupling limit

Finally we want that the dynamics of the system be determined by the lightest extended objects. In Hanany–Witten setups these are given by the smallest branes (the D4's in the example of section 3.2.1). We should identify a limit that decouples gravity, the massive string excitations (i.e. heavier bulk modes), the brane fluctuations along the 6 direction (see table 3.3), and leaves behind an interacting field theory.

Such a limit can be obtained by enforcing the following conditions:

$$l_\mathrm{s} \to 0 \ (M_\mathrm{s} \to \infty) \ , \ M_\mathrm{Pl} \to \infty \ , \ L \to 0 \ , \ g^2_\mathrm{YM} = \frac{l_\mathrm{s}^{p-3} g_\mathrm{s}}{L} \ \text{fixed} \ , \tag{3.19}$$

$p$ being the number of spacelike dimensions of the smallest branes in the setup, and $L$ the length of the finite interval they span.

All the above arguments can be repeated for higher dimensional branes. As we saw in figure 3.3, any D$p$-brane with $p \leq 6$ can end on an NS5-brane, and any D$p$-brane can end on a D$(p+2)$-brane. We are thus led to the conclusion that the Hanany–Witten setup with $N_\mathrm{c}$ color D$p$-branes and $N_\mathrm{f}$ flavor D$(p+2)$-branes engineers a $p$-dimensional interacting[11] field theory (with SU gauge and flavor groups) in the decoupling limit (3.19).

## 3.3. NS5-D6-D8 brane systems

We can now conveniently summarize the lessons of the previous section, and at the same time generalize the arguments to any D$p$-brane.

- The general Hanany–Witten setup engineering $p$-dimensional SYM with eight $Q$ supercharges requires suspending a bunch of color D$p$-branes between two consecutive NS5-branes, separated by a distance $L$ along the 6 direction.
- The SYM gauge coupling is given by $g^2_\mathrm{YM} = \frac{l_\mathrm{s}^{p-3} g_\mathrm{s}}{L}$, and remains finite in the limit (3.19) needed to decouple gravity and massive string modes.
- We can add flavors to the system by having the color D$p$'s end on a stack of flavor D$(p+2)$-branes, or by letting the former stretch to infinity beyond the NS5's.

restrict our attention to the good ones obeying the constraint. We thank S. Giacomelli for discussion on this and related issues.

[11] Even without the extra matter given by the flavor hypermultiplets this would be an interacting theory, as is always the case for the nonabelian Yang–Mills interaction at finite coupling.

| | $x^0$ | $x^1$ | $x^2$ | $x^3$ | $x^4$ | $x^5$ | $x^6$ | $x^7$ | $x^8$ | $x^9$ |
|---|---|---|---|---|---|---|---|---|---|---|
| NS5$_i$ | × | × | × | × | × | × | $x^6_i$ | · | · | · |
| D$p$ | | | $p-1$ times × | | | | × | · | · | · |
| D$(p+2)$ | | | $p-1$ times × | | | | $x^6_{\mathrm{D}p+2}$ | × | × | × |

**Table 3.5.:** Directions spanned by NS5, D$p$, D$(p+2)$-branes in a general Hanany–Witten configuration engineering a $p$-dimensional field theory. The D$p$-branes span a finite interval of length $L$ along direction 6.

- The field theory dynamics is governed by the lightest objects in the system, that is the lowest-dimensional branes.
- At finite SYM gauge coupling the D$p$'s ending on the NS5 will exert a pull on the latter, bending them. This is a one-loop effect in field theory.
- The boundary conditions introduced by the presence of the flavor D$(p+2)$'s on which the color D$p$'s end (respecting the ordering constraint) imply Nahm's equations (3.18). The simplest solution describes the "expansion" of a D$p$-brane into a fuzzy sphere whose radius diverges at the location of the D$(p+2)$. The system is thus better described by a fuzzy funnel.

### 3.3.1. D6

Let us now focus on the $p=6$ case in type IIA, namely the case of D6-branes suspended between NS5-branes. It will prove useful to be slightly more general, and to imagine having a sequence of many NS5-branes, not just two, as in figure 3.9. Since the system preserves eight $Q$ supercharges, the only possible algebra in six dimensions is the chiral $\mathcal{N}=(1,0)$, which can be enhanced to a superconformal algebra. This is reflected in the very nature of the field theory we are about to engineer. Moreover, the ten-dimensional Lorentz group is now broken to:

$$\mathrm{SO}(1,9) \to \mathrm{SO}(1,5) \times \mathrm{SO}(3) \cong \mathrm{SO}(1,5) \times \mathrm{SU}(2) \ . \tag{3.20}$$

The last factor in (3.20) is responsible for rotations in the $7,8,9$ plane in which the D6-branes are localized. The isometry group SU(2) is the correct R-symmetry group of the six-dimensional $\mathcal{N}=(1,0)$ chiral super(conformal) algebra.

There are a few differences with respect to the general recipe we just outlined. Let us spell them out.

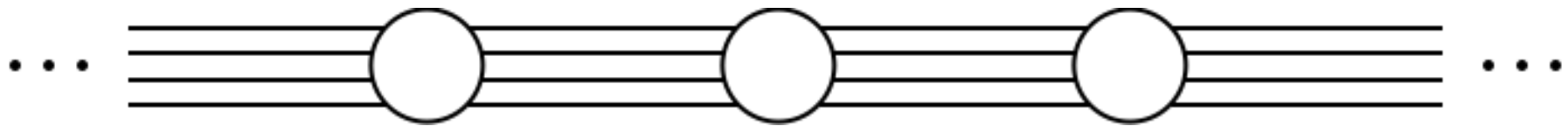

**Figure 3.9.:** A chain of NS5-branes with suspended D6-branes. The semi-infinite D6's to the left and right of the leftmost and rightmost NS5-brane (respectively) contribute flavor hypermultiplets. The finite segments all contain the same number $N_c$ of color D6-branes, giving rise to a chain of SU gauge groups connected by bifundamental hypermultiplets.

### No-bending condition

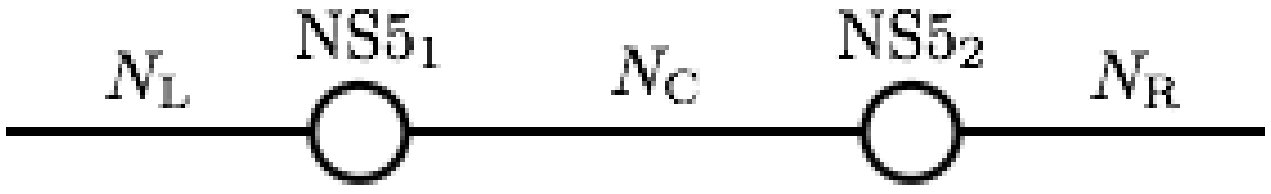

**Figure 3.10.:** Three segments of D6-branes, one finite (the D6's are suspended between the NS5's) with $N_C$ color branes, and two semi-infinite with $N_{L,R}$ flavor branes.

First of all, the D6's do not bend the NS5 they end on. Following Witten [42], it can been argued that, if there is a bending effect, then: it should be proportional to the net D-brane charge induced on the NS5 worldvolume, that is $N_L - N_R$; the scalars describing the embedding of the former into the latter should satisfy a Laplace equation (namely, they should satisfy a sort of minimal-area principle in order for the embedding to be supersymmetric); the bending should only depend on the D$p$ worldvolume directions transverse to the NS5.

All in all, this qualitative remarks suggest that we need to solve the Laplace equation in $6-p$ dimension. Clearly, for $p=6$ the transverse space is zero-dimensional and there is no Laplace equation to be solved to start with: No bending can occur. This has the very important consequence that *the net number of D6-branes ending on a certain NS5* (from the left and right) *must be zero*:[12]

$$N_{D6,L} = N_{D6,R} \ . \tag{3.21}$$

The same conclusion can be reached either by solving the appropriate Bianchi identity in supergravity with D6-brane sources (as we will see in section 3.3.3 in a slightly different way) or by requiring gauge anomaly cancellation in the six-dimensional field theory (i.e. its quantum consistency). In particular, the latter requirement imposes $N_f = 2N_c$ [2, 32].[13] We know by now that the easiest way to introduce flavors in a Hanany–Witten setup is to let the color D$p$-branes extend past the NS5-branes to infinity. In figure 3.10 we have three D6 segments; the no-bending condition imposes

$$N_\mathrm{L} = N_\mathrm{C} \ , \quad N_\mathrm{C} = N_\mathrm{R} \ \Rightarrow \ N_\mathrm{L} = N_\mathrm{R} = N_\mathrm{C} \ . \tag{3.22}$$

[12] As we will see later on, this statement needs to be modified in presence of a *Romans mass*, sourced by D8-branes.

[13] In $d=4$ $\mathcal{N}=2$ theories the same constraint guarantees the conformal invariance of the quantum theory, rather than gauge anomaly cancellation [47]. Notice also that in six dimensions there can be chiral flavors. The hyperino in a $(1,0)$ hypermultiplet is a chiral

The total number of flavors is $N_\mathrm{f} = N_\mathrm{L} + N_\mathrm{R} = 2N_\mathrm{C} = 2N_\mathrm{c}$. (The number of finite D6-branes in the central segment, $N_\mathrm{C}$, is the number of colors, $N_\mathrm{c}$.)

*The NS5-D6 Hanany–Witten setups automatically engineer anomaly-free* $(1,0)$ *superconformal theories in six dimensions.*

#### Tensor multiplets from NS5's

Second, we cannot neglect the presence of the NS5-branes anymore, not even at very small $g^2_\mathrm{YM} \sim g_\mathrm{s}$ coupling (when they become very heavy). The NS5's are now the smallest objects in the setup and will be completely contained inside the D6 worldvolume; see table 3.5 with $p = 6$. This simple observation is what makes most of the difference between six- and lower-dimensional (i.e. $d \leq 4$) field theories (engineered via Hanany–Witten setups). In fact the worldvolume theory of the type IIA NS5-brane supports a tensor multiplet, not a vector one (which would contain a more familiar gauge field, rather than a self-dual two-form). In the limit in which all NS5's are brought on top of each other, we cannot expect an enhanced gauge symmetry as is the case for "normal" D-branes; rather we will have a theory of so-called *tensionless strings* in six dimensions, sourcing the self-dual two-forms! This is believed to be a superconformal field theory (SCFT henceforth) describing the propagation of tensionless (i.e. massless) two-strings in $\mathrm{Mink}_6$ (spanned by directions $0,\ldots,5$ in table 3.5).

#### Dynamical gauge coupling

Third, the positions of the NS5-branes along the 6 direction are not fixed once and for all, as they were for suspended D$p$-branes with $p < 6$. The tensor multiplet of the $\mathcal{N} = (1,0)$ superalgebra contains a self-dual two-form and one real scalar $\Phi$ (as well as a "tensorino") [48]. The scalar provides (via its vev) the position $\phi_i$ of the $i$-th NS5 along the 6 direction. More precisely, if we had $N$ NS5-branes, $N-1$ such scalars would control the relative positions $\phi_{i+1} - \phi_i$ of the NS5-branes, while the $N$-th scalar would give the center-of-mass motion of the whole system along the 6 direction. In the easiest setup with just fermion and it has opposite chirality with respect to the gaugino in a $(1,0)$ vector multiplet. Therefore flavors contribute to the gauge anomalies, and can be used to balance off the contribution from the gaugini upon imposing $N_\mathrm{f} = 2N_\mathrm{c}$ [32].

one pair, the relative position of two consecutive NS5's, that is the interval's length $L = \phi_2 - \phi_1$, is related to $g^2_{\mathrm{YM}}$ via (3.12) as dictated by the Kaluza–Klein reduction. This means that *the vev of the dynamical scalar field in six dimensions governing the distance between the NS5's plays the role of gauge coupling "constant" in front of the YM term.* From (3.12) and $\left[g^{-2}_{\mathrm{YM}}\right] = \left[M^{d-4}\right]$ in $d$ dimensions we have:

$$\left[\frac{1}{g^2_{\mathrm{YM}6d}}\right] = \left[\frac{L}{g^2_{\mathrm{YM}7d}}\right] = \left[M^{-1+7-4}\right] = \left[M^2\right] \ , \tag{3.23}$$

which is exactly the canonical mass dimension of a scalar $\Phi$ in six spacetime dimensions.[14] Thus, the value of the gauge coupling is given by the vev of a (dimensionful) scalar field.

For each D6 segment suspended between the $i$-th and $i+1$-th NS5-brane, the SYM gauge kinetic term becomes

$$(\Phi_{i+1} - \Phi_i) \operatorname{Tr} F^2 \ , \tag{3.24}$$

and the effective gauge coupling is given by the difference of the scalar vevs:

$$\frac{1}{g^2_{\mathrm{YM}6d}} = \frac{\phi_{i+1} - \phi_i}{l_{\mathrm{s}}^3 g_{\mathrm{s}}} = \langle \Phi_{i+1} - \Phi_i \rangle \ . \tag{3.25}$$

From the latter equation it is clear that the limit in which all NS5-branes coincide ($\phi_i = \phi_{i+1}$, $\forall i$) gives the strongly-coupled fixed point: The coupling blows up and the gauge theory description becomes inadequate. The theory of coincident NS5's should be a superconformal one. Two main pieces of evidence support this hypothesis. First, the NS5-D6 brane setup preserves the right amount of $Q$ supercharges, eight. The only algebra with eight supercharges in six dimensions is the $\mathcal{N} = (1, 0)$ one, which is actually superconformal. Second, the continuum of massless states typical of a conformal theory sould be provided by tensionless strings between coincident NS5-branes. These are easily accounted for by massless D2-branes ending on the NS5's: They are string-like objects on the NS5 worldvolume, and they couple to the self-dual two-forms.

### D2 instantons

It is well known [49] that, in a suitable decoupling limit, the D$p$-branes in a $k$ D$p$ – $N$ D$(p + 4)$ system describe codimension-four solitons in the U$(N)$ gauge theory realized on the common D$(p+4)$ worldvolume. These solitons are actually (supersymmetric) instanton configurations at $k$ instanton-number, as can be argued from an analysis of the RR charges present in the system and the amount of preserved supersymmetry [35]. $k$ is a topological charge (the so-called winding number), thus the D$p$-brane charge is smeared on the D$(p+4)$ worldvolume: The D$p$'s are delocalized on the worldvolume of the former.

[14] In $d$ dimensions we would have $[\Phi] = \left[M^{(d-2)/2}\right]$.

In the case we are considering here, D2-branes parallel to the D6's (stretching along directions $0, 1, 6$) should act as instantons on the D6 worldvolume. Furthermore, in type IIA, a D2 is allowed to end on an NS5-brane, and couples to the self-dual two-form in the tensor multiplet. This is because it acts like a string on the NS5 worldvolume.[15] Thus the D2's become tensionless when the NS5's coincide (their tension goes like $\frac{1}{g_s} \sim \frac{1}{g^2_{\mathrm{YM},i}}$ – see (3.1) – for the $i$-th gauge group supported on the $i$-th D6 segment). In a six-dimensional superconformal theory, the BPS objects making up the massless spectrum are not particles, but rather instantonic strings, exactly those provided by tensionless D2's [19].

As we will see in the next section, the D2-NS5 system can be lifted to M-theory, where the so-called *M-strings* [50] (an M2 ending on an M5, i.e. a string-like object) can be counted by means of their *elliptic genus*. Upon applying the F/M-theory duality, the tensionless M2 on M5 system turns into a singular elliptically fibered Calabi–Yau. One can then apply techniques of topological strings to count the same BPS states [51].

### 3.3.2. Reduction from M-theory

Recall that to engineer the six-dimensional $\mathcal{N} = (2, 0)$ theory (of type $A_N$) one can start with $N$ flat M5-branes in M-theory on $\mathrm{Mink}_{11}$ [11]. The presence of just one kind of branes breaks in half the supersymmetry of the background (flat space), preserving sixteen (chiral) $Q$ supercharges. The algebra of the worldvolume theory is $\mathcal{N} = (2, 0)$. The R-symmetry is SO(5), corresponding to rotations in the five transverse directions to the worldvolume of the M5-branes, filling up the $0, \ldots, 5$ directions.

The superconformal theory is obtained in the limit in which all branes are brought on top of each other; the tensionless objects are the *M-strings* [50] coming from massless M2-branes stretching between coincident M5-branes. The theory is that of $N$ $(2,0)$ tensor multiplets. Each of them contains a self-dual two-form, five real scalars and two tensorini (left Weyl spinors).

[15] From figure 3.3 we know that a D1 can end on a type IIB NS5-brane: It acts as a point-like particle on its worldvolume. As such it is a source for the U(1) gauge potential in the six-dimensional $(1, 1)$ SYM multiplet supported on the NS5-brane. This picture is necessary for the D1-NS5 interaction to be consistent. Indeed the NS5-brane must be able to carry away the D1 charge (via a Chern–Simons, or Wess–Zumino, term) smeared on its worldvolume. T-dualizing once along a direction inside the type IIB NS5 worldvolume, we get a D2 on a type IIA NS5. The former is now a string in six dimensions and it sources a two-form, the self-dual tensor in a $(2, 0)$ tensor multiplet. Supersymmetry is reduced to $(1, 0)$ if the NS5 is nontrivially embedded in the spacetime, as is the case for an NS5-D6 Hanany–Witten setup. This does not change the coupling between the D2 and the tensor we just discussed though.

It is well known [35] that M-theory compactified on a circle $S^1_{\mathrm{M}}$ of radius $R_{\mathrm{M}}$ gives perturbative type IIA string theory with coupling $g_{\mathrm{s}} = \frac{R_{\mathrm{M}}}{l_{\mathrm{s}}}$: For $g_{\mathrm{s}} \to 0$ the radius is very small and we are effectively in ten dimensions; conversely, for big string coupling the circle "decompactifies" and we gain an extra dimension (the eleventh).[16] Moreover, from string dualities, one has that the eleven-dimensional Planck mass is related to the ten-dimensional type IIA string coupling and Planck length:

$$M_{\mathrm{Pl}\,(11)} = g_{\mathrm{s}}^{-\frac{1}{3}} l_{\mathrm{s}}^{-1} \ . \tag{3.26}$$

The tensions of the M2 and M5-branes in eleven dimensions are exactly equal to those of the type IIA D2 and NS5-branes respectively [35]:

$$T_{\mathrm{M2}} \sim M^3_{\mathrm{Pl}\,(11)} = g_{\mathrm{s}}^{-1} l_{\mathrm{s}}^{-3} = T_{\mathrm{D2}} \ , \quad T_{\mathrm{M5}} \sim M^6_{\mathrm{Pl}\,(11)} = g_{\mathrm{s}}^{-2} l_{\mathrm{s}}^{-6} = T_{\mathrm{NS5}} \ . \tag{3.27}$$

The $(2,0)$ tensor multiplet supported on the worldvolume of an M5-brane contains five real scalars. The same multiplet is also supported on the worldvolume of a type IIA NS5-brane. Hence, four out of five will descend to scalars describing the position of the NS5 in the four-dimensional transverse space.

The M2-brane can end on an M5-brane, much as a D2-brane can end on an NS5. Both smaller branes act as strings on the worldvolume of the latter. If we suspend an M2-brane between two M5's separated by a distance $L$, its tension will be given by $L\,T_{\mathrm{M2}}$: In the coincident limit $L \to 0$, the strings coming from suspended M2-branes become tensionless, and we expect to engineer a superconformal theory with $\mathcal{N} = (2,0)$ supersymmetry.

This is not a usual string theory though. To keep an interacting field theory while decoupling gravity we can send $M_{\mathrm{Pl}\,(11)} \to \infty$; we can then bring the M5-branes on top of each other, $L \to 0$ hence $T_{\mathrm{M2}} \to 0$. We are thus left with strings in six dimensions coupling to self-dual two-forms, and without any "string coupling" to play with (to set up perturbation theory as we would do with $g_{\mathrm{s}}$ in ten dimensions, for instance).

We have by now understood that the six-dimensional $\mathcal{N} = (2,0)$ theory can be equivalently engineered by coincident NS5's or M5's, connected by tensionless D2's and M2's respectively. The D2 on NS5 system is just a reduction of the M2 on M5 one to ten dimensions.

[16]Further nontrivial evidence for this is lent by the following fact [35]. The type IIA D0-branes, carrying a tension $T_{\mathrm{D0}} = (g_{\mathrm{s}} l_{\mathrm{s}})^{-1}$, become infinitely light at strong string-coupling, $g_{\mathrm{s}} \to \infty$, and are interpreted as the momentum states obtained by Kaluza–Klein reducing M-theory on a circle of radius $R_{\mathrm{M}} = T^{-1}_{\mathrm{D0}}$.

The new ingredient needed to reduce supersymmetry and be left with an interacting $\mathcal{N} = (1,0)$ theory, rather than its more supersymmetric $(2,0)$ cousin, are clearly D6-branes. In the type IIA setup summarized in table (3.5) with $p = 6$, they impose a further projection on the supercharges, (3.7), leaving only eight behind. The same mechanism can be understood in M-theory by lifting $k$ D6-branes: The background is given by M-theory on a $k$-center Taub-NUT space (also called ALF space) [52, 53]. When the D6-branes coincide, the Taub-NUT space becomes singular and locally resembles the $A_k$ orbifold singularity, i.e. $\mathbb{C}^2/\mathbb{Z}_{k+1}$. More precisely, the Taub-NUT space admits a canonical $S^1$ fibration (with radius $R$) over $\mathbb{R}^3$; the fibration is allowed to degenerate over $k$ points in $\mathbb{R}^3$. When $R$ is taken to infinity, the ALF space turns into the ALE space $A_k$. The Taub-NUT metric in M-theory describes a system of $k$ *Kaluza-Klein monopoles* [35]; when they all coincide we get an enhanced $\mathrm{SU}(k)$ gauge symmetry. The latter is recovered on the worldvolume of $k$ D6-branes, obtained by dualizing the $k$ coincident monopoles via tpye IIA – M-theory duality, that is upon identifying the Taub-NUT $S^1$ with the M-theory $S^1_{\mathrm{M}}$. The Taub-NUT space is part of the five-dimensional space transverse to the M5-branes, and the full background preserves half of the $Q$ supercharges of flat M5-branes in $\mathrm{Mink}_{11}$, that is eight.

*The NS5-D6 Hanany–Witten setup engineering six-dimensional* $(1,0)$ *theories is equivalent to M5-branes at an* $A_k$ *singularity in eleven dimensions.*

In this case, the $(2,0)$ tensor multiplet reduces to a $(1,0)$ tensor multiplet (containing one real scalar) plus a $(1,0)$ hypermultiplet (containing four real scalars). As we have seen, the vev of the tensor multiplet scalar controls the position of the NS5-brane, while giving a vev to the hypermultiplet scalars would correspond to moving along the Higgs branch of the theory.

## 3.3.3. D8

As we know from our discussion in section 3.2.2, the introduction of flavors in the type IIA brane system can be achieved by having the color D$p$-branes end on a stack of flavor D$(p+2)$-branes. In our present case $p = 6$, thus the flavor branes will be the type IIA D8-branes.

### The Romans mass

D8-branes are rather special. Usually a D$p$-brane couples electrically to a bulk RR potential of type II strings, that is a $(p+1)$-form $C_{p+1}$, and magnetically to a $(7-p)$-form (in ten spacetime dimensions). This is because a $p$-brane will sit at a point inside the $10-(p+1)=(9-p)$-dimensional space transverse to its worldvolume; we can measure its "electric" charge $\mu_p$ by surrounding this point with a $9-p-1=(8-p)$-dimensional sphere and computing the flux of the RR potential through the sphere. This is nothing but Gauss's law:

$$\mu_p = \int_{S^{8-p}} *_{10} F_{p+2} \ , \tag{3.28}$$

$*_{10}$ being the Hodge star in ten dimensions and $dC_{p+1} = F_{p+2}$ away from sources. The electric charge of the magnetically dual brane (that is the magnetic charge of the D$p$-brane) is given as usual by:

$$\mu_{6-p} = \int_{S^{p+2}} F_{p+2} \ . \tag{3.29}$$

In ten dimensions, a $(p+2)$-dimensional sphere surrounds (in the above sense) a $6-p$-brane. Furthermore, electric and magnetic charges have to satisfy Dirac's quantization condition [35]

$$\mu_p \, \mu_{6-p} \in 2\pi\mathbb{Z} \ . \tag{3.30}$$

Now, a D8-brane obviously has $p = 8$ so it is surrounded by a 0-sphere. Usually, whenever a system of D-branes and O-planes (of the same dimension) is put inside a compact space, the total D-brane charge of the system needs to add up to zero. This is because the flux lines of the relevant RR potential cannot escape to infinity, hence the total RR charge must vanish. One can obtain this result (the so-called *tadpole condition*) by integrating the appropriate Bianchi identity in supergravity with D-brane and O-plane sources.

However, this argument does not apply for D8-branes, which are surrounded by a 0-dimensional space: There is no need to have vanishing D8-charge in a compact space. The D8's couple to the RR nine-form potential $C_9$, which has a nondynamical ten-form as field strength, $F_{10}$. The latter is in fact proportional to the volume form of the ten-dimensional spacetime, the constant of proportionality being a zero-form, i.e. a scalar in ten dimensions. For this reason sometimes the latter is referred to as a *cosmological constant* [54, 33].[17] Schematically:

$$dC_9 = F_{10} = F_0 \, \mathrm{vol}_{10} \ , \quad F_0 = *_{10} F_{10} \ , \quad \mu_8 = F_0 \ . \tag{3.31}$$

[17]Another curious fact is the following: The D8's do not have magnetically dual branes (which would be $-2$-branes, $-1$-dimensional objects which make no sense physically). This is reflected by the fact that the "flux" $F_0$ is not the curvature of any $-1$-form RR potential $C_{-1}$.

$F_0$ is called *Romans mass* (for historical reasons) [54], and its presence defines *massive type IIA string theory*. Its Bianchi identity, away from sources, simply says $dF_0 = 0$. However, at the location of a D8-brane we have

$$dF_0 = n_{\rm D8}\,\delta(x - x_{\rm D8}) \Leftrightarrow F_0 = n_{\rm D8}\,\theta(x - x_{\rm D8})\ , \tag{3.32}$$

that is $F_0$ is piecewise constant. Here $x$ parameterizes the direction transverse to the D8 worldvolume inside $\text{Mink}_{10}$. (3.32) says that a stack of $n_{\rm D8}$ D8-branes manifests itself as the locus where the Romans mass jumps from 0 to $n_{\rm D8}$. Hence the D8's act as *domain walls* between regions with different "cosmological constant". Flux quantization then requires this constant to be an integer (as is the number of D8's in a certain region):

$$F_0 = \frac{n_0}{2\pi}\ , \quad n_0 \in \mathbb{Z}\ . \tag{3.33}$$

Notice that it is not known how to mimic such a behavior in M-theory, where the only scalar at our disposal is the eleven-dimensional Planck mass (which cannot be changed arbitrarily). For this reason, the M-theory origin of D8-branes (more generally the reduction from M-theory to massive type IIA) remains mysterious [55].[18]

## D8's as flavors

Let us now see what happens when we include flavor D8-branes to the NS5-D6 Hanany–Witten setup providing chiral $(1,0)$ tensor multiplets. Consider first the simple setup of figure 3.11.

First, the no-bending condition of the NS5-branes, $N_{\rm D6,L} = N_{\rm D6,R}$, needs to be properly adjusted due the presence of the D8's. The Bianchi identity for the (net) D6-brane charge in the presence of $k$ semi-infinite branes ending on the NS5-brane is:

$$dF_2 - HF_0 = k\,\theta(x^6)\,\delta^{(3)}(\mathbf{y} - \mathbf{y}_{\rm D6})\ . \tag{3.34}$$

Here $\mathbf{y}_{\rm D6}$ is a three-vector giving the location of the $k$ D6's in the $7, 8, 9$ plane parameterized by $\mathbf{y}$, and $x^6$ is the location of the NS5-brane along the 6 direction.

[18] It is known from the work of Hořava and Witten [56, 57] that strongly coupled $E_8 \times E_8$ heterotic string theory is dual to M-theory compactified on $S^1_{\rm M}/\mathbb{Z}_2$ (a segment), much like the strongly coupled type IIA string theory is better understood as M-theory on $S^1_{\rm M}$. This gives a duality between the heterotic string and type IIA. In particular, at each end of the M-theory segment we have a corresponding (8 D8)-O8 wall in type IIA. At strong coupling (that is in M-theory) we cannot move the D8's away from the O8, and the transverse motion of the D8's becomes a Wilson loop for the gauge theory on each boundary in M-theory. This gives a partial understanding (in a rather special situation) of the lift of a D8.

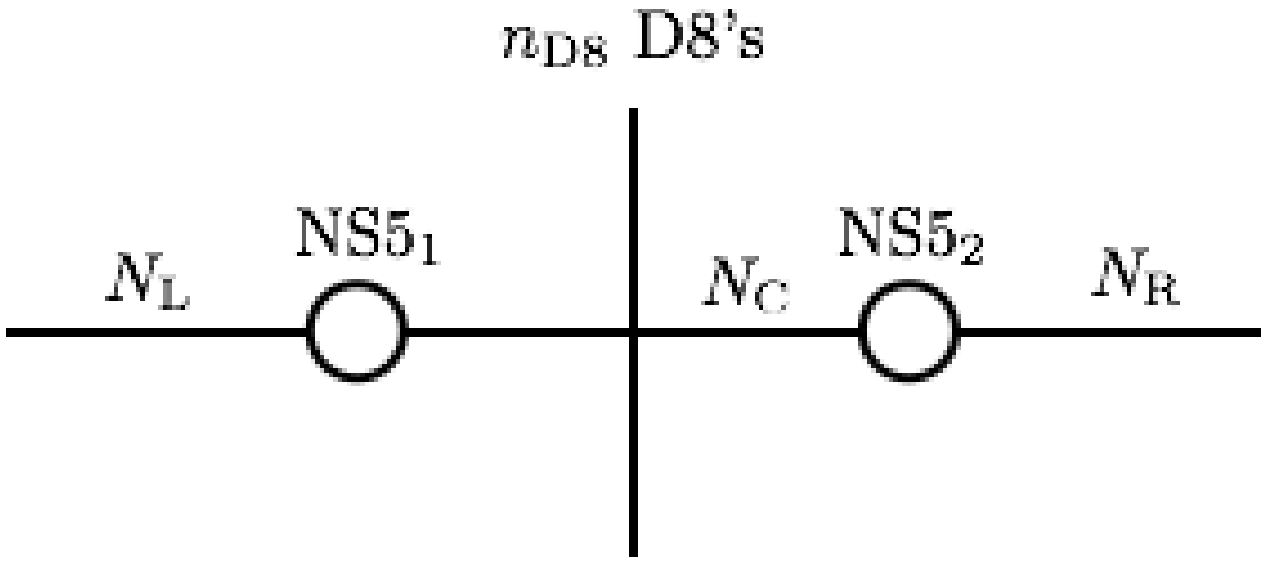

**Figure 3.11.:** The finite D6 segment is "crossed" by $n_{\mathrm{D8}}$ branes, raising the "cosmological constant" from $n_0$ (the value of the Romans mass to their left) to $n_0 + n_{\mathrm{D8}}$ (the value of the Romans mass to their right). The D8's contribute hypermultiplets in the $\mathbf{N}_{\mathrm{C}}$ fundamental representation of the SU($N_{\mathrm{C}}$) gauge group realized on the finite D6 worldvolume.

Integrating around the D6's (that is along the **y** plane) before and after the NS5 gives:

$$N_{\mathrm{D6,L}} - N_{\mathrm{D6,R}} = n_0 = 2\pi F_0 \ , \tag{3.35}$$

in a spacetime region with Romans mass given by $n_0$. That is, the D8's modify the condition imposing zero net D6-charge on an NS5 worldvolume, allowing the NS5 to absorb some of the D6-brane charge: The difference between the number of branes ending from the left and from the right equals the Romans mass. In turn, the Romans mass in two regions separated by a domain wall (i.e. a D8 stacks) jumps by the number of D8's in the stack. We summarize the situation of figure 3.11 in formulae:

$$N_{\mathrm{D6,L}} - N_{\mathrm{D6,C}} = n_0 \ , \quad N_{\mathrm{D6,C}} - N_{\mathrm{D6,R}} = n_0' = n_0 + n_{\mathrm{D8}} \ . \tag{3.36}$$

Once again, even in this more complicated setup, the total number of flavors is twice the number of colors, in agreement with anomaly cancellation in six dimensions:

$$N_{\mathrm{D6,L}} + N_{\mathrm{D6,R}} + n_{\mathrm{D8}} = 2N_{\mathrm{D6,C}} \ . \tag{3.37}$$

### D6-D8 Nahm poles

According to our discussion in section 3.2.3, we expect that the boundary conditions for D6's ending on a D8 be described by a Nahm pole of the kind (3.16):

$$X^i(x^6) = \frac{t^i}{x^6 - x^6_{\mathrm{D8}}} + \dots \ . \tag{3.38}$$

Basically, we expect that the D8's provide boundary conditions for the fields of the seven-dimensional SYM supported on the D6's. The position of D6's inside the three-dimensional space transverse to their common worldvolume is specified by three scalars $X^i$ (the R-symmetry is SO(3) ≅ SU(2)), obtained by dimensional reduction from the $d = 10$ $\mathcal{N} = 1$ SYM vector multiplet. As before, we allow these scalars to have a profile along the 6 direction. The simplest solution to Nahm's equation (3.15) (the covariant derivative is now computed with respect to $A_6$) is then (3.38). The $t^i$ generate an $\mathfrak{su}(2)$ subalgebra of $\mathfrak{su}(N_\mathrm{c})$.

The boundary data associated with the (nilpotent orbit of the) nilpotent element $\mu_\mathbb{C} \in \mathfrak{su}(N_\mathrm{D6})_\mathbb{C}$, which is specified by the embedding $\rho : \mathfrak{su}(2) \to \mathfrak{su}(N_\mathrm{D6})$, is interpreted as activating an operator vev in the six-dimensional field theory [22, 58]. Moreover, in the T-dual configuration[19] of color D7-branes ending on flavor D7's, the same nilpotent element defines so-called "T-brane data" [59]. Hence, in type IIB, the D6-D8 fuzzy funnel turns into a T-brane describing the recombination of a color D7 with a flavor one. The "T" in the name stands for triangular, and refers to the triangular (hence nilpotent) vev – along the adjoint representation of the color gauge group – that the complex "Higgs" scalar of the $d = 8$ SYM on the D7 worldvolume is given. (The Higgs field satisfies a so-called *Hitchin system* coupled to boundary defects; this is the T-dual version of Nahm's equations.) Whereas a diagonal vev would correspond to moving the branes in a stack apart (breaking the color group – hence the name of Higgs field), a nilpotent vev is not associated with movements of the branes, but rather with their *recombination* (also partially breaking the gauge group) which is not captured by the geometric moduli of type IIB. A better explanation would be the following: The complex structure deformation moduli of the *F-theory lift* of the D7-brane configuration do not capture the T-brane data. Once a proper treatment of this recombination process is identified in the F-theory language (see [60, 61, 62] for two complimentary approaches), it is also possible to study more general type IIB brane systems involving nonperturbative

[19] As is well known, type IIA/B string theory in the background of $N$ NS5-brane is T-dual to type IIB/A in the background of an $A_{N-1}$-tpye ALE space, a $\mathbb{C}^2/\mathbb{Z}_N$ singularity. The latter is a singular limit of an $N$-center Taub-NUT space, an $S^1$ fibration over $\mathbb{R}^3$ (that is $\mathbb{C}^2$, viewed as a cylinder of varying radius) collapsing over $N$ points in the base. The T-duality then amounts to exchanging the U(1) gauge fields $g_{\mu 9}$ and $B_{\mu 9}$ coming from the Kaluza–Klein reduction along the $S^1$ – parameterized by the periodic direction 9, say – of the ten-dimensional metric $g$ (of which the Taub-NUT solution is the KK monopole) and the B-field $B$ (of which the NS5 is the monopole) respectively. Performing such T-duality on the NS5-D6-D8 system in massive type IIA, we get a chain of D7-branes wrapping collapsed $\mathbb{P}^1$'s (the resolution 2-cycles of the $A_{N-1}$ singularity at zero Kähler size) intersecting each other according to the $A_{N-1}$ Dynkin diagram; the left- and rightmost D7's each end on a noncompact D7 – the T-dual of a D8 – responsible for the flavor symmetry. (Recall that $g^2_\mathrm{YM} \sim L^{-1}$, $L$ being the "length" of the interval along which the branes stratch. For infinite – noncompact – branes we get a gauge symmetry at zero coupling, i.e. a flavor symmetry.)

7-branes coupled to boundary defects (flavor 7-branes). These more general systems are needed to engineer $(1,0)$ theories that cannot be embedded in perturbative (massive) type IIA (like those coming from M-theory at an $E_{6,7,8}$ singularity). It is believed that the recent F-theory construction of [23] provides a full classification of six-dimensional $(1,0)$ (hence also $(2,0)$) theories.

However, a detailed explanation of these topics would take us to far, and we will have little to say about them in this thesis. For more details see [22].

### Ordering constraint on D8's

As we know from our discussion in section 3.2.3, the way we group the flavor branes in a stack is summarized by a partition $\mu$ of the number of color branes ending on them. The partition gives the dimensions of the various blocks of the Jordan canonical form of the nilpotent element $\mu_{\mathbb{C}}$. Each summand of the partition (i.e. the dimension of the corresponding Jordan block) gives the number of D6's ending on a certain D8.

Suppose we want to generalize a bit the configuration of figure 3.11 by inserting many D8-stacks crossing various finite D6 segments (suspended between consecutive NS5-branes). The number of D6-branes in the $i$-th finite segment is $n_i$ while $f_i$ is the number of D8-branes in the $i$-th stack (crossing the $i$-th D6 segment). The latter contribute $f_i$ flavor hypermultiplets in the fundamental $\mathbf{n}_i$, and raise the Romans mass by $f_i$ integer units. We need to impose gauge anomaly cancellation in the field theory, which translates into a generalization of formula (3.37). At the $i$-th NS5-brane we must have:

$$2n_i = n_{i-1} + n_{i+1} + f_i \ , \tag{3.39}$$

with $n_{i+1} - n_i$ being the net number of D6's ending on it. Clearly, the numbers $f_i$ can be obtained from the data specified by the partition, as we will see later on in section 3.4.

The simplest example is when every D6 ends on a different D8-brane. Consider for instance the configuration of figure 3.9, with $k$ semi-infinite D6's to the right and to the left of the rightmost and leftmost NS5-brane respectively. We know that, via a Hanany-Witten move, we can trade the semi-infinite D6's for finite D6's by having each of them end on a different D8-brane. This comes "at no cost"; it does not correspond to a Higgs branch deformation of the field theory. In field theory words, we are not activating any operator vev. Since a single D6 ends on a D8, both of the partitions (one for each D8 stack) will be given by:

$$\mu_{\mathrm{L,R}} = (\underbrace{1,\ldots,1}_{k}) \equiv \left(1^k\right) \ . \tag{3.40}$$

We have $k$ one-box rows, one for each D6 ending on a D8; equivalently, a column with depth $k$. See figure 3.12.

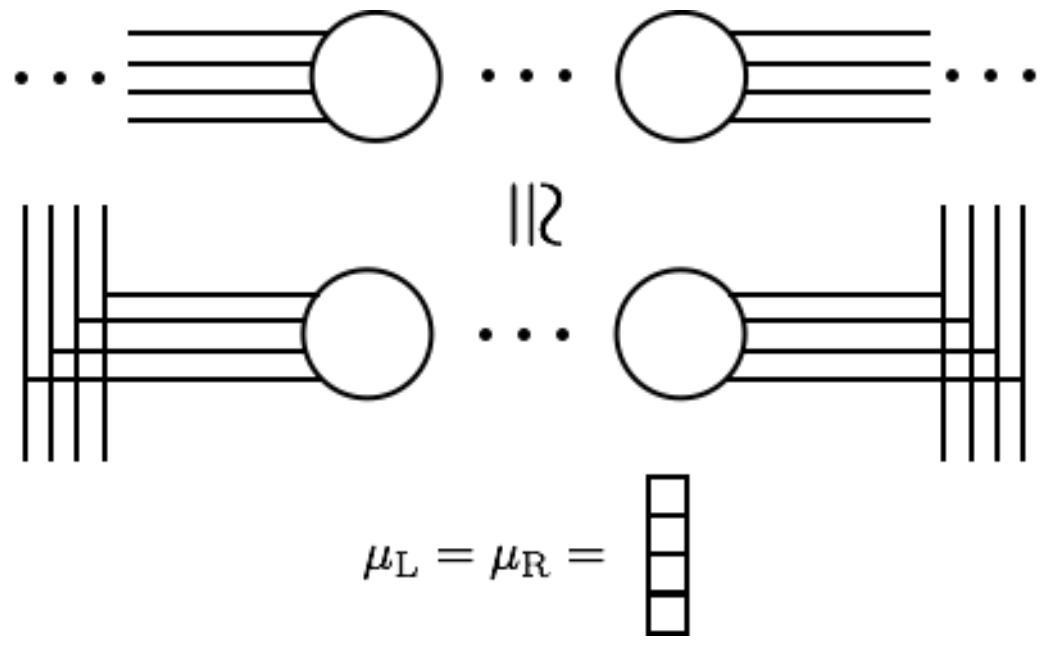

**Figure 3.12.:** We see the two brane configurations related by Hanany–Witten moves. On the top, semi-infinite D6-branes; on the bottom, finite D6's ending each on a different D8-brane. We also see the corresponding left and right partitions, given by (3.40) for $k = 4$.

Now, by activating Higgs branch deformations we can reach a configuration where all the D6's in the segment end on the same D8; see figure 3.13. In this case the partition is given by:

$$\mu_{\mathrm{L,R}} = (k) \ . \tag{3.41}$$

We have a single row with length $k$: $k$ D6's end on one D8. We will call this situation "full pole", following the nomenclature of [22].

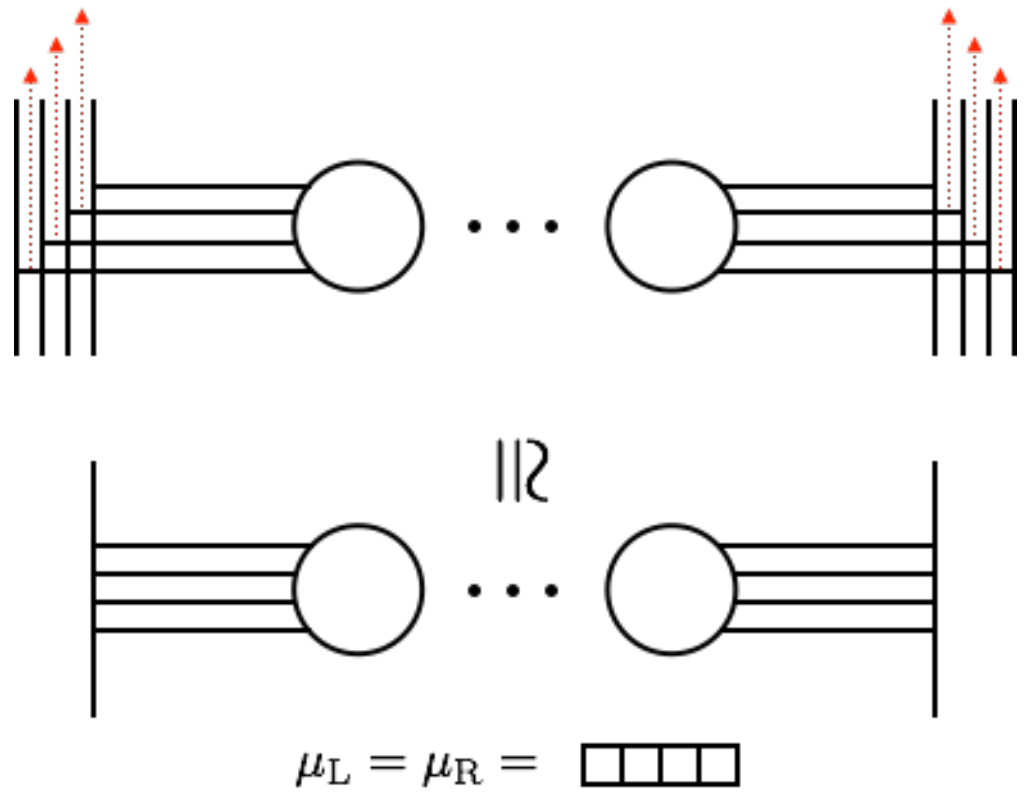

**Figure 3.13.:** To activate the Higgs branch deformation we give a vev to the hypermultiplet scalars parameterizing the position of the D6 along the $7, 8, 9$ plane. This way, we can slide the D6 segment trapped between consecutive D8's to infinity along the D8 worldvolume. We also see the left and right partitions, given by (3.41) for $k = 4$.

In the second configuration, the fuzzy sphere parameterized by the $X^i(x^6)$ is given by:

$$X^i(x^6)\,X^i(x^6) = \frac{\sum_i t^i t^i}{(x^6 - x^6_{\text{D8}})^2} = \frac{k(k-1)}{(x^6 - x^6_{\text{D8}})^2}\,1_2\ , \tag{3.42}$$

with radius

$$r(x^6) \underset{x^6 \to x^6_{\text{D8}}}{\sim} \frac{k}{x^6 - x^6_{\text{D8}}}\ , \tag{3.43}$$

which blows up at the location of the D8.

Notice that for $k = 1$, the Casimir of the $k$-dimensional irreducible representation of $\mathfrak{su}(2)$ vanishes: A single D6 ending on a D8 does not open up in a fuzzy funnel (its radius is zero, as can be seen by plugging $k = 1$ into (3.42)). This is in accordance with the fact the two configurations in figure 3.12 are connected by a Hanany–Witten move, which cannot keep track of such nontrivial effects. A single D8 imposes simple Dirichlet boundary conditions for the SYM gauge field on the D6 worldvolume, and Neumann for the scalars $X^i$.

*On the contrary, when $k > 1$ D6's end on a D8 we do have a nontrivial solution to Nahm's equation, with a pole at the location of the latter.* Therefore, for generic configurations, described by the partition $\mu$, we expect Nahm pole boundary conditions for the $X^i$, and the fuzzy sphere will have radius $\frac{\mu_i}{x^6 - x^6_{\text{D8}}}$.

Another way of describing the brane system is to say that we have fuzzy funnels protruding out of the leftmost (or rightmost) NS5-brane, to its left (or right). The $i$-th funnel represents the $i$-th D8-brane with $\mu_i$ D6-brane charge smeared on its worldvolume. Funnels with less D6-brane charge (a smaller summand in the partition $\mu$) will fit into larger funnels with more D6-brane charge (the ordering constraint requires that the partition be ordered). Moreover, as will be apparent from the supergravity solutions dual to the NS5-D6-D8 brane systems, funnels with less D6 charge will be pushed farther away from the central NS5-brane stack.

### 3.3.4. O8

#### O-planes

As is well known [63], on the worldvolume of $N_\text{c}$ coincident D$p$-branes a $\text{U}(N_\text{c})$ gauge group is realized.[20] To obtain the other classical groups one needs to place the stack "on top of a spacetime singularity", for instance that induced

[20] For $p = 6$, the center-of-mass U(1) will be decoupled from the low-energy spectrum due to an effect discussed in section 3.4. Hence we will be left with $\text{SU}(N_\text{c})$ gauge factors.

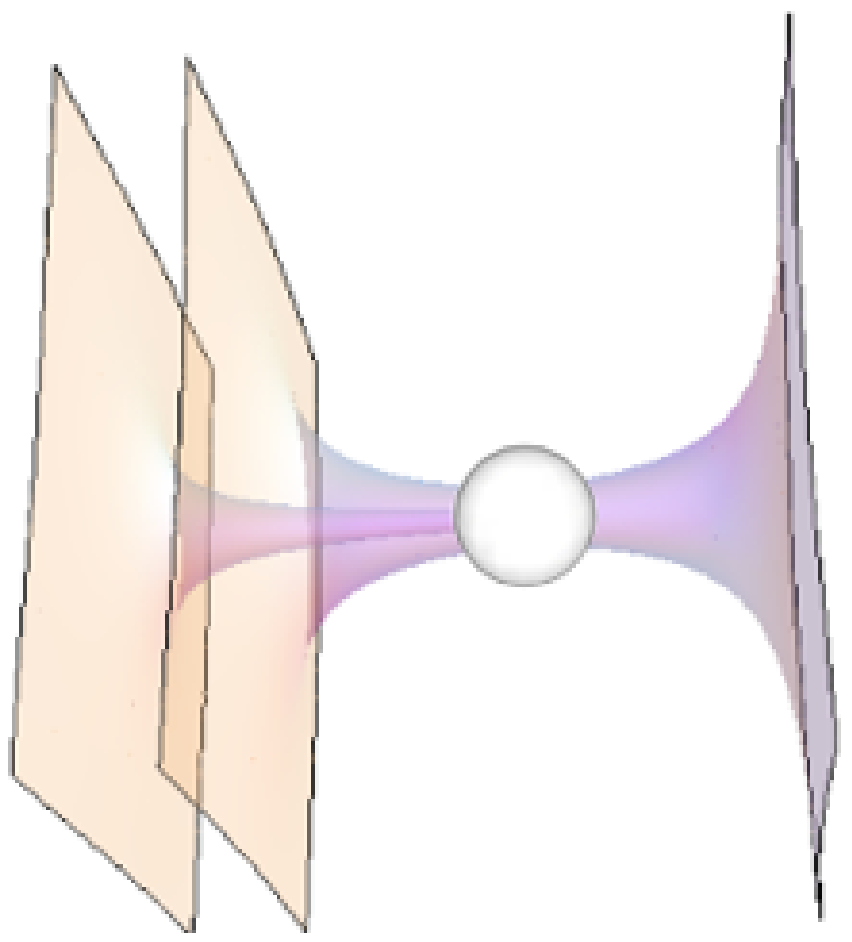

**Figure 3.14.:** An artist's impression of the D6-D8 fuzzy funnels protruding out of the central NS5-brane stack (when the NS5's are put on top of each other). Figure taken from [24]. A funnel created by less D6-branes (say $\mu_i$ of them) will fit into one created by more (say $\mu_j$, with $\mu_j \geq \mu_i$ for $i < j$). This reflects the ordering constraint imposed on the flavor D8-branes. A similar effect takes place in the dual supergravity solutions, see section 5.1.

by a $\mathbb{Z}_2$ *orbifold* acting on, say, $r$ out of ten spacetime directions $x^i$ as follows:

$$\sigma : x^i \mapsto -x^i \ , \quad i = 0, \ldots, r < 10 \ . \tag{3.44}$$

One can then combine this action with the so-called *worldsheet parity* reversal $\Omega$ (reversing the orientation of an open string and thus exchanging left- with right-movers). The fixed locus of such a combined action is called O$p$-plane, and is a $p+1 = (10-r)$-dimensional object.[21] Since it does not carry any new dynamical degrees of freedom (hence no field theory of massless excitations, not even in the limit $l_{\text{s}} \to 0$), it should not be seen as a physical object, but rather as a special extended locus in spacetime [38]. Nevertheless, such loci carry tension and charge. An O$p$-plane has opposite tension with respect to a

[21] In this very brief introduction to O-planes we have neglected details about the so-called GSO projection in a theory of superstrings, which are not crucial to our discussion. Moreover, the two actions $\Omega$ and $\sigma$ can also be combined with other discrete symmetries of the type IIA/B background, typically depending on the "internal compactification space", yielding more complicated actions on the type II spectrum. To be more concrete, in perturbative type IIB string theory the left-moving fermion number operator $(-1)^{F_{\text{L}}}$ provides another $\mathbb{Z}_2$ symmetry, which can be combined with $\Omega$ and $\sigma$ into $(-1)^{F_{\text{L}}}\Omega\sigma$. Its fixed locus is given by an O9, O7, O5 or O3-plane, according to the number of coordinates reflected by $\sigma$: zero, two, four or six respectively.

D$p$-brane and $2^{p-4}$ units of D$p$-brane charge [35]:[22]

$$q_{\mathrm{O}p^{\pm}} = \pm 2^{p-4} q_{\mathrm{D}p} \ . \tag{3.45}$$

As we can see, there exist both O-planes with positive and negative charge. We will call them O$p^{\pm}$ respectively.

When a stack of color branes is placed on top of an O$^{\mp}$-plane (together with its $\mathbb{Z}_2$ mirror) the combined $\Omega\sigma$ operator acts on the Chan–Paton indices of open strings connecting one brane to another, leading to an SO$(2N_{\mathrm{c}})$, respectively USp$(2N_{\mathrm{c}})$, gauge group realized on the common worldvolume.

## O8

In the present case we are interested in $p = 8$, and we will place flavor D8-branes on top of O8-planes. This way, not only the flavor group realized on the common D8 worldvolume is projected to either SO or USp, but also the gauge group realized on the D6 segment crossed by the D8-O8 system will suffer from such projections. This is due to the open strings connecting the D6's to their mirror D6's [19, 29].

In fact, each type of brane will get a mirror image under the $\mathbb{Z}_2$ orbifold action $\sigma : x^6 \to -x^6$ along the 6 direction ($r = 1$ so that $p + 1 = 9$, which is appropriate for an O8-plane in type IIA).

More precisely, an O8$^-$ (carrying $-16$ units of D8 charge) projects the SU flavor group on the D8 worldvolume to SO, the gauge SU on the D6 segment to USp; an O8$^+$ (carrying $+16$ units of D8 charge) does the opposite: flavor USp on the D8 worldvolume and gauge SO on the D6 one.

Following [19], we have to analyze a few possible configurations. Consider first the case in which an O8$^{\mp}$ is inserted between two consecutive NS5-branes.

If the O8$^{\mp}$ sits at the origin of the 6 direction, the rightmost NS5 will sit at $x^6_{\mathrm{NS5}} > 0$ and its image at $-x^6_{\mathrm{NS5}}$. A stack of $k$ D6-branes will connect the NS5 and its image, and will stretch along the interval $\left[0, x^6_{\mathrm{NS5}}\right]$, while its image will stretch from 0 to $-x^6_{\mathrm{NS5}}$. The gauge group is projected down to USp$(2k)$ (SO$(2k)$). We also have a single $(1,0)$ tensor multiplet and a decoupled hypermultiplet, both coming from the NS5. If we introduce a stack of D8's between the O8 and the NS5 to the right we also get fundamental hypermultiplets. More generally, we can introduce many NS5-branes sitting at $x^6_i > x^6_{\mathrm{NS5}}$ and connected by $k'$ D6 branes. The segments stretching along the intervals $\left[x^6_i, x^6_{i+1}\right]$ (with $x^6_{i+1} > x^6_i \ \forall i$) are not crossed by any O-plane, so no projection on the Chan–Paton indices takes place and the gauge groups are identified with their images supported on the "mirror" intervals $\left[-x^6_{i+1}, -x^6_i\right]$ (so that we get as many gauge factors as half the total number of D6 segments not crossed by the O8). This way we get vector multiplets and bifundamental hypermultiplets from strings going from one segment to itself and to the next, respectively. The total gauge group is given by:

$$\text{O8}^- \quad : \quad \text{USp}(2k) \times \prod_i \text{SU}(2k + 8i) \ ; \tag{3.46a}$$

$$\text{O8}^+ \quad : \quad \text{SO}(2k) \times \prod_i \text{SU}(2k - 8i) \ . \tag{3.46b}$$

[22] In conventions in which a D-brane and its $\mathbb{Z}_2$ mirror "on the other side of the O-plane" count as separate physical objects.

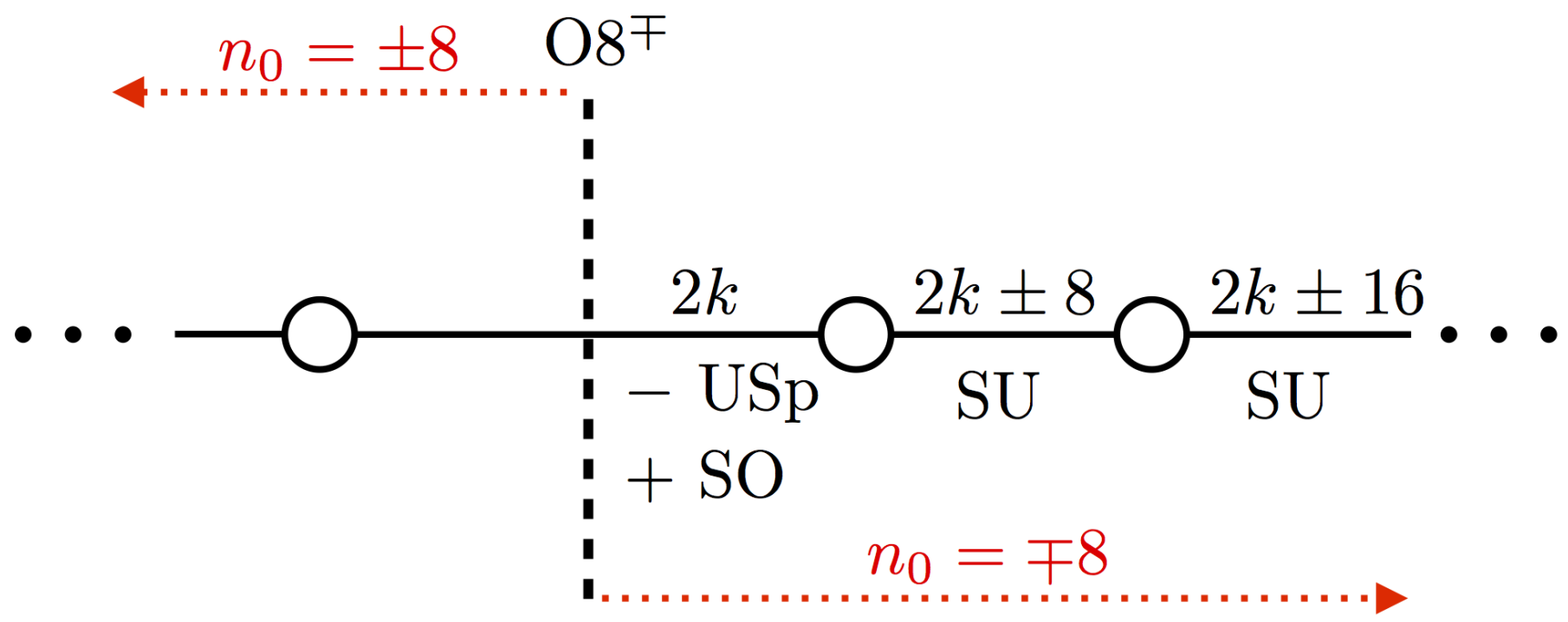

**Figure 3.15.:** An O8$^\mp$-plane crosses a D6 segment with $k$ branes, whose degrees of freedom suffer from the orientifold projection. All other branes (NS5's and finite D6 segments) on the two sides of the O-plane are identified. The first NS5 is located at $x^6_{\text{NS5}} > 0$ along the 6 direction. The O8 is located at $x^6 = 0$.

Let us see how the increase in the ranks of the SU gauge factors comes about, $k' = 2k \pm 8i$ for $i \geq 1$. Given the presence of the middle O8, the values of the Romans mass in the two regions (before and after it) – that we can think of as being generated by far away D8's – must be equal and opposite. Let us choose $n_0 = \pm 8$ before the O8$^\mp$, so that $n_0' = n_0 \mp 16 = \mp 8$. Starting from the central segment (the one crossed by the O-plane) of $k$ image D6-brane pairs, imposing (3.35) we get that the number $k'$ of D6's in the next segment to the right will be:

$$2k - k' = \mp 8 \Leftrightarrow k' = 2k \pm 8 \ ; \quad k' - k'' = \mp 8 \Leftrightarrow k'' = k' \pm 8 = 2k \pm 2 \cdot 8 \ , \tag{3.47}$$

and so on. This explains (3.46). The configuration is summarized in figure 3.15. One has to keep in mind that these relations will be modified in the obvious way if one includes stacks of D8's between the $i$-th and $i+1$-th NS5.

Next is the case in which $x^6_{\text{NS5}} = 0$, that is the NS5 is stuck at the O8$^{\mp}$. See figure 3.16.

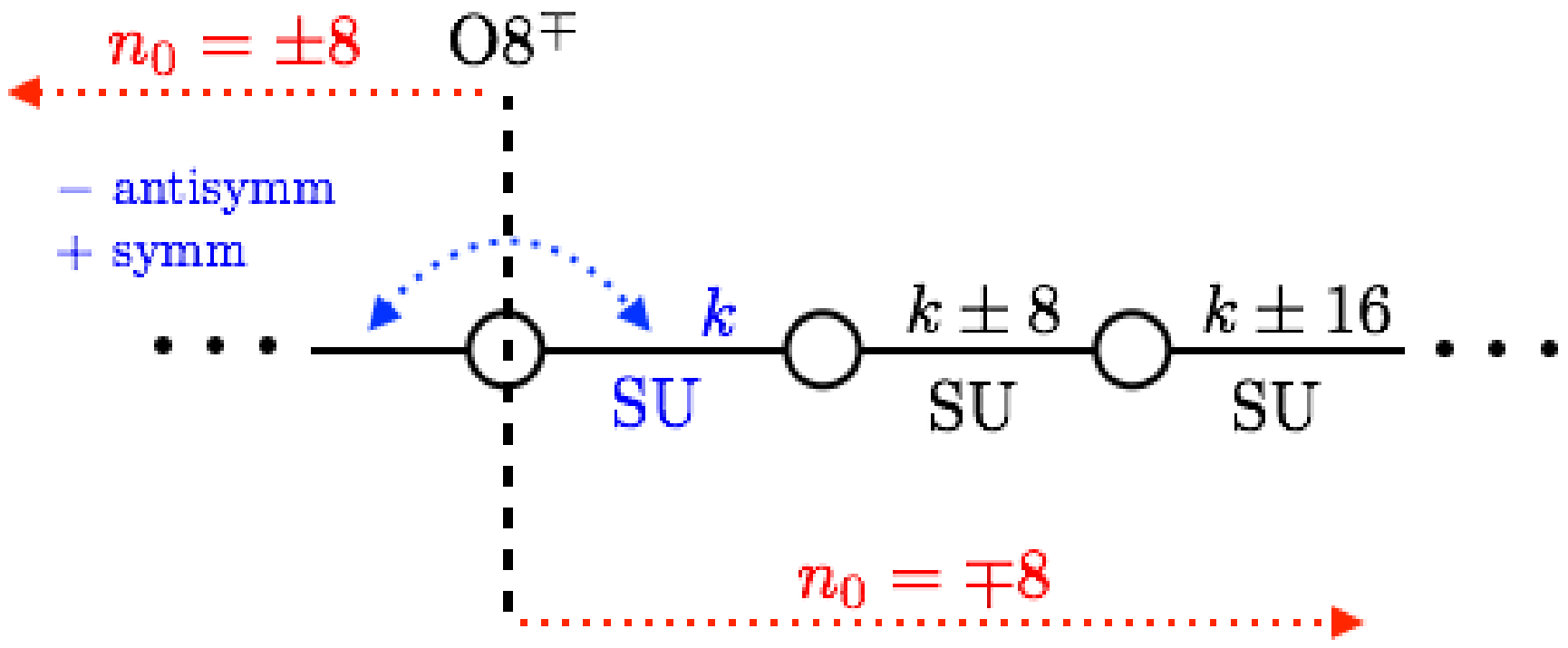

**Figure 3.16.:** An O8$^{\mp}$-plane with an NS5-brane stuck onto it; both are located at $x^6 = 0$. All other branes (NS5's and finite D6 segments) on the two sides of the O-plane are identified.

Consider suspending $k$ D6 between the former and the next NS5 at $x^6_1 > 0$. First of all, by construction there is only one dynamical tensor multiplet (the one associated with the NS5 to the right of the origin, identified with its image at $-x^6_1$). The orientifold action identifies the suspended D6's to the right with those to the left of the stuck NS5, so no projection on the Chan–Paton indices of strings going from the former D6's to the latter takes place (that is, we get an SU($k$) gauge factor). However, the "bifundamental" hypermultiplet these strings contribute is now in the antisymmetric (symmetric) representation of SU($k$). A computation along the lines of (3.47) gives:

$$\text{O8}^{\mp} \ : \quad \prod_i \text{SU}(k \pm 8i) \ . \tag{3.48}$$

Strings going from one D6 segment to the next will contribute bifundamental hypermultiplets.

One has to worry about the Nahm pole data coming from the D6-D8 bound states, given that the O8-plane at the origin of $x^6$ identifies all the branes to its right with those to its left. In particular, for each stack of D8's crossing the $i$-th D6 segment, in a way specified by the Young tableau $\mu_i$, we need to have the same number of D8's crossing the mirror segment on the other side of the O-plane. We then identify the partition $\mu_i$ at $x^6_{\text{D8}}$ with its mirror at $-x^6_{\text{D8}}$, and finally mod out by the action that swaps them [22].

### 3.3.5. O6

We can also add $O6^{\pm}$-planes parallel to the D6-branes. According to (3.45), each of them carries $\pm 4$ units of D6-brane charge, and projects the SU gauge factor realized on the D6 worldvolume down to USp (SO), by acting on the Chan–Paton indices of strings going from the D6 stack to itself. We also need to keep in mind an effect first discussed in [64]: When an O6 crosses the type IIA NS5-brane, it changes its sign (i.e. its D6 charge gets reversed).

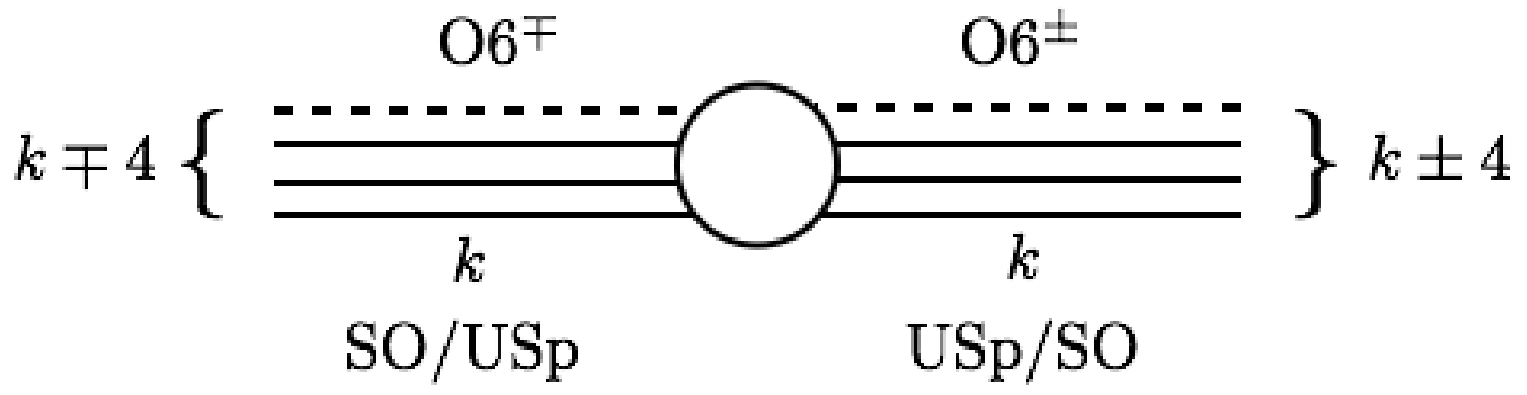

**Figure 3.17.:** An NS5-brane with $k$ semi-infinite D6's ending on it from the left and right. An $O6^{\mp}$-plane has been placed on top of the left segment, while an $O6^{\pm}$-plane has been placed on top of the right one. The charge of the O6 flips sign when crossing an NS5-brane, according to an effect first discussed in [64]. The number of D6-branes in the right stack is effectively increased (decreased) by 4 units, while the number of D6-branes in the left one is effectively decreased (increased) by 4 units.

Suppose we have an NS5-brane on which two semi-infinite D6 segments end, from the left and right. An O6-plane has been put on top of each segment. See 3.17. Recalling (3.21), the no-bending condition (with vanishing Romans mass) gives:

$$(N_{\mathrm{D6,L}} \mp 4) - (N_{\mathrm{D6,R}} \pm 4) = 0 \Leftrightarrow N_{\mathrm{D6,L}} - N_{\mathrm{D6,R}} = \pm 8 \;, \tag{3.49}$$

which is similar to (3.35), albeit with an "effective Romans mass" not generated by D8-branes.

If we consider a string of NS5's with suspended D6's in between, and an O6 on top of each D6 segment, the total gauge group will be a product of USp and SO gauge groups, whose ranks have to satisfy condition (3.49) evaluated at each NS5-brane. Schematically:

$$\ldots \times \mathrm{SO}(2k) \times \mathrm{USp}(2k-8) \times \mathrm{SO}(2k) \times \mathrm{USp}(2k-8) \times \mathrm{SO}(2k) \times \ldots \;. \tag{3.50}$$

We have bifundamental hypermultiplets from strings going from one segment to the next, as many tensor multiplets as gauge factors, possible fundamental hypermultiplets from semi-infinite D6 segments to the far right and far left of the above string. See figure 3.18.

We can also add an $O8^{\pm}$-plane to the above configuration. The only constraint is that we cannot have an NS5 stuck at the O8, since the O6 to its left would have opposite sign with respect to that to its right, whereas they should be identified. Let us put the O8 at the origin of the 6 direction, the NS5's at $x_i^6 > 0$, connected by $k$ D6 segments on top of which we lay an O6$^{\mp}$ (giving rise to $k$ pairs of image branes, rather than $k$ branes). Notice that to have a nonempty gauge group the O8 and the "middle" O6 should have opposite charge. Then the middle gauge factor will be SU($k$) (due to the common action of O6$^{\pm}$ and O8$^{\mp}$), and the next factors will be either SO or USp, with ranks obtained by applying the no-bending condition in presence of a nonvanishing Romans mass, (3.35), and with the number of D6's to the right and left of an NS5 effectively increased (decreased) by $\pm 4$.

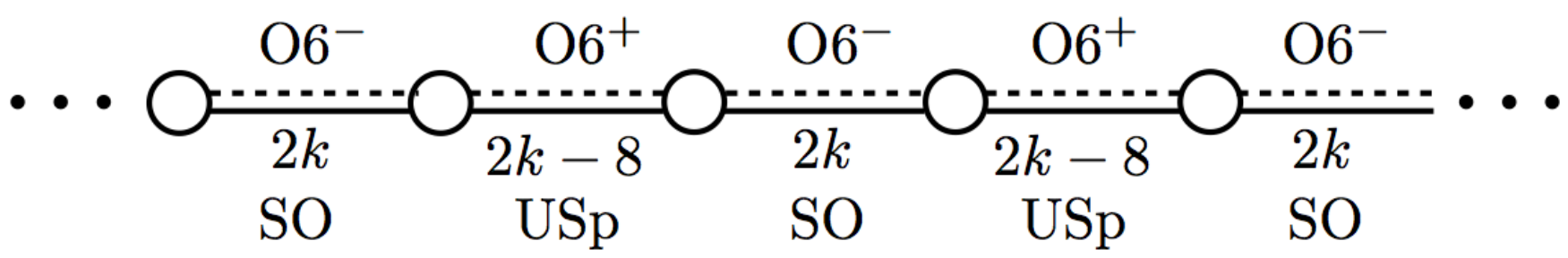

**Figure 3.18.:** A configuration of suspended D6-branes and many NS5-branes. O6-planes are put on top of each finite segment; their signs and the effective numbers of D6-branes in each finite segment – $2k$, $2k-8$ – respect the conditions of figure 3.17 and (3.49).

### M-theory lift

We will closlely follow the discussion in [22]. In section 3.3.2 we explained how to lift a system of $k$ D6-branes suspended between $N$ consecutive NS5-branes: We get $N$ M5-branes probing the $A_{k-1}$-type ALF space. The obvious generalization involves having the M5's probe a $D_k$-type ALF space, for a suitable $k$ [65, 66]. Intuitively, the O6 action should lift to the extra generator of the $D$ discrete group (which is not present in $A$). When all the $k$ centers of the ALF space coincide, we get the $D_k$ singularity (the ALE space), which corresponds to an SO($2k$) gauge group. We would thus be led to conclude that, in M-theory, $N$ M5's probing the $D_k$ ALF space give rise to $N$ SO gauge groups: Every "segment" of ALF space connecting two consecutive M5's supports one such factor, and the branes act as "domain walls" among them (contributing tensor multiplets). However, for odd $N$ this picture cannot be right, as it contrasts with what we would get in IIA. Due to the subtlety noted above regarding the sign of the O6 (when crossing an NS5), for odd $N$ the last gauge factor would be USp rather than SO. Turning the argument around, we conclude that there must be an even number of NS5's, $2N$, lifting to $2N$ "$\frac{1}{2}$-M5-branes" (one for each kind of domain wall connecting two consecutive gauge factors, USp $\times$ SO or SO × USp). The phenomenon of fractionalization of M5-branes probing a $D_k$ singularity has been observed in [67], and further generalized to the $E_k$ cases in [22].

## 3.4. Constructing linear quiver theories

In the previous discussion we have casually (and intentionally) hinted at the possibility to add several NS5-branes, connected by many D6 segments, possibly crossed by stacks of infinitely extended D8-branes. We also know that, as long as the NS5's are kept separated (that is, giving a vev to the scalars in the tensor multiplets), the instanton strings coming from D2's parallel to the D6's and stretching from one NS5 to the next will carry a finite tension. This means that the much feared tensionless strings in this phase are actually massive, and decouple from the massless spectrum. Thus they will not contribute to the low energy physics. The six-dimensional $(1,0)$ superconformal theory is said to be on its tensor branch, and we can give it an *effective linear quiver description*. This means that, away from the origin of the tensor branch, where the interacting fixed point of the RG flow is located, we can effectively describe the physics by means of a gauge theory, whose fields are summarized in a graphic bookkeeping device known as *quiver*.[23]

A quiver is a collection of nodes and arrows (as the name suggests), and it proves very useful in describing the matter content of a supersymmetric field theory (possibly with superpotential). The use of this language in physics is by now a venerable subject, and we will not attempt to give a full introduction here. Rather, we will just point out the main differences between the many (perhaps more common) lower-dimensional examples present in the literature and the six-dimensional theories we are really interested in.

Having introduced all possible ingredients of six-dimensional Hanany–Witten setups, semi-infinite D6's, O6's, D8's, O8's – besides the obvious NS5-D6 building block – we are finally ready to describe the ensuing $(1,0)$ theories on the tensor branch as linear quivers. Since we have been careful in applying the no-bending condition at each NS5-branes, the theories will be automatically anomaly-free (namely, $N_\mathrm{f} = 2N_\mathrm{c}$ for each gauge group) [19].

Consider a stack of $N$ NS5's and separate the branes in the stack fully, by introducing finite D6 segments suspended between two consecutive NS5's. To the right (left) of the rightmost (leftmost) NS5-brane we let D6-branes extend to infinity; see figure 3.9. By a Hanany–Witten move, the latter can be replaced by a stack of finite D6's, where each of the D6-branes ends on a different D8-brane; see figure 3.12. By activating a Higgs branch deformation (see section 3.4.1) we can slide to infinity trapped D6-segments (see figure 3.13). We will end up with D8's crossing various D6 segments towards the left and right tails of the NS5 chain; the "way" they intersect the D6's is dictated by an ordered partition of the number of (formerly) semi-infinite D6's. We have one partition for each tail. We can also add O8-planes in the D8 stacks, as well as O6-planes on top of each D6 segment, keeping in mind the sign flipping condition for the latter discussed in section 3.3.1. The presence of D8 charge changes the no-bending condition in a way dictated by (3.37). This affects the number of color D6-branes in each segment. This number is also affected by the O6's, in a way dictated by (3.49).

[23] In fact we can even do more, and write down a classical pseudo-Lagrangian [32], by specializing the “tensor hierarchy” actions of [4, 5].

All in all, we get the field content summarized in table 3.6. All multiplets are in massless representations of the $\mathcal{N} = (1,0)$ superconformal algebra. We use the notation of [31].

| multiplet | field content | quantity | rep. | strings |
|---|---|---|---|---|
| vector | $(A_\mu, \lambda_\alpha, D)_i$ | $i = 1, \dots, N-1$ | adjoint of $\mathrm{U}(r_i)$ | $6_i$-$6_i$ |
| hyper | $(h, \psi_{\dot\alpha})_i$ | $i = 1, \dots, N-2$ | bifund. $\overline{\mathbf{r}_i} \otimes \mathbf{r}_{i+1}$ | $6_i$-$6_{i+1}$ |
| hyper | $(\tilde{h}^{a_i}, \tilde{\psi}^{a_i}_{\dot\alpha})_i$ | $a_i = 1, \dots, f_i$<br>$\forall i = 1, \dots, N-1$ | fund. $\mathbf{r}_i$ | $8_{a_i}$-$6_i$ |
| tensor | $(\Phi, \chi_\alpha, B_{\mu\nu})_i$ | $i = 1, \dots, N$ | singlet | NS5 |
| linear | $((\boldsymbol{\pi}, C), \xi_{\dot\alpha})_i$ | $i = 1, \dots, N$ | singlet | NS5 |

**Table 3.6.:** *$D$ is an auxiliary field; $\lambda$ is a right Weyl spinor. $h$ and $\tilde{h}$ each represent four real scalars; $\psi$ ($\tilde{\psi}$) is a left Weyl spinor. $\Phi$ is one real scalar; $\chi$ is one left Majorana spinor; $B$ is a two-form with self-dual field strength: $H = *_6 H$. $\boldsymbol{\pi}$ is an $\mathrm{SU}(2)_R$ triplet of scalars; $C$ is an $\mathrm{SU}(2)_R$ singlet periodic scalar; $\xi$ is a left Weyl spinor. In the last column we have indicated the extended object(s) contributing the corresponding multiplet. The number $f_i$ of flavor hypermultiplets for the $i$-th gauge group is given by (3.55).*

### Massive U(1)'s

In the previous sections we have intentionally kept saying that the various gauge factors realized on the D6 segments are SU groups. This is strictly incorrect. The worldvolume gauge symmetry would be U. However there is a subtlety, that we will explain momentarily, due to which the center-of-mass U(1) in each U gauge factor becomes massive, thus decoupling from the low energy physics we are truly after. We will closely follow the discussion in [19, 31, 32].

The position $(x^7_i, x^8_i, x^9_i)$ of the $i$-th NS5 along the 7, 8, 9 plane is parameterized by the triplet $\boldsymbol{\pi}_i$, while its position $x_i^{10}$ along the M-theory circle is given by the periodic scalar $C_i$ (in absence of D8-branes). (Remember that the $\mathrm{SU}(2) \cong \mathrm{Sp}(1)$ R-symmetry of a $(1,0)$ six-dimensional theory engineered by one of our Hanany–Witten setups is realized by rotations in the $7,8,9$ plane.) The former acts as a Fayet–Iliopoulos term for the center-of-mass $\mathrm{U}(1)$ in the $i$-th U gauge factor (realized on the D6 segment suspended between the $i$-th and $i+1$-th NS5):

$$\boldsymbol{\pi}_i = \frac{\mathbf{x}_{i+1} - \mathbf{x}_i}{l_\mathrm{s}^3} \ , \quad \mathbf{x}_i \equiv (x_i^7, x_i^8, x_i^9) \ . \tag{3.51}$$

The latter acts as a $\theta$ angle in six dimensions, giving the coefficient of $\mathrm{Tr}(F_i \wedge F_i \wedge F_i)$ in the bosonic action ($F_i$ being the $\mathrm{U}(r_i)$ gauge field strength):

$$C_i = \frac{x_{i+1}^{10} - x_i^{10}}{l_\mathrm{s}^3} \ . \tag{3.52}$$

Both terms are given by dynamical fields that the NS5's contribute.

Now, the center-of-mass $\mathrm{U}(1)$'s suffer from an anomaly [68, 69, 70], believed to be canceled by a Green–Schwarz–West–Sagnotti mechanism which involves $C_i$ and renders them massive via a Stückelberg term. So, the $\mathrm{U}(1)$'s become nonanomalous and massive in one fell swoop. This justifies using $\mathrm{SU}(r_i)$ gauge factors rather than $\mathrm{U}(r_i)$.

### Constructing the quiver

Let us focus on the case without O-planes. The quiver diagram describing the theory on its tensor branch, is constructed as follows.

- For each gauge group $\mathrm{SU}(r_i)$ coming from a finite D6 segment we add a round node labeled by its rank $r_i$; this takes care of the vector multiplets.
- We connect two nodes by a link representing a bifundamental hypermultiplet $(h, \psi)$ and a tensor multiplet $(\Phi, \chi, B)$. The left- and rightmost NS5's also contribute one tensor multiplet each (which is not represented by any link).
- For each flavor group $\mathrm{SU}(f_i)$ coming from a D8 stack (crossing a D6 segment) we add a square node labeled by the number $f_i$ of hypermultiplets in the fundamental $\mathbf{r}_i$ of the gauge group $\mathrm{SU}(r_i)$.
- Gauge anomaly cancellation in six dimensions dictates that $N_\mathrm{f} = 2N_\mathrm{c}$. This constraint is implemented by requiring that (3.39) be satisfied at each node:

$$2r_i - r_{i+1} - r_{i-1} = f_i \ . \tag{3.53}$$

This is a direct generalization of (3.37) to the case of multiple D8 (sub-)stacks.

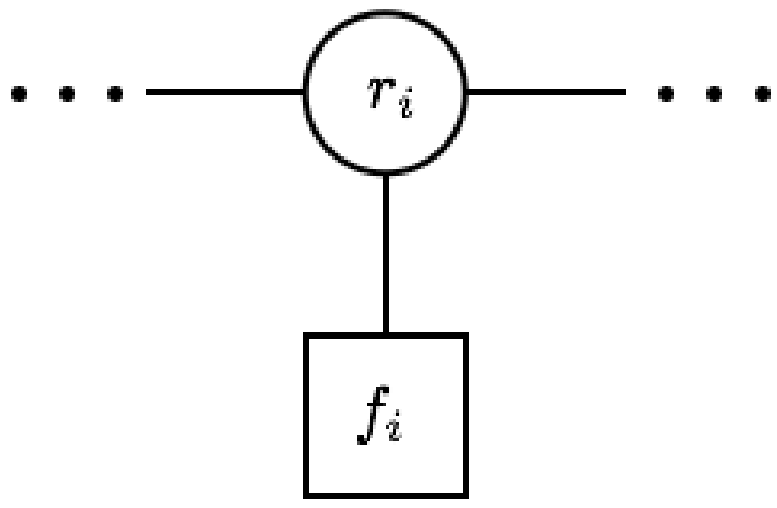

**Figure 3.19.:** We depict the basic node of a six-dimensional linear quiver gauge theory.

- As we will see in section 3.4.1, starting with one D8 stack to the far left (and one to the far right), we can parameterize the ways the D8's will be organized in sub-stacks crossing the D6 segments towards the tails of the quiver by specifying the left and right partition, $\mu_{\rm L}$ and $\mu_{\rm R}$.

- From the (transpose) partitions $\mu^t_{\rm L}$ and $\mu^t_{\rm R}$ we can directly read off the ranks of the various SU groups realized on finite D6 segments crossed by D8's:
$$\mu^t_i = r_i - r_{i-1} \quad \text{with } i \geq 1 \quad \text{and} \quad r_0 \equiv 0 \ . \tag{3.54}$$
The fact that $\mu^t$ is an ordered partition guarantees that $r_i \geq r_{i-1}$. See figure 3.6. This automatically implements the ordering constraint depicted in figure 3.7 (for the case of D3-D5 branes). The $i$-th "crossed" segment comes before the $i-1$-th going from right to left (that is from the central region where all the NS5's are initially placed), and has higher gauge group rank. Recalling (3.53) we also have:
$$f_i = 2r_i - r_{i+1} - r_{i-1} = (r_i - r_{i-1}) - (r_{i+1} - r_i) = -(\mu^t_{i+1} - \mu^t_i) \geq 0 \ . \tag{3.55}$$
Hence the partition $\mu^t$ (left or right) gives a discrete second derivative of the number of colors as the number $f_i$ of flavors. Each summand in the partition, i.e. the number of boxes in each row, gives the number of D6's ending on a certain D8 in the left (or right) flavor stack. The difference between two summands of the transpose partition (or Young tableau) $\mu^t$ gives the change of slope of a piecewise linear function (namely, $i \mapsto r_i$). Astonishingly, in [31] it was found that the graph of the latter gives a discretization of the graph of a continuum function which governs the metric of the *AdS holographic duals* of the linear quiver theories.

- The flavor Lie algebra which is left unbroken by the Nahm pole boundary conditions (3.38) is given by [58, 71]:

$$\mathfrak{s}\left(\bigoplus_i \mathfrak{u}(m_i)\right) , \tag{3.56}$$

  with $m_i$ the multiplicity of the summand $\mu_i$ in the (left or right) partition $\mu$. In general we will have many $\mathfrak{u}(1)$ factors.

The whole procedure is more easily explained by means of an example, which we now illustrate in detail.

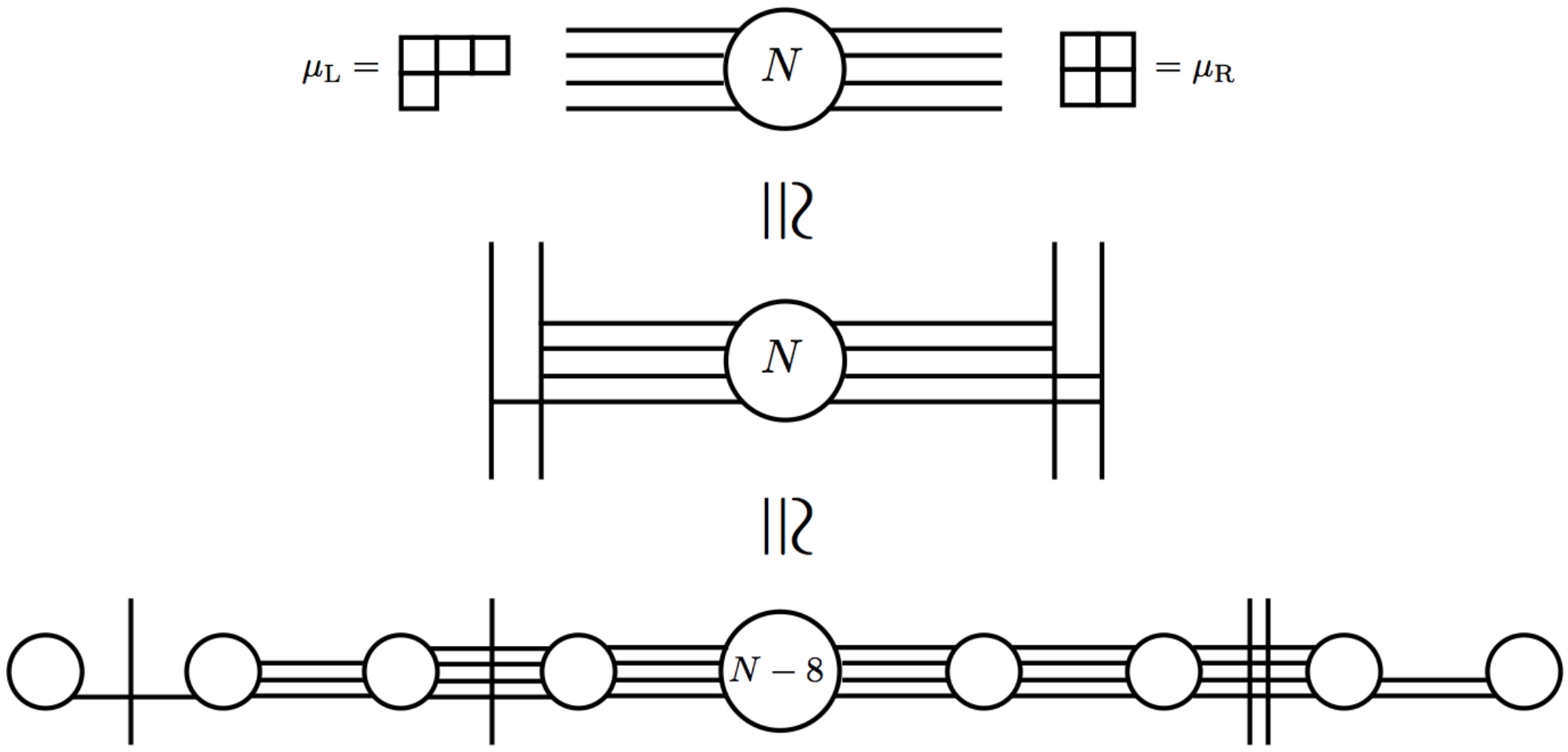

**Figure 3.20.:** Top: We enforce the boundary conditions summarized by the partitions $\mu_{\rm L}$ and $\mu_{\rm R}$ to an NS5-D6 brane system, with $N$ coinciding NS5's and 4 semi-infinite D6 segments ($N \gg 4$). The latter are replaced by the same amount of finite D6's, with each of them ending on a different D8-brane. Middle: We then modify the way we group the D8's by declaring how many D6's end on each of them. This corresponds to a Higgs branch deformation in field theory, by means of which we flow to a new theory. Note that, while moving along the Higgs branch, we stay at the origin of the tensor branch (all NS5's on top of each other). The new theory, on the left tail, has 3 D6's ending on the first D8 (a stack of 1 brane), 1 D6 ending on the second (again a stack of 1 D8-brane), thus the partition is given by $\mu_{\rm L} = (3, 1)$. On the right tail, it has 2 D6's ending on the first D8 (a stack of 1 brane), and 2 D6 ending on the second (a stack of 1 D8-brane), thus the partition is given by $\mu_{\rm R} = (2^2)$. Bottom: We perform various Hanany–Witten moves to reach this brane configuration from the middle one.

Let us see how to read off the quiver theory from both the partitions $\mu_{\rm L}$, $\mu_{\rm R}$

and the bottom configuration in figure 3.20. First, we have

$$\mu_{\rm L} = (3,1) \quad \Rightarrow \quad \mu^t_{\rm L} = (\mu^t_1, \mu^t_2, \mu^t_3) = (2,1,1) \ , \tag{3.57}$$
$$\mu_{\rm R} = (2^2) \quad \Rightarrow \quad \mu^t_{\rm R} = (\mu^t_1, \mu^t_2) = (2,2) \ . \tag{3.58}$$

By solving the set of equations (3.54) and (3.55) for both $\mu_{\rm L}$ and $\mu_{\rm R}$ we get:

$$\text{L} \ : \quad (r_1, r_2, r_3, r_{i\geq 4}) = (2,3,4,4) \ , \quad (f_1, f_2, f_3, f_{i\geq 4}) = (1,0,1,0) \ ; \tag{3.59}$$
$$\text{R} \ : \quad (r_1, r_2, r_{i\geq 3}) = (2,4,4) \ , \quad (f_1, f_2, f_{i\geq 3}) = (0,2,0) \ . \tag{3.60}$$

The unbroken flavor symmetry associated with the left and right partitions is given by:

$$\mathfrak{f}_{\rm L} = \mathfrak{s}\left(\mathfrak{u}(1) \oplus \mathfrak{u}(1)\right) \ , \quad \mathfrak{f}_{\rm R} = \mathfrak{s}\left(\mathfrak{u}(2)\right) = \mathfrak{su}(2) \ . \tag{3.61}$$

We can now construct the quiver; see figure 3.21.

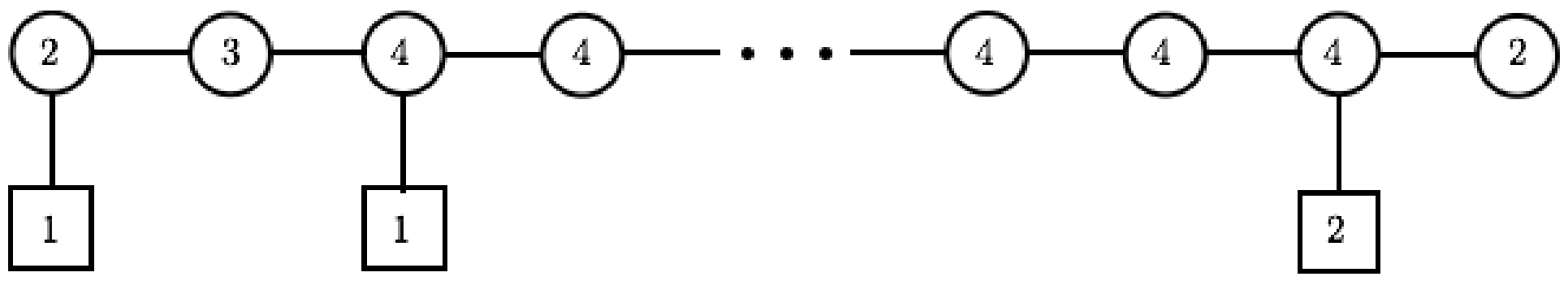

**Figure 3.21.:** A generic six-dimensional $(1,0)$ theory on its tensor branch. We show the two tails of the quiver, containing the data (3.59). The flavor symmetry factors are due to the presence of D8-branes. The ellipsis is meant to be a chain of SU(4) groups (realized on the worldvolume of $N-5$ finite D6 segments obtained by fully separating the NS5 stack with $N-4$ branes of figure 3.20, bottom image). The quiver can be read off directly from the bottom image in figure 3.20; the rightmost D8 sub-stack contains 2 branes, giving rise to the rightmost flavor group in the quiver.

### 3.4.1. Flows to interesting theories

In [24] it was argued that, starting from the Hanany–Witten setup with $k$ semi-infinite D6's sticking out of the outermost NS5's (a reduction of M5's at the $A_k$ ALF space), one can engineer more interesting superconformal theories with $(1,0)$ supersymmetry by triggering Higgs branch deformations. This argument had already been successfully applied in [30, 45]. From the field theory perspective, this means activating certain operator vev's and thus going to special loci in the Higgs branch of the moduli space of the theory. To access the Higgs branch we first go to the origin of the tensor branch.

This is easier to understand from the brane perspective. Consider first the configuration in which all the NS5's are on top of each other. We have two semi-infinite D6 stacks sticking out of the NS5 stack to the left and right. The D8's are then introduced in the system (at no cost)[24] by a Hanany–Witten move that replaces the semi-infinite D6's at the two tails with finite stacks ending on the D8's (basically by bringing in one D8 for every semi-infinite D6 from infinity). Entering the Higgs branch of the field theory requires sliding to infinity a D6 segment suspended between two D8's along the D8 worldvolume (i.e. along the $7, 8, 9$ directions). This is possible only when there are D6's "trapped" in between two (or more) D8's along the 6 direction. We thus have to modify the Dirichlet and Neumann boundary conditions imposed on the D6 fields by the D8's (see discussion in paragraph 3.3.3 and equation (3.40)). This is done by specifying two nontrivial partitions $\mu_{\rm L}$ and $\mu_{\rm R}$, one for each tail of the brane system. We then trigger Higgs deformations and reach a completely new Hanany–Witten setup. At the bottom of the flow we hope to find a new conformal field theory. We move away from the origin of the tensor branch by separating the NS5's fully, taking care of possible Hanany–Witten effects. We read off the effective gauge theory description (i.e. the quiver), and the new brane setup, precisely from the partitions $\mu_{\rm L}$ and $\mu_{\rm R}$ we started with.

The theory will be labeled by the number $N$ of NS5, $k$ of initial semi-infinite D6's, and the two partitions $\mu_{\rm L}$ and $\mu_{\rm R}$. We will call it

$$\mathcal{T}(N, k, \mu_{\rm L}, \mu_{\rm R}, G) \ . \tag{3.62}$$

In our case $G = A_k$, the SU($k$) gauge group engineered by the singularity in M-theory and descending onto the worldvolume of the $k$ D6-branes at the beginning of our procedure. It is possible to consider $G = D_{k+4}$ (see section 3.3.5) and $E_k$ [22] as well, but we will not pursue this here. Henceforth we will drop the label $G$ from (3.62).

The core of the procedure is the decoupling of the seven-dimensional SYM degrees of freedom (on the D6's), which is achieved thanks to the introduction of the D8-branes, enforcing nontrivial boundary conditions on the former.

### 3.4.2. The holographic limit

Now that the defining parameters of the $\mathcal{T}(N, k, \mu_{\rm L}, \mu_{\rm R})$ theories have been identified, and their string theory origin explained, we could try to find a *holographic limit* of the former. According to the AdS/CFT correspondence, in such a limit the physics of the field theory can be equally described by a *dual* AdS vacuum of (massive) type IIA string theory. This arises by taking the *near-*

[24]This does not trigger a flow in the field theory.

*horizon limit* of the closed string background generated by the Hanany–Witten setup that engineers the field theory.

However, implementing this near-horizon procedure requires a detailed knowledge of the string background, which is extremely hard to extract from arbitrarily complicated Hanany–Witten setups. A shortcut to solve this problem will be presented in chapter 5, where all $\mathrm{AdS}_7$ vacua of massive type IIA supergravity will be classified. There we will argue that they arise from the Hanany–Witten setups considered in this chapter. The holographic limit is then more easily found in the supergravity regime as the one that suppresses stringy ($g_\mathrm{s}$) and $\alpha'$ ($l_\mathrm{s}$) corrections that would modify the classical vacua.

In this section we will not explain how the correct holographic limit is found; we will simply point out what are its requirements on the field theory parameters.

Whereas in the more familiar instances of the AdS/CFT correspondence the holographic limit entails taking the ranks of the gauge group(s) to infinity, for the six-dimensional superconformal theories we should instead take *the number of gauge groups to infinity*. Since a gauge factor is supported on a finite D6 segment suspended between consecutive NS5-branes, this translates into taking *the number of NS5-branes to infinity*:

$$N \to \infty \ . \tag{3.63}$$

However, this limit would be too crude by itself, as it would wash away all the information associated with the partitions $\mu_{\mathrm{L,R}}$: Intuitively, it pushes back to infinity the D8 stacks (at $x^6 \to \mp\infty$ respectively) by extending the NS5 chain indefinitely.

In section 5.1 we will see that the correct limit that keeps track of that information is instead given by a refinement of (3.63), namely [31, 24]:

$$N \to \infty \ , \quad \frac{\mu_i}{N} \quad \text{fixed} \ . \tag{3.64}$$

# Chapter 4

# Ten-dimensional supergravity solutions

## 4.1. Holography

In the previous chapter we have seen that, in the $l_s \to 0$ limit, the worldvolume theory living on a D$p$-brane reduces to a usual field theory, whose field content is that of a super-Yang–Mills multiplet in $p+1$ dimensions. This is not that surprising if one remembers that D-branes are solitonic objects in string theory. As such, they should be described by collective coordinates. For a D$p$-brane in flat space, these are $9-p$ scalars (parameterizing its motion in the space transverse to its worldvolume) plus other fields, altogether making up a vector multiplet (SYM) in $p+1$ dimensions. The latter is the only multiplet with sixteen supercharges in any dimension, precisely the amount of supersymmetry preserved by a flat D-brane in $\text{Mink}_{10}$ (see (3.7) and discussion around it).

This observation can be strengthened by a string theory computation. Upon considering D-branes as fixed loci in spacetime open strings can end on, the quantized spectrum of massless excitations of an open string ending on a D-brane yields a SYM multiplet. (Massive modes will have masses of the order $M^2 \sim \alpha'^{-1}$, so that they decouple for $\alpha' \to 0$.)

Note that the worldvolume theory *does not* contain any graviton, although string theory *is* a theory of gravity. In other words, the open string sector of string theory does not see gravity.

In this chapter we are going to look at D-branes from a different perspective. Namely, as extended objects carrying tension (hence modifying the surrounding spacetime) and RR charges (hence acting like sources for the bulk RR potentials of type II string theory). Said differenty, D-branes can couple to *closed strings*. The closed string spectrum of string theory *does contain* gravity (in the form of a spin-two graviton).

This dual description of D-branes has led Maldacena [25] to propose a very interesting duality, whose details will be spelled out momentarily. The proposal, originally formulated for D3-branes in flat space and four-dimensional $\mathcal{N} = 4$ SYM, is the following:

*Type IIB string theory on* $\mathrm{AdS}_5 \times S^5$ *with* $N$ *quanta of* $F_5$ *is* equivalent *to* $\mathcal{N} = 4$ $\mathrm{U}(N)$ *SYM in four dimensions.*

This means that the (quantum) partition functions of the two theories are believed to be *equal*.

Taken at face value, this statement sounds rather surprising since the proposed *duality* relates a string theory, a theory of extended objects and gravity, to a usual (superconformal) field theory, with no gravity at all. However, once the duality is assumed to hold it becomes very helpful since it allows us to compute quantities in one theory and extract information on the other. This way of thinking is especially helpful when one of the two theories is in a regime of parameters that makes it incalculable (typically the strong-coupling regime).

The origin of the duality can be traced back to the gauge theory – brane worldvolume correspondence discussed in chapter 3. However, that correspondence allows one to match the two theories (and their parameters) only in a certain limit ($l_\mathrm{s} \to 0$). The statement entailed by the *holographic duality*, or *correspondence*, is much stronger, in that it requires that the two theories match at any value of their parameters.

Some quantities can be matched on the two sides without much effort. The match of more complicated quantities (observables) requires "precision tests" of holography, which constitute the subject of many years of recent research.

### 4.1.1. The D3 case

The prototypical example [25] of the holographic correspondence involves taking $N$ coincident, flat D3-branes that probe $\mathrm{Mink}_{10}$ in type IIB string theory. The gauge theory on the common worldvolume is just $d = 4$ $\mathcal{N} = 4$ SYM with gauge group $\mathrm{U}(N)$, and the closed string backround generated by the branes is also known [72].

The observation leading to the proposal of the *holographic*, or *AdS/CFT*, *correspondence* is neatly summarized in figure 4.1.

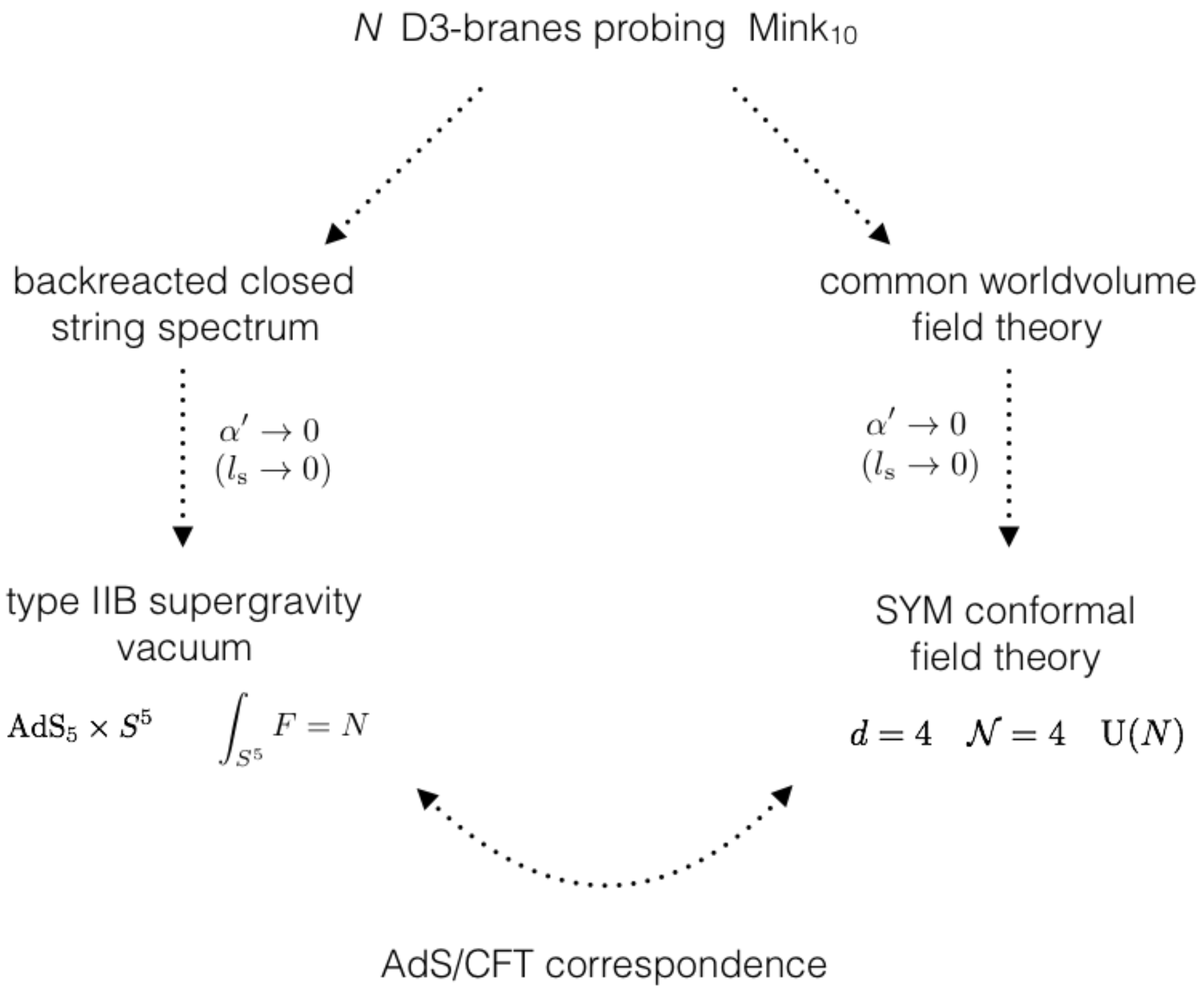

**Figure 4.1.:** The basic setup for the proposed AdS/CFT correspondence.

### Field theory

From the field theory perspective, the well-known *'t Hooft limit*,

$$N \to \infty \ , \quad g_{\rm YM}^2 \to 0 \ , \quad \lambda \equiv g_{\rm YM}^2 N \quad \text{fixed} \ , \tag{4.1}$$

considerably simplifies perturbation theory [73]: It allows us to organize it in inverse powers of $N$, and each order gives a finite contribution. (Essentially, only the so-called *planar diagrams* – a very special subclass of Feynman diagrams of the theory – contribute.) Two observations are in order:

- the limit requires taking *the rank of the gauge group to infinity*. This is often referred to in the literature as the *large $N$ limit*;
- the *classical limit* of the theory corresponds to taking $\lambda \to 0$. In other words, this means that the theory is weakly-coupled.

### String theory

The closed string background generated by $N$ D3-branes probing $\mathrm{Mink}_{10}$ was given in [72]. It is a background generated by tensionful threebranes coupling to the RR potential $C_4$ (hence sourcing the self-dual five-form flux $F_5$ according to (3.28)). The isometries of the solutions must be $\mathrm{SO}(1,3) \times \mathrm{SO}(6) \cong \mathrm{SO}(1,3) \times \mathrm{SU}(4)_R$.

$$e^{\phi} = g_{\mathrm{s}} \ , \tag{4.2a}$$

$$ds^2_{10} = H^{-\frac{1}{2}} ds^2_{\mathrm{Mink}_4} + H^{\frac{1}{2}} (dr^2 + r^2 ds^2_{S^5}) \ , \tag{4.2b}$$

$$H(r) = 1 + \frac{R^4}{r^4} \ , \quad R^4 = g_{\mathrm{s}} N \, l_{\mathrm{s}}^4 \ . \tag{4.2c}$$

$\mathrm{Mink}_4$ is the flat space filled by the D3 worldvolume. $r$ is a coordinate parameterizing the distance from the D3 stack inside the transverse space. The branes are located at $r = 0$. The five-form flux $F_5$ is proportional to the volume forms of $S^5$ and of the remaining space, with coefficient $\frac{N l_{\mathrm{s}}}{R^5}$. Its quantization requires $N \in \mathbb{Z}$, as is obvious from the setup.

Notice that, far away from the D3-branes, at $r \to \infty$, the ten-dimensional metric reduces to flat space (since $H \to 1$), so their backreaction is completely negligible and the space is regular. At $r = 0$ we have a coordinate singularity, signaling the presence of a horizon.

### The near-horizon limit

To switch off the interaction between the D3-branes and the closed string spectrum they backreact on, one needs to take a limit whereby the only remaining physics is the one dictated by the common worldvolume theory. We know that this limit is $l_{\mathrm{s}} \to 0$, which implements the gauge theory – brane worldvolume correspondence of the previous chapter. However, since we are currently dealing with a stack of D-branes, not just one, we should be careful with the $W$ boson masses coming from tensionful strings stretching from one brane to another. These masses go like $\frac{r_{i+1} - r_i}{\alpha'}$, where $r_i$ are the positions of the various branes along the $r$ direction, in the phase where they are all separated (the Higgsed phase). In order to keep such masses finite (possibly zero), we take the limit

$$l_{\mathrm{s}} \to 0 \ (\alpha' \to 0) \ , \quad g_{\mathrm{s}} \text{ fixed} \ , \quad N \text{ fixed} \ , \quad \frac{r_i}{\alpha'} \text{ fixed } \forall i \ . \tag{4.3}$$

This is the *decoupling limit* taken in a careful way. It leaves behind a $d = 4$ $\mathcal{N} = 4$ $\mathrm{U}(N)$ SYM theory decoupled from the bulk string modes.

Let us see what the same limit implies on the closed string background (4.2). A naive calculation would give $H \to 1$ as $\alpha' \to 0$; this is however wrong as the masses $\frac{r_i}{\alpha'} \sim \frac{r}{\alpha'}$ are required to stay finite. Hence we must also take $r \to 0$. This washes away the $1 + \dots$ part in $H$, and we are left with:

$$ds^2_{10} \underset{r\to 0}{\sim} R^2 \left(ds^2_{\mathrm{AdS}_5} + ds^2_{S^5}\right) \ , \quad ds^2_{\mathrm{AdS}_5} = \frac{dr^2}{r^2} + r^2 ds^2_{\mathrm{Mink}_4} \ . \tag{4.4}$$

Starting from an asymptotically flat metric we have gained a curved metric.

## Matching the parametric regimes

We can now try to match parameters on the two sides. By the gauge theory – brane worldvolume correspondence we know that for D3-branes $g^2_{\mathrm{YM}} = g_{\mathrm{s}}$ (see equation (3.3) with $p = 3$). Hence

$$\lambda = g_{\mathrm{s}} N \ \Leftrightarrow \ g_{\mathrm{s}} = \frac{\lambda}{N} \ . \tag{4.5}$$

At fixed $\lambda$ the large $N$ limit implies $g_{\mathrm{s}} \to 0$, so it is equivalent to suppressing the higher string loop corrections.

*The large $N$ limit in field theory corresponds to the suppression of $g_s$ corrections on the string theory side.*

To see what we have to do in order to suppress $\alpha'$ corrections (needed to neglect the massive bulk excitations), let us consider the quantity

$$\frac{\alpha'}{R^2} = \frac{l_{\mathrm{s}}^2}{R^2} = \frac{1}{\sqrt{\lambda}} \ . \tag{4.6}$$

Taking $\alpha' \to 0$ (that is $l_{\mathrm{s}}$ much smaller than the *compactification scale* $R$) requires $\lambda \to \infty$.

*The limit that recovers ordinary two-derivative supergravity is the strongly-coupled limit in field theory.* This limit is the opposite of the classical limit in field theory.

We now understand the usefulness of the holographic duality. When the field theory is weakly-coupled, we can perform computations (scattering amplitudes, correlators etc.) by reliably using perturbation theory in the 't Hooft limit; when it is strongly-coupled, we can use the *dual* weakly-coupled supergravity AdS vacuum to perform computations.

### Checks involving symmetries

Some obvious checks involve the matching of symmetries on the two sides of the correspondence.

- The SYM coupling $g_{\mathrm{YM}}$ can be promoted to a complex one $\tau \equiv \frac{i}{g_{\mathrm{YM}}^2} + \theta$ by the introduction of a theta angle $\theta$. The Montonen–Olive duality of SYM (also called strong-coupling / weak-coupling duality), $g_{\mathrm{YM}} \mapsto g_{\mathrm{YM}}^{-1}$, is then understood as the S-duality transformation, $\tau \mapsto -1/\tau$ at zero $\theta$ angle. The latter is a self-duality of type IIB string theory, enjoying the full quantum invariance under $\mathrm{SL}(2,\mathbb{Z})$. We can see the origin of the $\theta$ angle in string theory. Upon expanding the DBI plus WZ action (3.2) for a D3-brane in $l_{\mathrm{s}}$ (equivalently in $\alpha'$) up to two derivatives, we obtain an expression of the form

  $$\int_{\mathrm{D3}} d^{3+1}x \left( e^{-\phi} \mathcal{F}_{\mu\nu}\mathcal{F}^{\mu\nu} + C_0\, \epsilon_{\mu\nu\rho\sigma}\mathcal{F}^{\mu\nu}\mathcal{F}^{\rho\sigma} \right) \ . \tag{4.7}$$

  The first integrand is the standard kinetic term for the SYM gauge potential $A$, the second is precisely a $\theta$ angle contribution. The identification with the SYM complexified coupling gives

  $$\tau = \frac{i}{g_{\mathrm{s}}} + C_0 \equiv \frac{i}{g_{\mathrm{YM}}^2} + \theta \ . \tag{4.8}$$

- $\mathcal{N}=4$ SYM is believed to be conformal at the quantum level (the gauge coupling $\beta$ function vanishes up to three-loop order, and is believed to vanish at all orders in perturbation theory). Therefore the superalgebra is enhanced to a superconformal one. We have sixteen new supercharges, called $S$, that together with the other sixteen $Q$ supercharges sum up to thirty-two. This is the same amount of supersymmetry preserved by the background (4.2) of type IIB supergravity. (Without any branes! We have already taken into account the brane backreaction on the string background.) The $S$ charges are needed to close the superconformal algebra, $[K,Q]=S$, $K$ being the generator of special conformal transformations.

- The bosonic part $\mathrm{SO}(4,2)$ of the superconformal group $\mathrm{SU}(2,2|4)$ is realized on the supergravity side by the isometries of $\mathrm{AdS}_5$.

- The R-symmetry group $\mathrm{SU}(4)_R$ of $\mathcal{N}=4$ is realized by the isometries of the internal space $S^5$.

- There is a caveat with the gauge group of the SYM theory. The correspondence should work with its $\mathrm{SU}(N)$ version. The center-of-mass $\mathrm{U}(1)$ is actually decoupled in the brane picture (it is IR free in field theory), hence its dynamics cannot be captured by the duality [28].

For an account on "precision tests" of holography in the $\mathcal{N} = 4$ case (e.g. operator-state correspondence in AdS/CFT), other instances with less or more supersymmetry and different dimensions we refer the reader to the comprehensive review [28] and references therein.

### 4.1.2. The NS5-D6-D8 case

The string theory embedding of the field theories discussed in chapter 3 proves their existence, and gives us a concrete way to identify their dynamical degrees of freedom. However, our ultimate goal is to be able to compute quantities in field theory, either by exact methods (e.g. localization, indices, and the like) or via approximations. Moreover, all the "information" contained in a conformal field theory should be given by correlators of its operators. It is not an easy task to extract such information from the brane system.

A very useful tool that allows us to perform (approximate) computations should be holography. We will see a clear example of how this works when we will compute the $a$ coefficient of the *conformal anomaly* of six-dimensional $(1, 0)$ theories in section 4.2.

We can now try to generalize the lessons learned in the prototypical case of flat D3-branes to obtain a philosophy that is applicable to more complicated brane constructions. In brief, the procedure should be the following.

1. Start with a brane setup engineering a certain field theory via the gauge theory – brane worldvolume correspondence (in the sense of Hanany–Witten).

2. The brane setup gives rise to a certain closed string background since it backreacts onto the NSNS and RR bulk fields of type II string theory.

3. Focus on the supergravity regime by neglecting possible $\alpha'$ and $g_s$ corrections to the background.

4. Compute the near-horizon limit of the supergravity background. This typically entails understanding which limit(s) of the free parameters in the theory precisely suppresses string loop and $\alpha'$ corrections, enabling us to reliably compute quantities in the (classical) supergravity regime of string theory.

5. The supergravity background thus obtained should be an AdS vacuum with bosonic content specified by the (near-horizon limit of the) closed string background.

However, in realistic situations, going from point 1. to point 5. in an algorithmic way is far from being easy. The biggest obstruction to the implementation of such an algorithm comes from the difficulty of computing the bosonic spectrum of an arbitrary configuration of intersecting branes. Localized metrics are very hard to find (same exceptions can be found in [74, 75, 76]) and, even when they are known, the correct near-horizon limit is difficult to identify.

A possible way out is to jump directly to point 5. and look for AdS vacua of type II supergravity of the appropriate dimension and with the correct amount of supersymmetry to allow for dual field theories engineered by Hanany–Witten setups. One could then try to prove (a posteriori) that they descend from near-horizon limits of the latter. How to find such a *holographic limit* is far from obvious, but a strong suggestion comes from identifying a limit of the parameters of the AdS vacua that suppresses $\alpha'$ and $g_{\rm s}$ corrections, thus justifying our use of supergravity instead of its full-fledged quantum parent (string theory). Since these parameters are also present in the Hanany–Witten setup, one can translate them into parameters in field theory, and see what this limit corresponds to. Finally, if we believe in the holographic correspondence, we can rely on the supergravity vacua to compute quantities in the classical regime of string theory (that we can trust) and extract information about the strongly-coupled regime of the dual field theory.

With this philosophy in mind, in section 5.1 we will justify the holographic limit for six-dimensional $(1,0)$ theories engineered by NS5-D6-D8 Hanany–Witten setups that we proposed in section 3.4.2. We will do so from the supergravity perspective of the dual $\text{AdS}_7$ vacua of massive type IIA.

## 4.2. The holographic conformal anomaly of $(1,0)$ theories

A very interesting observable that can be computed holographically is the so-called *conformal anomaly* of an SCFT. This computation can be either matched to the field theory result, if the latter is known, or used as a prediction, if it is not.

In AdS/CFT, a field propagating in the AdS vacuum of type II supergravity, a *bulk field*, corresponds to an operator in the superconformal theory. This is the *state-operator correspondence*. Which fields correspond to which operators is a nontrivial question in general, and the answer typically depends on the details of the theory we are considering.

However, one can often identify a set of operators in the SCFT which will have a clear counterpart in supergravity. This observation is purely due to symmetry reasons. For instance, the stress-energy tensor $T_{\mu\nu}$ of the field theory couples to the bulk metric $g_{\mu\nu}$. Without entering into the details, this has as important consequence the fact that, when the SCFT is defined on a curved background, its conformal invariance is broken [77]:

$$\langle T^{\mu}_{\mu} \rangle \neq 0 \ . \tag{4.9}$$

In fact, tracelessness of the stress-energy tensor means the associated conformal current is conserved.

In an even number $d$ of dimensions, whenever the trace of $T$ is nonvanishing it can be used to define the so-called *Weyl* or *conformal anomaly.* It is constrained by general covariance and by the specific number $d$ to be of the form

$$\langle T^{\mu}_{\mu} \rangle \propto a\, E_{(d)} + \sum_i c_i I_i \ , \tag{4.10}$$

where $E_{(d)}$ is the Euler density in $d$ dimensions and $I_i$ are Weyl invariants of weight $d$ (their number depends on $d$ – there are three in $d = 6$). Both can be computed from the Riemann tensor specified by the metric of the space the field theory is defined on [77].

In particular, $a$ is of central importance in the formulation of so-called *a-theorems* in even dimensions. It has been proven that, in $d = 2$[1] [36] and $d = 4$ [78, 79], $a$ is monotonic along the RG flow, namely:

$$\Delta a = a_{\text{UV}} - a_{\text{IR}} > 0 \ . \tag{4.11}$$

Moreover

$$a \geq 0 \ , \tag{4.12}$$

and the bound is saturated if and only if the theory has no local degrees of freedom [36, 80]. This, and the fact that the number of degrees of freedom reduces in going from the UV to the IR, are usually taken as strong indications that $a$ serves as a good estimate of such number.

What we said so far applies to conformal field theories, not necessarily supersymmetric. In supersymmetric theories $a$ can also be related to other anomalies. For example, in $d = 4$ $\mathcal{N} = 1$ theories $a$ is related to the 't Hooft anomaly for the U(1) R-symmetry, which allows to reliably compute the former in many strongly-coupled cases.

[1] In two dimensions $a$ coincides with $c$, the central charge of the (S)CFT, and is usually denoted $c$ for this reason. There are no $c$ anomalies in the sense of (4.10).

In six dimensions, the 't Hooft anomalies for the $(2,0)$ theories of type $A_{N-1}$ were computed long ago. Their $a$ anomaly has been computed holographically in [77]. In the large $N$ limit (i.e. large number of M5-branes), the leading order scales like $N^3$. An exact computation in field theory for all $(2,0)$ theories was recently presented in [81], and a six-dimensional $a$ theorem thus established.

For $(1,0)$ theories [82] found that $a$ is related to the R-symmetry and diffeomorphism anomalies; similar results hold for the $c_i$ [83]. In turn, R-symmetry and diffeomorphism anomalies can be computed on the tensor branch, where the superconformal theories are described by linear quiver gauge theories. This allows to also extract $a$ by invoking an analog of the 't Hooft anomaly matching principle [84, 85].

### 4.2.1. A sketch of the computation

Recently, the field-theoretic computation of the $a$ anomaly of [84, 85] was extended to all linear quiver gauge theories engineered by the Hanany–Witten setups we presented in chapter 3 [31]. The field theory computation was then matched (in the holographic limit) to the one performed in the the dual AdS vacua. As is typical for theories admitting a holographic dual, $a \propto c_i$ and $a \sim N^3$, as for their $(2,0)$ cousins.

In this thesis, we are going to show the details of the holographic computation of $a$ for some $(1,0)$ theories engineered by particular Hanany–Witten setups; see section 6.8.8. These theories are dual to the AdS$_7$ vacua presented in section 5.6. The original computation has first appeared in [III, IV].

Here is how it works. Following the original argument of [77] (performed in Einstein frame), we have, for an SCFT in $d$ dimensions:

$$\langle T^\mu_\mu \rangle \propto \frac{l^{d-1}}{G_{\rm N}^{(d+1)}} \cdot \ldots \; . \tag{4.13}$$

The ellipsis represents the Weyl anomaly (i.e. a polynomial in $E_{(d)}$ and $I_i$); it is zero for odd $d$. The quantity $l$ is related to the AdS$_{d+1}$ cosmological constant by $-\frac{4}{d-1}\Lambda = \frac{2d}{l^2}$; it is the AdS radius. $G_{\rm N}^{(d+1)}$ is Newton's constant in $d+1$ dimensions.

The details of the supergravity AdS compactification enter in $G_{\rm N}$, since the inverse of the latter is simply proportional to the volume of the internal compactification space: $G_{\rm N}^{(d+1)} = g_{\rm s}^{-2}\,{\rm vol}_{10-(d+1)}$. Hence for a six-dimensional SCFT we would have to compute:

$$l^5\,{\rm vol}_{M_3}\ . \tag{4.14}$$

In our vacua, however, both the dilaton $g_{\rm s} = e^{\phi(y)}$ and the AdS "radius" $e^{A(y)}$[2] have a nontrivial profile and depend on the coordinates $y$ of the internal space $M_3$. In particular, $l$ should be understood as the integral of $e^A$ over $M_3$. We should then translate everything from Einstein to string frame, which we use throughout the thesis. The ten-dimensional metrics in the two frames are related by $g^{\rm E}_{10} = e^{-\frac{\phi}{2}} g^{\rm s}_{10}$.

All in all, we get:

$$a \propto \int_{M_3} e^{5A-2\phi}\,{\rm vol}_3\ . \tag{4.15}$$

The exact proportionality factor is fixed by comparison with the explicit known result for the $(2,0)$ theory of type $A_N$ given in [77]. The latter has as holographic dual the background ${\rm AdS}_7 \times S^4$ in eleven-dimensional supergravity, which reduces to the ${\rm AdS}_7 \times M_3$ vacuum presented in section 5.6.1. This way, we can "propagate" the coefficient found in [77] to our massive type IIA vacua.

We are going to use (4.15) in section 6.8.8 to estimate the $a$ anomaly for some $(1,0)$ theories and for their *compactifications* on a Riemann surface, i.e. four-dimensional $\mathcal{N}=1$ theories admitting a holographic dual.

It is satisfactory to notice that, in the ($N \to \infty$, $\frac{\mu}{N}$ fixed) holographic limit proposed in section 3.4.2 (and justified in section 4.1.2), the $a$ anomaly of $\mathcal{T}(N,k,\mu_{\rm L},\mu_{\rm R})$ theories computed in field theory matches with that coming out of formula (4.15) applied to their AdS duals [31].

## 4.3. The ten-dimensional system equivalent to supersymmetry

In this section we will introduce some elements of type II supergravity needed to understand the results presented in part II of this thesis. We will use the system of *supersymmetry equations* derived here to find AdS vacua of type II supergravity, which are holographically dual to superconformal theories in the sense explained in section 4.1.

[2] In our case the radius is given by the warping function $A(y)$. See (4.23) and recall that $ds^2_{{\rm AdS}_{d+1}}$ is the unit radius metric on AdS, as explained in footnote 6.

### 4.3.1. Elements of type II supergravity

#### Field content

Ten-dimensional type II supergravity is the low energy limit of type II string theory: It is an effective description whereby only the massless fields in the string spectrum are retained. The theory enjoys maximal $\mathcal{N} = 2$ supersymmetry in ten dimensions (hence the II in the name), that is thirty-two supercharges. The supermultiplet contains the following fields, organized in "sectors" (for reasons we will not explain here). The NSNS sector contains the dilaton $\phi$, the graviton $g$ and the $B$ field. (The field strength of the latter is a three-form, locally given by $H = dB$, and it is called NSNS flux.) The RR sector contains the usual $p$-form potentials $C_p$ (odd numbers $p = 1, \ldots, 9$ in IIA, even numbers $p = 0, \ldots, 8$ in IIB). The R–NS and NS–R sectors each contain a gravitino $\psi$ (spin $\frac{3}{2}$ Majorana–Weyl spinor) and a dilatino $\lambda$ (spin $\frac{1}{2}$ Majorana–Weyl spinor), so that we have two of each in total. In type IIA the two gravitini (as well as the two dilatini) have opposite chirality; in type IIB they all have the same chirality, which we choose to be positive.

There are two free parameters in the theory: the characteristic string length $l_\mathrm{s}$ (with $l_\mathrm{s}^2 = \alpha'$ and $l_\mathrm{s}^{-1} = M_\mathrm{s}$) and the string coupling $g_\mathrm{s} = e^{\langle\phi\rangle}$ (the vev of the dilaton which is arbitrary, since no potential is specified by the Lagrangian of the theory). We can *correct* the theory by higher-derivative terms weighted by powers of $\alpha'$ (obtained by integrating out the massive string excitations of the full string theory spectrum) and by quantum "stringy" terms weighted by powers of $g_\mathrm{s}$. String theory provides a clear prescription to compute all such corrections.

The RR fluxes $F_{p+1} = dC_p - H \wedge C_{p-1}$ (away from D-brane sources) satisfy a self-duality condition $F_p = (-)^{\lfloor \frac{p}{2} \rfloor} *_{10} F_{10-p}$, simply due to dimensional reasons.[3] If we define the formal sums (often called "polyforms")

$$C = \sum_p C_p \ , \quad F = \sum_p F_p \ , \tag{4.16}$$

for appropriate values of $p$ in IIA/B, then the total flux $F$ and its self-duality condition can be re-written as

$$F = d_H C \ , \quad F = \lambda *_{10} F \ . \tag{4.17}$$

$d_H = d - H\wedge$ is the ordinary exterior derivative twisted by the NSNS three-form

[3] For $p = 5$ this implements the famous self-duality of the five-form flux in type IIB, that cannot be included in the supergravity Lagrangian in a standard way. A possible solution has been proposed by Pasti–Sorokin–Tonin [6] and, more recently, by Sen [86].

$H$,[4] and $\lambda$ is an operator acting on a $p$-form as follows:

$$\lambda F_p = (-)^{\lfloor \frac{p}{2} \rfloor} F_p \ . \tag{4.18}$$

## Vacua

Let us focus for the time being on a *compactification* of type II theory on a *four-dimensional external space.* This means that the ten-dimensional spacetime $M_{10}$ the theory is defined on will be split in two parts, $\text{Mink}_4$, or $\text{AdS}_4$, and a $10 - 4 = 6$-dimensional *internal* space, which is often taken to be compact.[5]

A *vacuum* of type II supergravity is then a solution to its equations of motion that does not contain any particle in the four-dimensional spacetime. Since we usually require that such a solution be supersymmetric, we should impose that the supersymmetry variations of the various fields in the massless spectrum vanish. Furthermore, since we are only interested in classical solutions of the theory, we should set the expectation values of the fermions to zero. In turn, the latter requirement guarantees that the variations of the bosonic NSNS and RR fields automatically vanish (as they always contain one or more terms depending on the fermionic fields). Thus what we should really impose by hand is that the terms depending on bosonic fields in the supersymmetry variations of the fermionic fields be zero.

For type IIA/B supergravity this requirement reads ($M = 0, \ldots, 9$):

$$\delta\psi^1_M = \left(D_M - \frac{1}{4} H_M\right) \epsilon_1 + \frac{e^\phi}{16} F\, \Gamma_M \Gamma_{(10)}\, \epsilon_2 = 0 \ , \tag{4.19a}$$

$$\delta\psi^2_M = \left(D_M + \frac{1}{4} H_M\right) \epsilon_2 - \frac{e^\phi}{16} \lambda(F) \Gamma_M \Gamma_{(10)}\, \epsilon_1 = 0 \ , \tag{4.19b}$$

$$\Gamma^M \delta\psi^1_M - \delta\lambda^1 = \left(D - \partial\phi - \frac{1}{4} H\right) \epsilon_1 = 0 \ , \tag{4.19c}$$

$$\Gamma^M \delta\psi^2_M - \delta\lambda^2 = \left(D - \partial\phi + \frac{1}{4} H\right) \epsilon_2 = 0 \ . \tag{4.19d}$$

The two Killing spinors $\epsilon_a$ in ten dimensions are the *infinitesimal generators of the supersymmetry variations.* Their independent components are the thirty-two supercharges of type II theory. In IIA $\epsilon_1$ has positive chirality while $\epsilon_2$ has

[4] It is a differential:

$$d_H^2 = \frac{1}{2} \{d_H, d_H\} = \frac{1}{2} \{d, d\} - \{d, H\wedge\} + \frac{1}{2} \{H\wedge, H\wedge\} = -(dH)\wedge = 0 \ .$$

[5] $\text{dS}_4$ is incompatible with unbroken supersymmetry in four dimensions and will not be considered in the following.

negative chirality. In IIB they both have positive chirality. $\Gamma^M$ are a basis of gamma matrices in ten dimensions; $\Gamma_{(10)}$ is the chiral gamma matrix. We also defined $D = \Gamma^M D_M$ and $\partial\phi = \Gamma^M \partial_M \phi$. Moreover, in (4.19) we have written the RR and NSNS fluxes in their *slashed* form, that is as *bispinors*. To make a bispinor out of a differential form, we can use the so-called *Clifford map*. On a generic $p$-form in $d$ dimensions:

$$F_p = \frac{1}{p!} F_{i_1 \dots i_p}\, dx^{i_1} \wedge \dots \wedge dx^{i_p} \;\longleftrightarrow\; \not{F}_p \equiv \frac{1}{p!} F_{i_1 \dots i_p}\, \gamma_{\alpha\beta}^{i_1 \dots i_p} \ . \tag{4.20}$$

$\gamma_{(d)}^{i_1 \dots i_p}$ is the total antisymmetrization (of the product) of $p$ gamma matrices in $d$ dimensions. It carries two spinorial indices $\alpha$ and $\beta$, hence the name "bispinor". In writing (4.19) we also used the partially slashed notation $H_M \equiv \not{H}_M = \frac{1}{6} H_{MNP} \Gamma^{NP}$, while the slash over $F$ has been dropped (as for other quantities), implying all differentials $dx^{i_j}$ have been turned into gamma matrices $\gamma^{i_j}$.

The first-order differential system (4.19) will be referred to as the *supersymmetry system*: A solution to its equations is automatically a solution to the second-order equations of motion of type II supergravity if we also impose the *Bianchi identities* for the RR and NSNS fluxes on top of the former [87, 88]. These identities, away from D-brane sources, read:

$$d_H F = 0 \ , \quad dH = 0 \ . \tag{4.21}$$

As we said at the beginning of this paragraph, a vacuum of the theory should not contain any "particles" in the four-dimensional space we are compactifying the theory on. This type of compactification requires fibering $\text{Mink}_4$, or $\text{AdS}_4$, on the six-dimensional internal space $M_6$:

$$\text{Mink}_4 \hookrightarrow M_{10} \to M_6 \ , \quad \text{AdS}_4 \hookrightarrow M_{10} \to M_6 \ . \tag{4.22}$$

However, such a fibration would involve a connection. Since we want the solution to be a vacuum, the former should enjoy maximal symmetry in four dimensions (the Poincaré group $\text{ISO}(3,1)$ in $\text{Mink}_4$ or the conformal one, $\text{SO}(3,2)$, in $\text{AdS}_4$). A connection would necessarily be a vector in the external space, selecting a preferred direction inside it. Hence we have to set it to zero. This kills possible off-diagonal components of the ten-dimensional metric. The external metric must be that of $\text{Mink}_4$ or $\text{AdS}_4$, but we can still allow that their "size" be a function of the internal space. This size is controlled by a *warping function* $A_6$:[6]

$$ds^2_{10} = e^{2A_6(y)} ds^2_{\text{Mink}_4,\, \text{AdS}_4} + ds^2_{M_6} \ . \tag{4.23}$$

[6] In (4.23), $ds^2_{\text{AdS}_4}$ is the unit radius AdS metric. Therefore, the radius of the external space is actually controlled by its size, given by the function $A_6$ which depends on the generic point of internal space $M_6$.

We should also restrict the support of all bosonic fields to the internal space, which is parameterized by local coordinates $y$, not to select any particular direction in the external one (thus breaking some of its isometries). For the scalars this poses no difficulties: $A_6 = A_6(y)$, $\phi = \phi(y)$. For the $p$-form fluxes, we should require that they have no legs on the external space. For $H$, this amounts to requiring that it have only purely internal indices. For the polyform $F_{(10)} \equiv F$ in (4.17), keeping in mind its self-duality, we can impose the following condition:

$$F_{(10)} = F + e^{4A_6}\,\mathrm{vol}_4 \wedge *_6\lambda F \ . \tag{4.24}$$

Here $F$ is the sum of the "purely internal" $F_p$ RR fluxes: $p \leq 6$. They are allowed to depend only on $y$. The "external" fluxes are determined via self-duality by (4.24).

### Decomposing the supersymmetry parameters

We should now proceed to decompose the supersymmetry variation parameters $\epsilon_{1,2}$ according to the Ansatz (4.23). This requires a decomposition of the gamma matrices in ten dimensions (forming a representation of the Clifford algebra $\mathrm{Cl}(9,1)$), which can be represented as tensor products of the ones in four and six dimensions:

$$\Gamma_\mu = e^{A_6}\gamma_\mu \otimes 1_6 \ , \quad \Gamma_m = \gamma_{(4)} \otimes \gamma_m \ . \tag{4.25}$$

The $e^{A_6}$ in (4.25) is needed to close the Clifford algebra $\mathrm{Cl}(3,1)$ (of which the $\gamma_\mu$, $\mu = 0, \ldots, 3$, form a representation in Lorentzian signature) with respect to $e^{2A_6}ds_4^2$ in (4.23). ($\gamma_{(4)}$ is the chiral gamma in four dimensions.) The $\gamma_m$ form a representation of $\mathrm{Cl}(6,0)$, $m = 4, \ldots, 9$ (in Euclidean signature). This decomposition acts on the tensor product of the spaces of spinors in four and six dimensions.

The supersymmetry variations parameters $\epsilon_a$ are chiral (Weyl) spinors which also satisfy the Majorana condition ($\epsilon_a^* = \epsilon_a$). Specializing to the minimal $\mathcal{N} = 1$ supersymmetric case in ten dimensions (sixteen supercharges) and exploiting the decomposition (4.25), it is possible to write $\epsilon_a$ as

$$\epsilon_1 = \zeta_+ \otimes \eta^1_+ + \zeta_- \otimes \eta^1_- \ , \quad \epsilon_2 = \zeta_+ \otimes \eta^2_\mp + \zeta_- \otimes \eta^2_\pm \ . \tag{4.26}$$

(The upper sign in $\eta^2$ is for IIA, the lower sign for IIB.) $\zeta$ and $\eta^{1,2}$ are Majorana–Weyl spinors in four ($\zeta_\pm = \zeta_\mp^*$) and six dimensions ($\eta^{1,2}_\pm = (\eta^{1,2}_\mp)^*$), respectively.

We require that the latter be a solution to (4.19) for every "external" spinor $\zeta$ of chirality $\pm$: Choosing a particular $\zeta$ would in fact be equivalent to selecting a preferred direction in $\mathrm{Mink}_4$, or $\mathrm{AdS}_4$, which we should avoid according to our general philosophy presented in section 4.3.1. At this point, the only difference between the two external spaces are the Killing spinor equations that the $\zeta$ are required to satisfy:

$$D_\nu \zeta_\pm = 0 \ , \ \text{in Mink}_4 \ ; \quad D_\nu \zeta_\pm = \frac{1}{2}\mu\gamma_\nu\zeta_\mp \ , \ \text{in AdS}_4 \ . \tag{4.27}$$

$\mu$ is a complex number related to the cosmological constant $\Lambda$ in AdS by $\Lambda = -|\mu|^2$.

Using the outlined decomposition of bosonic and fermionic fields, directly inspired by the definition of a type II supersymmetric vacuum, it is now possible to transform the system of equations (4.19) into:[7]

$$\left(D_m - \frac{1}{4}H_m\right)\eta^1_+ \mp \frac{e^\phi}{8}F\,\gamma_m\eta^2_\mp = 0 \ , \tag{4.28a}$$

$$\left(D_m + \frac{1}{4}H_m\right)\eta^2_\mp - \frac{e^\phi}{8}\lambda(F)\,\gamma_m\eta^1_+ = 0 \ , \tag{4.28b}$$

$$\mu e^{-A_6}\eta^1_+ + \partial A_6\,\eta^1_+ - \frac{e^\phi}{4}F\eta^2_\mp = 0 \ , \tag{4.28c}$$

$$\mu e^{-A_6}\eta^2_\pm + \partial A_6\,\eta^2_\mp - \frac{e^\phi}{4}\lambda(F)\eta^1_+ = 0 \ , \tag{4.28d}$$

$$2\mu e^{-A_6}\eta^1_- + D\eta^1_+ + \left(\partial(2A_6 - \phi) + \frac{1}{4}H\right)\eta^1_+ = 0 \ , \tag{4.28e}$$

$$2\mu e^{-A_6}\eta^2_\pm + D\eta^2_\mp + \left(\partial(2A_6 - \phi) - \frac{1}{4}H\right)\eta^2_\mp = 0 \ . \tag{4.28f}$$

(Remember that $F$ is the polyform given by the sum of purely internal RR fluxes.) The Bianchi identities (4.21) now read:

$$dH = 0 \ , \quad (d - H\wedge)F = 0 \ , \quad (d + H\wedge)(e^{4A_6} *_6 F) = 0 \ . \tag{4.29}$$

## 4.3.2. The system of equations

It is important to understand that the system (4.28) also determines the metric on $M_6$, once an explicit solution $(\eta^{1,2}, A_6, \phi, F, H)$ has been found. This is not obvious from the system though, since the internal metric $g_{mn}$ does not appear as an unknown. However, using the language of $G$ *structures* in complex geometry, one realizes that the data specified by the spinors $\eta^{1,2}$ and the internal metric $g_6$ is equivalent to a pair of forms $J$ and $\Omega$ that one can define on $M_6$ under certain assumptions (namely, the existence of a *complex structure*). Since it is often easier to work with forms than with spinors, it seems reasonable to try to recast (4.28) in terms of differential forms, and only then try to find a solution. (Notice that the Bianchi identities are already formulated purely in terms of forms, the RR and NSNS fluxes.) Unfortunately, we immediately face a new difficulty.

[7] The dependence of (4.19) on the external spinor $\zeta$ factors out, once we impose (4.26) and (4.27). The system should give a supersymmetric solution for *any* $\zeta$ (since we are interested in a four-dimensional vacuum), thus we can discard the factor depending on it from all the equations. Upon specializing $M$ first to $\mu$, then to $m$, we are left with (4.28).

In the language of $G$ structures, the group $G$ is the stabilizer of one spinor, say $\eta^1$, on the tangent bundle $TM_6$ of the internal manifold $M_6$. Since we have two independent spinors, the common stabilizer can take many forms, e.g. SU(2) or SU(3) in six dimensions, and it can even jump from one group to the other on different loci in $M_6$. This means that the reformulation of the spinorial system (4.28) in terms of differential forms will depend on the properties of the spinors $\eta^{1,2}$ on $M_6$ and their common stabilizer on $TM_6$. Since such properties crucially depend on the number of spacetime dimensions $d$, and on the parity of this number (see e.g. [35, Vol. 2, App. B]), it is particularly hard to generalize the supersymmetry system from $d$ to $d'$, or to a different external space (from Mink to AdS and viceversa) in a quick way. (Nonetheless, this approach has been pretty fruitful in finding Mink or AdS vacua in diverse dimensions with given $G$ structure. For an account on early results see [89, Intro.].)

A more recent (and powerful) approach uses the so-called *generalized structures* of *generalized complex geometry* first introduced in [90, 91]. It arises from the observation that, in six dimensions, the common stabilizer of $\eta^{1,2}$ on the *generalized tangent bundle* $TM_6 \oplus T^*M_6$ is *always* $\mathrm{SU}(3) \times \mathrm{SU}(3)$. One can then translate the spinors into differential forms, and provide a unified way of recasting (4.28) in terms of the latter which does not depend on the choice of $G$ structure on $TM_6$.

The application of the same language to the *generic ten-dimensional spacetime* $M_{10}$ of type II supergravity has been worked out in [89], where it was found that preserved $\mathcal{N} = 1$ supersymmetry in ten dimensions (i.e. sixteen supercharges) in type IIA as well as IIB is equivalent to the following system of differential equations,

$$d_H(e^{-\phi}\Phi) = -(\tilde{K} \wedge + \iota_K)F_{(10)} \ , \tag{4.30a}$$

$$L_K g = 0 \ , \quad d\tilde{K} = \iota_K H \ , \tag{4.30b}$$

and algebraic constraints,

$$\left(e_{+_1}\cdot\Phi\cdot e_{+_2},\Gamma^{MN}\left[\pm d_H(e^{-\phi}\Phi\cdot e_{+_2})+\frac{1}{2}e^{\phi}d^{\dagger}(e^{-2\phi}e_{+_2})\Phi-F_{(10)}\right]\right)=0\ , \tag{4.31a}$$

$$\left(e_{+_1}\cdot\Phi\cdot e_{+_2},\left[d_H(e^{-\phi}e_{+_1}\cdot\Phi)-\frac{1}{2}e^{\phi}d^{\dagger}(e^{-2\phi}e_{+_2})\Phi-F_{(10)}\right]\Gamma^{MN}\right)=0\ . \tag{4.31b}$$

$\Phi=\epsilon_1\otimes\overline{\epsilon_2}$ is the key ten-dimensional bispinor, i.e. a differential form under (4.20), obtained from the supersymmetry variation parameters $\epsilon_{1,2}$ ($\overline{\epsilon_2}\equiv\epsilon_2^{\dagger}\Gamma^0$). *The Clifford map implements the reformulation of the spinorial system in terms of differential forms.*

$g$ is the ten-dimensional metric, and $L_K g$ its Lie derivative along $K$. $d_H$ and $\iota_K$ are, respectively, the twisted exterior derivative we defined under (4.17) and and the interior product with respect to $K$. $K$ and $\tilde{K}$ are two ten-dimensional one-forms defined as follows:

$$K=\frac{1}{64}(\bar{\epsilon}_1\Gamma_M\epsilon_1+\bar{\epsilon}_2\Gamma_M\epsilon_2)\,dx^M\ ,\quad \tilde{K}=\frac{1}{64}(\bar{\epsilon}_1\Gamma_M\epsilon_1-\bar{\epsilon}_2\Gamma_M\epsilon_2)\,dx^M\ . \tag{4.32}$$

$e_{+_a}$ are two ten-dimensional vectors. They can be chosen quite freely (for more details see [89]), provided they satisfy the constraints

$$e_{+_a}^2=0\ ,\quad e_{+_a}\cdot K_a=\frac{1}{2}\ ,\quad a=1,2\ , \tag{4.33}$$

with $K=K_1+K_2$ and $\tilde{K}=K_1-K_2$. $(F_p,G_q)\equiv(F_p\wedge\lambda(G_q))_{\text{top}}$ is the Chevalley–Mukai pairing between two polyforms. Upon wedging, we only retain the top-form part.

In generalized complex geometry language, the system (4.30), (4.31) contains the differential form $\Phi$ and two sections $e_{+_a}$ of the *generalized tangent bundle* $TM_{10}\oplus T^*M_{10}$ of the ten-dimensional spacetime $M_{10}$. The polyform $\Phi$ defines a so-called *generalized* ISpin(7) *structure* on such bundle, containing two copies of the stabilizer on $TM_{10}$ of a single supersymmetry parameter $\epsilon$. However, $\Phi$ is not able to define a ten-dimensional metric $g$ and a $B$ field by itself. To this end, it has to be supplemented by the two sections $e_{+_1}$, $e_{+_2}$, although these introduce partly spurious degrees of freedom in the system (which can be exactly quantified [89]).

### Specialization of the ten-dimensional system to AdS$_d$

Equations (4.30), (4.31) are necessary and sufficient for minimal ($\mathcal{N}=1$) supersymmetry in ten dimensions to hold: For extended supersymmetry one should impose that there be several solutions with same bosonic content.

Equations (4.30) are necessary but not sufficient for preserved supersymmetry. For this reason they have to be supplemented by the two "pairing" equations (4.31), acting as additional algebraic constraints on $\Phi$. These look daunting at first, but they often turn out to be fully redundant. (This will always be the case for the AdS vacua that we will consider in this thesis; see the appendices A.3 and B.1.2 for a proof.) Thus we will not dwell on them. The most important equations are then (4.30a) and (4.30b).

Let us first discuss the first equation in (4.30b). It actually implies [89] that the vector $K$ is an isometry of the full ten-dimensional (closed) background: It implies that the Lie derivatives along $K$ of all NSNS and RR bosonic fields vanish. Moreover, it implies that $L_K \Phi = 0$: $K$ is a supersymmetric isometry, keeping the bispinor $\Phi$ (constructed upon the supersymmetry variation parameters $\epsilon_{1,2}$) invariant.

When we specialize $M_{10}$ to a compactification on an AdS external space, $\mathrm{AdS}_d \hookrightarrow M_{10} \to M_{10-d}$, the second equation in (4.30b) just fixes the norms of the internal Killing spinors $\eta$ to be equal ($\epsilon = \zeta \otimes \eta$, in a decomposition à la (4.26)). The last and most important equation, (4.30a), implies two classes of sub-equations: one class of *geometric equations*, determining the metric on the internal space $M_{10-d}$, and one class of *fluxed equations* that, when coupled to the Bianchi identities for RR and NSNS fluxes, fully determine the latter.

The power of the system should by now be clear. Not only can we specialize it very easily to AdS (or Mink) in different dimensions (simply by plugging an appropriate Ansatz $\epsilon = \zeta \otimes \eta$ into $\Phi$), but it also allows us to *determine* metric and fluxes of the ten-dimensional type II background once we solve a bunch of differential equations (which are typically easier to solve than many coupled equations for Killing spinors).

### The strategy to solve the system on AdS$_d$

We will now briefly outline the strategy to solve the system (4.30), assuming (4.31) are redundant. (The same strategy would apply even if the latter were not redundant.) From now on we will only be interested in AdS solutions (vacua of type II supergravity).

- Compactify $M_{10}$ as in $\mathrm{AdS}_d \hookrightarrow M_{10} \to M_{10-d}$, upon choosing $d$.
- Split the two supersymmetry variation parameters $\epsilon_a$ according to the Ansatz $\epsilon = \zeta \otimes \eta$; $\zeta$ is a spinor on AdS$_d$ satisfying (4.27), while $\eta$ is a spinor on $M_{10-d}$. The details of such a splitting depend on $d$, on the theory (IIA/IIB) and on the amount of preserved supersymmetry one wishes the system to be equivalent to.

- Plug the $\epsilon = \zeta \otimes \eta$ Ansatz into $\Phi = \epsilon_1 \otimes \overline{\epsilon_2}$. The latter then takes the (schematic) structure: external form $\wedge$ internal form (upon using Fierz-like identities), where both factors are polyforms defined on $\mathrm{AdS}_d$ and $M_{10-d}$ respectively.

- Plug the decomposed form of $\Phi$ into (4.30a). We can factor out the dependence on the external polyform (since the equation must be valid for any $\zeta$). This gives differential equations for the forms (of different degrees) that make up the internal polyform. The latter equations are necessary for preserved supersymmetry on $M_{10-d}$.

- We now focus on the $d = 5, 6, 7$ cases. Given the properties of the $\eta$ spinors on $M_{10-d}$ (whose norms are fixed to be equal by the second in (4.30b)), one can parameterize the internal forms in terms of a Vielbein on $M_{10-d}$ and a few "angles", acting as local coordinates on the space. See section 5.4.2 for $d = 7$, section 7.4 for $d = 6$, section 6.3.3 for $d = 6$.

- A subclass of the remaining necessary equations (the *geometric* ones) will determine the Vielbein – i.e. an internal metric (as a function of the angles introduced by the parameterization), the warping function, the dilaton (see section 5.5.1 for $d = 7$, sections 7.5.1 and 7.5.2 for $d = 6$, section 6.4.1 for $d = 5$). Another subclass (the *fluxed* equations), accompanied by the Bianchi identities for the RR and NSNS fluxes, will determine the latter (see section 5.5.2 for $d = 7$, section 7.5.3 for $d = 6$, section 6.4.2 for $d = 5$).

- So far we have used the term 'determine' in a generic sense. More precisely, at this stage we face two possibilities: Either the two subclasses of equations *fully* determine metric and fluxes explicitly (we have an explicit bosonic solution without the need of any Ansatz), or we are left with some residual (typically coupled) PDEs (for $d = 6$ see section 7.6) or sometimes ODEs (for $d = 7$ see section 5.5.3). To solve these, one might need to specify a further Ansatz, thus only determining a subclass of the full solution space (for a very interesting subclass in $d = 5$ see section 6.6).

# Part II

## Original contributions

# Chapter 5

# All AdS$_7$ solutions of type II supergravity

Sections 5.2 through 5.6 of this chapter are based on the published article [I]. Some observations are based on the published letter [IV].

## 5.1. Relation to six-dimensional Hanany–Witten setups

In this section we will provide evidence in favor of the interpretation of the solutions of section 5.6 as near-horizon limits of the NS5-D6-D8 Hanany–Witten setups introduced in chapter 3.

To this end, we will use details of the solutions that will only be introduced in later sections. We will refer to the relevant equations and paragraphs wherever needed. The reader can equivalently come back to this section after having read the rest of the chapter.

We will use observations first appeared in [I, 24, 31].

### A Myers-like effect and the ordering constraint

In supergravity language, D8-brane sources with different D6 charge (that is different summand $\mu_i$ in the left or right partitions $\mu_{\mathrm{L,R}}$ which order the D8 stacks) are stabilized by supersymmetry (i.e. as a consequence of the supersymmetry equations) at different values of the coordinate $r$ introduced above equation (5.52).

The quantity

$$q(r) = \frac{1}{4}\sqrt{1-x^2(r)}e^{A(r)-\phi(r)} = e^{-\phi}\,\mathrm{radius}(S^2_r) \tag{5.1}$$

fixes their position according to:

$$q(r_{\rm D8}) = \frac{1}{2}(-n_2 + \mu\, n_0) = \frac{1}{2}(-n_2' + \mu\, n_0') \ . \tag{5.2}$$

$n_2$ and $n_0$ are defined in (5.72) and (5.73) respectively, and they are the quanta of the $F_2 - F_0 B_2$ and $F_0$ (the Romans mass) fluxes. $\mu$ is the integer slope defined in (5.84) ($\mu = \frac{\Delta n_2}{\Delta n_0}$) and used in section 5.6.3; it represents the D6 charge of a D8-D6 bound state (that is a fuzzy D6 opening up into a D8). A prime indicates the flux integers on the other side of the D8.

In the supergravity solution, the D8-D6 bound states appear as "creases" on the compact internal space $M_3$, which is topologically an $S^3$. More precisely, it is an $S^2$ fibration over a finite interval parameterized by $r$; the fiber shrinks at the two endpoints of the interval, making the whole space compact. A crease is just an $S^2$ fiber, that we can imagine being wrapped by a D8-brane located at $r_{\rm D8}$ along $r$.

Formula (5.1) then tells us that the radius of the $S^2$ fiber divided by $g_{\rm s}$ is linear in $\mu$, the D6 charge of the D8-D6 bound state. In turn we have:

$$r_{\rm D8} \sim g_{\rm s}\, n_{\rm D6} \ . \tag{5.3}$$

At the endpoints of the finite interval parameterized by $r$ the $S^2$ fiber shrinks, its radius is zero. Since big D6 charge means big radius, the corresponding $S^2$ fiber will be located farther away from the endpoints; this is the meaning of (5.3).

Another, intuitive, way of understanding (5.3) is the following. More D6 charge means more electrostatic repulsion among the D6's smeared onto the D8 worldvolume, making the D8 puff up a bit more, towards the interior of $M_3$ (along $r$). This is very similar to what Myers considered in his famous setup [44], namley $N$ D2-branes with D0 charge smeared on their worldvolume in a constant $F_4$ background. Upon T-dualizing six times $F_4$ turns into $F_{10} = *_{10} F_0$, sourcing the D8's.

This Myers-like effect[1] in supergravity reflects what we found in section 3.3.3. A D8-D6 fuzzy funnel is the geometric interpretation of the field theory boundary condition (3.38). Those with smaller D6 charge (i.e. smaller summand $\mu_i$ in the partition $\mu_{\rm L,R}$) will fit into those with bigger D6 charge, without intersecting them (the radius of the noncommutative sphere is smaller). In supergravity, a crease on the $S^3$ created by a D8 with bigger D6 charge will be pushed farther away from the (left or right) pole of the sphere, towards the central region, and will not intersect another D8 with smaller D6 charge.

[1] Notice that in the original setup of Myers the branes probe flat space in a constant flux background, while here the space is curved, and the fluxes not constant.

In section 5.5.7 we will see that $F_0$ is monotonous throughout the compact internal space $M_3$. This is a consequence of the boundary conditions (5.61) one needs to impose on the system of ODEs (5.53) (equivalent to preserved supersymmetry) in order to have regular solutions. In the left (or northern) hemisphere $F_0 > 0$ and the D8's have positive D6 charge; in the right (or southern) hemisphere $F_0 < 0$ and the D8's have negative D6 charge (i.e. positive anti-D6 charge). Hence, e.g. in the northern one:

$$n_{\text{D8},i} = n_{0,i} - n_{0,i-1} < 0 \ , \quad n_{\text{D6},i} = n_{2,i} - n_{2,i-1} < 0 \ ; \tag{5.4}$$
$$2(q_{i+1} - q_i) = (-n_{2,i} + n_{0,i}\mu_i) - (-n_{2,i} + n_{0,i}\mu_{i-1}) = n_{0,i}(\mu_i - \mu_{i-1}) \ . \tag{5.5}$$

$i$ labels the D8-D6 bound state in supergravity, i.e. the one with $\mu_i$ D6 charge. The number of D8's is given by the jump of the Romans mass before and after it; similarly, the jump of the $n_2$ flux integer gives the induced D6 charge on the stack. From (5.53) one can easily obtain the condition

$$\partial_r q(r) = \frac{1}{4} F_0 e^{A(r)} > 0 \quad \text{for} \quad F_0 > 0 \quad \text{(northern hemisphere)} \ . \tag{5.6}$$

Therefore

$$q_{i+1} > q_i \Rightarrow \mu_i > \mu_{i-1} \ . \tag{5.7}$$

This is the supergravity implementation of the ordering constraint advocated for in section 3.3.3.[2] This observation first appeared in [24].

### The holographic limit revisited

In section 3.4.2 we have proposed a limit of the parameters of the $\mathcal{T}(N, k, \mu_{\text{L}}, \mu_{\text{R}})$ theories that would allow them to have holographic duals. Later, in section 4.1.2, we have described the philosophy in the context of the AdS/CFT correspondence that justifies such limit from the dual supergravity perspective. Typically, one is after a parametric regime in the supergravity AdS vacuum where stringy ($g_{\text{s}}$) and $\alpha'$ corrections become negligible. In particular, following [31], we proposed to take:

$$N \to \infty \ , \quad \frac{\mu_i}{N} \quad \text{fixed} \ . \tag{5.8}$$

In this section, we will see *why* such a limit actually suppresses possible unwanted corrections. We will also justify it by using the interpretation of the AdS$_7$ vacua as near-horizon geometries of the NS5-D6-D8 Hanany–Witten setups.

[2] Here we are using a slightly different convention whereby, if we literally interpreted the supergravity slopes $\mu_i$ as the summands of the partition $\mu$, $\mu_i$ would come "before" $\mu_{i-1}$ (i.e. it would represent an upper row in the Young tableau, with the bottom one being $\mu_0$).

The possibility of making the dilaton and the curvature small throughout the internal space naturally comes from a symmetry of the system of ODEs (5.5.3), equivalent to preserved supersymmetry. It was first discovered in [I, 24] and later discussed in greater detail in [31].

Consider the transformation (5.63) (with $\Delta A \equiv n$):

$$A \mapsto A + n \ , \quad \phi \mapsto \phi - n \ , \quad r \mapsto e^n r \ , \quad x \mapsto x \ . \tag{5.9}$$

It is a symmetry of the system of ODEs (5.53) and it also leaves the boundary conditions (5.61) invariant. In order to be a symmetry of the full solution it needs to be accompanied by the rescalings

$$n_{2,i} \mapsto e^{2n}\, n_{2,i} \ , \quad \mu_i \mapsto e^{2n}\, \mu_i \ , \quad n_0 \quad \text{invariant} \ , \tag{5.10}$$

so that (5.1), (5.2) stay invariant. The $B$ field (5.48) (or (5.95) in a massless $F_0 = 0$ region), and its flux $H$, should also be rescaled accordingly:

$$B_2 \mapsto e^{2n}\, B_2 \ , \quad N \mapsto e^{2n}\, N \ , \tag{5.11}$$

with $N = -\frac{1}{4\pi^2}\int H \in \mathbb{Z}$.

- The transformation (5.11) allows to make $N$ parametrically large. This realizes the first condition in (5.8).
- The symmetry (5.9) allows one to make $A$ as large as one wishes; the same holds for the "compactification scale" (i.e. the AdS radius) $R \equiv e^{2A} \mapsto e^{2A}e^{2n}$. In turn this guarantees that the curvature of the solution be arbitrarily small, according to (5.17) and (5.52). So $\frac{\alpha'}{R^2} \to 0$.
- We also have $e^{2\phi} \mapsto \frac{e^{2\phi}}{e^{2n}}$, so that $g_s$ can be made parametrically small.
- Finally, both $N$ and $\mu_i$ scale in the same way, so that $\frac{\mu_i}{N}$ remains invariant. This realizes the second condition in (5.8).

### The near-horizon limit interpretation

As we have seen in section 4.1.1, finding a limit suppressing $g_s$ and $\alpha'$ corrections might mean directly identifying the near-horizon limit that takes us from a closed string background to its AdS supergravity vacuum. In the present case, this is not immediately obvious, but we can still imagine how the near-horizon geometry could arise.

- First of all, as we suggested already in section 3.3.3, the summand $\mu_i$ in a partition $\mu$ should correspond to the number of D6-branes ending on the $i$-th D8 in the stack. Actually, we have $f_i$ D8's in the (left or right) stack on which $\mu_i$ D6's end. $f_i$ was defined in (3.55).

  In supergravity language, this $\mu_i$ should correspond to the slope defined in (5.84) and used in section 5.6.3. It gives the D6 charge of the $i$-th D8-D6 bound state; since it is an integer, it can be thought of as the integral of an $f_2$ worldvolume flux induced by D6-branes ending on the D8 (the first Chern class of a worldvolume bundle on the $S^2$ fiber wrapped by the D8).

- $N = -\frac{1}{4\pi^2}\int H$ should correspond to the number of NS5-branes. A single NS5-brane has the role of a large gauge transformation for the $B$ field in supergravity; see formula (5.68). A possible explanation is the following.

  The $B$ field is chosen to vanish at both poles of the internal space $M_3 \cong S^3$ (see dashed green line in figures 5.7(a) and 5.8(a)). However, its flux through $M_3$, $N$, is generically nonvanishing. So to pass from one pole to the other we have to make $N$ large gauge transformations (5.68) in the middle region (this takes care of $H$ flux quantization). Since such a large gauge transformation would change the flux integer $n_2$ (hence the number of D6's) in light of (5.48), (5.71), and (5.72), as well as $n_0$ in light of (5.74), we decide to perform all of them away from the region where $F_0 \neq 0$ (which is true at the poles in light of (5.61)). So we will distribute the D8s' to the left (northern hemisphere) and right (southern hemisphere) of the "middle" massless region (generically present) where we perform the $N$ large gauge transformations.

  Thus, the $N$ NS5's will be kept on top of each other and in the central massless region.

- We can now imagine zooming in close to the $N$ NS5 / $k$ D6 intersection, hoping that this forgets about the presence of the D8's, which are kept far away at first. As we have seen, the naive picture on the left side of figure 3.8 must be modified to take into account the Nahm pole, describing a D6 that opens up into a D8-brane. This field theory boundary condition should be related to the D8-D6 bound state of supergravity: The fuzzy sphere describing the D6 is not scaled away, and remains as a source in supergravity (the creases on the $S^3$). The NS5's then get dissolved into $H$ flux, and their position (along $r$) is unimportant. The Myers-like effect takes care of the ordering constraint: D8-brane sources in supergravity will wrap $S^2$ fibers far away from the poles if their D6-charges are big.

This is what remains of the fuzzy funnels protruding out of the central NS5 stack to the left and right in the original Hanany–Witten setup.

### 5.1.1. Supergravity solutions of brane setups

We will now present the Hanany–Witten brane constructions (their associated left and right partitions, and the linear quiver description) that should give rise to the (numerical) solutions of section 5.6: the unique $\mathrm{AdS}_7 \times M_3$ solution of massless type IIA supergravity summarized in figure 5.1; an $\mathrm{AdS}_7$ vacuum of massive type IIA supergravity (with nonzero Romans mass $F_0$ throughout the internal space $M_3$) without D8-brane sources summarized in figure 5.2; finally, an AdS7 vacuum of massive type IIA supergravity (with varying Romans mass and a central massless region) featuring the presence of D8-brane sources with D6 charge summarized in figure 5.3. These three classes of solutions will also be given an analytic definition in chapter 6.

## 5.2. Introduction

Interacting quantum field theories generally become hard to define in more than four dimensions. A Yang–Mills theory, for example, becomes strongly coupled in the UV. In six dimensions, a possible alternative would be to use a two-form gauge field. Its nonabelian formulation is still unclear, but string theory predicts that a $(2, 0)$ superconformal completion of such a field actually exists on the worldvolume of M5-branes. Understanding these branes is still one of string theory's most interesting challenges.

This prompts the question of whether other non-trivial six-dimensional theories exist. There are in fact several other string theory constructions [14, 20, 19, 92] that would engineer such theories. Progress has also been made (see for example [4, 5, 93]) in writing explicitly their classical actions.

Another way to establish the existence of superconformal theories in six dimensions is to look for supersymmetric $\mathrm{AdS}_7$ solutions in string theory. In this chapter, we classify such solutions. As we will review later, in M-theory, one only has $\mathrm{AdS}_7 \times S^4$ (which is holographically dual to the $(2, 0)$ theory) or an orbifold thereof. That leaves us with $\mathrm{AdS}_7 \times M_3$ in IIA with non-zero Romans mass $F_0 \neq 0$ (which cannot be lifted to M-theory) or in IIB.

Here we will show that, while there are no such solutions in IIB, many do exist in IIA with non-zero Romans mass $F_0$.

Our methods are reminiscent of the generalized complex approach for $\mathrm{Mink}_4 \times M_6$ or $\mathrm{AdS}_4 \times M_6$ solutions [94]. We start with a similar system [95] for $\mathrm{Mink}_6 \times M_4$, and we then use the often-used trick of viewing $\mathrm{AdS}_7$ as a warped product of $Mink_6$ with a line. This allows us to obtain a system valid for $AdS_7 \times M_3$. A similar procedure was applied in [96] to derive a system for $AdS_5 \times M_5$ from $Mink_4 \times M_6$. The system we derive is written in terms of differential forms satisfying some algebraic constraints; mathematically, these constraints mean that the forms define a generalized identity×identity structure on $TM_3 \oplus T^*M_3$. This fancy language, however, will not be needed here; we will give a parameterization of such structures in terms of a vielbein $\{e_a\}$ and some angles, and boil the system down to one written in terms of those quantities.

When one writes supersymmetry as a set of PDEs in terms of forms, they may have some interesting geometrical interpretation (such as the one in terms of generalized complex geometry in [94]); but, to obtain solutions, one usually needs to make some Ansatz, such as that the space is homogeneous or that it has cohomogeneity one. One then reduces the differential equations to algebraic equations or to ODEs, respectively.

The $AdS_7 \times M_3$ case is different. As we will see, the equations actually determine explicitly the vielbein $\{e_a\}$ in terms of derivatives of our parameterization function. This gives a local, explicit form for the metric, without any Ansatz. By a suitable redefinition we find that the metric describes an $S^2$ fibration over a one-dimensional space.

This is actually to be expected holographically. A $(1, 0)$ superconformal theory has an $Sp(1) \cong SU(2)$ R-symmetry group, which should appear as the isometry group of the internal space $M_3$. With a little more work, all the fluxes can also be determined, and they are also left invariant by the $SU(2)$ isometry group of our $S^2$ fiber. All the Bianchi identities and equations of motion are automatically satisfied, and existence of a solution is then reduced to a system of two coupled ODEs.[3] From this point on, our analysis is pretty standard: in order for $M_3$ to be compact, the coordinate $r$ on which everything depends should in fact parameterize an interval $[r_N, r_S]$, and the $S^2$ should shrink at the two endpoints of the interval, which we from now on will call "poles". This requirement translates into certain boundary conditions for the system of ODEs.

We have studied the system numerically. We can obtain regular[4] solutions if we insert brane sources. We exhibit solutions with D6's, and solutions with one or two D8 stacks, appropriately stabilized by flux. For example, in the solution with two D8 stacks, they have opposite D6 charge, and their mutual electric attraction is balanced against their gravitational tendency to shrink. (For D8-branes, there is no problem with the total D-brane charge in a compact space; usually such problems are found by integrating the flux sourced by the brane over a sphere surrounding the brane, whereas for a D8 such a transverse sphere is simply an $S^0$.) We think that there should exist generalizations with an arbitrary number of stacks.

[3] This is morally a hyper-analogue to the reduction performed in [96] along the generalized Reeb vector, although in our case the situation is so simple that we need not introduce that reduction formalism.

[4] On the loci where branes are present, the metric is of course not regular, but such singularities are as usual excused by the fact that we know that D-branes have an alternative definition as boundary conditions for open strings, and are thought to be objects in the full theory. The singularity is particularly mild for D8's, which manifest themselves as jumps in the derivatives of the metric and other fields — which are themselves continuous.

It is natural to think that our regular solutions with D8-branes might be related to D-brane configurations in [20, 19], which should indeed engineer six-dimensional $(1, 0)$ superconformal theories. Supersymmetric solutions for configurations of that type have actually been found [76] (see also [75]); non-trivially, they are fully localized. It is in principle possible that their results are related to ours by some limit. Such a relationship is not obvious, however, in part because of the SU(2) symmetry, that forces our sources to be only parallel to the $S^2$-fiber. It would be interesting to explore this possibility further.

We will begin our analysis in section 5.3 by finding the pure spinor system (5.22) relevant for supersymmetric $\mathrm{AdS}_7 \times M_3$ solutions. In section 5.4 we will then derive the parameterization (5.36) for the pure spinors in terms of a vielbein and some functions. In section 5.5 will then use this parameterization to analyze the system (5.22). As we mentioned, we will reduce the problem to a system of ODEs; regularity imposes certain boundary conditions on this system. Fluxes and metric are fully determined by a solution to the system of ODEs. Finally, in section 5.6 we study the system numerically, finding some regular examples, shown in figures 5.7 and 5.8.

## 5.3. Supersymmetry and pure spinor equations in three dimensions

In this section, we will derive a system of differential equations on forms in three dimensions that is equivalent to preserved supersymmetry for solutions of the type $\mathrm{AdS}_7 \times M_3$. We will derive it by a commonly-used trick: namely, by considering $\mathrm{AdS}_{d+1}$ as a warped product of $\mathrm{Mink}_d$ and $\mathbb{R}$. We will begin in section 5.3.1 by reviewing a system equivalent to supersymmetry for $\mathrm{Mink}_6 \times M_4$. In section 5.3.2 we will then translate it to a system for $\mathrm{AdS}_7 \times M_3$.

### 5.3.1. $\mathrm{Mink}_6 \times M_4$

Preserved supersymmetry for $\mathrm{Mink}_4 \times M_6$ was found [94] to be equivalent to the existence on $M_6$ of an $\mathrm{SU}(3) \times \mathrm{SU}(3)$ structure satisfying certain differential equations reminiscent of generalized complex geometry [90, 91].

Similar methods can be useful in other dimensions. For $\text{Mink}_6 \times M_4$ solutions, [95] found a system in terms of an $\text{SU}(2) \times \text{SU}(2)$ structure on $M_4$, described by a pair of pure spinors $\phi^{1,2}$. Similarly to the $\text{Mink}_4 \times M_6$ case, they can be characterized in two ways. One is as bilinears of the internal parts $\eta^{1,2}$ of the supersymmetry parameters in (A.2):[5]

$$\phi^1_\mp = e^{-A_4} \eta^1_+ \otimes \eta^{2\,\dagger}_\mp \ , \qquad \phi^2_\mp = e^{-A_4} \eta^1_+ \otimes \eta^{2c\,\dagger}_\mp \ , \tag{5.12}$$

where the warping function $A_4$ is defined by

$$ds^2_{10} = e^{2A_4} ds^2_{\text{Mink}_6} + ds^2_{M_4} \ . \tag{5.13}$$

The upper index in (5.12) is relevant to IIA, the lower index to IIB; so in IIA we have that $\phi^{1,2}$ are both odd forms, and in IIB that they are both even. One can also give an alternative characterization of $\phi^{1,2}$, as a pair of pure spinors which are *compatible*. This stems directly from their definition as an $\text{SU}(2) \times \text{SU}(2)$ structure, and it means that the corresponding generalized almost complex structures commute. This latter constraint can also be formulated purely in terms of pure spinors as $(\phi^1, \phi^2) = (\bar{\phi}^1, \phi^2)$.[6] This can be shown similarly to an analogous statement in six dimensions; see [97, App. A].

The system equivalent to supersymmetry now reads [95][7]

$$d_H\big(e^{2A_4-\phi}\text{Re}\phi^1_\mp\big) = 0 \ , \tag{5.14a}$$
$$d_H\big(e^{4A_4-\phi}\text{Im}\phi^1_\mp\big) = 0 \ , \tag{5.14b}$$
$$d_H\big(e^{4A_4-\phi}\phi^2_\mp\big) = 0 \ , \tag{5.14c}$$
$$e^\phi F = \mp 16 *_4 \lambda(dA_4 \wedge \text{Re}\phi^1_\mp) \ , \tag{5.14d}$$
$$(\overline{\phi^1_\pm}, \phi^1_\pm) = (\overline{\phi^2_\pm}, \phi^2_\pm) = \frac{1}{4} \ . \tag{5.14e}$$

Here, $\phi$ is the dilaton; $d_H = d - H\wedge$ is the twisted exterior derivative; $A_4$ was defined in (5.13); $F$ is the internal RR flux, which, as usual, determines the external flux via self-duality:

$$F_{(10)} \equiv F + e^{6A_4}\text{vol}_6 \wedge *_4 \lambda F \ . \tag{5.15}$$

[5] As usual, we are identifying forms with bispinors via the Clifford map $dx^{m_1} \wedge \ldots \wedge dx^{m_k} \mapsto \gamma^{m_1 \cdots m_k}$. $\mp$ denotes chirality, and $\eta^c \equiv B_4 \eta^*$ denotes Majorana conjugation; for more details see appendix A.1. The factors $e^{-A_4}$ are included for later convenience.

[6] As usual, the Chevalley pairing in this equation is defined as $(\alpha, \beta) = (\alpha \wedge \lambda(\beta))_{\text{top}}$; $\lambda$ is the sign operator defined on $k$-forms as $\lambda \omega_k \equiv (-)^{\lfloor \frac{k}{2} \rfloor} \omega_k$.

[7] We have massaged a bit the original system in [95], by eliminating $\text{Re}\phi^1_\mp$ from the first equation of their (4.11).

Actually, (5.14) contains an assumption: that the norms of the $\eta^i$ are equal. For a noncompact $M_4$, it might be possible to have different norms; (5.14) would then have to be slightly changed. (See [98, Sec. A.3] for comments on this in the $\mathrm{Mink}_4 \times M_6$ case.) As shown in appendix A.1, however, for our purposes such a generalization is not relevant.

With this caveat, the system (5.14) is equivalent to supersymmetry for $\mathrm{Mink}_6 \times M_4$. It can be found by direct computation, or also as a consequence of the system for $\mathrm{Mink}_4 \times M_6$ in [94]: one takes $M_6 = \mathbb{R}^2 \times M_4$, with warping $A_6 = A_4$, internal metric $ds^2_{M_6} = e^{2A_4}((dx^4)^2 + (dx^5)^2) + ds^2_{M_4}$, and, in the language of [98],

$$\Phi_1 = e^{A_4}(dx^4 + i dx^5) \wedge \phi^2_{\mp} \ , \qquad \Phi_2 = (1 + i e^{2A_4} dx^4 \wedge dx^5) \wedge \phi^1_{\mp} \ . \tag{5.16}$$

Furthermore, (5.14) can also be found as a consequence of the ten-dimensional system (4.30), (4.31). [95] also give an interpretation of the system in terms of calibrations, along the lines of [99].

### 5.3.2. $\mathbf{AdS_7} \times M_3$

As we anticipated, we will now use the fact that AdS can be used as a warped product of Minkowski space with a line. We would like to classify solutions of the type $\mathrm{AdS}_7 \times M_3$. These in general will have a metric

$$ds^2_{10} = e^{2A_3} ds^2_{\mathrm{AdS}_7} + ds^2_{M_3} \tag{5.17}$$

where $A_3$ is a new warping function (different from the $A_4$ in (5.13)). Since

$$ds^2_{\mathrm{AdS}_7} = \frac{d\rho^2}{\rho^2} + \rho^2 ds^2_{\mathrm{Mink}_6} \ , \tag{5.18}$$

(5.17) can be put in the form (5.13) if we take

$$e^{A_4} = \rho e^{A_3} \ , \qquad ds^2_{M_4} = \frac{e^{2A_3}}{\rho^2} d\rho^2 + ds^2_{M_3} \ . \tag{5.19}$$

A genuine $\mathrm{AdS}_7$ solution is one where not only the metric is of the form (5.18), but where there are also no fields that break its SO(6,2) invariance. This can be easily achieved by additional assumptions: for example, $A_3$ should be a function of $M_3$. The fluxes $F$ and $H$, which in section 5.3.1 were arbitrary forms on $M_4$, should now be forms on $M_3$. For IIA, $F = F_0 + F_2 + F_4$: in order not to break $\mathrm{SO}(6,2)$, we impose $F_4 = 0$, since it would necessarily have a leg along $\mathrm{AdS}_7$; for IIB, $F = F_1 + F_3$.

Following this logic, solutions to type II equations of motion of the form AdS$_7 \times M_3$ are a subclass of solutions of the form Mink$_6 \times M_4$. In appendix A.1, we also show how the AdS$_7 \times M_3$ supercharges get translated in the Mink$_6 \times M_4$ framework, and that the internal spinors have equal norm, as we anticipated in section 5.3.1. Using (A.10), we also learn how to express the $\phi^{1,2}$ in (5.12) in terms of bilinears of spinors $\chi_{1,2}$ on $M_3$:

$$\phi^1_\mp = \frac{1}{2}\left(\psi^1_\mp + ie^{A_3}\frac{d\rho}{\rho}\wedge\psi^1_\pm\right) , \qquad \phi^2_\mp = \mp\frac{1}{2}\left(\psi^2_\mp + ie^{A_3}\frac{d\rho}{\rho}\wedge\psi^2_\pm\right) , \quad (5.20)$$

with

$$\psi^1 = \chi_1 \otimes \chi_2^\dagger , \qquad \psi^2 = \chi_1 \otimes \chi_2^{c\,\dagger} . \quad (5.21)$$

As in section 5.3.1, we have implicitly mapped forms to bispinors via the Clifford map, and in (5.20) the subscripts $_\pm$ refer to taking the even or odd form part. (Recall also that $\phi^{1,2}_-$ is relevant to IIA, and $\phi^{1,2}_+$ to IIB; see (5.14).) The spinors $\chi_{1,2}$ have been taken to have unit norm.

$\psi^{1,2}$ are differential forms on $M_3$, but not just any forms. (5.21) imply that they should obey some algebraic constraints. Those constraints could be interpreted in a fancy way as saying that they define an identity$\times$identity structure on $TM_3 \oplus T^*M_3$. However, three-dimensional spinorial geometry is simple enough that we can avoid such language: rather, in section 5.4 we will give a parameterization that will allow us to solve all the algebraic constraints resulting from (5.21).

We can now use (5.20) in (5.14). Each of those equations can now be decomposed in a part that contains $d\rho$ and one that does not. Thus, the number of equations would double. However, for (5.14a), (5.14b) and (5.14c), the part that does not contain $d\rho$ actually follows from the part that does. The "norm" equation, (5.14e), simply reduces to a similar equation for a three-dimensional norm. Summing up:

$$d_H \mathrm{Im}(e^{3A_3-\phi}\psi^1_\pm) = -2e^{2A_3-\phi}\mathrm{Re}\psi^1_\mp , \quad (5.22a)$$

$$d_H \mathrm{Re}(e^{5A_3-\phi}\psi^1_\pm) = 4e^{4A_3-\phi}\mathrm{Im}\psi^1_\mp , \quad (5.22b)$$

$$d_H(e^{5A_3-\phi}\psi^2_\pm) = -4ie^{4A_3-\phi}\psi^2_\mp , \quad (5.22c)$$

$$\pm\frac{1}{8}e^\phi *_3 \lambda F = dA_3 \wedge \mathrm{Im}\psi^1_\pm + e^{-A_3}\mathrm{Re}\psi^1_\mp , \quad (5.22d)$$

$$dA_3 \wedge \mathrm{Re}\psi^1_\mp = 0 , \quad (5.22e)$$

$$(\psi^{1,2}_+, \overline{\psi^{1,2}_-}) = -\frac{i}{2} ; \quad (5.22f)$$

again with the upper sign for IIA, and the lower for IIB.

The system (5.22) is equivalent to supersymmetry for AdS$_7 \times M_3$. As we show in appendix A.1, a supersymmetric AdS$_7 \times M_3$ solution can be viewed as a supersymmetric Mink$_6 \times M_4$ solution, and for this the system (5.14) is equivalent to supersymmetry. (5.22) can also be obtained directly from the ten-dimensional system (4.30), (4.31), but extra work is needed to show that (4.31) are fully redundant.

In (5.22) the cosmological constant of AdS$_7$ does not appear directly, since we have taken its radius to be one in (5.18). We did so because a non-unit radius can be reabsorbed in the factor $e^{2A_3}$ in (5.17).

Before we can solve (5.22), we have to solve the algebraic constraints that follow from the definition of $\psi^{1,2}$ in (5.21); we will now turn to this problem.

## 5.4. Parameterization of the pure spinors

In section 5.3.2 we obtained a system of differential equations, (5.22), that is equivalent to supersymmetry for an AdS$_7 \times M_3$ solution. The $\psi^{1,2}$ appearing in that system are not arbitrary forms; they should have the property that they can be rewritten as bispinors (via the Clifford map $dx^{i_1} \wedge \ldots \wedge dx^{i_k} \mapsto \gamma^{i_1 \cdots i_k}$) as in (5.21). In this section, we will obtain a parameterization for the most general set of $\psi^{1,2}$ that has this property. This will allow us to analyze (5.22) more explicitly in section 5.5.

We will begin in section 5.4.1 with a quick review of the case $\chi_1 = \chi_2$, and then show in section 5.4.2 how to attack the more general situation where $\chi_1 \neq \chi_2$.

### 5.4.1. One spinor

We will use the Pauli matrices $\sigma_i$ as gamma matrices, and use $B_3 = \sigma_2$ as a conjugation matrix (so that $B_3 \sigma_i = -\sigma_i^t B_3 = -\sigma_i^* B_3$). We will define

$$\chi^c \equiv B_3 \chi^* \ , \qquad \overline{\chi} \equiv \chi^t B_3 \ ; \tag{5.23}$$

notice that $\chi^{c\,\dagger} = \chi^t B_3^\dagger = \overline{\chi}$.

We will now evaluate $\psi^{1,2}$ in (5.21) when $\chi_1 = \chi_2 \equiv \chi$; as we noted in section 5.3.2, $\chi$ is normalized to one. Notice first a general point about the Clifford map $\alpha_k = \frac{1}{k!} \alpha_{i_1 \ldots i_k} dx^{i_1} \wedge \ldots \wedge dx^{i_k} \mapsto \alpha\!\!\!/_k \equiv \frac{1}{k!} \alpha_{i_1 \ldots i_k} \gamma^{i_1 \cdots i_k}$ in three dimensions (and, more generally, in any odd dimension). Unlike what happens in even dimensions, the antisymmetrized gamma matrices $\gamma^{i_1 \cdots i_k}$ are a redundant basis for bispinors. For example, we see that the slash of the volume form is a number: $\mathrm{vol}\!\!\!/_3 = \sigma^1 \sigma^2 \sigma^3 = i$. More generally we have

$$\alpha\!\!\!/ = -i \, {*\!\!\!/}\lambda\alpha . \tag{5.24}$$

In other words, when we identify a form with its image under the Clifford map, we lose some information: we effectively have an equivalence $\alpha \cong -i * \lambda\alpha$. When evaluating $\psi^{1,2}$, we can give the corresponding forms as an even form, or as an odd form, or as a mix of the two.

Let us first consider $\chi \otimes \chi^\dagger$. We can choose to express it as an odd form. In its Fierz expansion, both its one-form part and its three-form part are a priori non-zero; we can parameterize them as

$$\chi \otimes \chi^\dagger = \frac{1}{2}(e_3 - i\text{vol}_3) \ . \tag{5.25}$$

(We can also write this in a mixed even/odd form as $\chi \otimes \chi^\dagger = \frac{1}{2}(1 + e_3)$; recall that the right hand sides have to be understood with a Clifford map applied to them.) $e_3$ is clearly a real vector, whose name has been chosen for later convenience. The fact that the three-form part is simply $-\frac{i}{2}\text{vol}_3$ follows from $||\chi|| = 1$. Notice also that

$$e_3\chi = \sigma_i \chi e_3^i = \sigma_i \chi\chi^\dagger\sigma^i\chi = \frac{1}{2}(-e_3 - 3i\text{vol}_3)\chi \quad \Rightarrow \quad e_3\chi = \chi \tag{5.26}$$

where we have used (5.25), and that $\sigma_i \alpha_k \sigma^i = (-)^k(3-2k)\alpha_k$ on a $k$-form. (5.26) also implies that $e_3$ has norm one.[8]

Coming now to $\chi \otimes \overline{\chi}$, we notice that the three-form part in its Fierz expansion is zero, since $\overline{\chi}\chi = \chi^t B_3 \chi = 0$. The one-form part is now a priori no longer real; so we write

$$\chi \otimes \overline{\chi} = \frac{1}{2}(e_1 + ie_2) \ . \tag{5.27}$$

Similar manipulations as in (5.26) show that $(e_1 + ie_2)\chi = 0$; using this, we get that

$$e_i \cdot e_j = \delta_{ij} \ . \tag{5.28}$$

In other words, $\{e_i\}$ is a vielbein, as notation would suggest.

## 5.4.2. Two spinors

We will now analyze the case with two spinors $\chi_1 \neq \chi_2$ (again both with norm one). We will proceed in a similar fashion as in [100, Sec. 3.1].

Our aim is to parameterize the bispinors $\psi^{1,2}$ in (5.21). Let us first consider their zero-form parts, $\chi_2^\dagger\chi_1$ and $\chi_2^{c\,\dagger}\chi_1$. The parameterization (5.26) can be applied to both $\chi_1$ and $\chi_2$, resulting in two one-forms $e_3^i$. (This notation is a bit inconvenient, but these two one-forms will cease to be useful very soon.) Using then (5.25) twice, we see that

$$|\chi_2^\dagger \chi_1|^2 = \chi_2^\dagger \chi_1 \chi_1^\dagger \chi_2 = \mathrm{Tr}(\chi_1 \chi_1^\dagger \chi_2 \chi_2^\dagger) = \frac{1}{4}\,\mathrm{Tr}\left((1+e_3^1)(1+e_3^2)\right)$$
$$= \frac{1}{2}(1 + e_3^1 \cdot e_3^2)\ . \tag{5.29}$$

[8] An alternative, perhaps more amusing, way of seeing this is to consider $\chi \otimes \chi^\dagger$ as a two-by-two spinorial matrix. It has rank one, which will be true if and only if its determinant is one. Using that $\det(A) = \frac{1}{2}(\text{Tr}(A)^2 - \text{Tr}(A^2))$ for 2×2 matrices, one gets easily that $e_3$ has norm one.

Similarly we have

$$|\chi_2^{c\,\dagger}\chi_1|^2 = \mathrm{Tr}(\chi_1 \chi_1^{c\,\dagger} \chi_2 \chi_2^{c\,\dagger}) = \frac{1}{4}\,\mathrm{Tr}\left((1+e_3^1)(1-e_3^2)\right) = \frac{1}{2}(1 - e_3^1 \cdot e_3^2)$$
$$= 1 - |\chi_2^\dagger \chi_1|^2\ . \tag{5.30}$$

Both $|\chi_2^\dagger \chi_1|^2$ and $|\chi_2^{c\,\dagger}\chi_1|^2$ are positive and $\leq 1$. Thus we can parameterize $\chi_2^\dagger \chi_1 = e^{ia}\cos(\psi)$, $\chi_2^{c\,\dagger}\chi_1 = e^{ib}\sin(\psi)$. (The name of this angle should not be confused with the forms $\psi^{1,2}$.) By suitably multiplying $\chi_1$ and $\chi_2$ by two phases, we can assume $a = -\frac{\pi}{2}$ and $b = \frac{\pi}{2}$; we will reinstate generic values of these phases at the very end. Thus we have

$$\chi_2^\dagger \chi_1 = -i\cos(\psi)\ , \qquad \chi_2^{c\,\dagger}\chi_1 = i\sin(\psi)\ . \tag{5.31}$$

Just as in [100, Sec. 3.1], we can now introduce

$$\chi_0 = \frac{1}{2}(\chi_1 - i\chi_2)\ , \qquad \tilde{\chi}_0 = \frac{1}{2}(\chi_1 + i\chi_2)\ . \tag{5.32}$$

In three Euclidean dimensions, a spinor and its conjugate form a (pointwise) basis of the space of spinors. For example, $\chi_0$ and $\chi_0^c$ are a basis. We can then expand $\tilde{\chi}_0$ on this basis. Actually, its projection on $\chi_0$ vanishes, due to (5.31): $\chi_0^\dagger \tilde{\chi}_0 = \frac{i}{4}(\chi_1^\dagger \chi_2 + \chi_2^\dagger \chi_1) = 0$. With a few more steps we get

$$\tilde{\chi}_0 = \frac{\chi_0^{c\,\dagger}\tilde{\chi}_0}{||\chi_0||^2}\chi_0^c = \tan\left(\frac{\psi}{2}\right)\chi_0^c\ . \tag{5.33}$$

We can now invert (5.32) for $\chi_1$ and $\chi_2$, and use (5.33). It is actually more symmetric-looking to define $\chi_0 \equiv \cos\left(\frac{\psi}{2}\right)\chi$, to get

$$\chi_1 = \cos\left(\frac{\psi}{2}\right)\chi + \sin\left(\frac{\psi}{2}\right)\chi^c\ , \quad \chi_2 = i\left(\cos\left(\frac{\psi}{2}\right)\chi - \sin\left(\frac{\psi}{2}\right)\chi^c\right)\ . \tag{5.34}$$

We have thus obtained a parameterization of two spinors $\chi_1$ and $\chi_2$ in terms of a single spinor $\chi$ and of an angle $\psi$. Let us count our parameters, to see if our result makes sense. A spinor $\chi$ of norm 1 accounts for 3 real parameters; $\psi$ is one more. We should also recall we have rotated both $\chi_{1,2}$ by a phase at the beginning of our computation, to make things easier. We have a grand total of 6 real parameters, which is correct for two spinors of norm 1 in three dimensions.

We can now use the parameterization (5.34), and the bilinears (5.25), (5.27) obtained in section 5.4.1:

$$\begin{aligned}\chi_1 \otimes \chi_2^\dagger = -i \bigg[&\cos^2\left(\frac{\psi}{2}\right)\chi\chi^\dagger - \sin^2\left(\frac{\psi}{2}\right)\chi^c\chi^{c\,\dagger} + \\ &+ \cos\left(\frac{\psi}{2}\right)\sin\left(\frac{\psi}{2}\right)(\chi^c\chi^\dagger - \chi\chi^{c\,\dagger})\bigg] \\ &= -\frac{i}{2}\left[e_3 - i\sin(\psi)e_2 - i\cos(\psi)\mathrm{vol}_3\right] \ . \end{aligned} \tag{5.35}$$

A computation along these lines allows us to evaluate $\chi_1 \otimes \overline{\chi_2}$ as well. We can also reinstate at this point the phases of $\chi_1$ and $\chi_2$, absorbing the overall factor $-i$. The bilinear in (5.35) is expressed as an odd form, but we will also need its even-form expression; this can be obtained by using (5.24). Recalling the definition (5.21), we get:

$$\psi_+^1 = \frac{e^{i\theta_1}}{2}\left[\cos(\psi) + e_1 \wedge (-ie_2 + \sin(\psi)e_3)\right] \ , \tag{5.36a}$$

$$\psi_-^1 = \frac{e^{i\theta_1}}{2}\left[e_3 - i\sin(\psi)e_2 - i\cos(\psi)\mathrm{vol}_3\right] \ , \tag{5.36b}$$

$$\psi_+^2 = \frac{e^{i\theta_2}}{2}\left[\sin(\psi) - (ie_2 + \cos(\psi)e_1) \wedge e_3\right] \ , \tag{5.36c}$$

$$\psi_-^2 = \frac{e^{i\theta_2}}{2}\left[e_1 + i\cos(\psi)e_2 - i\sin(\psi)\mathrm{vol}_3\right] \ . \tag{5.36d}$$

Notice that these satisfy automatically (5.22f).

Armed with this parameterization, we will now attack the system (5.22) for AdS$_7 \times M_3$ solutions.

## 5.5. General results

In section 5.3.2, we have obtained the system (5.22), equivalent to supersymmetry for AdS$_7 \times M_3$ solutions. The $\psi_\pm^{1,2}$ appearing in that system are not just any forms; they should have the property that they can be written as bispinors as in (5.21). In section 5.4.2, we have obtained a parameterization for the most general set of $\psi_\pm^{1,2}$ that fulfills that constraint; it is (5.36), where $\{e_i\}$ is a vielbein.

Thus we can now use (5.36) into the differential system (5.22), and explore its consequences.

### 5.5.1. Purely geometrical equations

We will start by looking at the equations in (5.22) that do not involve any fluxes. These are (5.22e), and the lowest-component form part of (5.22a), (5.22b) and (5.22c).

First of all, we can observe quite quickly that the IIB case cannot possibly work. (5.22a), (5.22b) and (5.22c) all have a zero-form part coming from their right-hand side, which, using (5.36), read respectively

$$\cos(\psi)\cos(\theta_1) = 0 \ , \qquad \cos(\psi)\sin(\theta_1) = 0 \ , \qquad \sin(\psi)e^{i\theta_2} = 0 \ . \tag{5.37}$$

These cannot be satisfied for any choice of $\psi$, $\theta_1$ and $\theta_2$. Thus we can already exclude the IIB case.[9]

Having disposed of IIB so quickly, we will devote the rest of the analysis to IIA. Actually, we already know that we can get something new only with non-zero Romans mass, $F_0 \neq 0$. This is because for $F_0 = 0$ we can lift to an eleven-dimensional supergravity solution AdS$_7 \times N_4$. There, we only have a four-form flux $G_4$ at our disposal, and the only way not to break the SO(6,2) invariance of AdS$_7$ is to switch it on along the internal four-manifold $N_4$. This is the Freund–Rubin Ansatz, which requires $N_4$ to admit a Killing spinor. This means that the cone $C(N_4)$ over $N_4$ admits a covariantly constant spinor; but in five dimensions the only manifold with restricted holonomy is $\mathbb{R}^5$ (or one of its orbifolds, of the form $\mathbb{C}^2/\Gamma \times \mathbb{R}$). Thus we know already that all solutions with $F_0 = 0$ lift to AdS$_7 \times S^4$ (or AdS$_7 \times S^4/\Gamma$) in eleven dimensions. (In fact we will see later how AdS$_7 \times S^4$ reduces to ten dimensions.) We will thus focus on $F_0 \neq 0$, and use the case $F_0 = 0$ as a control.

In IIA, the lowest-degree equations of (5.22a), (5.22b) and (5.22c) are one-forms; they are less dramatic than (5.37), but still rather interesting. Using (5.36), after some manipulations we get

$$\begin{aligned} e_1 &= -\frac{1}{4}e^A\sin(\psi)d\theta_2 \ , \qquad e_2 = \frac{1}{4}e^A(d\psi + \tan(\psi)d(5A-\phi)) \ , \\ e_3 &= \frac{1}{4}e^A\left(-\cos(\psi)d\theta_1 + \frac{\cot(\theta_1)}{\cos(\psi)}d(5A-\phi)\right) \ , \end{aligned} \tag{5.38}$$

[9] This quick death is reminiscent of the fate of AdS$_4 \times M_6$ with SU(3) structure in IIB. The system in [94] has a zero-form equation and two-form equation coming from the right-hand side of its fluxless equation, which look like $\cos(\theta) = 0 = \sin(\theta)J$, where $\theta$ is an angle similar to $\psi$ in (5.36). This is consistent with a no-go found with lengthier computations in [101].

and

$$xdx = (1 + x^2)d\phi - (5 + x^2)dA \ , \tag{5.39}$$

where

$$x \equiv \cos(\psi) \sin(\theta_1) \ , \tag{5.40}$$

and we have dropped the subscript $_3$ on the warping function: $A \equiv A_3$ from now on. Notice that (5.38) determine the vielbein. Usually (i.e. in other dimensions), the geometrical part of the differential system coming from supersymmetry gives the *derivative* of the forms defining the metric. In this case, the forms themselves are determined in terms of derivatives of the angles appearing in our parameterizations. This will allow us to give a more complete and concrete classification than is usually possible.

We still have (5.22e). Notice that (5.22a) allows to write it as $dA \wedge d(e^{3A-\phi}x) = 0$. Using also (5.39), we get

$$dA \wedge d\phi = 0 \ . \tag{5.41}$$

This means that $\phi$ is functionally dependent on $A$:[10]

$$\phi = \phi(A) \ . \tag{5.42}$$

(5.39) then means that $x$ too is functionally dependent on $A$: $x = x(A)$.

### 5.5.2. Fluxes

So far, we have analyzed (5.22e), and the one-form part of (5.22a), (5.22b) and (5.22c). Before we look at their three-form part too, it is convenient to look at (5.22d), which gives us the RR flux, for reasons that will become apparent.

First we compute $F_0$ from (5.22d):

$$F_0 = 4xe^{-A-\phi}\frac{3 - \partial_A\phi}{5 - 2x^2 - \partial_A\phi} \ . \tag{5.43}$$

The Bianchi identity for $F_0$ says that it should be (piecewise) constant. It will thus be convenient to use (5.43) to eliminate $\partial_A\phi$ from our equations.

Before we go on to analyze our equations, let us also introduce the new angle $\beta$ by

$$\sin^2(\beta) = \frac{\sin^2(\psi)}{1 - x^2} \ . \tag{5.44}$$

We can now use $x$ as defined in (5.40) to eliminate $\theta_1$, and $\beta$ to eliminate $\psi$. This turns out to be very convenient in the following, especially in our analysis of the metric in section 5.5.4 below (which was our original motivation to introduce $\beta$).

[10](5.42) excludes the case where $A$ is constant in a region. However, it is easy to see that this case cannot work. Indeed, in this case (5.39) can be integrated as $e^\phi \propto \sqrt{1 - x^2}$, which is incompatible with (5.43) below.

After these preliminaries, let us give the expression for $F_2$ as one obtains it from (5.22d):

$$F_2 = \frac{1}{16}\sqrt{1-x^2}e^{A-\phi}(xe^{A+\phi}F_0 - 4)\mathrm{vol}_{S^2} \ , \tag{5.45}$$

where

$$\mathrm{vol}_{S^2} = \sin(\beta)d\beta \wedge d\theta_2 \tag{5.46}$$

is formally identical to the volume form for a round $S^2$ with coordinates $\{\beta, \theta_2\}$. We will see later that this is no coincidence.

Finally, let us look at the three-form part of (5.22a), (5.22b) and (5.22c). One of them can be used to determine $H$:

$$H = \frac{1}{8}e^{2A}\sqrt{1-x^2}\,\frac{6 + xF_0e^{A+\phi}}{4 + xF_0e^{A+\phi}}\,dx \wedge \mathrm{vol}_{S^2} \ , \tag{5.47}$$

while the other two turn out to be identically satisfied.

Our analysis is not over: we should of course now impose the equation of motion, and the Bianchi identities for our fluxes. The equation of motion for $F_2$, $d * F_2 + H * F_0 = 0$, follows automatically from (5.22d), much as it happens in the pure spinor system for AdS$_4 \times M_6$ solutions [94]. We should then impose the Bianchi identity for $F_2$, which reads $dF_2 - HF_0 = 0$ (away from sources). This does not follow manifestly from (5.22d), but in fact it is a consequence of the explicit expressions (5.43), (5.45) and (5.47) above. When $F_0 \neq 0$, it also implies that the $B$ field such that $H = dB$ can be locally written as

$$B_2 = \frac{F_2}{F_0} + b \tag{5.48}$$

for a closed two-form $b$. Using a gauge transformation, it can be assumed to be proportional (by a constant) to $\mathrm{vol}_{S^2}$; we then have that it is a constant, $\partial_A b = 0$.

The equation of motion for $H$, which reads for us $d(e^{7A-2\phi} *_3 H) = e^{7A}F_0 *_3 F_2$ (again away from sources), is also automatically satisfied, as shown in general in [102]. Finally, since we have checked all the conditions for preserved supersymmetry, the Bianchi identities and the equations of motion for the fluxes, the equations of motion for the dilaton and for the metric will now follow [87].

### 5.5.3. The system of ODEs

Let us now sum up the results of our analysis of (5.22). Most of our equations determine some fields: (5.38) give the vielbein, and (5.43), (5.45), (5.47) give the fluxes. The only genuine differential equations we have are (5.39), and the condition that $F_0$ should be constant. Recalling that $\phi$ is functionally dependent on $A$, (5.42), these two equations can be written as

$$\partial_A \phi = 5 - 2x^2 + \frac{8x(x^2-1)}{4x - F_0 e^{A+\phi}} \,, \tag{5.49a}$$

$$\partial_A x = 2(x^2-1)\,\frac{x e^{A+\phi} F_0 + 4}{4x - F_0 e^{A+\phi}} \,. \tag{5.49b}$$

We thus have reduced the existence of a supersymmetric solution of the form AdS$_7 \times M_3$ in IIA to solving the system of ODEs (5.49). It might look slightly unsettling that we are essentially using at this point $A$ as a coordinate, which might not always be a wise choice (since $A$ might not be monotonic). For that matter, our analysis has so far been completely local; we will start looking at global issues in section 5.5.4, and especially 5.5.6.

Unfortunately we have not been able to find analytic solutions to (5.49), other than in the $F_0 = 0$ case (which we will see in section 5.6.1). For the more interesting $F_0 \neq 0$ case, we can gain some intuition by noticing that the system becomes autonomous (i.e. it no longer has explicit dependence on the "time" variable $A$) if one defines $\tilde{\phi} \equiv \phi + A$. The system for $\{\partial_A \tilde{\phi}, \partial_A x\}$ can now be thought of as a vector field in two dimensions; we plot it in figure 5.4.

We will study the system (5.49) numerically in section 5.6. Before we do that, we should understand what boundary conditions we should impose. We will achieve this by analyzing global issues about our setup, that we have so far ignored.

### 5.5.4. Metric

The metric

$$ds^2_{M_3} = e_a e_a \tag{5.50}$$

following from (5.38) looks quite complicated. However, it simplifies enormously if we rewrite it in terms of $\beta$ in (5.44):[11]

$$ds^2_{M_3} = e^{2A}(1-x^2)\left[\frac{16}{(4x - e^{A+\phi}F_0)^2}\, dA^2 + \frac{1}{16}\, ds^2_{S^2}\right] \,, \tag{5.51}$$

[11] In fact, the definition of $\beta$ was originally found by trying to understand the global properties of the metric (5.50). Looking at a slice $x =$const, one finds that the metric in $\{\theta_1, \theta_2\}$ has constant positive curvature; the definition of $\beta$ becomes then natural. Nontrivially, this definition also gets rid of non-diagonal terms of the type $dA d\theta_1$ that would arise from (5.38).

with $ds^2_{S^2} = d\beta^2 + \sin^2(\beta)d\theta_2^2$ the metric of round $S^2$. *Without any Ansatz*, the metric has taken the form of a fibration of a round $S^2$, with coordinates $\{\beta, \theta_2\}$, over an interval with coordinate $A$. Notice that none of the scalars appearing in (5.51) (and indeed in the fluxes (5.43), (5.45), (5.47)) were originally intended as coordinates, but rather as functions in the parameterization of the pure spinors $\psi^{1,2}$. Usually, one would then need to introduce coordinates independently, and to make an Ansatz about how all functions should depend on those coordinates, sometimes imposing the presence of some particular isometry group in the process.

Here, on the other hand, the functions we have introduced are suggesting themselves as coordinates to us rather automatically. Since so far our expressions for the metric and fluxes were local, we are free to take their suggestion. We will take $\beta$ to be in the range $[0, \pi]$, and $\theta_2$ to be periodic with period $2\pi$, so that together they describe an $S^2$ as suggested by (5.51), and also by the two-form (5.46) that appeared in (5.45), (5.47).[12]

It is not hard to understand why this $S^2$ has emerged. The holographic dual of any solutions we might find is a $(1, 0)$ CFT in six dimensions. Such a theory would have SU(2) R-symmetry; an SU(2) isometry group should then appear naturally on the gravity side as well. This is what we are seeing in (5.51).

The fact that the $S^2$ in (5.51) is rotated by R-symmetry also helps to explain a possible puzzle about IIB. Often, given a IIA solution, one can produce a IIB one via T-duality along an isometry. All the Killing vectors of the $S^2$ in (5.51) vanish in two points; T-dualizing along any such direction would produce a non-compact solution in IIB, but still a valid one. But the IIB case died very quickly in section 5.5.1; there are no solutions, not even non-compact or singular ones. Here is how this puzzle is resolved. Since the SU(2) isometry group of the $S^2$ is an R-symmetry, supercharges transform as a doublet under it (we will see this more explicitly in section 5.5.5). Thus even the strange IIB geometry produced by T-duality along a U(1) isometry of $S^2$ would not be supersymmetric.

Even though we have promoted $\beta$ and $\theta_2$ to coordinates, it is hard to do the same for $A$, which actually enters in the seven-dimensional metric (see (5.17)). We would like to be able to cover cases where $A$ is non-monotonic. One possibility would be to use $A$ as a coordinate piecewise. We find it clearer, however, to introduce a coordinate $r$ defined by $dr = 4e^A \frac{\sqrt{1-x^2}}{4x - e^{A+\phi}F_0} dA$, so that the metric now reads

$$ds^2_{M_3} = dr^2 + \frac{1}{16} e^{2A}(1 - x^2) ds^2_{S^2} \; . \tag{5.52}$$

[12] A slight variation is to take $\mathbb{RP}^2 = S^2/\mathbb{Z}_2$ instead of $S^2$; this will not play much of a role in what follows, except for some solutions with O6-planes that we will mention in sections 5.6.1 and 5.6.2.

In other words, $r$ measures the distance along the base of the $S^2$ fibration. Now $A$, $x$ and $\phi$ have become functions of $r$. From (5.49) and the definition of $r$ we have

$$\begin{aligned} \partial_r \phi &= \frac{1}{4}\frac{e^{-A}}{\sqrt{1-x^2}}(12x + (2x^2-5)F_0 e^{A+\phi}) \ , \\ \partial_r x &= -\frac{1}{2}e^{-A}\sqrt{1-x^2}(4 + xF_0 e^{A+\phi}) \ , \\ \partial_r A &= \frac{1}{4}\frac{e^{-A}}{\sqrt{1-x^2}}(4x - F_0 e^{A+\phi}) \ . \end{aligned} \tag{5.53}$$

We have introduced a square root in the system, but notice that $-1 \leq x \leq 1$ already follows from requiring that $ds^2_{M_3}$ in (5.51) has positive signature. (We choose the positive branch of the square root.)

Let us also record here that the NS three-form also simplifies in the coordinates introduced in this section:

$$H = -(6e^{-A} + xF_0 e^{\phi})\text{vol}_3 \ , \tag{5.54}$$

where vol$_3$ is the volume form of the metric $ds^2_{M_3}$ in (5.52) or (5.51).

We have obtained so far that the metric is the fibration of an $S^2$ (with coordinates $(\beta, \theta_2)$) over a one-dimensional space. The SU(2) isometry group of the $S^2$ is to be identified holographically with the R-symmetry group of the $(1,0)$ superconformal dual theory. For holographic applications, we would actually like to know whether the total space of the $S^2$ fibration can be made compact. We will look at this issue in section 5.5.6. Right now, however, we would like to take a small detour and see a little more clearly how the R-symmetry SU(2) emerges in the pure spinors $\psi^{1,2}$.

### 5.5.5. SU(2) covariance

We have just seen that the metric takes the particularly simple form (5.52) in coordinates $(r, \beta, \theta_2)$; the appearance of the $S^2$ is related to the SU(2) R-symmetry group of the $(1,0)$ holographic dual.

Since these coordinates are so successful with the metric, let us see whether they also simplify the pure spinors $\psi^{1,2}$. We can start by the zero-form parts of (5.36), which read

$$\psi^1_0 = ix + \sqrt{1-x^2}\,\cos(\beta) \ , \qquad \psi^2_0 = \sqrt{1-x^2}\,\sin(\beta)e^{i\theta_2} \ . \tag{5.55}$$

Recalling that $(\beta, \theta_2)$ are the polar coordinates on the $S^2$ (see the expression of $ds^2_{S^2}$ in (5.51)), we recognize in (5.55) the appearance of the $\ell = 1$ spherical

harmonics

$$y^\alpha = \{\sin(\beta)\cos(\theta_2), \sin(\beta)\sin(\theta_2), \cos(\beta)\} \ . \tag{5.56}$$

Notice that $y^3$ appears in $\psi^1 = \chi_1 \otimes \chi_2^\dagger$, while $y^1 + iy^2$ appears in $\psi^2 = \chi_1 \otimes \chi_2^{c\,\dagger}$. This suggests that we introduce a 2×2 matrix of bispinors. From (A.4) we see that for IIA $\binom{\chi_1}{\chi_1^c}$ and $\binom{\chi_2}{-\chi_2^c}$ are both SU(2) doublets, so that it is natural to define

$$\Psi = \begin{pmatrix} \chi_1 \\ \chi_1^c \end{pmatrix} \otimes (\chi_2^\dagger \, , -\chi_2^{c\,\dagger}) = \begin{pmatrix} \psi^1 & \psi^2 \\ (-)^{\rm deg}(\psi^2)^* & -(-)^{\rm deg}(\psi^1)^* \end{pmatrix} \ , \tag{5.57}$$

where $(-)^{\rm deg}$ acts as $\pm$ on a even (odd) form. The even-form part can then be written as

$$\Psi_+^{ab} = i{\rm Im}\psi_+^1 \, 1_2 + \left({\rm Re}\psi_+^2 \sigma_1 - {\rm Im}\psi_+^2 \sigma_2 + {\rm Re}\psi_+^1 \sigma_3\right) \ , \tag{5.58a}$$

where $\sigma_\alpha$ are the Pauli matrices while the odd-form part is

$$\Psi_-^{ab} = {\rm Re}\psi_-^1 \, 1_2 + i\left({\rm Im}\psi_-^2 \sigma_1 + {\rm Re}\psi_-^2 \sigma_2 + {\rm Im}\psi_-^1 \sigma_3\right) \ . \tag{5.58b}$$

(5.58) shows more explicitly how the R-symmetry SU(2) acts on the bispinors $\Psi^{ab}$, which split between a singlet and a triplet. If we go back to our original system (5.22), we see that (5.22a), (5.22d), (5.22e) each behave as a singlet, while (5.22b), (5.22c) behave as a triplet — thanks also to the fact that the factor $e^{5A-\phi}$ appears in both those equations.

More concretely, (5.55) can now be written as

$$\Psi_0^{ab} = ix \, 1_2 + \sqrt{1-x^2}\, y^\alpha \sigma_\alpha \ ; \tag{5.59a}$$

the one-form part reads

$$\Psi_1^{ab} = \sqrt{1-x^2} dr \, 1_2 + i\left[xy^\alpha dr + \frac{1}{4} e^A \sqrt{1-x^2}\, dy^\alpha\right]\sigma_\alpha \ . \tag{5.59b}$$

The rest of $\Psi^{ab}$ can be determined by (5.24): $\Psi_3^{ab} = -i *_3 \Psi_0^{ab} = -i\Psi_0^{ab}\,{\rm vol}_3$, $\Psi_2^{ab} = -i *_3 \Psi_1^{ab}$. (The three-dimensional Hodge star can be easily computed from (5.52).)

We will now turn to the global analysis of the metric (5.52).

### 5.5.6. Topology

We now wonder whether the $S^2$ fibration in (5.51) can be made compact.

One possible strategy would be for $r$ to be periodically identified, so that the topology of $M_3$ would become $S^1 \times S^2$. This is actually impossible: from (5.53) we have

$$\partial_r(xe^{3A-\phi}) = -2\sqrt{1-x^2}e^{2A-\phi} \le 0\ . \tag{5.60}$$

This can also be derived quickly from (5.22a) using the singlet part of (5.59). Now, $xe^{3A-\phi}$ is continuous;[13] for $r$ to be periodically identified, $xe^{3A-\phi}$ should be a periodic function. However, thanks to (5.60), it is nowhere-increasing. It also cannot be constant, since $x$ would be $\pm 1$ for all $r$, which makes the metric in (5.51) vanish. Thus $r$ cannot be periodically identified.

We then have to look for another way to make $M_3$ compact. The only other possibility is in fact to shrink the $S^2$ at two values of $r$, which we will call $r_{\rm N}$ and $r_{\rm S}$; the topology of $M_3$ would then be $S^3$. The subscripts stand for "north" and "south"; we can visualize these two points as the two poles of the $S^3$, and the other, non-shrunk copies of $S^2$ over any $r \in (r_{\rm N}, r_{\rm S})$ to be the "parallels" of the $S^3$. Of course, since (5.53) does not depend on $r$, we can assume without any loss of generality that $r_{\rm N} = 0$.

We will now analyze this latter possibility in detail.

## 5.5.7. Local analysis around poles

We have just suggested to make $M_3$ compact by having the $S^2$ fiber over an interval $[r_{\rm N}, r_{\rm S}]$, and by shrinking it at the two endpoints. In this case $M_3$ would be homeomorphic to $S^3$.

To realize this idea, from (5.52) we see that $x$ should go to 1 or $-1$ at the two poles $r_{\rm N}$ and $r_{\rm S}$. To make up for the vanishing of the $\sqrt{1-x^2}$'s in the denominators in (5.53), we should also make the numerators vanish. This is accomplished by having $e^{A+\phi} = \pm 4/F_0$ at those two poles (which is obviously only possible when $F_0 \neq 0$). We can now also see that $\partial_r x \sim -4e^{-A}\sqrt{1-x^2} \le 0$ around the poles. Since, as we noticed earlier, $-1 \le x \le 1$, $x$ should actually be 1 at $r_{\rm N}$, and $-1$ at $r_{\rm S}$. Summing up:

$$\left\{x = 1,\ e^{A+\phi} = \frac{4}{F_0}\right\} \ \text{ at } r = r_{\rm N}\ , \quad \left\{x = -1,\ e^{A+\phi} = -\frac{4}{F_0}\right\} \ \text{ at } r = r_{\rm S}\ . \tag{5.61}$$

Since we made both numerators and denominators in (5.53) vanish at the poles, we should be careful about what happens in the vicinity of those points. We want to study the system around the boundary conditions (5.61) in a power-series approach. (The same could also be done directly with (5.49).) Let us first expand around $r_{\rm N}$. As mentioned earlier, thanks to translational invariance in $r$ we can assume $r_{\rm N} = 0$ without any loss of generality. We get

$$\begin{aligned}
\phi &= -A_0^+ + \log\left(\frac{4}{F_0}\right) - 5e^{-2A_0^+}r^2 + \frac{172}{9}e^{-4A_0^+}r^4 + O(r)^6 \ , \\
x &= 1 - 8e^{-2A_0^+}r^2 + \frac{400}{9}e^{-4A_0^+}r^4 + O(r)^6 \ , \\
A &= A_0^+ - \frac{1}{3}e^{-2A_0^+}r^2 - \frac{4}{27}e^{-4A_0^+}r^4 + O(r)^6 \ .
\end{aligned} \tag{5.62}$$

[13] This might not be fully obvious in presence of D8-branes, but we will see later that it is true even in that case, basically because $\phi$ is a physical field, and $A$ and $x$ appear as coefficients in the metric.

$A_0^+$ here is a free parameter. The way it appears in (5.62) is explained by noticing that (5.53) is symmetric under

$$A \to A + \Delta A \ , \qquad \phi \to \phi - \Delta A \ , \qquad x \to x \ , \qquad r \to e^{\Delta A} r \ . \tag{5.63}$$

Applying (5.62) to (5.52), and setting for a moment $r_{\rm N} = 0$, we find that the metric has the leading behavior

$$ds^2_{M_3} = dr^2 + r^2 ds^2_{S^2} + O(r)^4 = ds^2_{\mathbb{R}^3} + O(r)^4 \ . \tag{5.64}$$

This means that the metric is regular around $r = r_{\rm N}$. The expansion of the fluxes (5.45), (5.47) is

$$F_2 = -\frac{10}{3}F_0 e^{-A_0^+} r^3 {\rm vol}_{S^2} + O(r)^5 \ , \quad H = -10 e^{-A_0^+} r^2 dr \wedge {\rm vol}_{S^2} + O(r)^3 \ . \tag{5.65}$$

As for the $B$ field, recall that it can be written as in (5.48). (5.65) shows that around $r = r_{\rm N} = 0$, the term $F_2/F_0$ is regular as it is, without the addition of $b$; this suggests that one should set $b = 0$. To make this more precise, consider the limit

$$\lim_{r\to 0} \int_{\Delta_r} H = \lim_{r \to 0} \int_{S^2_r} B_2 \tag{5.66}$$

where $\Delta_r$ is a three-dimensional ball such that $\partial \Delta_r = S^2_r$. In (5.48), the first term goes to zero because $x \to 1$; so the limit is equal to $\int_{S^2} b$, which is constant. This constant signals a delta in $H$. So we are forced to conclude that

$$b = 0 \tag{5.67}$$

near the pole. (However, we will see in section 5.5.8 that $b$ can become non-zero if one crosses a D8 while going away from the pole.)

To be more precise, (5.67) should be understood up to gauge transformations. $B$ is not a two-form, but a 'connection on a gerbe', in the sense that it transforms non-trivially on chart intersections: on $U \cap U'$, $B_U - B_{U'}$ can be a 'small' gauge transformation $d\lambda$, for $\lambda$ a 1-form, or more generally a 'large' gauge transformation, namely a two-form whose periods are integer multiples of $4\pi^2$. In our case, if we cover $S^3$ with two patches $U_{\rm N}$ and $U_{\rm S}$, around the equator we can have $B_{\rm N} - B_{\rm S} = N\pi \mathrm{vol}_{S^2}$. In this case $\int_{S^3} H = B_{\rm N} - B_{\rm S} = N\pi\mathrm{vol}_{S^2} = (4\pi^2)N$, in agreement with flux quantization for $H$. Thus $b = 0$ is also gauge equivalent to any integer multiple of $\pi\mathrm{vol}_{S^2}$. In practice, however, we will prefer to work with $b = 0$ around the poles, and perform a gauge transformation whenever

$$\hat{b}(r) \equiv \frac{1}{4\pi}\int_{S^2_r} B_2 \tag{5.68}$$

gets outside the "fundamental region" $[0,\pi]$. In other words, we will consider $\hat{b}$ to be a variable with values in $[0,\pi]$, and let it begin and end at 0 at the two poles. $\hat{b}$ will then wind an integer number $N$ of times around $[0,\pi]$, and this will make sure that $\int_{S^3} H = (4\pi^2)N$, thus taking care of flux quantization for $H$.

So far we have discussed the expansion around the north pole; a similar discussion holds for the expansion around the south pole $r_{\rm S}$. The expressions that replace (5.62), (5.64), (5.65) can be obtained by using the symmetry of (5.53) under

$$x \to -x\ , \quad F_0 \to -F_0\ , \quad r \to -r\ . \tag{5.69}$$

The free parameter $A_0^+$ can now be changed to a possibly different free parameter $A_0^-$.

We have hence checked that the boundary conditions (5.61) are compatible with our system (5.53), and that they give rise to a regular metric at the poles.

## 5.5.8. D8-branes

There is one more ingredient that we will need in section 5.6 to exhibit compact solutions: brane sources. In presence of branes the metric cannot be called regular: their gravitational backreaction will give rise to a singularity. A random singularity would call into question the validity of a solution, since the curvature and possibly the dilaton[14] would diverge there, making the supergravity approximation untrustworthy. We are however sure of the existence of D-branes, in spite of the singularities in their geometry, because we have an open string realization for them.

D8-branes in particular are even more benign, in a way, because the singularity manifests itself simply as a discontinuity in the derivatives of the coefficients in the metric. In general relativity, such a discontinuity would be subject to the so-called Israel junction conditions [103], which are a consequence of the Einstein equations. As we mentioned earlier, in our case, however, supersymmetry guarantees that the equations of motion for the dilaton and metric are automatically satisfied [87]. Hence, the conditions on the first derivatives will follow from imposing continuity of the fields and supersymmetry.

[14] In presence of Romans mass, the string coupling is bounded by the inverse radius of curvature in string units: $e^\phi \lesssim \frac{l_s}{R_{\rm curv}}$, and is actually generically of the order of the bound [55].

Let us be more concrete. We will suppose we have a stack of $n_{\rm D8}$ D8-branes, possibly with a worldvolume gauge field-strength $f_2$ (not to be confused with the RR field-strength $F_2$), which induces a D6-brane charge distribution on it. The Bianchi identity for such an object reads

$$d_H F = \frac{1}{2\pi} n_{\rm D8} e^{\mathcal{F}} \delta \quad \Rightarrow \quad d\tilde{F} = \frac{1}{2\pi} n_{\rm D8} e^{2\pi f_2} \delta \quad (\delta \equiv dr \delta(r)) \ . \tag{5.70}$$

As usual $\mathcal{F} = B_2 + 2\pi f_2$; recall from section 5.3.2 that $F = F_0 + F_2$; and likewise we have defined

$$\tilde{F} \equiv e^{-B_2} F = F_0 + (F_2 - B_2 F_0) \ . \tag{5.71}$$

In other words, $\tilde{F} = F_0 + \tilde{F}_2$, with $\tilde{F}_2 = F_2 - B_2 F_0$. Since $\tilde{F}_2$ is closed away from sources, it makes sense to define

$$n_2 \equiv \frac{1}{2\pi} \int_{S^2} \tilde{F}_2 \ . \tag{5.72}$$

Flux quantization then requires $n_2$ to be an integer, and that

$$F_0 = \frac{n_0}{2\pi} \ , \tag{5.73}$$

with $n_0$ an integer. (We are working in string units where $l_s = 1$.) Integrating now (5.70) across the magnetized stack of D8's gives

$$\Delta n_0 = n_{\rm D8} \ , \quad \Delta \tilde{F}_2 = f_2 \Delta n_0 \ . \tag{5.74}$$

All physical fields should be continuous across the D8 stack. For example, $\Delta\phi = 0$. Also, the coefficients of the metric should not jump; in particular, from (5.17), we see that $\Delta A = 0$. Also, since $x$ appears in front of $ds^2_{S^2}$ in (5.52), we should have $\Delta x = 0$.

Imposing that the $B$ field does not jump is trickier. A first caveat is that $B$ would actually be allowed to jump by a gauge transformation, as discussed in section 5.5.7. However, we find it less confusing to put the intersection between the charts $U_{\rm N}$ and $U_{\rm S}$ away from the D8's, and to treat $\int_{S^2} B_2$ as a periodic variable as described in section 5.5.7.

Thus we will simply impose that $B$ does not jump. First, recall that it can be written as in (5.48), when $F_0 \neq 0$. The $b$ term was shown in (5.67) to be vanishing near the pole, but we will soon see that this conclusion is not valid between D8's. In fact, it is connected to the flux integer $n_2$ defined in (5.72): from (5.48) we have

$$\tilde{F}_2 = -F_0 b \; ; \tag{5.75}$$

integrating this on $S^2$, we get $2\pi n_2 = -F_0 \int_{S^2} b$, or in other words

$$b = -\frac{n_2}{2F_0} \mathrm{vol}_{S^2} \; . \tag{5.76}$$

We can use our result (5.45) for $F_2$; for this section, it will be convenient to define

$$p \equiv \frac{1}{16} x\sqrt{1-x^2} e^{2A} \; , \quad q \equiv \frac{1}{4}\sqrt{1-x^2} e^{A-\phi} \; , \tag{5.77}$$

so that

$$F_2 = (pF_0 - q)\mathrm{vol}_{S^2} \; . \tag{5.78}$$

From this and (5.76) we now have

$$B_2 = \left(p - \frac{q}{F_0} - \frac{n_2}{2F_0}\right) \mathrm{vol}_{S^2} \; . \tag{5.79}$$

Let us call $n_0$, $n_2$ the flux integers on one side of the D8 stack, and $n_0'$, $n_2'$ the fluxes on the other side. Let us at first assume that both $n_0$ and $n_0'$ are non-zero. Then, equating $B$ on the two sides, we see that $p$ cancels out, and we get

$$\frac{1}{n_0}\left(q + \frac{1}{2} n_2\right) = \frac{1}{n_0'}\left(q + \frac{1}{2} n_2'\right) \; , \tag{5.80}$$

or in other words

$$q|_{r=r_{\rm D8}} = \frac{n_2' n_0 - n_2 n_0'}{2(n_0' - n_0)} \; , \tag{5.81}$$

with $q$ as defined in (5.77). Notice that, in (5.48), the term $F_2/F_0$ and $b$ can both separately jump, while the whole $B_2$ is staying continuous. For this reason, as we anticipated in section 5.5.7, the conclusion $b = 0$ (which implies $n_2 = 0$ by (5.76)) will hold near the poles, but can cease to hold after one crosses a D8. (5.81) is also satisfying in that it is symmetric under exchange $\{n_0, n_2\} \leftrightarrow \{n_0', n_2'\}$. Notice also that, under a gauge transformation for the $B$ field, $n_2 \to n_2 + n_0 \Delta B$, $n_2' \to n_2' + n_0' \Delta B$, and (5.81) remains unchanged.

A constraint on the discontinuity should also come from the $F_2$ Bianchi identity (5.70). Using (5.78), we see that the only discontinuities are coming from the jump in $F_0$, so that we get

$$d_H F = \Delta F_0 (1 + p\,\mathrm{vol}_{S^2})\delta = \Delta F_0 \, \exp(p\,\mathrm{vol}_{S^2})\delta \; . \tag{5.82}$$

Comparing this with (5.70) we see that $\mathcal{F} = p\,\mathrm{vol}_{S^2}$. It also follows that

$$d\tilde{F}_2 = \Delta F_0(-B_2 + p\,\mathrm{vol}_{S^2})\delta = \frac{\Delta F_0}{F_0}\left(q + \frac{1}{2}n_2\right)\mathrm{vol}_{S^2}\ . \tag{5.83}$$

The expression on the right-hand side is not ambiguous thanks to (5.78). Comparing (5.83) with (5.70) again, we see that $f_2 = \frac{1}{F_0}\left(q + \frac{n_2}{2}\right)$. Going back to (5.74), we learn that

$$\frac{\Delta n_2}{\Delta n_0} = \frac{1}{n_0}\left(q + \frac{1}{2}n_2\right)\ . \tag{5.84}$$

This is actually nothing but (5.81) again.

(5.83) shows that our D8 is actually also charged under $F_2$, and thus that it is actually a D8-D6 bound state.

In fact, we should mention that it also acts as a source for $H$. This should not come as a surprise: it comes from the fact that $B$ appears in the DBI brane action. The simplest way to see this phenomenon for us is to notice that $H$ in (5.54) contains $F_0$. Since $F_0$ jumps across the D8, so does $H$, and its equation of motion now gets corrected to

$$d(e^{7A-2\phi} * H) - e^{7A}F_0 * F_2 = -xe^{7A-\phi}\Delta F_0\delta\ . \tag{5.85}$$

The localized term on the right hand side is exactly what one obtains by varying the DBI action $-\int_{S^2} e^{7A-\phi}\sqrt{\det(g+\mathcal{F})}$: the variation for a single D8 is $-e^{7A-\phi}\frac{\mathcal{F}}{\det(g+\mathcal{F})}\delta = -xe^{7A-\phi}\delta$. This was guaranteed to work: the equation of motion for $H$ was shown in [102] to follow in general from supersymmetry even in presence of sources. (The CS term $\int Ce^{\mathcal{F}}$ does not contribute, as remarked below [102, (B.7)].)

Yet another check one could perform is whether the D8 source is now BPS — namely, whether the supersymmetry variation induced on its worldvolume theory can be canceled by an appropriate $\kappa$-symmetry transformation. This check is made simpler by the fact that brane calibrations are actually the same forms that appear in the bulk supersymmetry conditions (as first noticed in [99] for compactifications to four dimensions). In our case, we see from [95, Table 1] that the appropriate calibration for a space-filling brane is $e^{6A_4-\phi}\mathrm{Re}\phi^1_-$; for our AdS$_7$ case, we should pick in (5.20) its part along $d\rho$. So our brane calibration is

$$e^{7A-\phi}\mathrm{Im}\psi^1_+\ . \tag{5.86}$$

The condition that a single brane should be BPS boils down to demanding that the pull-back of the form $e^{\mathcal{F}}\mathrm{Im}\psi^1_+$ equal the generalized volume form $\sqrt{\det(g+\mathcal{F})}$ on the brane. Alternatively, this is equivalent to demanding that the pullback on the brane of

$$e^{\mathcal{F}}\mathrm{Re}\psi_+^1 \ , \quad e^{\mathcal{F}}\psi_+^2 \tag{5.87}$$

vanish. We checked explicitly that this condition holds precisely if (5.81) does.

We should be a bit more careful, however, about what happens for multiple branes. In that case, (5.87) become non-abelian, because they both contain the worldsheet field $f_2$. Satisfying this condition now requires $\mathcal{F}$ to be proportional to the identity, and this in turn requires that the D6-brane charge $\Delta n_2$ should be an integer multiple of $n_{\rm D8} = \Delta n_0$. In other words, a bunch of D8-branes should be made of magnetized branes which all have the same induced D6-brane charge.

Finally, in our analysis so far we have left out the case where $F_0$ is zero on one of the sides of the D8 stack, say the right side, so that $n_0' = 0$. This time we cannot apply (5.79) on the right side of the D8. An expression for $B$ in this case will be given in (5.95) below. Imposing continuity of $B$ this time does not lead to (5.81), but to a different condition in terms of the integration constants appearing in (5.95). However, the Bianchi identity for $F_2$ can still be applied on the left side of the D8, where $F_0 \neq 0$; this still leads to (5.81). In other words, in this case we have (5.81) plus an extra condition imposing continuity of $B$. This will be important in our example with two D8's in section 5.6.3.

Let us summarize the results of this section. We have obtained that one can insert D8's in our setup, provided their position $r_{\rm D8}$ is such that the condition (5.81) is satisfied. When $F_0$ is non-zero on both sides of the D8, this ensures that the Bianchi for $F_2$ is satisfied, and that $B$ is continuous. In the special case where $F_0 = 0$ on one side, continuity of $B$ has to be imposed independently.

### 5.5.9. Summary of this section

Supersymmetric solutions of the form AdS$_7 \times M_3$ cannot exist in IIB. In IIA we have reduced the problem to solving the system of ODEs (5.49) (or (5.53)). Given a solution to this system, the flux is given by (5.43), (5.45) and (5.47), and the metric is given by (5.51) (or (5.52)). This describes an $S^2$ fibration over a segment; the space is compact if the $S^2$ fiber shrinks at the endpoints of the segment, giving a topology $M_3 = S^3$. This imposes the boundary conditions (5.61) on the system (5.53). D8-branes can be inserted along the $S^2$, at values $r = r_{\rm D8}$ that satisfy (5.81).

We now turn to a numerical study of the system, which will show that nontrivial solutions do indeed exist.

## 5.6. Explicit solutions

We will now show some explicit $\mathrm{AdS}_7 \times M_3$ solutions, by solving the system (5.53). We will start in section 5.6.1 by looking briefly at the massless solution, which is in a sense unique; it has a D6-brane and an anti-D6 at the two poles. In section 5.6.2 we will switch on Romans mass, and we will obtain a solution with a D6 at one pole only. In section 5.6.3 we will then obtain regular solutions with D8-branes.

### 5.6.1. Warm-up: review of the $F_0 = 0$ solution

We will warm up by reviewing the solution one can get for $F_0 = 0$.

As we remarked in section 5.5.1, in the massless case one can always lift to eleven-dimensional supergravity, and there we can only have $\mathrm{AdS}_7 \times S^4$ (or an orbifold thereof). The metric simply reads

$$ds^2_{11} = R^2 \left( ds^2_{\mathrm{AdS}_7} + \frac{1}{4} ds^2_{S^4} \right) , \tag{5.88}$$

being $R$ an overall radius. Let us now have a look at how this reduces to IIA. It is not obvious whether the reduction will preserve any supersymmetry; but, as we will now see, this can be arranged.

To reduce, we have to choose an isometry. Since $S^4$ has Euler characteristic $\chi = 2$, like any even-dimensional sphere, any vector field has at least two zeros, and so our reduction will have at least two points where the dilaton goes to zero; we expect some other strange feature at those two points, and as we will see this expectation is borne out.

How should we choose the isometry? We can think about U(1) isometries on $S^d$ as rotations in $\mathbb{R}^{d+1}$. The infinitesimal generator $v$ is an element of the Lie algebra $\mathfrak{so}(d+1)$, namely an antisymmetric $(d+1) \times (d+1)$ matrix $v$. Moreover, two such elements $v_i$ that can be related by conjugation, $v_1 = O v_2 O^t$, for $O \in \mathrm{SO}(d+1)$, can be thought of as equivalent. Any antisymmetric matrix can be put in a canonical block-diagonal form where every block is of the form $\left(\begin{smallmatrix} 0 & a \\ -a & 0 \end{smallmatrix}\right)$, with $a$ an angle. For even $d$, this implies that there is at least one zero eigenvalue, which corresponds to the fact that there is no vector field without zeros on the sphere. For $d = 4$, we have two angles $a_1$ and $a_2$. Our solution can be reduced along any of these vector fields, but we also want the reduction to preserve some supersymmetry. The infinitesimal spinorial action of the vector field we just described is proportional to $a_1 \gamma_{12} + a_2 \gamma_{34}$. If we demand that this matrix annihilates at least one spinor $\chi$ (so that, at the finite level, $\chi$ is kept invariant), we get either $a_1 = a_2$ or $a_1 = -a_2$.

To make things more concrete, let us introduce a coordinate system on $S^4$ adapted to the isometry we just found:

$$ds^2_{S^4} = d\alpha^2 + \sin^2(\alpha) ds^2_{S^3} = d\alpha^2 + \sin^2(\alpha)\left(\frac{1}{4}ds^2_{S^2} + (dy + C_1)^2\right) \ , \quad (5.89)$$

with $dC_1 = \frac{1}{2}\mathrm{vol}_{S^2}$ and $\alpha \in [0, \pi]$. We have written the $S^3$ metric as a Hopf fibration over $S^2$; the $1/4$ is introduced so that all spheres have unitary radius. The reduction will now proceed along the vector

$$\partial_y \ . \quad (5.90)$$

We can actually generalize this a bit by considering the orbifold $S^4/\mathbb{Z}_k$, where $\mathbb{Z}_k$ is taken to be a subgroup of the U(1) generated by $\partial_y$. This is equivalent to multiplying the $(dy + C_1)^2$ term in (5.89) by $\frac{1}{k^2}$.

We can now reduce the eleven-dimensional metric (5.88), quotiented by the $\mathbb{Z}_k$ we just mentioned, using the string-frame reduction $ds^2_{11} = e^{-\frac{2}{3}\phi}ds^2_{10} + e^{\frac{4}{3}\phi}(dy + C_1)^2$. We obtain a metric of the form (5.17), with

$$e^{2A} = R^2 e^{\frac{2}{3}\phi} = \frac{R^3}{2k}\sin(\alpha) \ , \quad ds^2_{M_3} = \frac{R^3}{8k}\sin(\alpha)\left(d\alpha^2 + \frac{1}{4}\sin^2(\alpha)ds^2_{S^2}\right) \ . \quad (5.91)$$

We could now also reduce the Killing spinors on $S^4$, which are given in appendix A.2 in our coordinates. There are indeed two of them which can be reduced, confirming our earlier arguments. This would allow us to compute directly the $\psi^{1,2}$. We will instead proceed by using the equations we derived in section 5.5. It is actually more convenient, in this case, to work directly with the system (5.49), that can be more easily solved explicitly:

$$x = \sqrt{1 - e^{4(A - A_0)}} \ , \quad \phi = 3A - \phi_0 \quad (5.92)$$

where $A_0$ and $\phi_0$ are two integration constants. This can be seen to be the same as (5.91) by taking

$$x = \cos(\alpha) \ , \quad A_0 = \frac{1}{2}\log\left(\frac{R^3}{2k}\right) \ , \qquad \phi_0 = 3\log R \ . \quad (5.93)$$

The fluxes can now be computed from (5.45) and (5.47):

$$F_2 = -\frac{1}{2}k\mathrm{vol}_{S^2} \ , \quad H = -\frac{3}{32}\frac{R^3}{k}\sin^3(\alpha)d\alpha \wedge \mathrm{vol}_{S^2} \ ; \quad (5.94)$$

the $B$ field then can be written as

$$B_2 = \frac{3}{32}\frac{R^3}{k}\left(x - \frac{x^3}{3}\right)\mathrm{vol}_{S^2} + b \quad (5.95)$$

where again $b$ is a closed two-form. The simple result for $F_2$ in (5.94) could be expected from the fact that the metric (5.89) is an $S^1$ fibration over $S^2$ with Chern class $c_1 = -k$.

However, (5.91) might appear problematic for two reasons. First of all, the warping function goes to zero at the two poles $\alpha = 0$, $\alpha = \pi$.[15] Second, $ds^2_{M_3}$ would be singular at the poles even if it were not multiplied by an overall factor $e^{2A} = \frac{R^3}{2k}\sin(\alpha)$, because of the $1/4$ in front of $ds^2_{S^2}$. Indeed, when we expand it around, say, $\alpha = 0$, we find $d\alpha^2 + \frac{\alpha^2}{4} ds^2_{S^2}$; this would be regular without the $1/4$, but as it stands it has a conical singularity.

However, these singularities at the poles have the behavior one expects near a D6. Near the north pole $\alpha = 0$, $ds^2_{M_3}$ in (5.91) looks like $ds^2_{M_3} \sim \alpha\left(d\alpha^2 + \frac{1}{4}\alpha^2 ds^2_{S^2}\right)$. In terms of the $r$ variable we used in (5.52), this looks like

$$ds^2_{M_3} \sim dr^2 + \left(\frac{3}{4}r\right)^2 ds^2_{S^2} \ . \tag{5.96}$$

Near the ordinary flat-space D6-brane metric, $ds^2_{M_3} \sim \rho^{-1/2}(d\rho^2 + \rho^2 ds^2_{S^2})$, which also looks like (5.96) with $r = \frac{4}{3}\rho^{3/4}$.

The presence of D6's could actually be inferred more directly. First of all, we know that D6-branes result from loci where the size of the eleventh dimension goes to zero; this indeed happens at the two poles. Moreover, from the expression of $F_2$ in (5.94), the integral of $F_2$ over the $S^2$ is constant and equal to $-2\pi k$. We can take the $S^2$ close to the north or the south pole, where it signals the presence of D6-brane charge. More precisely, there are $k$ anti-D6-branes at the north pole and $k$ D6-branes at the south pole.

One crucial difference with the usual D6 behavior, however, is the presence of the NS three-form $H$. From (5.94) we see that it does not vanish near the D6. Rather, it diverges: near the anti-D6 at $r = r_{\rm N} = 0$,[16]

$$H \sim r^{-1/3}\mathrm{vol}_3 \ . \tag{5.97}$$

This can also be inferred directly from eleven-dimensional supergravity, using the reduction formula $G_4 = e^{\phi/3} H \wedge e^{11}$. Since $\phi \sim r$, the three-form energy density diverges as $e^{-2\phi}H^2 \sim (r_{\rm N} - r)^{-8/3}$. We should remember, in any case, that this solution is non-singular in eleven dimensions; the diverging behavior

[15] The warping function also goes to zero at the equator of the AdS$_6 \times S^4$ solution [104], recently shown [105] to be the only AdS$_6$ solution in massive IIA. This solution can also be T-dualized, without breaking supersymmetry, both using its non-abelian and the more usual abelian isometries [106], differently from what we saw for AdS$_7$ in section 5.5.4.

[16] It is interesting to ask what happens in the Minkowski limit. From (5.54) we see that $H = -6e^{-A}\mathrm{vol}_3$; taking $R \to \infty$, $e^{-A}$ tends to zero except than in a region $\alpha \ll R^{-1/3}$, which gets smaller and smaller in the limit.

in (5.97) is cured by M-theory, just like the divergence of the curvature of (5.96) is.

The simultaneous presence of D6's and anti-D6's in a BPS solution might look unsettling at first, since in flat space they cannot be BPS together. It is true that the conditions imposed on the supersymmetry parameters $\epsilon_i$ by a D6 and by an anti-D6 brane are incompatible. But in flat space the $\epsilon_i$ are constant, while in our present case they are not. The condition changes from the north pole to the south pole; so much so that an anti-D6 is BPS at the north pole, and a D6 is BPS at the south pole. Although we have not been working explicitly with spinors in this chapter, but rather with forms, we can see this by performing a brane probe analysis in the language of calibrations, as we did for D8-branes at the end of section 5.5.8. The relevant polyform is again (5.86); for a D6 we should use its zero-form part, which from (5.36) is simply $\cos(\theta_1)\sin(\psi) = x$. For a D6 or anti-D6, this should be equal to plus or minus the internal volume form of the D6, which is $\pm 1$; this happens precisely at the north and south pole.

In figure 5.5 we show some parameters for the solution as a function of the $r$ defined in (5.52), for uniformity with latter cases. We also show the radius of the transverse sphere, which near the poles has the angular coefficient 3/4 of (5.96).

We have obtained this massless IIA solution by reducing the M-theory solution AdS$_7 \times S^4/\mathbb{Z}_k$, but other orbifolds would be possible as well. One could for example have quotiented by the $\hat{D}_{k-2}$ groups, which would have resulted in IIA in an orientifold by the action of the antipodal map on the $S^2$. The transverse $S^2$ would have been replaced by an $\mathbb{RP}^2$; at the poles we would have had O6's together with the $k$ D6's/anti-D6's of the $A_k$ case.

We will see in section 5.6.3 solutions with $F_0 \neq 0$ and without any D6-branes. But we will at first try in the next subsection to introduce $F_0$ without any D8-branes.

### 5.6.2. Massive solution without D8-branes

In section 5.6.1 we reviewed the only solution for $F_0 = 0$, related to AdS$_7 \times S^4$ by dimensional reduction; it has a D6 and an anti-D6 at the poles of $M_3 \cong S^3$.

We now start looking at what happens in presence of a non-zero Romans mass, $F_0 \neq 0$. We saw in section 5.5.7 that in this case it is possible for the poles to be regular points. It remains to be seen whether those boundary conditions can be joined by a solution of the system (5.53).

We can for example impose the boundary condition (5.61) at $r = r_{\rm N}$, and evolve numerically towards positive $r$ using (5.53). The procedure is standard: we use the approximate power-series solution (5.62) from $r = r_{\rm N} = 0$ to a very small $r$, and then use the values of $A$, $\phi$, $x$ thus found as boundary conditions for a numerical evolution of (5.53). One example of solution is shown in figure 5.6(a). It stops at a finite value of $r$, where it resembles there the south pole behavior of the massless case in figure 5.5; for example, $e^A$ goes to zero at the right extremum.

This is actually easy to understand already from the system, both in (5.49) and in (5.53). As $A$ and $\phi$ get negative, they suppress the terms containing $F_0$, and the system tends to the one for the massless case.

An alternative, and perhaps more intuitive, understanding can be found using the form (5.49) of the system, which we drew in figure 5.4 as a vector field flow on the space $\{A + \phi, x\}$. The green circle in that figure represents the point $\{A + \phi = \log(4/F_0), x = 1\}$, which is the appropriate boundary condition for the north pole in (5.61). In that figure the 'time' variable is $A$. From (5.62), we see that $A$ has a local maximum at $r = r_{\rm N}$. So the stream in figure 5.4 has to be followed backwards, starting from the green circle at the top. We can see that the integral curve asymptotically approaches $x = -1$, but does not get there in finite 'time'; in other words, $A \to -\infty$. The flow corresponding to the solution in figure 5.6(a) is shown in figure 5.6(b).

In the massless case, we saw in section 5.6.1 that the singularities at the poles are actually D6-branes. In this case too we have D6's at the south pole. This is confirmed by considering the integral of $F_2$ along a sphere $S^2$ in the limit where it reaches the south pole: it gives a non-zero number. By tuning $A_0^+$, this can be arranged to be $2\pi$ times an integer $k$, where $k$ is the number of D6-branes at the south pole. The presence of these D6-branes without any anti-D6 is not incompatible with the Bianchi identity $dF_2 - HF_0 = k\delta_{D6}$, because integrating it gives $-F_0 \int H = k$. In other words, the flux lines of the D6's are absorbed by $H$-flux, as is often the case for flux compactifications. Notice also that these D6's are calibrated; the computation runs along similar lines as the one we presented for the massless solution in section 5.6.1.

To be more precise, the singularity is not the usual D6 singularity, in that there is also a NS three-form $H$ diverging as in (5.97). This is consistent with the prediction in [107, Eq. (4.15)] (given there in Einstein frame), and in general with the analysis of [108, 109], which found that it is problematic to have ordinary D6-brane behavior in a massive AdS$_7 \times S^3$ setup precisely like the one we are considering here. (In the language of [108], the parameter $\alpha$ of our solution goes to a negative constant; this enables the solution to exist and to evade the global no-go they found, but at the cost of the diverging $H$ in (5.97), [107, Eq. (4.15)].) More precisely, the asymptotic behavior we find is the one discovered in [109, Eq. (3.4)].

Thus the singularity at the south pole in figure 5.6 is the same we found in the massless case we saw in section 5.6.1. In that case, the singularity is cured by M-theory. In the present case, the non-vanishing Romans mass prevents us from doing that. However, we still think it should be interpreted as the appropriate response to a D6; for this reason we think it is a physical solution.

So far we have examined what happens when we impose that the north pole is regular. It is also possible to have a D6 and anti-D6 singularity at both poles, as in the previous section, or an O6 at one of the poles (keeping D6's at the other pole). Roughly speaking, this corresponds to a trajectory similar to the one in figure 5.6(b), in which one "misses" the green circle to the left or to the right, respectively. As we have seen, the D6 solution is very similar to the massless one. The O6 solutions also turn out to be very similar to their massless counterpart:[17] near the pole, their asymptotics is $e^A \sim r^{-1/5}$, $e^\phi \sim r^{-3/5}$, $x \sim 1 - r^{4/5}$. This leads to the same asymptotics for the metric as in the massless O6 solution near the critical radius $\rho_0 = g_s l_s$. Once again, however, in the massive case we have a diverging NS three-form; this time $H \sim r^{-3/5}\text{vol}_3$. Finally, in such a case the $S^2$ is replaced by an $\mathbb{RP}^2$ because of the orientifold action.

### 5.6.3. Regular massive solution with D8-branes

We will now examine what happens in presence of D8-branes.

The first possibility that comes to mind is to put all of them together in a single stack. The idea is the following. We once again use the power-series expansion (5.62) from $r = r_\text{N} = 0$ to a small $r$, and use the resulting values of $A$, $\phi$ and $x$ as boundary conditions for a numerical evolution of (5.53). This time, however, we should stop the evolution at a value of $r$ where (5.81) is satisfied. At this point $F_0$ will change, and (5.53) will change as well. Generically, the evolution on the other side of the D8 will lead to a D6 or an O6 singularity, as discussed in section 5.6.2. However, if $F_0$ is negative, according to (5.61), the point $\{x = -1,\, e^{A+\phi} = -\frac{4}{F_0}\}$ leads to a regular South Pole. Fortunately, our solution still has a free parameter, namely $A_0^+ = A(r_\text{N})$. By fine-tuning this parameter, we can try to reach $\{x = -1,\, e^{A+\phi} = -\frac{4}{F_0}\}$ and obtain a regular solution.

Alternatively, after stopping the evolution from the North Pole to the D8, one can look for a similar solution starting from the South Pole, and then match the two — in the sense that one should make sure that $A$, $\phi$, and $x$ are continuous. One combination of them, namely $q$, will already match by construction. It is then enough to match two variables, say $A$ and $x$; this can be done by adjusting $A_0^+$ and $A_0^-$.

[17] In the different setup of [110], an O6 in presence of $F_0$ gets modified in such a way that its singularity disappears. This does not happen here.

Naively, however, we face a problem when we try to choose the flux parameters on the two sides of the D8's. We concluded in (5.67) that near the poles we should have $b = 0$; this seems to imply, via (5.76), that $n_2 = 0$ on both sides of the D8. (5.81) would then lead to $q = 0$ on the D8, which can only be true at the poles $x = \pm 1$.

This confusion is easily cleared once we remember that $B$ can undergo a large gauge transformation that shifts it by $k\pi \text{vol}_{S^2}$, as we explained towards the end of section 5.5.7. We saw there that we can keep track of this by introducing the variable $\hat{b}$ in (5.68). We now simply have to make sure that $\hat{b}$ winds an integer amount of times $N$ around the fundamental domain $[0, \pi]$; this can be interpreted as the presence of $N$ large gauge transformations, or as the presence of a non-zero quantized flux $N = \frac{1}{4\pi^2}\int H$.

We still face one last apparent problem. It might seem that making sure that $\hat{b}$ winds an integer amount of times requires a further fine-tuning on the solution; this we cannot afford, since we have already used both our free parameters $A_0^{\pm}$ to make sure all the variables are continuous, and that the poles are regular.

Fortunately, such an extra fine-tuning is in fact not necessary. Let us call $(n_0, n_2)$ the flux parameters before the D8, and $(n_0', n_2')$ after it. For simplicity let us also assume $n_2' = 0$, so that no large gauge transformations are needed on that side. As we remarked at the end of section 5.5.8, $\Delta n_2 = n_2' - n_2 = -n_2$ should be an integer multiple of $\Delta n_0 = n_0' - n_0 = n_{\text{D8}}$: $\Delta n_2 = \mu \Delta n_0$, $\mu \in \mathbb{Z}$. To take care of flux quantization, it is enough to also demand that $n_2 = N n_0$ for $N$ integer. Indeed, from (5.73), (5.76), (5.68), we see that in that case at the North Pole we get $\hat{b} = -\pi N$; since this is an integer multiple of $\pi$, it can be brought to zero by using large gauge transformations. Together, the conditions we have imposed determine $n_0' = n_0 \left(1 - \frac{N}{\mu}\right)$.

All this gives a strategy to obtain solutions with one D8 stack. We show one concrete example in figure 5.7. One might find it intuitively strange that the D8-branes are not "slipping" towards the South Pole. The branes back-react on the geometry, bending the $S^3$, much as a rubber band on a balloon. This by itself, however, would not be enough to prevent them from slipping. Rather, we also have to take into account the Wess–Zumino term in the brane action. This term, which takes into account the interaction of the branes with the flux, balances with the gravitational DBI term to stabilize the D8's. The formal check of this is that the branes are calibrated, something we have already seen in section 5.5.8 (see discussion around (5.86), (5.87)). The D8 stack is made of $n_{\text{D8}} = 50$ D8-branes; each of these D8's has worldsheet flux $f_2$ such that $\int_{S^2} f_2 = -2\pi$, which means that it has an effective D6-brane charge equal to $-1$. A single D8/D6 bound state probe with this charge is calibrated exactly at $r = r_{\text{D8}}$, and thus will not slip to the South Pole. The solution can perhaps be thought of as arising from the one in figure 5.6 via some version of Myers' effect.[18]

We can also look for a configuration with two stacks of D8-branes, again with regular poles. The easiest thing to attempt is a symmetric configuration where the two stacks have the same number of D8's, with opposite D6 charge. As for the solution with one D8, (5.61) implies $F_0$ at the north pole and negative $F_0$ at the south pole. For our symmetric configuration, these two values will be opposite, and there will be a central region between the two D8 stacks where $F_0 = 0$.

We show one such solution in figure 5.8. As for the previous solution with one D8, we have started from the North Pole and South Pole; now, however, we did not try to match these two solutions directly, but we inserted a massless region in between. From the northern solutions, again we found at which value of $r = r_{\rm D8}$ it satisfies (5.81). We then stopped the evolution of the system there, evaluated $A$, $\phi$, $x$ at $r_{\rm D8}$, and used them as a boundary condition for the evolution of (5.53), now with $F_0 = 0$. Now we matched this solution to the southern one; namely, we found at which values of $r = r_{\rm D8'}$ their $A$, $\phi$ and $x$ matched. This requires translating the southern solution in $r$ by an appropriate amount, and picking $A_0^- = A_0^+$. Given the symmetry of our configuration, this is not surprising: the southern solution is related to the northern one under (5.69). Moreover, matching a region with $F_0 \neq 0$ to the massless one means imposing an extra condition, namely the continuity of $B$ in $r_{\rm D8}$, as we mentioned at the end of 5.5.8.

The parameter $A_0^+ = A_0^- = A_0$ would at this point be still free. However, one still has to impose flux quantization for $H$. As we recalled above, this is equivalent to requiring that the periodic variable $\hat{b}$ starts and ends at zero. Unlike the case with one D8 above, this time we do need a fine-tuning to achieve this, since the expression for $B$ is not simply controlled by the massive expression (5.79). Fortunately we can use the parameter $A_0$ for this purpose. The solution in the end has no moduli.

As for the solution with one D8 stack we saw earlier, in this case too the D8-branes are not "slipping" towards the North and South Pole because of their interaction with the RR flux: each of the two stacks is calibrated. In this case, intuitively this interaction can be understood as the mutual electric attraction between the two D8 stacks, which indeed have opposite charge under $F_2$; the balance between this attraction and the "elastic" DBI term is what stabilizes the branes.

Let us also remark that for both solutions (the one with one D8 stack, and the one with two) it is easy to make sure, by taking the flux integers to be large enough, that the curvature and the string coupling $e^{\phi}$ are as small as one wishes, so that we remain in the supergravity regime of string theory. In figures 5.7 and 5.8 they are already rather small (moreover, in the figure we use some rescalings for visualization purposes).

[18]We thank I. Bena, S. Kuperstein, T. Van Riet and M. Zagermann for very useful conversations about this point and about the existence of solutions with a single D8. These solutions are consistent with the analysis in [111].

Thus we have found regular solutions, with one or two stacks of D8-branes. It is now in principle possible to go on, and to add more D8's. We have found examples with four D8 stacks, which we are not showing here. We expect that generalizations with an arbitrary number of stacks should exist, especially if there is a link with the brane configurations in [20, 19]. Another possibility that might also be realized is having an O8-plane at the equator of the $S^3$.

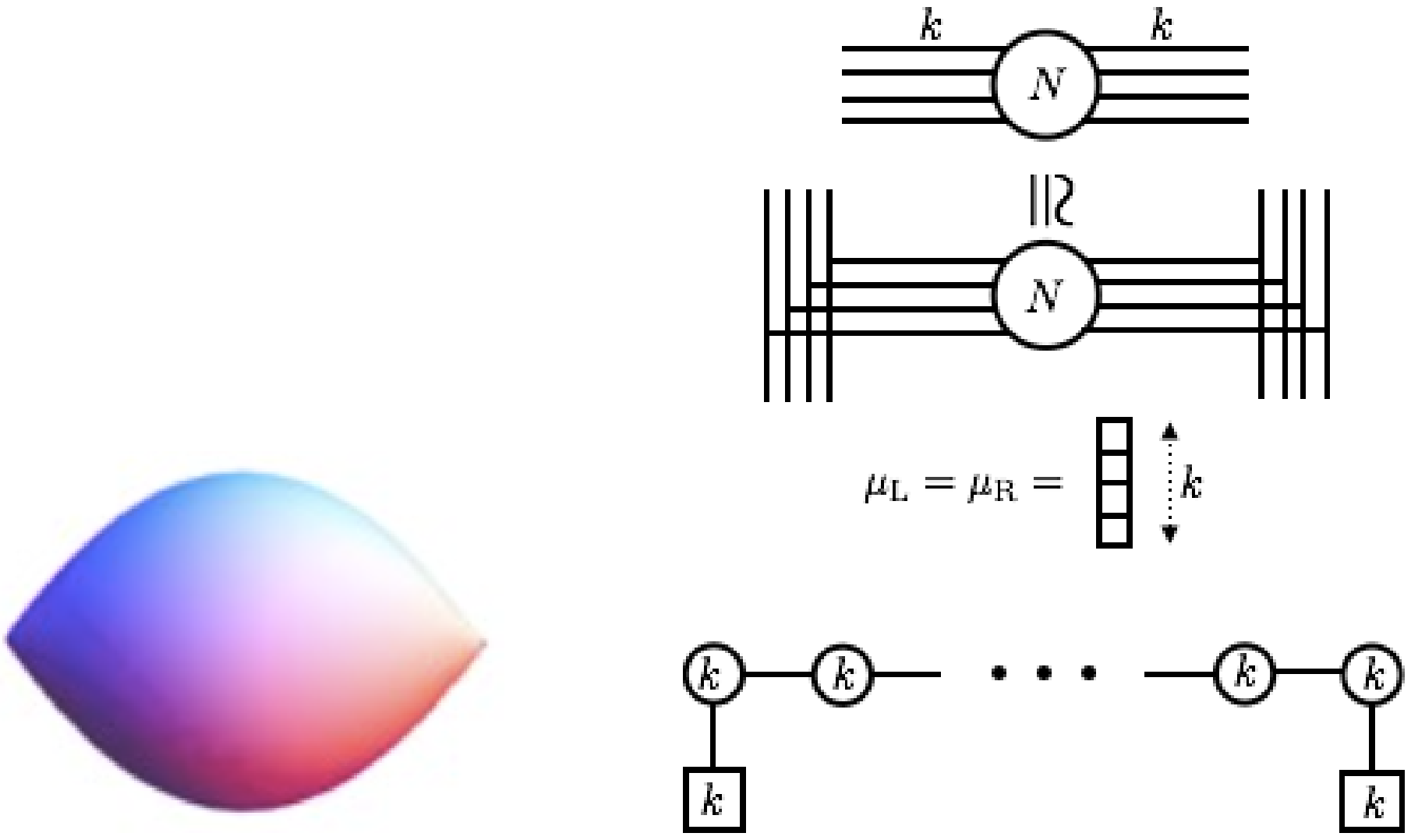

(a) (b)

**Figure 5.1.:** In figure 5.1(b) we see a Hanany–Witten NS5-D6 brane setup ($N$ NS5-branes and $k$ D6-branes) and an equivalent NS5-D6-D8 setup, obtained by bringing in from infinity a D8-brane for each semi-infinite D6-brane (the NS5's sit in the middle massless region). We also see the left and right partitions, from which we can read off the quiver on the bottom line. Every D8 has a D6 ending on its worldvolume. However, since this setup is expected give rise to the massless solution of section 5.6.1, we should perform a limit that pushes back to infinity the D8's. This is achieved by requiring $N \to \infty$, so that the massless region where the NS5's sit eats up all the space inside $M_3$. In this limit the data specified by the partitions is washed away (see discussion around (3.63)) and, in supergravity, the D8-brane sources with slope $\mu = \pm 1$ (i.e. D6 charge) become indistinguishable from the (anti-)D6-brane sources found in the analysis of section 5.6.1. In fact, the latter kind of sources is what one finds upon reducing from eleven-dimensional supergravity. In figure 5.1(a) (taken from [31]) we see an artist's impression of the internal space $M_3$, an $S^2$ fibration over a finite interval. Topologically, it is an $S^3$; however, the northern and southern poles are not regular points for the metric (due to the presence of D6-brane sources). These singularities get resolved once we go back to eleven-dimensional supergravity (implying that the configuration of NS5-D6 branes at strong coupling is better described by M5-branes probing an $A_{k-1}$ ALF space). Since there are no D8-brane sources, we do not see the appearance of any creases on $M_3$.

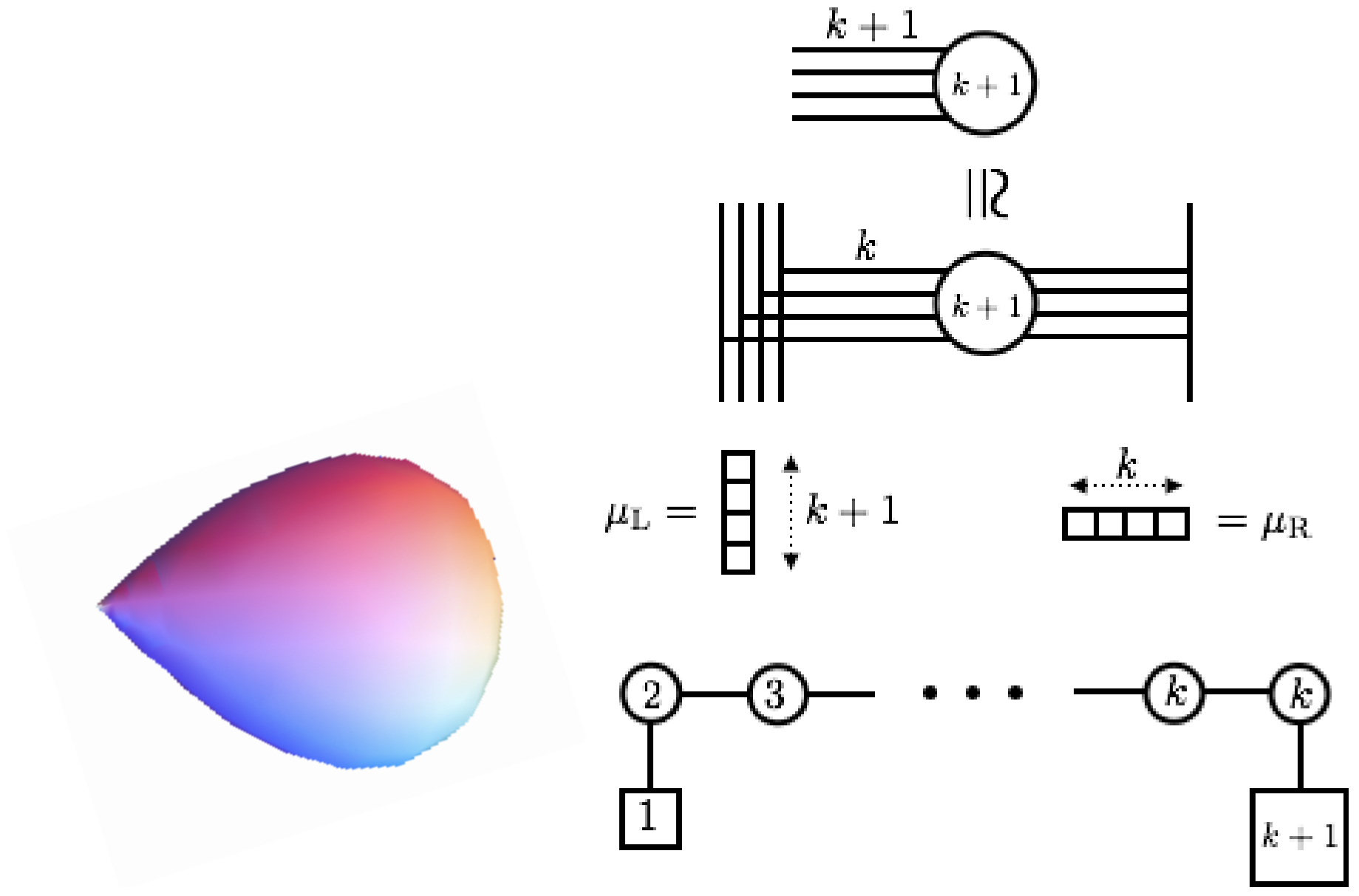

(a) (b)

**Figure 5.2.:** In figure 5.2(b) we see an NS5-D6-D8 brane setup in massive type IIA string theory. $F_0$ is nonvanishing throughout the internal space $M_3$ but constant. We also see an equivalent setup obtained via Hanany–Witten moves, allowing us to easily read off the two partitions, hence the quiver theory. In the holographic limit, the D8 stack with one D6 ending on each D8 is pushed back to infinity, and is indistinguishable from a D6-brane source in supergravity (becoming a singular point for the metric, see figure 5.2(a) – figure taken from [31]). The other pole of the internal $M_3 \cong S^3$ is a regular point for the metric, as there are no brane sources modifying its asymptotics (recall that the NS5-branes turn into pure $H$ flux in the near-horizon limit).

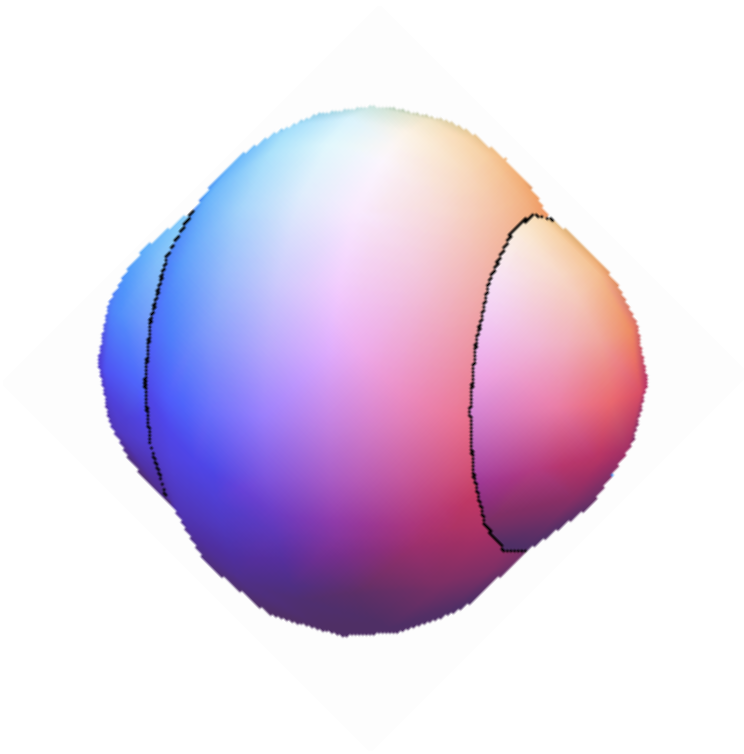

(a)

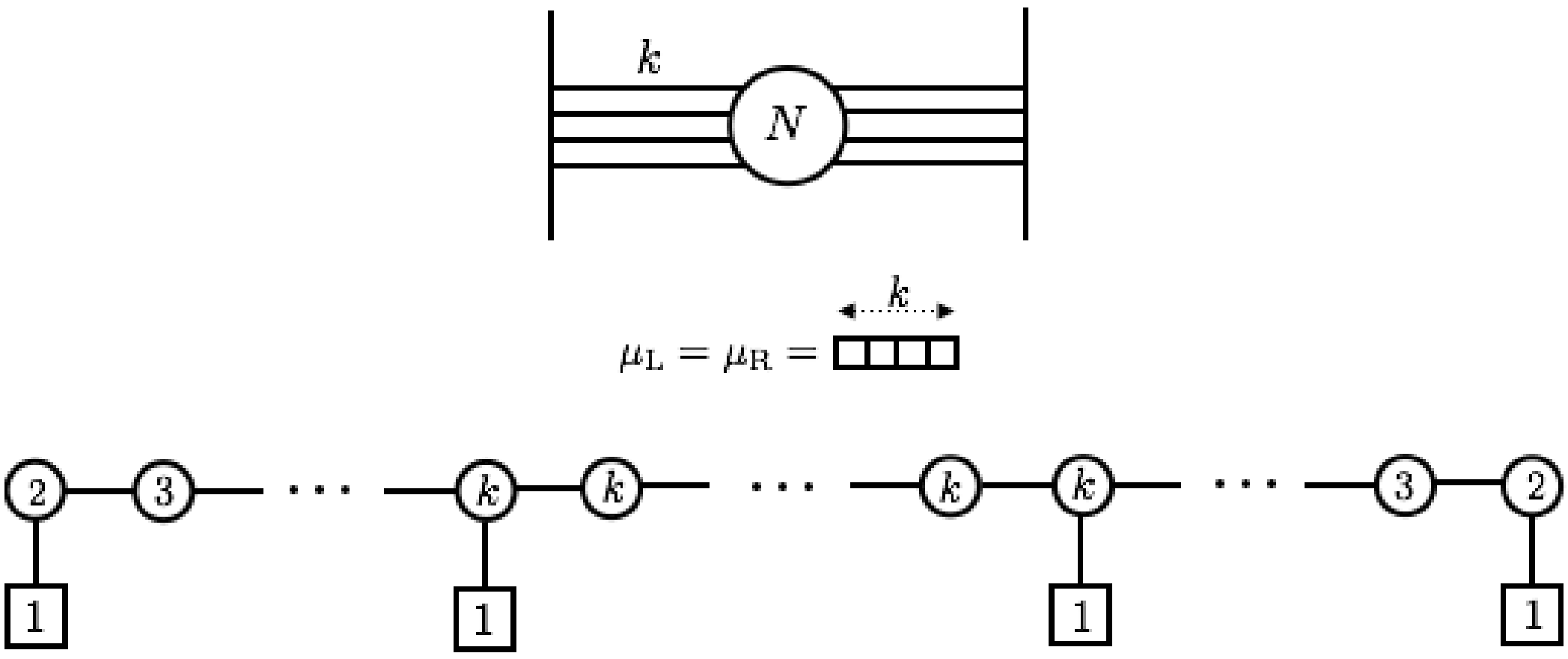

(b)

**Figure 5.3.:** In figure 5.3(b) we see an NS5-D6 brane setup in massive type IIA string theory with nonvanishing $F_0$ throughout the internal space $M_3$. To perform the holographic limit, this time we scale both $N$ and $\mu$, so that $\frac{\mu}{N}$ stays finite (see equations (5.10) and (5.11) and discussion around it). In figure 5.3(a) (taken from [31]) we see an artist's impression of the internal space. Both poles of $M_3 \cong S^3$ are regular points for the internal metric. The presence of D8-brane sources (represented by the black creases) with nonzero D6 charge (given by the supergravity "slope" $\mu = \pm k$) "shields" the singularity we would have in presence of D6-brane sources. This is due to a Myers-like effect.

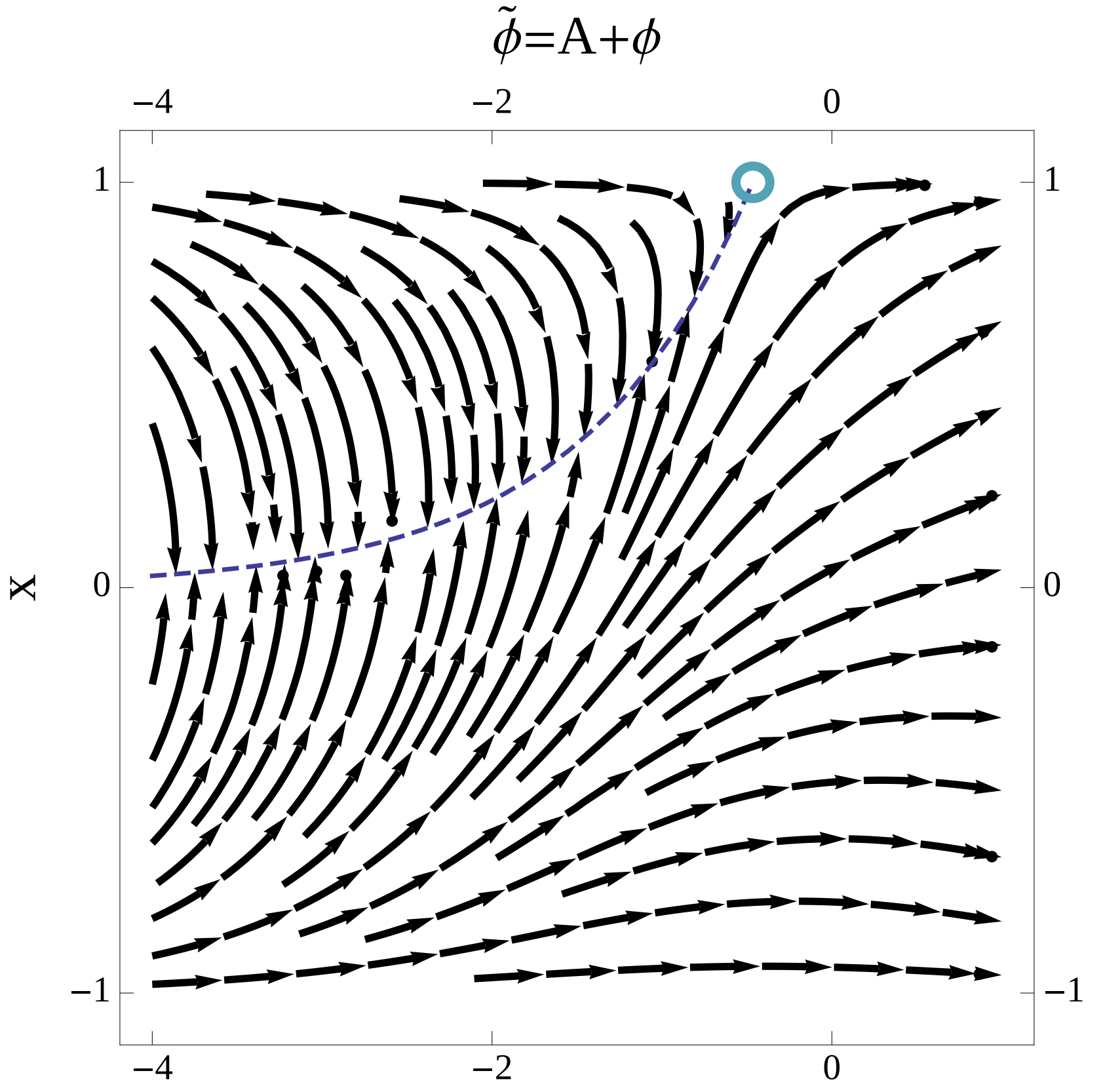

**Figure 5.4.:** A plot of the vector field induced by (5.49) on $\{\tilde{\phi} \equiv \phi + A, x\}$, for $F_0 = 40/2\pi$ (in agreement with flux quantization, (5.73) below). The green circle represents the point $\{\phi + A = \log(4/F_0), x = 1\}$, whose role will become apparent in section 5.5.7. The dashed line represents the locus along which the denominators in (5.49) vanish.

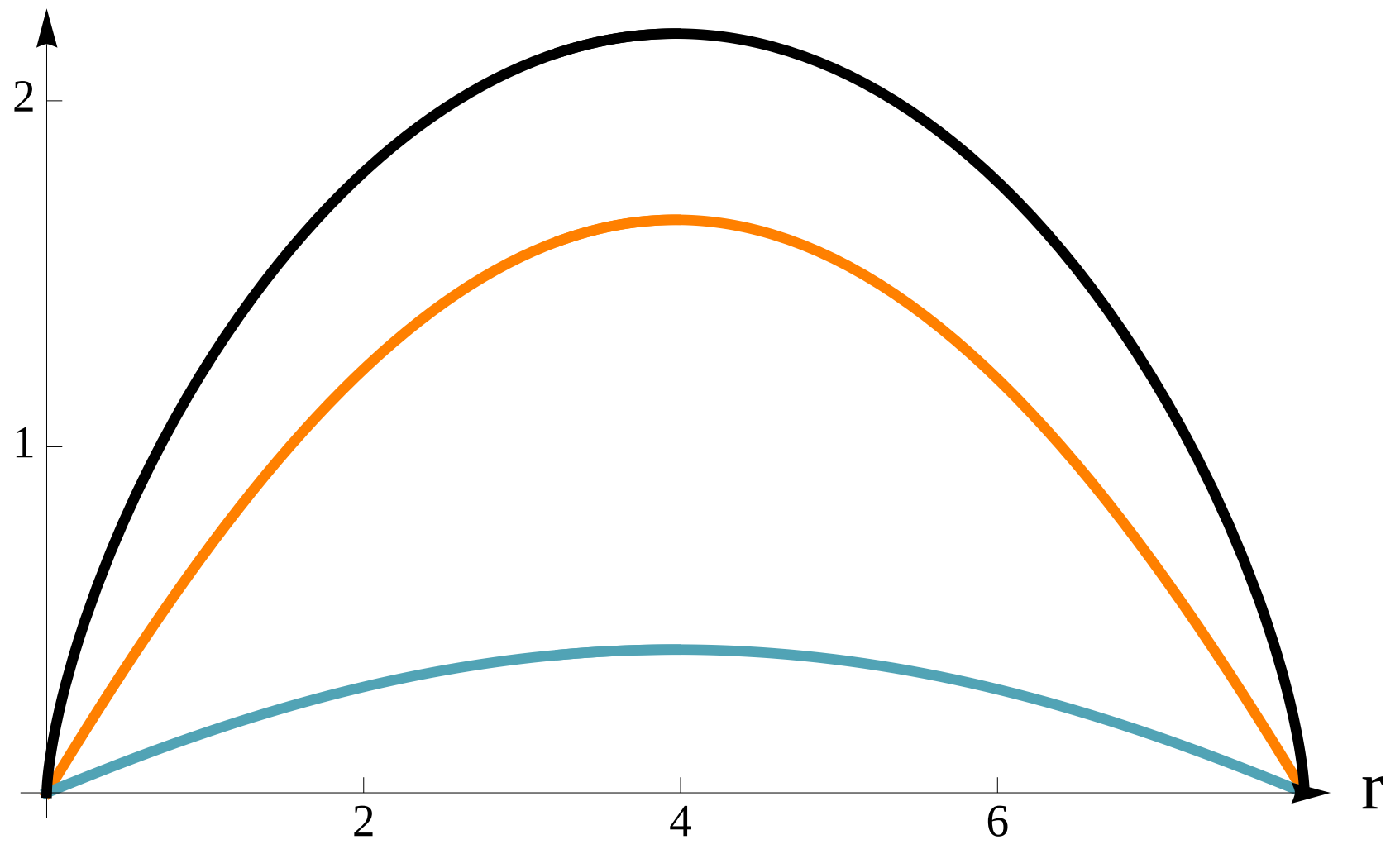

**Figure 5.5.:** Massless solution in IIA. We show here the radius of the $S^2$ (orange), the warping factor $e^{2A}$ (black; multiplied by a factor 1/20), and the string coupling $e^{\phi}$ (green; multiplied by a factor 5). We see that the warping goes to zero at the two poles. The angular coefficient of the orange line can be seen to be 3/4 as in (5.96). The two singularities are due to $k$ D6 and $k$ anti-D6 (in this picture, $k = 20$).

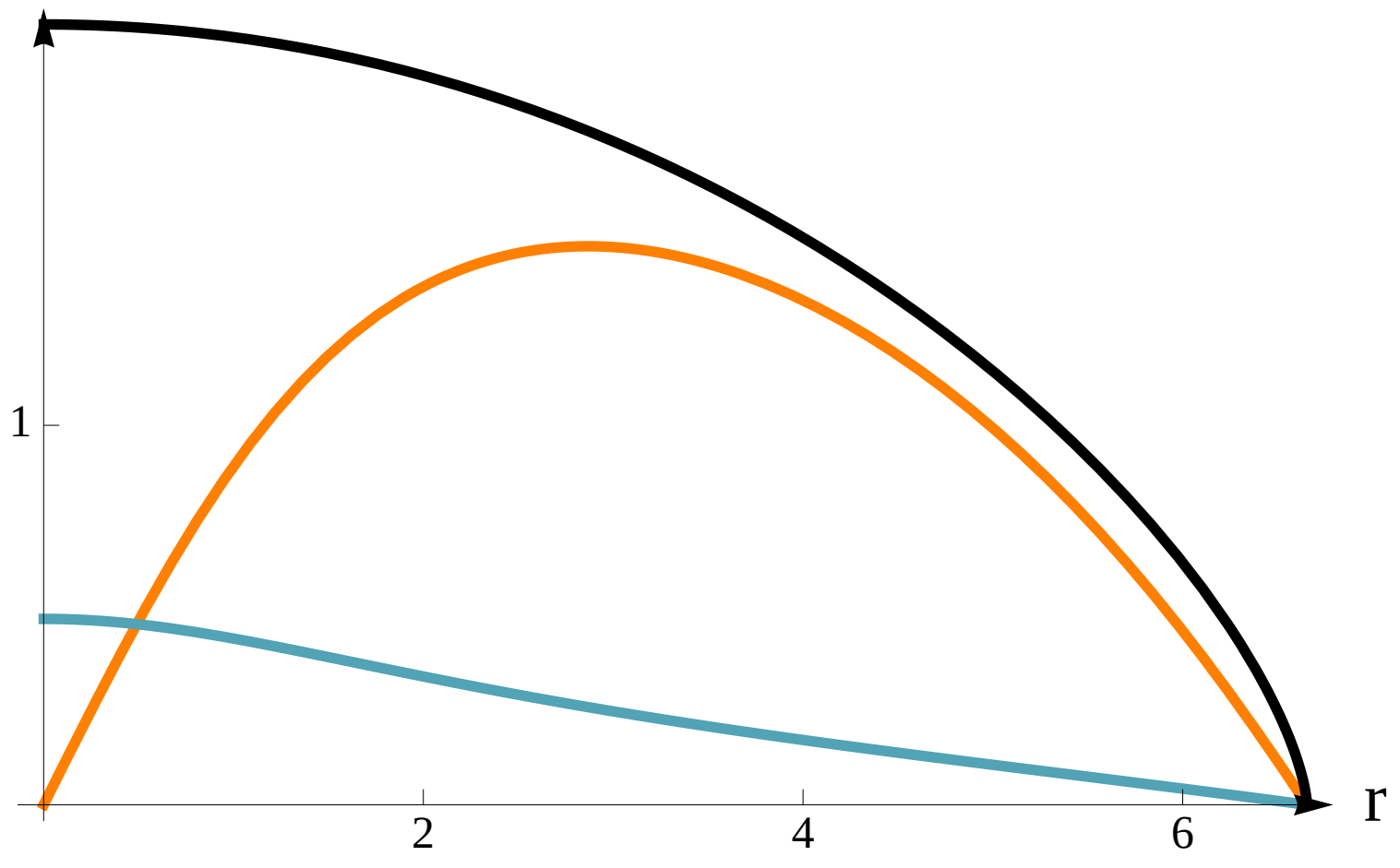

(a)

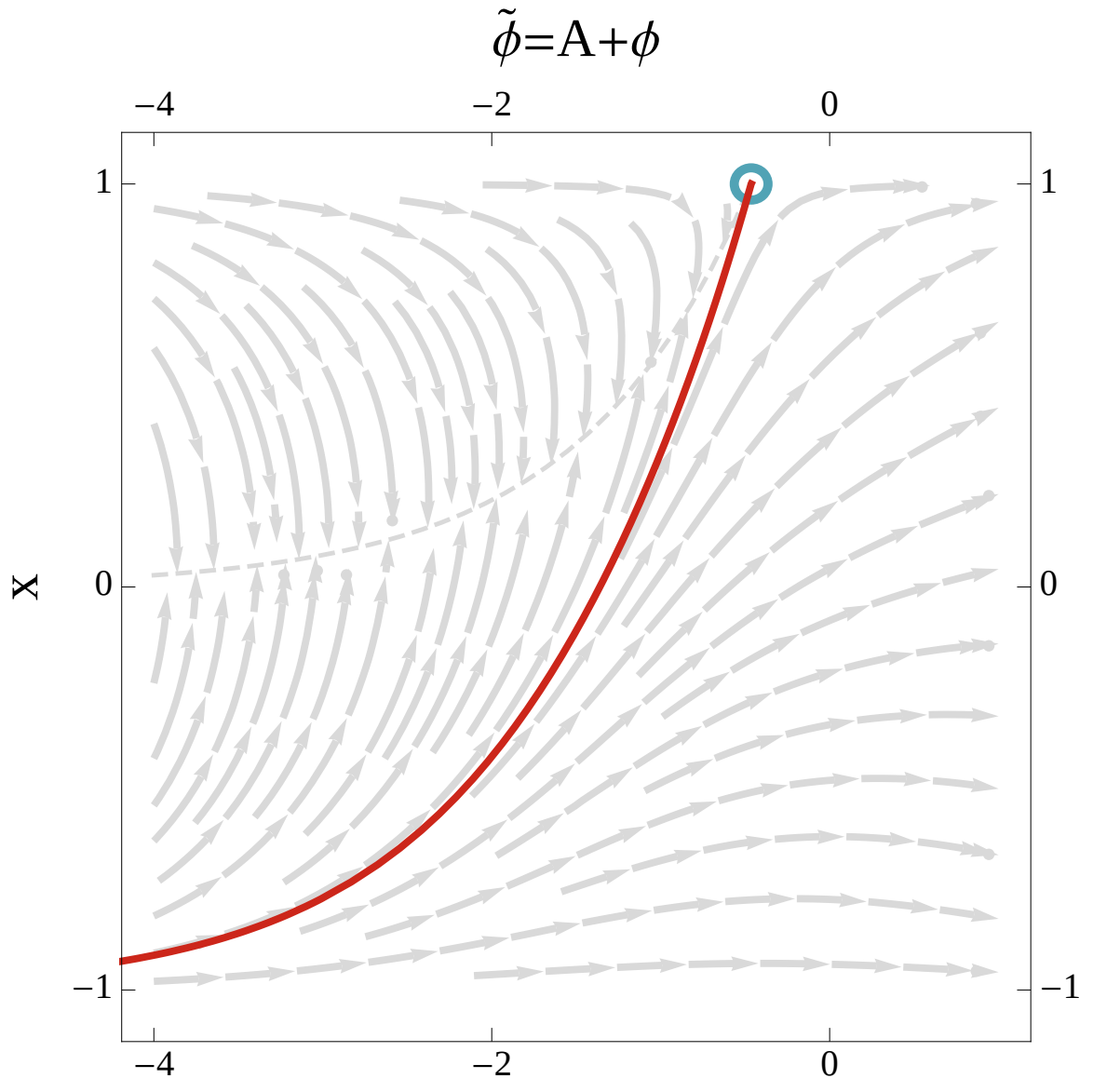

(b)

**Figure 5.6.:** Solution for $F_0 = 40/2\pi$. We imposed regularity at the north pole, and evolved towards positive $r$. In (a) we again plot the radius of the $S^2$ (orange), the warping factor $e^{2A}$ (black; multiplied by a factor 1/20), and the string coupling $e^{\phi}$ (green; multiplied by a factor 5). With increasing $r$, the plot gets more and more similar to the one for the massless case in figure 5.5. There is a stack of D6's at the south pole (in this picture, $k = 112$ of them), as in the massless case, although this time it also has a diverging NS three-form $H$. Notice that the size of the $S^2$ goes linearly near both poles, but with angular coefficients 1 near the north pole (appropriate for a regular point) and 3/4 for the south pole (appropriate for a D6, as seen in (5.96)). In (b), we see the path described by the solution in the $\{A + \phi, x\}$ plane, overlaid to the vector field shown in figure 5.4.

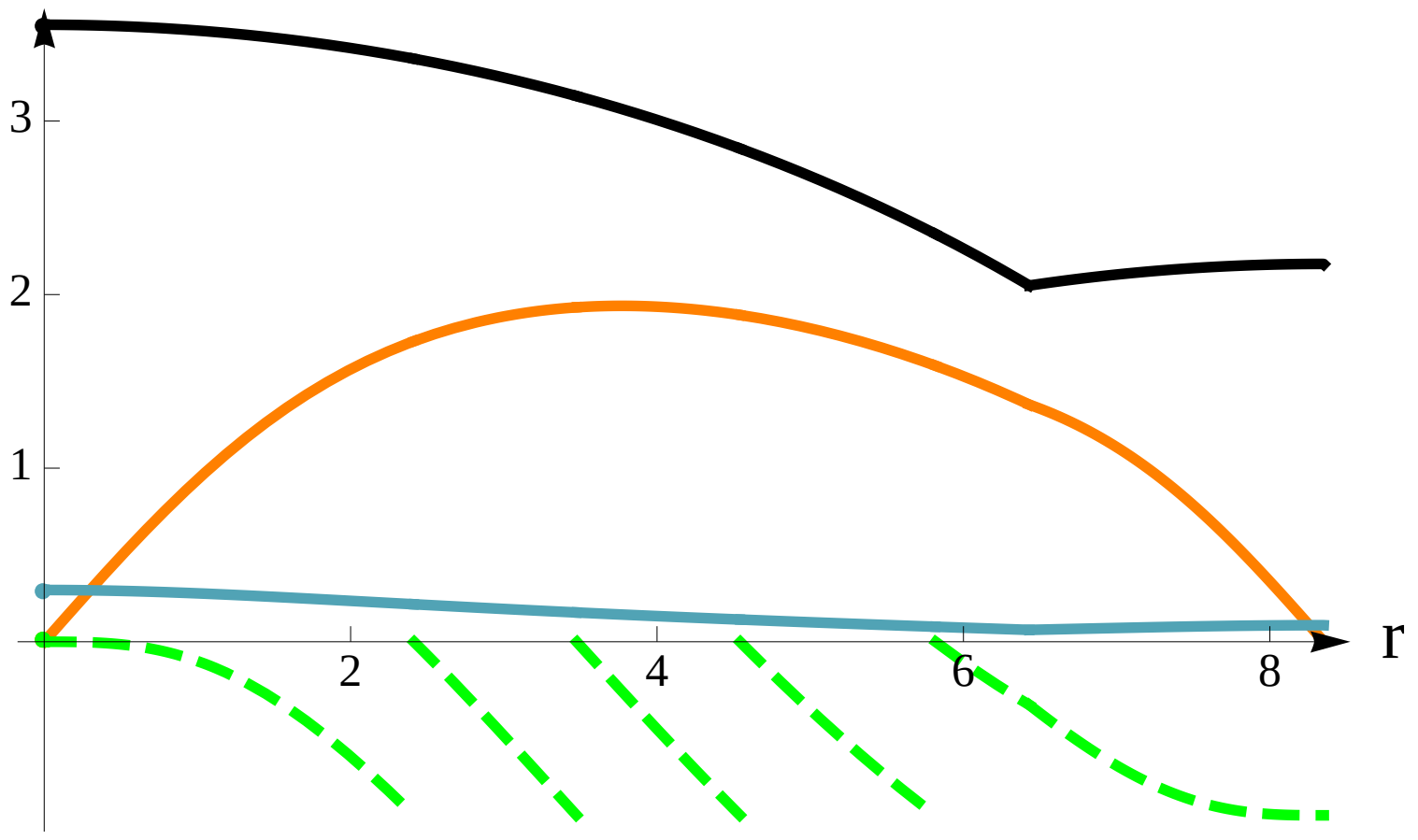

(a)

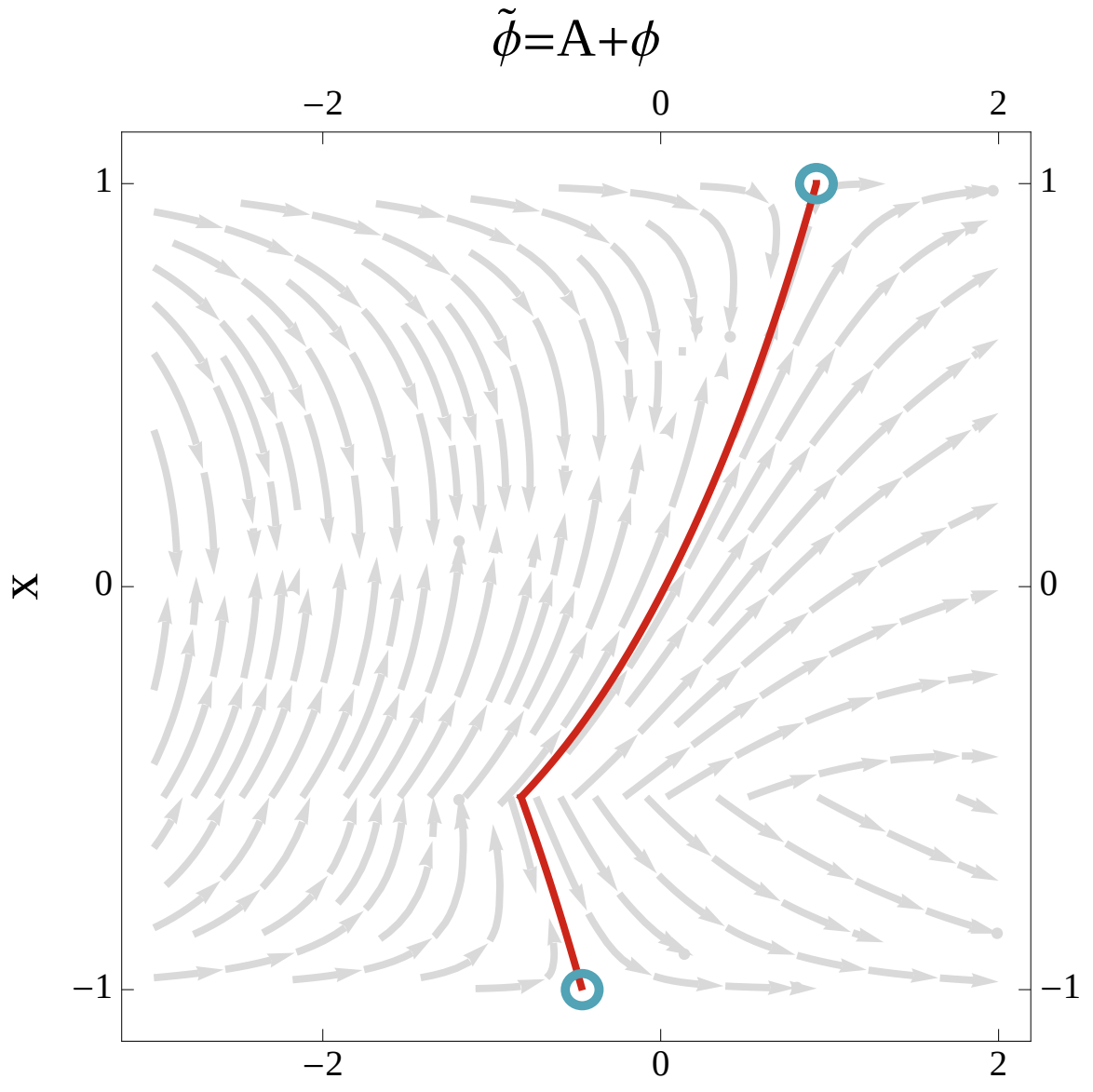

(b)

**Figure 5.7.:** Regular solution with one D8 stack. Its position can be seen in the graph as the value of $r$ where the derivatives of the functions jump; it is fixed by (5.81). In (a) we again plot the radius of the $S^2$ (orange), the warping factor $e^{2A}$ (black; rescaled by a factor 1/20), and the string coupling $e^{\phi}$. We also plot $\frac{1}{\pi}\hat{b}(r) = \frac{1}{4\pi^2}\int_{S^2_r} B_2$ (dashed, light green); to guide the eye, we have periodically identified it as described in section 5.5.7. (The apparent discontinuities are an artifact of the identification.) The fact that it starts and ends at $\hat{b} = 0$ is in compliance with flux quantization for $H$; we have $\frac{1}{4\pi^2}\int H = -5$. The flux parameters are $\{n_0, n_2\} = \{10, -50\}$ on the left (namely, near the north pole), $\{-40, 0\}$ on the right (near the south pole). In (b), we see the path described by the solution in the $\{A + \phi, x\}$ plane, overlaid to the relevant vector field, that this time changes with $n_0$.

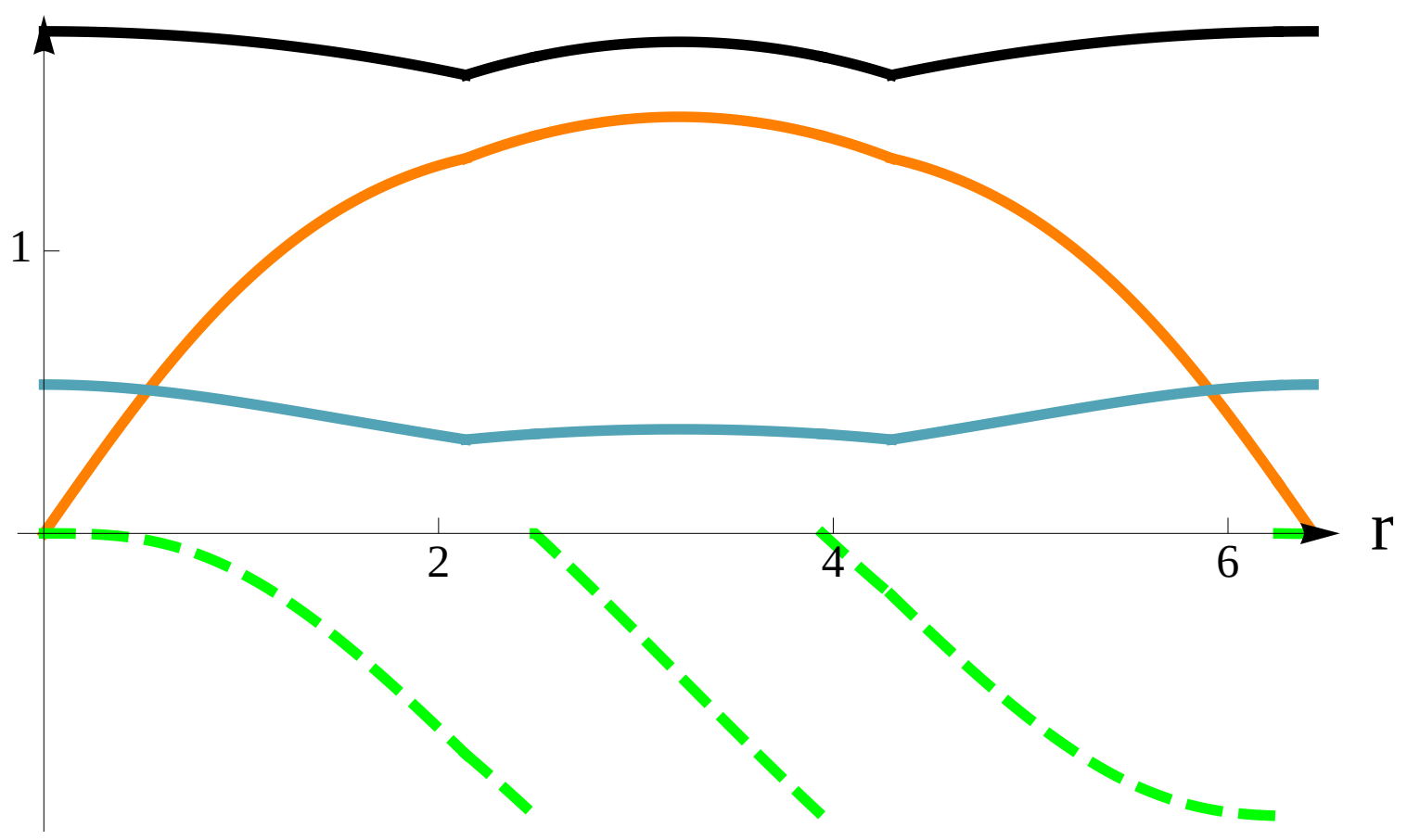

(a)

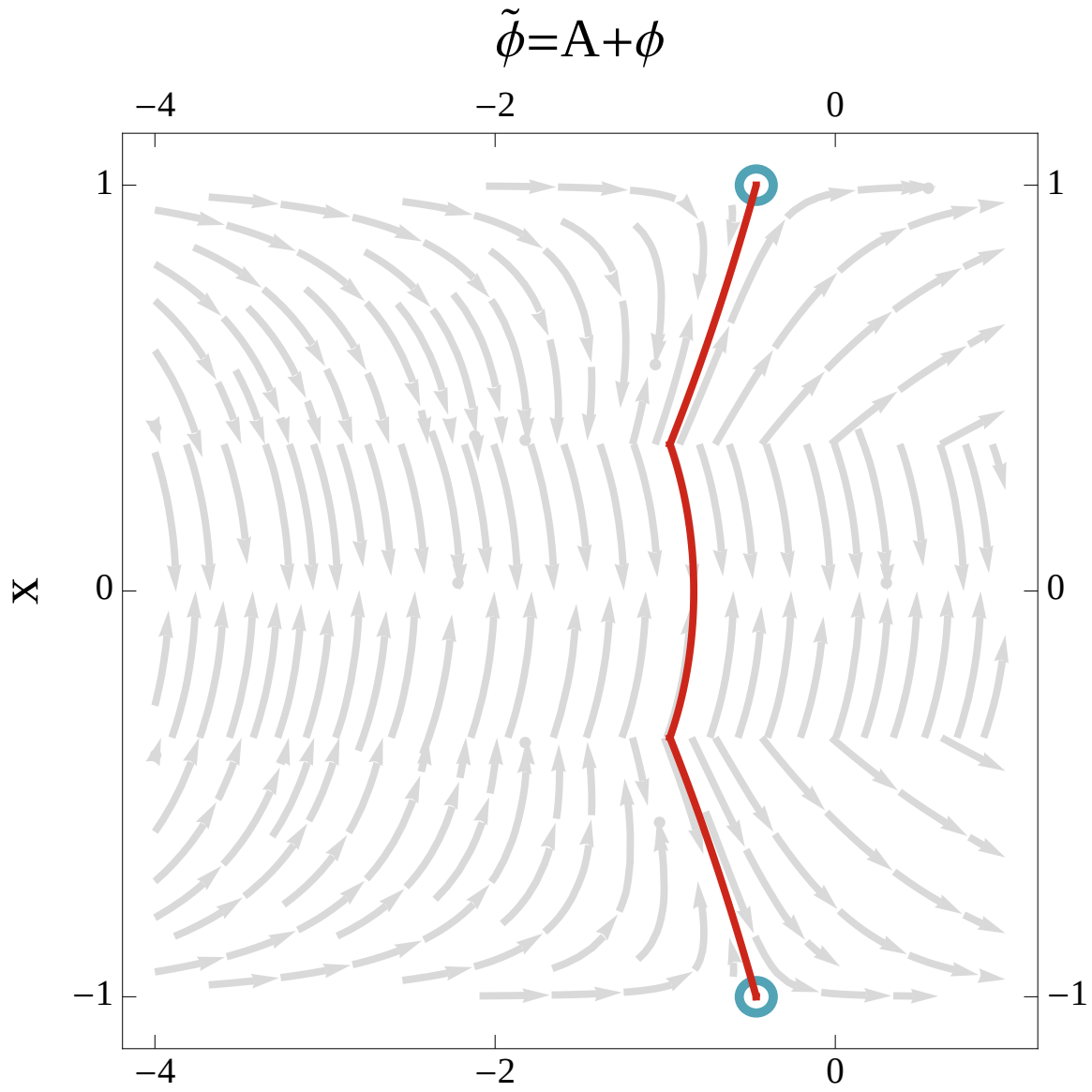

(b)

**Figure 5.8.:** Regular solution with two D8 stacks. As in figure 5.7, their positions are the two values of $r$ where the derivatives of the functions jump. In (a) we again plot the radius of the $S^2$ (orange), the warping factor $e^{2A}$ (black; rescaled by a factor 1/20), and the string coupling $e^{\phi}$ (green; rescaled by a factor 5), and $\hat{b}$ (as in figure 5.7; this time $\frac{1}{4\pi^2}\int H = -3$). The flux parameters are: $\{n_0, n_2\} = \{40, 0\}$ on the left (namely, near the north pole); $\{0, -40\}$ in the middle; $\{-40, 0\}$ on the right (near the south pole). The region in the middle thus has $F_0 = 0$; it is indeed very similar to the massless case of figure 5.5. In (b), we see the path described by the solution in the $\{A + \phi, x\}$ plane, overlaid to the relevant vector field, that again changes with $n_0$.

# Chapter 6

# AdS$_5$ solutions of massive type IIA supergravity

Sections 6.2 through 6.9 of this chapter are based on the published article [III] and letter [IV].

## 6.1. Relation to six-dimensional Hanany–Witten setups

In this chapter we are going to present an infinite class of analytic AdS$_5$ vacua of massive type IIA supergravity with eight supercharges (they are half-BPS with respect to $\mathcal{N} = 1$ supersymmetry in ten dimensions). They are related by a bijective correspondence to the AdS$_7$ ones of chapter 5. In particular, the internal space of the former is given by $M_3 \hookrightarrow M_5 \to \Sigma$: It is a fibration of a compact three-dimensional manifold $M_3$ over a Riemann surface $\Sigma$ of genus $g > 1$. $M_3$ turns out to be a deformation of the internal space of the AdS$_7$ vacua.

The natural interpretation of this (explained in section 6.8.8) is that the AdS$_5$ vacua are the holographic duals of four-dimensional $\mathcal{N} = 1$ superconformal theories obtained by compactifying the $(1, 0)$ theories on $\Sigma$. The compactification is twisted: The SU$(2)_R$ R-symmetry of $(1, 0)$ theories is mixed with an SO(2) group of local transformations on $\Sigma$. The commutant is just U(1), which is exactly the R-symmetry of $d = 4$ $\mathcal{N} = 1$ theories. It is realized on the supergravity side by a U(1) isometry of $M_3$.

We are thus led to conjecture that it should be possible to engineer these $\mathcal{N} = 1$ theories by compactifying the NS5-D6-D8 Hanany–Witten setups of chapter 3 on a Riemann surface, spanning two directions wrapped by all branes.

Supersymmetry is reduced according to the topological twist induced by the Riemann surface.

Prominent examples of this kind of construction are Gaiotto's class $\mathcal{S}$ theories [112], obtained by compactifying the $A_N$-type $(2,0)$ theory on a Riemann surface, i.e. the worldvolume theory of $N$ coincident M5-branes. The compactified theory enjoys $\mathcal{N}=2$ supersymmetry in four dimensions, half of the sixteen $Q$ supercharges of the $\mathcal{N}=(2,0)$ algebra. Similarly, the compactification of the $\mathcal{T}(N,k)$ theory (with trivial partitions $\mu_{\rm L}=\mu_{\rm R}=(1^k)$) on a Riemann surface produces a class of four-dimensional $\mathcal{N}=1$ theories, called class $\mathcal{S}_k$ [113]. The former is the theory of $N$ M5's at the $A_k$ ALF space; its eleven-dimensional supergravity dual is just $\mathrm{AdS}_7\times S^4/\mathbb{Z}_k$, which reduces to the massless vacuum $\mathrm{AdS}_7\times M_3$ without D8-brane sources of section 5.6.1 ($F_0=0$ throughout $M_3$).

By a similar token, our $\mathrm{AdS}_5$ vacua should be dual to compactifications on $\Sigma$ of more complicated $\mathcal{T}(N,k,\mu_{\rm L},\mu_{\rm R})$ theories with nontrivial partitions. The latter reflect the "insertion" of D8 stacks in the type IIA Hanany–Witten setup, rendering the M-theory description inapplicable. Instead, we should use massive type IIA string theory.

## 6.2. Introduction

The study of supersymmetric conformal field theories (CFT) in four dimensions using holography is by now a venerable subject. Their holographic duals are $\mathrm{AdS}_5$ solutions in either IIB supergravity or M-theory. A comprehensive analysis of supersymmetric $\mathrm{AdS}_5$ solutions of IIB supergravity was carried out in [88]; these include the Freund–Rubin compactifications and the Pilch–Warner solution [114]. Analogous studies were performed for $\mathcal{N}=1$ [115] and $\mathcal{N}=2$ [116] supersymmetric $\mathrm{AdS}_5$ backgrounds of M-theory, where new analytic solutions were found. $\mathrm{AdS}_5$ solutions arising in M-theory usually have a higher-dimensional origin: they are compactifications ("twisted" in a certain way) of CFT's in six dimensions. Actually this latter CFT is essentially always the $(2,0)$ theory living on the worldvolume of M5-branes, as in [117] (and in the more recent examples [118, 119]).

$\mathrm{AdS}_7$ solutions in type II supergravity were classified in chapter 5. A new infinite class of solutions was found in massive IIA: the internal space $M_3$ is always topologically an $S^3$, but its shape is not round — rather, it is a fibration of a round $S^2$ over an interval.[1] Both D6's and D8's can be present (and, a bit more exotically, O6's and O8's). The CFT duals of these solutions are $(1,0)$ supersymmetric theories, which were argued in [24] to be the ones obtained in [19, 20] from NS5-D6-D8 configurations (see also [12, 13] for earlier related theories). A similar class of (1,0) theories can be found in F-theory [21, 22].

[1]This Ansatz was also considered in [108, 107, 109, 111], also in a non-supersymmetric setting.

This prompts the question of whether these (1,0) theories, when compactified on a Riemann surface, can also give rise to CFTs in four dimensions. If so, their duals should be AdS$_5$ solutions in massive IIA.

In this chapter we classify AdS$_5$ solutions of massive IIA, and we find many analytic examples. The new (and physically sensible) ones are in bijective correspondence with the AdS$_7$ solutions; this strongly suggests that their dual CFT$_4$ are indeed twisted compactifications of the (1,0) CFT$_6$. The correspondence is via a simple universal map, which was directly inspired by the map in [120] from AdS$_4$ to AdS$_7$ solutions. At the level of the metric it reads

$$e^{2A}(ds^2_{\mathrm{AdS}_5} + ds^2_{\Sigma_g}) + dr^2 + e^{2A}v^2 ds^2_{S^2} \;\to\; \sqrt{\frac{4}{3}}\left(\frac{4}{3}e^{2A}ds^2_{\mathrm{AdS}_7} + dr^2 + \frac{v^2}{1+3v^2}e^{2A}ds^2_{S^2}\right) , \tag{6.1}$$

where $A$, $v$ are functions of $r$ and $\Sigma_g$ is a Riemann surface of genus $g \geq 2$. This map is so simple that it also allows us to find analytic expressions for the AdS$_7$ solutions. For example, the simplest massive AdS$_5$ solution has metric

$$ds^2 = \sqrt{\frac{3}{4}\frac{n_2}{F_0}}\left(\sqrt{\tilde{y}+2}\,(ds^2_{\mathrm{AdS}_5} + ds^2_{\Sigma_g}) + \frac{d\tilde{y}^2}{4(1-\tilde{y})\sqrt{\tilde{y}+2}} + \right. \\ \left. + \frac{1}{9}\frac{(1-\tilde{y})(\tilde{y}+2)^{3/2}}{2-\tilde{y}}ds^2_{S^2}\right) \tag{6.2}$$

with $\tilde{y} \in [-2, 1]$. Its AdS$_7$ "mother", obtained via the map (6.1), reads on the other hand

$$ds^2 = \frac{n_2}{F_0}\left(\frac{4}{3}\sqrt{\tilde{y}+2}\,ds^2_{\mathrm{AdS}_7} + \frac{d\tilde{y}^2}{4(1-\tilde{y})\sqrt{\tilde{y}+2}} + \frac{1}{3}\frac{(1-\tilde{y})(\tilde{y}+2)^{3/2}}{8-4\tilde{y}-\tilde{y}^2}ds^2_{S^2}\right) . \tag{6.3}$$

Both these solutions have a stack of $n_2$ D6-branes at $\tilde{y} = 2$, and are regular elsewhere. The D6's can also partially or totally be replaced by several D8-branes, much like in a Myers effect [44]. (In a way, these solutions realize the vision of [121].) Such more complicated solutions are obtained by gluing together copies of (6.3), or sometimes also of a more complicated metric that we will see later on.

We start our analysis in complete generality. We use the time-honored trick of reducing the study of AdS$_5$ solutions to that of Minkowski$_4$ solutions whose internal space $M_6$ has a conical isometry. One can then use the general classification of [94], which uses generalized complex geometry on $M_6$. Due to the conical structure of $M_6$, the "pure spinor equations" of [94] become a certain new set of equations on $M_5$. (The idea of applying the pure spinor equations to AdS$_5$ solutions in this way goes back to [96], where it was applied to IIB solutions.) It is immediately seen that the only possibility that leads to solutions is that of an SU(2) structure on $M_6$ (where the pure spinors are of so-called type 1 and type 2), which means in turn that there is an identity structure on $M_5$.

The practical consequence of this is that we can determine the metric on $M_5$ in full generality. It is a fibration of a three-dimensional fiber $M_3$ over a two-dimensional space $\mathcal{C}$. The three-dimensional fiber also has a Killing vector, which is holographically dual to R-symmetry on the field theory side. The fluxes are also fully determined. The independent functions (one function $a_2$ in the metric, the warping $A$, and the dilaton $\phi$) have to satisfy a total of six PDEs.

The problem simplifies dramatically once we impose what we will call the "compactification Ansatz". This consists in imposing that: 1) The metric of $\mathcal{C}$ is conformally related to that of a surface $\Sigma$, which does not depend on the coordinates of the three-dimensional space orthogonal to $\mathcal{C}$ inside $M_5$. The conformal factor is equal to the warping function $e^{2A}$ in front of the AdS$_5$ metric; 2) neither $A$, nor the dilaton $\phi$, nor the function $a_2$ entering the metric and fluxes, depend on the coordinates of $\Sigma$. Under this Ansatz, $\Sigma$ has constant curvature[2] (and we can compactify it to produce a compact Riemann surface $\Sigma_g$); the PDEs reduce to only three. Moreover, these PDEs are all polynomial in one of the local coordinates on $M_3$. Thus they can be in fact reduced to a set of ODEs. At this point the analysis branches out in several possibilities; for each of those, only one ODE survives. In the massless case, there is a "generic case", which is the reduction to IIA of the Bah–Beem–Bobev–Wecht solution [123, 119], and two special cases being the reduction of the $\mathcal{N} = 1$ Maldacena–Núñez solution [117] and the Itsios–Núñez–Sfetsos–Thompson solution [124]. In the massive case, we get new solutions. Again there is a generic case and two special cases. In the generic case, we find no solutions. The first special case, with positive curvature on $\Sigma_g$, has singularities which we cannot interpret physically. The second, with constant negative curvature[3] on $\Sigma_g$, leads to physically sensible solutions. These latter ones are the main result of this chapter.

Solving the ODE produces several solutions, of which (6.2) is the simplest. Without D8's, the most general solution has either two D6 stacks (unlike (6.2), which has one), or one D6 stack and one O6. As we already mentioned, there is also the possibility of introducing D8's, which can be done by gluing together copies of (6.2), of the Maldacena–Núñez solution, and possibly also of the more complicated solution we just mentioned. As we also anticipated, the map (6.1) can then be used to produce analytical expressions for all the $AdS_7$ solutions in [I, 24].

[2] For compactifications of (2,0) theories, the fact that $\Sigma$ has constant curvature was explained in [122].

[3] Compactifying on $T^2$ the NS5–D6–D8 configurations of [19, 20] and T-dualizing twice should lead to the NS5–D4–D6 system of [125]; the holographic dual to those solutions was found in [126].

All these new explicit solutions are begging further investigation, particularly regarding their field theory interpretation. This might be the beginning of a correspondence between $CFT_6$ and $CFT_4$ similar to the celebrated class $\mathcal{S}$ theories [112] (although notice that we do not discuss Riemann surfaces with punctures here, as was done in [118]). A feature that those theories also had is that (at the supergravity level) the ratio of the number of degrees of freedom in four and six dimensions is proportional to $g-1$, just like for [118] (and for [119]); this is a simple consequence of the map (6.1). We compute the central charges for the $CFT_6$ in a couple of simple cases; for example, for a symmetric solution with two D8's. Along with the NSNS flux integer $N$, there is also another flux integer $\mu$, which is basically the D6 charge of the D8's; the number of degrees of freedom is a simple cubic polynomial in $N$ and $\mu$, and agrees with an earlier approximate computation in [24]. It would be interesting to also compute contributions from stringy corrections, which we have not done here.

This chapter is organized as follows. In section 6.3, we write the system of pure spinor equations relevant for supersymmetry. In section 6.4, we analyze the system: we determine the metric and fluxes in terms of a few functions, subject to a set of PDEs, which we summarize in subsection 6.4.4. In section 6.6 we introduce the compactification Ansatz, for which we are able to give a complete list of cases. One of these classes (apparently the only physically sensible one which was not already known) is then analyzed in 6.8 in more detail. The highlights of that analysis are the correspondence to $AdS_7$ in section 6.8.2, the explicit solutions in sections 6.8.5, 6.8.6, 6.8.7, and the preliminary field theory considerations in section 6.8.8. In appendix C.1 we provide the proof of the existence of a Killing vector on $M_5$. In section 6.5 we consider an Ansatz simpler than the one in section 6.6; it reproduces a certain solution of [115]. Finally, in section 6.7, we summarize the already known solutions which we recovered in our analysis.

## 6.3. The conditions for supersymmetry

In this section, we will derive a system of differential equations on forms in five dimensions that is equivalent to preserved supersymmetry for solutions of the type $AdS_5 \times M_5$. We will derive it by considering $AdS_5$ as a warped product of Mink$_4$ and $\mathbb{R}$. We will begin in section 6.3.1 by reviewing a system equivalent to supersymmetry for Mink$_4 \times M_6$. In section 6.3.2 we will then translate it to a system for AdS$_5 \times M_5$.

### 6.3.1. Mink$_4 \times M_6$

Preserved supersymmetry for Mink$_4 \times M_6$ was found [94] to be equivalent to the existence on $M_6$ of an $\mathrm{SU}(3) \times \mathrm{SU}(3)$ structure satisfying a set of differential equations. The system is described by a pair of pure spinors

$$\phi_- \equiv e^{-A_6} \chi_1^+ \otimes \chi_2^{-\,\dagger} \ , \qquad \phi_+ \equiv e^{-A_6} \chi_1^+ \otimes \chi_2^{+\,\dagger} \ , \tag{6.4}$$

where the warping function $A_6$ is defined by

$$ds^2_{10} = e^{2A_6} ds^2_{\mathrm{Mink}_4} + ds^2_{M_6} \ , \tag{6.5}$$

and the $\pm$ superscripts indicate the chirality of $\chi_1$ and $\chi_2$. The pure spinors $\phi_-$ and $\phi_+$ can be expressed as a sum of odd and even forms respectively, via application of the Fierz expansion and the Clifford map

$$dx^{m_1} \wedge \cdots \wedge dx^{m_k} \to \gamma^{m_1 \ldots m_k} \ . \tag{6.6}$$

The system of differential equations equivalent to supersymmetry for type IIA supergravity reads:

$$d_H\big(e^{2A_6-\phi}\mathrm{Re}\phi_-\big) = -\frac{c_-}{16} F \ , \tag{6.7a}$$

$$d_H\big(e^{3A_6-\phi}\phi_+\big) = 0 \ , \tag{6.7b}$$

$$d_H\big(e^{4A_6-\phi}\mathrm{Im}\phi_-\big) = -\frac{c_+ e^{4A_6}}{16} *_6 \lambda F \ . \tag{6.7c}$$

Here, $\phi$ is the dilaton, $d_H = d - H\wedge$ is the twisted exterior derivative and $c_\pm$ are constants such that

$$\|\chi_1\|^2 \pm \|\chi_2\|^2 = c_\pm e^{\pm A_6} \ . \tag{6.8}$$

$F$ is the internal Ramond-Ramond flux which determines the external flux via self-duality:

$$F_{(10)} \equiv F + e^{6A_6} \mathrm{vol}_4 \wedge *_6 \lambda F \ . \tag{6.9}$$

$\lambda$ is an operator acting on a $p$-form $F_p$ as $\lambda F_p = (-1)^{[\frac{p}{2}]} F_p$, where square brackets denote the integer part.

### 6.3.2. AdS$_5$ $\times$ $M_5$

As we anticipated, we will now use the fact that anti-de Sitter space can be treated as a warped product of Minkowski space with a line. We would like to classify solutions of the type AdS$_5 \times M_5$. These in general will have a metric[4]

$$ds^2_{10} = e^{2A} ds^2_{\mathrm{AdS}_5} + ds^2_{M_5} \ . \tag{6.10}$$

Since

$$ds^2_{\mathrm{AdS}_5} = \frac{d\rho^2}{\rho^2} + \rho^2 ds^2_{\mathrm{Mink}_4} \ , \tag{6.11}$$

$ds^2_{10}$ in equation (6.10) can be put in the form of equation (6.5) if we take

$$e^{A_6} = \rho e^{A} \ , \qquad ds^2_{M_6} = \frac{e^{2A}}{\rho^2} d\rho^2 + ds^2_{M_5} \ . \tag{6.12}$$

In order to preserve the SO$(4,2)$ invariance of AdS$_5$, $A$ should be a function of $M_5$. In addition, the fluxes $F$ and $H$, which in subsection 6.3.1 were arbitrary forms on $M_6$, should now be forms on $M_5$. For IIA, $F = F_0 + F_2 + F_4 + F_6$; in order not to break SO$(4,2)$, we impose $F_6 = 0$.

Following the decomposition of the geometry of $M_6$ we wish to decompose the system of equations (6.7) so as to obtain the system equivalent to preserved supersymmetry for AdS$_5 \times M_5$. We start by decomposing the generators of the Clifford algebra Cl$(6,0)$ as

$$\gamma^{(6)}_\rho = \frac{e^A}{\rho} 1 \otimes \sigma_1 \ , \qquad \gamma^{(6)}_m = \gamma_m \otimes \sigma_2 \ , \qquad m = 1, \ldots, 5 \tag{6.13}$$

where $\sigma_1$, $\sigma_2$ are the Pauli matrices and $\gamma_m$ generate Cl$(5,0)$. Accordingly, the chirality matrix $\gamma^{(6)}_7 = 1 \otimes \sigma_3$ and the chiral spinors $\chi^+_1$, $\chi^-_2$ are decomposed in terms of Spin$(5)$ spinors $\eta_1$, $\eta_2$ as

$$\chi^+_1 = \sqrt{\frac{\rho}{2}} \eta_1 \otimes \begin{pmatrix} 1 \\ 0 \end{pmatrix} \ , \qquad \chi^-_2 = \sqrt{\frac{\rho}{2}} \eta_2 \otimes \begin{pmatrix} 0 \\ 1 \end{pmatrix} \ . \tag{6.14}$$

$\phi_-$ and $\phi_+$ now read

$$\phi_- = \frac{1}{2} \left( \frac{e^A}{\rho} d\rho \wedge \psi^1_+ + i \psi^1_- \right) \ , \qquad \phi_+ = \frac{1}{2} \left( -i \frac{e^A}{\rho} d\rho \wedge \psi^2_- + \psi^2_+ \right) \ , \tag{6.15}$$

where

$$\psi^1 \equiv e^{-A} \eta_1 \otimes \eta_2^\dagger \ , \qquad \psi^2 \equiv e^{-A} \eta_1 \otimes \overline{\eta_2} \ . \tag{6.16}$$

[4] Here $ds^2_{\mathrm{AdS}_5}$ is the unit radius metric on AdS$_5$. Moreover from now on $A_5 \equiv A$.

The bar is defined as $\overline{\eta} \equiv (\eta^c)^\dagger \equiv (B\eta^*)^\dagger = -\eta^t B$, where $B$ is a conjugation matrix that in five Euclidean dimensions can be taken to satisfy $B^* = B$, $B^t = -B$, $B^2 = BB^* = -1$. The subscripts plus and minus on $\psi^1$, $\psi^2$ refer to taking the even and odd form part respectively, in their expansion as forms. One should keep in mind here a comment about odd dimensions: the Clifford map (6.6) is not injective. Rather, a form $\omega$ and its cousin $*\lambda\omega$ are mapped to the same bispinor (recall the definition of $\lambda$ right after (6.9)). Thus a bispinor can always be expressed both as an even and as an odd form, and in particular we have

$$\psi^{1,2}_- = *\lambda\psi^{1,2}_+ \ . \tag{6.17}$$

Applying the decomposition (6.15) to equations (6.7) we obtain a *necessary and sufficient* system of equations for supersymmetric AdS$_5 \times M_5$ solutions:

$$d_H\big(e^{3A-\phi}\mathrm{Re}\psi^1_+\big) + 2e^{2A-\phi}\mathrm{Im}\psi^1_- = 0 \ , \tag{6.18a}$$

$$d_H\big(e^{4A-\phi}\psi^2_-\big) - 3ie^{3A-\phi}\psi^2_+ = 0 \ , \tag{6.18b}$$

$$d_H\big(e^{4A-\phi}\mathrm{Re}\psi^1_-\big) = 0 \ , \tag{6.18c}$$

$$d_H\big(e^{5A-\phi}\mathrm{Im}\psi^1_+\big) - 4e^{4A-\phi}\mathrm{Re}\psi^1_- = \frac{c_+}{8}e^{5A} * \lambda F \ . \tag{6.18d}$$

We also obtain the condition $c_- = 0$; it follows that the relation $\|\chi_1\|^2 \pm \|\chi_2\|^2 = c_\pm e^{\pm A_6}$ becomes

$$\|\eta_1\|^2 = \|\eta_2\|^2 = \frac{1}{2}c_+ e^A \ . \tag{6.19}$$

Henceforth, without loss of generality, we set $c_+ = 2$.

The stabilizer group $\mathcal{G} \in$ Spin(5) of $\eta_1$ and $\eta_2$ can be either SU(2) or the identity group. In the next section we parametrize $\psi^1$, $\psi^2$ in terms of these structures. We will see however that only the identity case leads to supersymmetric solutions. An identity structure is actually a choice of Vielbein; so we will end up parameterizing the $\psi^1$ and $\psi^2$ in terms of a Vielbein.

### 6.3.3. Parameterization of $\psi^1$, $\psi^2$ and the identity structure

We first consider the case where there is only one spinor, $\eta_1 = \eta_2$ of norm $e^{\frac{A}{2}}$. In five dimensions it defines an SU(2) structure. This can be read off from the Fierz expansions of $\eta_1 \otimes \eta_1^\dagger$ and $\eta_1 \otimes \overline{\eta_1}$, which as remarked in (6.17) can be written both as even and as odd forms:

$$\begin{aligned} \psi^1_+ &= \frac{1}{4}e^{-ij} \ , \qquad & \psi^2_+ &= \frac{1}{4}\omega \ , \\ \psi^1_- &= \frac{1}{4}v \wedge e^{-ij} \ , \qquad & \psi^2_- &= \frac{1}{4}v \wedge \omega \ . \end{aligned} \tag{6.20}$$

Application of Fierz identities yields

$$v\eta_1 = \eta_1 \tag{6.21}$$

and the following set of algebraic constraints on the 1-form $v$ and 2-forms $j$ and $\omega$:

$$\begin{aligned} \iota_v v = 1\ , \qquad & \iota_v j = \iota_v \omega = 0 \\ j\wedge\omega = 0\ , \qquad & \omega\wedge\omega = 0\ , \qquad \omega\wedge\overline{\omega} = 2j\wedge j = \text{vol}_4, \end{aligned} \tag{6.22}$$

where $\text{vol}_4$ is the volume form on the four-dimensional subspace orthogonal to $v$. This set of forms and constraints define precisely an SU(2) structure in five dimensions.

In this case, however, the two-form part of (6.18b) tells us $\psi^2 = 0$, which is only possible for $\eta_1 = 0$. Hence, there are no supersymmetric AdS$_5 \times M_5$ solutions in type IIA supergravity with an SU(2) structure on $M_5$.

Let us then consider the case of two spinors $\eta_1$ and $\eta_2$, which as mentioned earlier define an identity structure. We can expand $\eta_2$ in terms of $\eta_1$ as

$$\eta_2 = a\eta_1 + a_0\eta_1^c + \frac{1}{2}b\overline{w}\,\eta_1\ , \tag{6.23}$$

where $a$, $a_0 \in \mathbb{C}$, $b \in \mathbb{R}$ and $w$ is a complex vector that we normalize such that $w\cdot\overline{w} = 2$ (so that Re$w$ and Im$w$ are orthogonal and have norm 1). Also, by redefining if necessary $a \to a + \frac{b}{2}\overline{w}\cdot v$, $w \to w - (w\cdot v)v$ (which leaves (6.23) invariant, upon using (6.21)), we can assume

$$w\cdot v = 0\ . \tag{6.24}$$

Now (6.19) implies

$$|a|^2 + |a_0|^2 + b^2 = 1\ . \tag{6.25}$$

The identity structure is then spanned by $v$, $w$ and

$$u \equiv \frac{1}{2}\iota_{\overline{w}}\,\omega\ , \tag{6.26}$$

in terms of which

$$\omega = w\wedge u\ , \qquad -ij = \frac{1}{2}(w\wedge\overline{w} + u\wedge\overline{u})\ . \tag{6.27}$$

From (6.22) we now see that $u$ is also orthogonal to $v$, as well as to $w$ and $\overline{w}$; moreover, it satisfies $u\cdot\overline{u} = 2$. In other words,

$$\{v, \text{Re}w, \text{Im}w, \text{Re}u, \text{Im}u\} \tag{6.28}$$

are a Vielbein.

We can now expand $\psi^1$ and $\psi^2$ in terms of this Vielbein. We separate out their even and odd parts:

$$\begin{aligned}
\psi^1_+ &= \frac{\overline{a}}{4}\exp\left[-ij + \frac{w}{\overline{a}} \wedge (\overline{a_0}u - bv)\right] \ , \\
\psi^1_- &= \frac{1}{4}(\overline{a}v + bw) \wedge \exp\left[-ij + \frac{w}{\overline{a}} \wedge (\overline{a_0}u - bv)\right] \ , \\
\psi^2_+ &= -\frac{a_0}{4}\exp\left[-ij + \frac{u}{a_0} \wedge (aw - bv)\right] \ , \\
\psi^2_- &= -\frac{1}{4}(a_0 v + bu) \wedge \exp\left[-ij + \frac{u}{a_0} \wedge (aw - bv)\right] \ .
\end{aligned} \tag{6.29}$$

## 6.4. Analysis of the conditions for supersymmetry

Having obtained the expansions (6.29) of $\psi^1$, $\psi^2$ in terms of the identity structure on $M_5$, we can proceed with the study of the system (6.18). In section 6.4.1 we study the constraints imposed on the geometry of $M_5$ while in section 6.4.2 we obtain the expressions of the fluxes in terms of the geometry. The analysis in 6.4.1 is *local*.

### 6.4.1. Geometry

The equations of the system (6.18) which constrain the geometry of $M_5$ are (6.18a), (6.18b) and (6.18c) with the exception of the three-form part of (6.18a) which determines $H$. In the following study of these constraints, it is convenient to introduce the notation

$$a \equiv a_1 + i a_2 \ , \qquad k_1 \equiv \overline{a}v + bw \ , \qquad k_2 \equiv -bv + aw \ . \tag{6.30}$$

The zero form part of (6.18b), the one-form part of (6.18a), the two-form part of (6.18c) and the two-form part (6.18b) yield the following set of equations:

$$a_0 = 0 \ , \tag{6.31a}$$
$$d\big(e^{3A-\phi}a_1\big) + 2e^{2A-\phi}\mathrm{Im}k_1 = 0 \ , \tag{6.31b}$$
$$d\big(e^{4A-\phi}\mathrm{Re}k_1\big) = 0 \ , \tag{6.31c}$$
$$d\big(e^{4A-\phi}bu\big) - 3ie^{3A-\phi}u \wedge k_2 = 0 \ . \tag{6.31d}$$

It can then be shown that the higher-form parts of (6.18a), (6.18b) and (6.18c) follow from the above equations.

(6.31a) simplifies quite a bit (6.29), which now becomes

$$\begin{aligned} \psi_+^1 &= \frac{1}{4}\bar{a}\exp\left[-ij + \frac{b}{\bar{a}}v\wedge w\right] \ , \quad \psi_+^2 = \frac{1}{4}(aw - bv)\wedge u \wedge e^{-ij} \ , \\ \psi_-^1 &= \frac{1}{4}(\bar{a}v + bw)\wedge e^{-ij} \ , \quad \psi_-^2 = -\frac{1}{4}bu\wedge\exp\left[-ij - \frac{a}{b}v\wedge w\right] \ . \end{aligned} \tag{6.32}$$

It is also interesting to see what the pure spinors $\phi_\pm$ on $M_6$ look like:

$$\phi_+ = \frac{1}{4}E_1\wedge E_2\wedge\exp\left[\frac{1}{2}E_3\wedge\overline{E_3}\right] \ , \tag{6.33}$$

$$\phi_- = E_3\wedge\exp\left[\frac{1}{2}(E_1\wedge\overline{E_1} + E_2\wedge\overline{E_2})\right] \ , \tag{6.34}$$

where

$$E_1 \equiv ie^A b\frac{d\rho}{\rho} + aw - bv \ , \quad E_2 \equiv u \ , \quad E_3 \equiv e^A\bar{a}\frac{d\rho}{\rho} + i(\bar{a}v + bw) \ . \tag{6.35}$$

(6.33) are the canonical forms of a type 1 – type 2 pure spinor pair (where the "type" of a pure spinor is the lowest form appearing in it); or, in other words, of a pure spinor pair associated with an SU(2) structure on $M_6$ (although remember that the structure on $M_5$ is the identity). It would be interesting to push this further, and to start an analysis similar to the one in [96]: in that paper, the language of generalized complex geometry is used to set up a generalized reduction procedure, which eventually leads to a set of four-dimensional equations.

Let us now go back to (6.31). Given (6.31a), equation (6.25) becomes

$$a_1^2 + a_2^2 + b^2 = 1 \ . \tag{6.36}$$

Equations (6.31b) and (6.31c) can be integrated by introducing local coordinates $y$,

$$y = -\frac{1}{2}e^{3A-\phi}a_1 \ , \tag{6.37}$$

and $x$ such that

$$\mathrm{Im}k_1 = e^{-2A+\phi}dy \ , \qquad \mathrm{Re}k_1 = e^{-4A+\phi}dx \ . \tag{6.38}$$

$M_5$ possesses an abelian isometry generated by the Killing vector

$$\xi \equiv \frac{1}{2}(\eta_1^\dagger\gamma^m\eta_2 - \eta_2^\dagger\gamma^m\eta_2)\partial_m = -e^A b(\mathrm{Re}k_2)^\sharp \tag{6.39}$$

where $m = 1, \ldots, 5$ and the $\sharp$ superscript denotes the vector dual to the one-form it acts on. A straightforward way to show that $\xi$ is a Killing vector is to work directly with the supersymmetry variations (see appendix C.1) which yield $\nabla_{(m}\,\xi_{\nu)} = 0$ and $L_\xi \phi = L_\xi A = 0$, where $\nabla$ is the Levi-Civita connection and $L_\xi$ is the Lie derivative with respect to $\xi$. It would be interesting to show this directly using the language of generalized complex geometry, and to make contact with the analysis in [96].

Expressing $w$, $v$ in terms of $\mathrm{Re}k_2$, $\mathrm{Re}k_1$, and $\mathrm{Im}k_1$ we can write the metric on $M_5$ as

$$ds^2_{M_5} = ds^2_{\mathcal{C}} + (\mathrm{Re}k_2)^2 + \frac{e^{-4A+2\phi}}{b^2}\Big[(b^2+a_2^2)e^{-4A}dx^2 + (b^2+a_1^2)dy^2 + \\ + 2a_1a_2e^{-2A}dxdy\Big] \ , \quad (6.40)$$

where $ds^2_{\mathcal{C}} = u\bar{u}$, and $\mathcal{C}$ denotes the two-dimensional subspace spanned by $u$.

Let us introduce local coordinates $x^I$, $I = 1,2,3$ such that

$$ds^2_{\mathcal{C}} + (\mathrm{Re}k_2)^2 = g_{IJ}(x^I, x, y)dx^I dx^J \ . \quad (6.41)$$

$\phi$, $A$ and $a_2$ are in principle functions of $x^I$, $x$ and $y$. Given the fact that $L_\xi \mathrm{Re}k_1 = L_\xi \mathrm{Im}k_1 = 0$,[5] we can further introduce a coordinate $x^3 \equiv \psi$ adapted to the the Killing vector

$$\xi = 3\partial_\psi \ , \quad (6.42)$$

in terms of which

$$\mathrm{Re}k_2 = -\frac{1}{3}e^A bD\psi \ , \qquad D\psi \equiv d\psi + \rho \ , \qquad \rho = \rho_i(x^i, x, y)dx^i \ . \quad (6.43)$$

where $x^i$, $i = 1,2$ are local coordinates on $\mathcal{C}$. Thus

$$g_{IJ}(x^I, x, y)dx^I dx^J = (g_{\mathcal{C}})_{ij}(x^i, x, y)dx^i dx^j + \frac{1}{9}e^{2A}b^2 D\psi^2. \quad (6.44)$$

In addition, since $\xi$ is a Killing vector and $L_\xi \phi = L_\xi A = 0$, $A$, $\phi$ and $a_2$ are independent of $\psi$.

The exterior derivative on $M_5$ can be decomposed as

$$d = d_2 + d\psi \wedge \partial_\psi + dx \wedge \partial_x + dy \wedge \partial_y \ , \quad (6.45)$$

[5] Deduced from $\iota_\xi \mathrm{Re}k_1 = \iota_\xi \mathrm{Im}k_1 = 0$ and equation (6.38)

where $d_2$ is the exterior derivative on $\mathcal{C}$. We can thus further refine equation (6.31d) as follows:

$$d_2 u = i\rho_0 \wedge u \ , \tag{6.46a}$$
$$\partial_\psi u = iu \ , \tag{6.46b}$$
$$\partial_x u = f_1 u \ , \tag{6.46c}$$
$$\partial_y u = f_2 u \ , \tag{6.46d}$$

where

$$\rho_0 \equiv \rho + *_2 d_2 \log\big(b e^{4A-\phi}\big) \ , \tag{6.47a}$$
$$f_1(x^i, x, y) \equiv -\partial_x \log\big(e^{4A-\phi} b\big) + \frac{3e^{-5A+\phi} a_2}{b^2} \ , \tag{6.47b}$$
$$f_2(x^i, x, y) \equiv -\partial_y \log\big(e^{4A-\phi} b\big) + \frac{3e^{-3A+\phi} a_1}{b^2} \ . \tag{6.47c}$$

$*_2$ is the Hodge star defined by $g_{\mathcal{C}}$, such that $*_2 u = -iu$. Integrability of equations (6.46) yields the constraints

$$\partial_y f_1 = \partial_x f_2 \tag{6.48}$$

and

$$\partial_x \rho_0 = - *_2 \, d_2 f_1 \ , \tag{6.49a}$$
$$\partial_y \rho_0 = - *_2 \, d_2 f_2 \ . \tag{6.49b}$$

We can write $ds^2_{\mathcal{C}}$ as

$$ds^2_{\mathcal{C}} = e^{2\varphi(x^i,x,y)}(dx_1^2 + dx_2^2) \ . \tag{6.50}$$

The Gaussian curvature or one-half the scalar curvature of $\mathcal{C}$, $\ell(x^i, x, y)$, is

$$\ell(x^i, x, y) = -e^{-2\varphi}(\partial^2_{x_1} + \partial^2_{x_2})\varphi \ . \tag{6.51}$$

Equations (6.46b) and (6.46c), (6.46d) are solved by

$$u = e^{\varphi + i\psi}(dx_1 + i dx_2) \ , \qquad \partial_x \varphi = f_1 \ , \qquad \partial_y \varphi = f_2 \ . \tag{6.52}$$

Equation (6.46a) then yields

$$\rho_0 = \partial_{x_2}\varphi dx_1 - \partial_{x_1}\varphi dx_2 \ , \tag{6.53}$$

and thus

$$d_2 \rho_0 = \ell(x^i, x, y) \mathrm{vol}_{\mathcal{C}} \ . \tag{6.54}$$

Compatibility of (6.54) with (6.49a), (6.49b) requires that $\ell$ obey the equations

$$\partial_x \ell + 2 f_1 \ell = \Delta_2 f_1 \ , \tag{6.55a}$$
$$\partial_y \ell + 2 f_2 \ell = \Delta_2 f_2 \ , \tag{6.55b}$$

where $\Delta_2 \equiv {d_2}^\dagger d_2 + d_2 {d_2}^\dagger$. The last two equations also follow from (6.51), bearing in mind that $\Delta_2 \varphi = -e^{-2\varphi}(\partial^2_{x_1} + \partial^2_{x_2})\varphi$.

### 6.4.2. Fluxes

In this section we give the expressions for the fluxes in terms of the geometry of $M_5$. In the following expressions we employ the notation

$$\begin{aligned} \zeta_1 &\equiv \mathrm{Re}(a k_1)^\sharp = -2y e^{A} \partial_x - a_2 e^{2A-\phi} \partial_y \ , \\ \zeta_2 &\equiv \frac{1}{b^2} \mathrm{Im}(a k_1)^\sharp = a_2 e^{4A-\phi} \partial_x - 2y e^{-A} \partial_y \ . \end{aligned} \tag{6.56}$$

The NSNS three-form flux $H$ is given by the three-form part of equation (6.18a):

$$\begin{aligned} H = d\left( \frac{1}{6y} dx \wedge D\psi + \frac{1}{3} e^{A} \mathrm{Re}(a k_1) \wedge D\psi + \frac{e^{3A-\phi} a_2}{2y} \mathrm{vol}_{\mathcal{C}} \right) + \\ - \frac{1}{6y^2} dx \wedge dy \wedge D\psi + \frac{e^{-2A}}{y} dx \wedge \mathrm{vol}_{\mathcal{C}} + \frac{e^{3A-\phi} a_2}{2y^2} dy \wedge \mathrm{vol}_{\mathcal{C}} \ , \end{aligned} \tag{6.57}$$

where $\mathrm{Re}(a k_1) = -2y e^{-7A+2\phi} dx - a_2 e^{-2A+\phi} dy$.

The RR fluxes can be computed from equation (6.18d):

$$F_0 = -4e^{2A-2\phi}b^2\partial_y A - e^{-A}\iota_{\zeta_1} d\big(e^{A-\phi}a_2\big) \,, \tag{6.58a}$$

$$\begin{aligned} F_2 = & \big[-4e^{-A-\phi}a_2 + 4e^{4A-2\phi}\partial_x A - e^{-5A}\iota_{\zeta_2} d\big(e^{5A-\phi}a_2\big)\big]\,\mathrm{vol}_{\mathcal{C}} + \\ & + \frac{1}{3} d\big(e^{A-\phi}a_2\big)\wedge D\psi + F_0\frac{1}{3}e^{A}\mathrm{Re}(ak_1)\wedge D\psi + \\ & - \frac{e^{-A}}{b^2} *_2 d_2\big(e^{A-\phi}a_2\big)\wedge \mathrm{Im}(ak_1) + 4e^{-4A} *_2 d_2 A\wedge dx \,, \end{aligned} \tag{6.58b}$$

$$\begin{aligned} F_4 = & \frac{1}{3}\big[e^{-6A}\partial_y\big(e^{5A-\phi}a_2\big)dx - e^{-2A}\partial_x\big(e^{5A-\phi}a_2\big)dy + \\ & - 4e^{-2A}dy\big]\wedge d\psi\wedge \mathrm{vol}_{\mathcal{C}} + \\ & - \frac{1}{3}\big[4e^{-\phi}a_2 + e^{-4A}\iota_{\zeta_2} d\big(e^{5A-\phi}a_2\big)\big]\,\mathrm{Re}(ak_1)\wedge d\psi\wedge \mathrm{vol}_{\mathcal{C}} + \\ & - \frac{1}{3}e^{-10A+2\phi} *_2 \big[d_2\big(e^{5A-\phi}a_2\big)\big]\wedge dx\wedge dy\wedge D\psi \,, \end{aligned} \tag{6.58c}$$

where $\mathrm{Im}(ak_1) = a_2 e^{-4A+\phi}dx - 2ye^{-5A+2\phi}dy$.

The fluxes can also be computed from the expression

$$F = \mathcal{J}_+ \cdot d_H(e^{-\phi}\mathrm{Im}\phi_-) \tag{6.59}$$

on $M_6$ [97]. The operator $\mathcal{J}_+\cdot$ is associated with the pure spinor $\phi_+$, which can be found in (6.33):

$$\mathcal{J}_+\cdot = \frac{i}{2}\sum_{i=1}^{2}(E_i\wedge \overline{E_i}\llcorner - \overline{E_i}\wedge E_i\llcorner) + \frac{i}{2}(E_3\llcorner\overline{E_3}\llcorner + E_3\wedge\overline{E_3}\wedge) \,. \tag{6.60}$$

The degree of difficulty of computing the fluxes from (6.59) is proportional to the degree of the flux. The opposite is true for computing the fluxes from (6.18d).

### 6.4.3. Bianchi identities

In order to have a complete supersymmetric AdS$_5 \times M_5$ solution, apart from the conditions for supersymmetry (which imply the equations of motion [87]) the Bianchi identities of the fluxes need to be imposed. In this section we study the latter and the extra constraints that follow from their application.

We start with the Bianchi identity of $H$ i.e. $dH = 0$. We find that it determines

$$d_2\rho = e^{-2A}\big[6 + 12y(\partial_y A - f_2) - 6e^{5A-\phi}a_2(\partial_x A - f_1) + 3\partial_x\big(e^{5A-\phi}a_2\big)\big]\mathrm{vol}_{\mathcal{C}} \,. \tag{6.61}$$

Next, we turn to the Bianchi identities of the RR fluxes. The Bianchi identity of $F_0$ just says that it is a constant. The Bianchi identity of $F_2$ is

$$dF_2 - F_0\, H = 0\ . \tag{6.62}$$

The non-zero components on the left-hand side are the $dx \wedge \mathrm{vol}_{\mathcal{C}}$ and $dy \wedge \mathrm{vol}_{\mathcal{C}}$ components and imposing that they vanish yields the equations:

$$\begin{aligned} \partial_x \mathcal{Q} + 2 f_1 \mathcal{Q} - \left[\frac{1}{3}\partial_x\big(e^{A-\phi}a_2\big) - \frac{F_0}{6y}\right] *_2 d_2\rho - F_0 \frac{e^{-2A}}{y} + \\ + \Delta_2\big(e^{A-\phi}a_2\big)\frac{e^{-5A+\phi}a_2}{b^2} + \Delta_2(e^{-4A}) + \\ - d_2\big(e^{A-\phi}a_2\big)\cdot d_2\left(\frac{2e^{-5A+\phi}a_2}{b^2}\right) = 0\ , \end{aligned} \tag{6.63a}$$

$$\begin{aligned} \partial_y \mathcal{Q} + 2 f_2 \mathcal{Q} - \frac{1}{3}\partial_y\big(e^{A-\phi}a_2\big) *_2 d_2\rho - F_0 \frac{e^{3A-\phi}a_2}{2y^2} + \\ - \Delta_2\big(e^{A-\phi}a_2\big)\frac{2e^{-6A+2\phi}y}{b^2} + \\ + d_2\big(e^{A-\phi}a_2\big)\cdot d_2\left(\frac{4e^{-6A+2\phi}y}{b^2}\right) = 0\ . \end{aligned} \tag{6.63b}$$

where

$$\mathcal{Q}(x^i,x,y) \equiv -4e^{-A-\phi}a_2 + 4e^{4A-2\phi}\partial_x A - e^{-5A}\iota_{\zeta_2} d\big(e^{5A-\phi}a_2\big) - F_0\frac{e^{3A-\phi}a_2}{2y}\ . \tag{6.64}$$

Finally, the Bianchi identity of $F_4$

$$dF_4 - H \wedge F_2 = 0\ , \tag{6.65}$$

is automatically satisfied.

### 6.4.4. Summary so far

So far, we have analyzed the constraints imposed by supersymmetry and the Bianchi identities without any Ansatz; let us summarize what we have obtained.

First of all, we have already determined the local form of the metric: (6.40), (6.43). Most notably, we see the emergence of a Killing vector $\xi$ generating a U(1) isometry, and of a two-dimensional space $\mathcal{C}$. The geometry of $\mathcal{C}$ is constrained by (6.46). The $S^1$ upon which the U(1) acts is fibered over $\mathcal{C}$ with $\rho$ being the connection of the fibration. The curvature of the connection is given by (6.61).

In fact the U(1) isometry is a symmetry of the full solution as it also leaves invariant the fluxes; the latter can be verified by computing the Lie derivative with respect to $\xi$ of the fluxes' expressions as presented in section 6.4.2. This symmetry was to be expected: it is a U(1) R-symmetry corresponding to the R-symmetry of the dual $\mathcal{N} = 1$ field theory. The surface $\mathcal{C}$ is of less immediate interpretation, but already at this stage it seems to suggest that the field theory should be a compactification on $\mathcal{C}$ of a six-dimensional field theory. We will see later that this expectation is indeed borne out for the explicit solutions we will find.

We have also reduced the task of finding solutions to a set of partial differential equations on three functions: $a_2$, the dilaton $\phi$, and the warp factor $A$, which in general depend on four variables i.e. the coordinates $x^i, x, y$. Supersymmetry equations alone give us (6.48), (6.55a), (6.55b). Moreover, the fluxes should satisfy the relevant Bianchi identities, which away from sources give the further equations (6.58a), (6.63a), (6.63b). Thus we have a total of six partial differential equations. Solving all of them might seem a daunting task, but we will see in the next section that they simplify dramatically with a simple Ansatz. This will allow us to find many explicit solutions.

## 6.5. A simple Ansatz

We assume that $\phi$ and $A$ are functions of $y$ only and that $g_{\mathcal{C}}$ is independent of $x$ i.e. $f_1 = 0$. From equation (6.47b) it follows that $a_2 = 0$. The metric becomes

$$ds^2_{M_5} = ds^2_{\mathcal{C}} + \frac{1}{9}e^{2A}b^2 D\psi^2 + e^{-8A+2\phi}dx^2 + \frac{e^{-4A+2\phi}}{b^2}dy^2 \ , \tag{6.66}$$

where now $b^2 = 1 - a_1^2$. Equation (6.48) is satisfied trivially while equations (6.49a) and (6.49b) yield $\partial_x \rho = \partial_y \rho = 0$ (in the present Ansatz $\rho_0 = \rho$). $A(y)$ and $\phi(y)$ are subject to the differential equations coming from the Bianchi identities of $F_0$ and $F_2$, and equation (6.55b),

$$\partial_y \ell + 2 f_2 \ell = 0 \ . \tag{6.67}$$

$\ell$ is determined by (6.54) and (6.61) to be $\ell = 6e^{-2A} + 12e^{-2A}y(\partial_y A - f_2)$.

We first look at the Bianchi identity of $F_2$; it yields:

$$F_0(\partial_y A - f_2) = 0 \ , \tag{6.68}$$

so either $F_0 = 0$ or $f_2 = \partial_y A$. We consider the two cases $F_0 = 0$ and $F_0 \neq 0$ separately.

### 6.5.1. $F_0 = 0$

In this case, from the expression (6.58a) for $F_0$ we conclude that $\partial_y A = 0$ i.e. $A$ is constant which without loss of generality we set to zero. $F_2$ is zero, as can be seen from its expression (6.58b). We thus need to solve equation (6.67). This yields the ODE

$$\partial_y f_2 + 2 f_2^2 = 0 \ , \tag{6.69}$$

which is solved by

$$f_2 = \frac{1}{2}\frac{c}{cy - k} \ , \qquad c,\, k = \text{const.} \ . \tag{6.70}$$

Recalling the definition (6.47c) of $f_2$, equation (6.70) is in turn solved for

$$e^{2\phi} = \frac{k - cy}{2(c_1 - ky^2)} \ , \qquad c_1 = \text{const.} \ . \tag{6.71}$$

Equations (6.46) are then solved by

$$u = e^{i\psi}\sqrt{2(k - cy)}\,\widehat{u}(x_1, x_2) \ . \tag{6.72}$$

Substituting (6.70) and (6.72) in (6.61) yields

$$d_2\rho = 12k\,\text{vol}_\Sigma \ , \tag{6.73}$$

where $\Sigma$ is the surface spanned by $\widehat{u}$. Its Gaussian curvature is thus $12k$.

This solution was first discovered by Gauntlett, Martelli, Sparks and Waldram [115] (see appendix 6.7.1), and it is the T-dual of the AdS$_5 \times Y^{p,q}$ solution in type IIB supergravity.

### 6.5.2. $F_0 \neq 0$

In this case $f_2 = \partial_y A$; equations (6.46) are solved by

$$u = e^{i\psi}e^{A}\,\widehat{u}(x_1, x_2) \ . \tag{6.74}$$

$\ell = 6e^{-2A}$ obeys equation (6.67) automatically and

$$d_2\rho = 6\text{vol}_\Sigma \ . \tag{6.75}$$

Substituting $f_2$ in (6.47b) gives

$$e^{4A} = -\frac{1}{12y}\partial_y\beta \ , \tag{6.76}$$

where $\beta(y) \equiv e^{10A-2\phi}b^2$. The Bianchi identity of $F_0$ becomes then an ODE for $\beta$:

$$e^{12A}F_0 = -\beta\,\partial_y e^{4A} \ . \tag{6.77}$$

The situation appears promising: we have reduced the problem to the ODE (6.77). However, as we will now see, one cannot obtain physical compact solutions to this system.

Let us introduce the coordinate $\tilde{y}$ by $d\tilde{y} = \frac{e^{-2A+\phi}}{b}dy$, so that the metric (6.66) contains $d\tilde{y}^2$. (6.76) now reads

$$F_0 = 16e^{-\phi}b^2\partial_{\tilde{y}}A \ . \tag{6.78}$$

In order to obtain a compact solution, we should have the factor in front of the $S^1$ in (6.66), namely $e^A b$, go to zero for some $y = y_0$. For a regular point, this is impossible: since $A$ and $\phi$ should go to constant at $y_0$, we should have $b$ go to zero; but from (6.78) we see that this is in contradiction with $F_0 \neq 0$. We might think of having a singularity corresponding to a brane, but since only an $S^1$ would shrink at $y = y_0$, such a brane would be codimension-2; there are no such objects in IIA supergravity.

## 6.6. A compactification Ansatz

We have reduced the general classification problem to a set of six PDEs. To simplify the problem, we will now make an Ansatz.

We assume that $A$, $\phi$ and $a_2$ are functions of $x$ and $y$ only, and that

$$ds^2_{\mathcal{C}} = e^{2A}ds^2_{\Sigma}(x_1, x_2) \ . \tag{6.79}$$

In other words, the ten-dimensional metric becomes $ds^2_{10} = e^{2A}(ds^2_{\mathrm{AdS}_5} + ds^2_{\Sigma}) + ds^2_{M_3}$. It will soon follow that $\Sigma$ has constant curvature; from now on we will assume it to be a compact Riemann surface $\Sigma_g$. For $g \geq 1$ this involves a quotient by a discrete subgroup, but since no functions depend on its coordinates, this presents no difficulty.

This Ansatz is motivated by the fact that most known solutions in eleven-dimensional supergravity (and hence in massless IIA) are of this type. We also have in mind our original motivation for this chapter: finding solutions dual to twisted compactifications of CFT$_6$. If one wants to study a CFT$_6$ on $\mathbb{R}^4 \times \Sigma_g$ rather than on $\mathbb{R}^6$, one needs to replace $ds^2_{\mathrm{AdS}_7} = \frac{d\rho^2}{\rho^2} + \rho^2 ds^2_{\mathbb{R}^6}$ with $\frac{d\rho^2}{\rho^2} + \rho^2(ds^2_{\mathbb{R}^4} + ds^2_{\Sigma})$ in the UV, and then look for a solution that represents the flow to the IR. Our Ansatz is basically that in the IR fixed point this metric is only modified in the $\rho^2$ term multiplying $ds^2_\Sigma$, which drops out and becomes a constant.

Whatever its origin, we will now see that this Ansatz is remarkably effective at simplifying the system of PDEs: we will be able to completely classify the resulting solutions. One particular case will be source of many solutions, which will be analyzed in section 6.8.

### 6.6.1. Simplifying the PDEs

(6.79) implies

$$f_1 = \partial_x A \ , \qquad f_2 = \partial_y A \ . \tag{6.80}$$

The integrability condition (6.48) is then satisfied trivially, while equations (6.49a) and (6.49b) yield $\partial_x \rho = \partial_y \rho = 0$ (in the present Ansatz $\rho_0 = \rho$).

Equations (6.54), (6.61) yield $\ell = e^{-2A}[6 + 3\partial_x(e^{5A-\phi}a_2)]$. (6.55a), (6.55b) are then solved by

$$e^{5A-\phi}a_2 = cx + \epsilon \quad \ c = \text{const.} \ , \tag{6.81}$$

where $\epsilon = \epsilon(y)$ is a function of $y$ only. It follows that

$$\ell = e^{-2A}(6+3c) \tag{6.82}$$

i.e. the Gaussian curvature of $\Sigma_g$ is equal to $6 + 3c$.

Given the definitions (6.47b), (6.47c), the equations $f_1 = \partial_x A$ and $f_2 = \partial_y A$ become

$$\partial_x \left(e^{10A-2\phi} b^2\right) = 6e^{5A-\phi} a_2 \ , \tag{6.83a}$$

$$\partial_y \left(e^{10A-2\phi} b^2\right) = 6e^{7A-\phi} a_1 \ . \tag{6.83b}$$

Recall that $a_1 = -2ye^{-3A+\phi}$ and $b^2 = 1 - a_1^2 - a_2^2$. Using (6.81) we can solve these for

$$e^{10A-2\phi} - 4y^2 e^{4A} = c(c+3)x^2 + 2(c+3)\epsilon x + \beta \ , \tag{6.84a}$$

$$e^{4A} = -\frac{\epsilon'}{2y}x - \frac{1}{12y}\left(\beta' - 2\epsilon\epsilon'\right) \ , \tag{6.84b}$$

where $\beta = \beta(y)$ is a function of $y$ only, and a prime denotes differentiation with respect to $y$.

So far we have solved the differential equations imposed by supersymmetry; we now need to impose the Bianchi identities. First, the expression for $F_0$, (6.58a), becomes

$$e^{12A}F_0 = -[c(c+3)x^2 + 2(c+3)\epsilon x + \beta](e^{4A})' + e^{4A}\partial_y(cx+\epsilon)^2 + 2e^{8A}cy \ . \tag{6.85}$$

Recalling (6.84b), we see that this equation is polynomial in $x$, of degree 3. In other words, we can view it as a set of four ODEs in $y$.

The Bianchi identities for $F_2$, (6.63), become

$$\partial_x^2(e^{6A-2\phi}) = 0 \ , \tag{6.86a}$$

$$\partial_y\partial_x(e^{6A-2\phi}) + F_0\frac{\epsilon'}{2y} = 0 \ . \tag{6.86b}$$

Substituting equations (6.81) and (6.84) in (6.86a) yields the differential equation

$$36(\epsilon')^2\beta = -(c+3)(\beta' - 2\epsilon\epsilon')\left[c\beta' - 2(c+6)\epsilon\epsilon'\right] \ . \tag{6.87}$$

Notice that the $x$ dependence has dropped out from this equation. Concerning (6.86b), just as for (6.85), it can be written as a polynomial in $x$ of degree 3, and viewed as four ODEs in $y$.

So we appear to have reduced the problem to four ODEs from (6.85), one from (6.86a) (which becomes (6.87)), and four from (6.86b), for a total of nine ODEs in $y$. However, many of these ODEs actually happen not to be independent from each other. For example, the $x^3$ component of both (6.85) and (6.86b) gives

$$4c(c+3)\left(\frac{\epsilon'}{y}\right)' + F_0\left(\frac{\epsilon'}{y}\right)^3 = 0 \ , \tag{6.88}$$

as well as the $x^2$ component of (6.86b).

To analyze the remaining ODEs, as a warm-up we will first look at the case $F_0 = 0$, where we will reproduce several known solutions. We will then look at the case $F_0 \neq 0$, which we will further split into a generic case where $\epsilon' \neq 0$, and a special case where $\epsilon' = 0$; both will give rise to new solutions.

## 6.6.2. $F_0 = 0$

For $F_0 = 0$, (6.88) becomes $c(c+3)(\epsilon' - y\epsilon'') = 0$. We can then have either $\epsilon' = y\epsilon''$, $c = -3$, or $c = 0$. In the $c = 0$ case, actually the $x^2$ coefficient of (6.85) gives again $\epsilon' = y\epsilon''$. So this case becomes a subcase of the $\epsilon' = y\epsilon''$ case.

- **Case 1:** $\epsilon' = y\epsilon''$. In this case we have

$$\epsilon = \frac{1}{2}c_1y^2 + c_2 \ , \qquad c_1,\, c_2 = \text{const.} \ . \tag{6.89}$$

The $x^3$ component of (6.85) is (6.88), which we just looked at. The $x^2$ and $x^1$ components both require

$$\left(\frac{\beta'}{y}\right)' = 2\frac{c+3}{c+6}c_1y \ . \tag{6.90}$$

The solution to this ODE is

$$\beta = \frac{c+6}{c+3}\frac{1}{4}c_1^2 y^4 + \frac{1}{2}c_3 y^2 + c_4 \qquad c_3,\, c_4 = \text{const.}\ . \tag{6.91}$$

The $x^0$ component of (6.85) then gives

$$(2c_1c_2 - c_3)(2(c+6)c_1c_2 - cc_3) + \frac{36}{(c+3)}c_1^2 c_4 = 0\ . \tag{6.92}$$

Generically this can be solved for $c_4$. In this case, the transformation

$$x \to x + \frac{\delta}{c}\ , \qquad c_2 \to c_2 - \delta\ , \qquad \beta \to \beta + \frac{(3+c)(\delta^2 - 2\delta\epsilon)}{c} \tag{6.93}$$

leaves the solution invariant and $\delta$ can be chosen such that

$$\beta = \frac{c+6}{c+3}\epsilon^2. \tag{6.94}$$

This branch reproduces the solution obtained from reduction to ten dimensions of the Bah–Beem–Bobev–Wecht AdS$_5$ solution of M-theory [119], as described in appendix 6.7.4.

This however does not cover the case $c_1 = 0$. Treating this separately, we find that (6.92) leads to $c = 0$. This branch reproduces the Itsios–Núñez–Sfetsos–Thompson solution [124], discussed in subsection 6.7.2.

- **Case 2:** $c = -3$. In this case, the $x^2$ component of (6.85) gives $\epsilon' = 0$. With this, the whole of (6.85) gives

$$2\beta(\beta' - y\beta'') + y\beta'^2 = 0\ . \tag{6.95}$$

This equation is nonlinear, but if one defines $z = y^2/2$ it becomes $2\beta\partial_z^2\beta = (\partial_z\beta)^2$, which is easily solved by the square of a linear function; in other words, by

$$\beta = c_2(y^2 + 4c_1)^2\ , \qquad c_1,\, c_2 = \text{const.}\ . \tag{6.96}$$

This case reproduces the solution obtained from reduction to ten dimensions of the Maldacena–Núñez AdS$_5$ solution of M-theory [117], described in subsection 6.7.3.

### 6.6.3. $F_0 \neq 0$

We will divide the analysis in the generic case, where $c \neq 0$ and $-3$, and two special cases $c = 0$ or $-3$. Let us note that from (6.87), we see that $\epsilon' = 0$ implies either $c = 0$ or $-3$; in other words, if $c \neq 0$ and $-3$, then $\epsilon' \neq 0$. On the other hand, from (6.88), we see that $\epsilon' \neq 0$ implies $c \neq 0$ and $-3$; in other words, if $c = 0$ or $-3$, then $\epsilon' = 0$.

### Generic case

We begin by analyzing (6.85) with the aid of (6.87) and (6.88). In particular, combining the last two and on the condition that

$$\beta' \neq \frac{c+3}{c} 2\epsilon\epsilon' \ , \tag{6.97}$$

we obtain the following expression for the derivative of (6.84b):

$$(e^{4A})' = \frac{(\epsilon')^2}{8c(c+3)y^3} \left[ F_0 \epsilon' x + \frac{1}{6} F_0 (\beta' - 2\epsilon\epsilon') - 4cy^2 \right] \ . \tag{6.98}$$

Substituting $(e^{4A})'$ and the expression for $\beta$ provided by (6.87) in (6.85), we find that the latter gives

$$\beta' = \frac{c+3}{c} 2\epsilon\epsilon' \ ; \tag{6.99}$$

this is incompatible with the assumption (6.97).

We thus proceed with the case $c\beta' = (c+3)2\epsilon\epsilon'$. Equation (6.87) fixes

$$\beta = \frac{c+3}{c}\epsilon^2 \ , \tag{6.100}$$

while equation (6.86b) follows from (6.87) and (6.88); the latter can be solved by quadrature. The solution is

$$\epsilon = -\frac{2\sqrt{2c(c+3)}}{3F_0^2}(F_0 y - 2c_1)\sqrt{F_0 y + c_1} + c_2 \ , \qquad c_1, \, c_2 = \text{const.} \ . \tag{6.101}$$

Upon substituting the derived expressions for $\beta$ and $\epsilon$, (6.85) becomes

$$\frac{3y(c+3)(cx+\epsilon)^2}{c(c_1 + F_0 y)} = 0 \ , \tag{6.102}$$

which cannot hold for $c \neq 0, -3$.

We conclude that there are no solutions in the generic case.

### Special cases

- For $c = 0$, equation (6.88) is trivially satisfied, while the $x^1$ component of (6.85) yields $\epsilon = 0$. Then $a_2 = 0$ and this leads to the (unphysical) massive solution of section 6.5.2.

- For $c = -3$, (6.88) is again trivially satisfied, while (6.85) yields the following ODE for $\beta$:

$$e^{12A} F_0 = -\beta \, (e^{4A})' - 6e^{8A} y \; . \tag{6.103}$$

Using (6.84b) we see $e^{4A} = -\frac{\beta'}{12y}$. This ODE is nonlinear, and a little tougher than the ones we saw so far in this subsection. Hence we defer its further analysis to the next section. We will see there that it leads to many new AdS$_5$ solutions.

## 6.7. Recovered solutions

In this section we discuss a set of known, supersymmetric AdS$_5 \times M_5$ solutions of type IIA supergravity with zero Romans mass, which we recovered in our analysis. Two of them descend from AdS$_5$ solutions of M-theory, whose reduction to ten dimensions we present. We focus on the geometry of the solutions, as the fluxes are determined by it. We aim to adhere to the notation of the original papers; whenever there is overlap with notation used in the main body of the chapter, we add a hat $\hat{}$ .

There are more supersymmetric AdS$_5$ solutions in IIA [126, 127, 128, 129] that should be particular cases of our general classification of section 6.4. These are outside the compactification Ansatz of section 6.6.

### 6.7.1. The Gauntlett–Martelli–Sparks–Waldram solution

The metric on $M_5$ reads

$$ds^2_{M_5} = \frac{k - cy}{6m^2} ds^2_{C_k} + e^{-6\lambda} \sec^2 \zeta + \frac{1}{9m^2} \cos^2 \zeta D\psi^2 + e^{-6\lambda} dx_3^2 \; , \tag{6.104}$$

where

$$e^{6\lambda} = \frac{2m^2(\hat{a} - ky^2)}{k - cy} \; , \qquad \cos^2 \zeta = \frac{\hat{a} - 3ky^2 + 2cy^2}{\hat{a} - ky^2} \; . \tag{6.105}$$

The dilaton is given by $e^{-2\phi} = e^{6\lambda}$.

$\hat{a}$, $c$ are constants, $k = 0, \pm 1$ and $m^{-1}$ is the radius of AdS$_5$. $C_k$ is a Riemann surface of unit radius; it is a sphere $S^2$, a torus $T^2$ or a hyperbolic space $H^2$ for $k = 1, 0$ or $-1$ respectively. The GMSW solution is the reduction to ten dimensions of an AdS$_5 \times M_6$ solution of M-theory, where $M_6$ is a fibration of $S^2$ over $C_k \times T^2$ and the reduction is along an $S^1 \subset T^2$.

The solution is the one recovered in subsection 6.5.1. The constants $c$ and $k$ are identified with the corresponding of 6.5.1, while $\hat{a} = c_1$. The coordinate $x_3$ is related to $x$ via $x_3 = -x$; a minus is introduced for matching the expressions of the fluxes. Finally, in 6.5.1 $m = 1$.

### 6.7.2. The Itsios–Núñez–Sfetsos–Thompson solution

This solution was discovered in [124] by nonabelian T-dualizing the AdS$_5 \times T^{1,1}$ solution in type IIB supergravity. The metric on $M_5$ reads

$$ds^2_{M_5} = \lambda_1^2 ds^2_{S^2} + \frac{\lambda_2^2\lambda^2}{\Delta} x_1^2 D\psi^2 + \frac{1}{\Delta}\left[(x_1^2 + \lambda^2\lambda_2^2)dx_1^2 + (x_2^2 + \lambda_2^4)dx_2^2 + \right.$$
$$\left. + 2x_1x_2dx_1dx_2\right] \; , \qquad (6.106)$$

where

$$\Delta = \lambda_2^2 x_1^2 + \lambda^2(x_2^2 + \lambda_2^4) \; , \qquad \lambda_1^2 = \lambda_2^2 = \frac{1}{6} \; , \qquad \lambda^2 = \frac{1}{9} \; , \qquad (6.107)$$

and

$$ds^2_{S^2} = d\theta_1^2 + \sin^2\theta_1 \phi_1^2 \; , \qquad \rho = \cos\theta_1 d\phi_1 \; . \qquad (6.108)$$

The dilaton is given by $e^{-2\phi} = \Delta$.

The solution fits into the $c_1 = 0$ branch of the first case of subsection 6.6.2 for $c_3 = -12$ (achieved by setting the constant warp factor to zero) and $\epsilon = c_2 = \lambda\lambda_2^2$. $\Sigma_g$ is $S^2$ of radius $\frac{1}{\sqrt{6}}$. The coordinate transformation relating $x_1, x_2$ to $x, y$ is:

$$x_1^2 = -36y^2 + 36\epsilon x + 6c_4 - 6\epsilon^2 \; , \qquad x_2 = 6y \; . \qquad (6.109)$$

### 6.7.3. The Maldacena–Núñez solution

We write the metric of the $\mathcal{N} = 1$ Maldacena–Núñez solution [117] in the form presented in [115]:

$$e^{-2\lambda} ds^2_{11} = ds^2_{\mathrm{AdS}_5} + \frac{1}{3} ds^2_{H^2} + e^{-6\lambda}\sec^2\zeta dy^2 + \frac{1}{9m^2}\cos^2\zeta\left((d\psi + \tilde{P})^2 + ds^2_{S^2}\right) \; , \qquad (6.110)$$

where

$$e^{6\lambda} = \hat{a} + y^2 \; , \qquad \cos^2\zeta = \frac{\hat{a} - 3y^2}{\hat{a} + y^2} \; , \qquad (6.111)$$

and $m^{-1}$ is the radius of AdS$_5$. The metrics on $H^2$ and $S^2$ are

$$ds^2_{H^2} = \frac{dX^2 + dY^2}{Y^2} \; , \qquad ds^2_{S^2} = d\theta^2 + \sin^2\theta d\nu^2 \; , \qquad (6.112)$$

while the connection of the fibration of $\psi$ is

$$\tilde{P} = -\cos\theta d\nu - \frac{dX}{Y} \; . \qquad (6.113)$$

### Reduction to ten dimensions

We reduce the Maldacena–Núñez solution to ten dimensions, along $\nu$. In order to do so, we rewrite the part of $ds^2_{M_6}$ involving $d\psi$ or $d\nu$ as

$$\frac{1}{9m^2}\cos^2\zeta\left[(d\nu+A_1)^2+\sin^2\theta D\psi^2\right] \ , \tag{6.114}$$

where

$$A_1=-\cos\theta D\psi \ , \qquad \rho=-\frac{dX}{Y} \ . \tag{6.115}$$

Reducing along $d\nu$ yields then

$$ds^2_{10}=e^{2A}ds^2_{\text{AdS}_5}+ds^2_{M_5} \ , \tag{6.116}$$

where

$$e^{-2A}ds^2_{M_5}=\frac{1}{3}ds^2_{H^2}+e^{-6A+2\phi}\sec^2\zeta dy^2+\frac{1}{9m^2}\cos^2\zeta\left(d\theta^2+\sin^2\theta D\psi^2\right) \ . \tag{6.117}$$

Furthermore,

$$\phi=\frac{3}{4}\log\left(\frac{1}{9m^2}e^{2\lambda}\cos^2\zeta\right) \ , \qquad A=\lambda+\frac{1}{3}\phi \ . \tag{6.118}$$

The reduced Maldacena–Núñez solution fits into the second case of section 6.6.2, for $\epsilon=0$ (achieved by by a $x\to x+\frac{\epsilon}{3}$ shift), $c_1=-\frac{\hat{a}}{12}$ and $c_2=1$. $\Sigma_g$ is $H^2$ of radius $\frac{1}{\sqrt{3}}$. In our conventions $m=1$. The coordinate transformation relating $x$ to $y,\theta$ is:

$$x=-\frac{1}{9}(\hat{a}-3y^2)\cos\theta \ . \tag{6.119}$$

### AdS$_7$ variables

For our discussion in the main text, it is useful to also include two parameters $R$ and $k$ which are usually set to one. If we use the slightly awkward-looking

$$\beta=\frac{4}{k^2}\left(y^2-\frac{3^4}{2^{10}}R^6\right)^2 \tag{6.120}$$

the corresponding solution, using (6.141) and (6.144), is

$$ds^2_{M_5}=e^{2A}ds^2_{\Sigma_g}+\frac{1}{3^{3/2}k}\frac{64dy^2}{\sqrt{9^2R^6-32^2y^2}}+\frac{(9^2R^6-32^2y^2)^{3/2}}{16(3^5R^6+32^2y^2)} \ , \tag{6.121}$$

$$e^{4A}=\frac{9^2R^6-32^2y^2}{3\cdot 2^8k^2} \ , \qquad e^{4\phi}=\frac{(9^2R^6-32^2y^2)^3}{2\cdot 6^3k^6(3^5R^6+32^2y^2)^2} \ . \tag{6.122}$$

These again look messy, but upon using the map (6.153) and defining an angle $\alpha$ via

$$\cos\alpha \equiv \frac{32}{9R^3}y \qquad (6.123)$$

turn into the expressions for the metric, $A$ and $\phi$ of the massless AdS$_7$ solution, obtained by reducing AdS$_7 \times S^4/\mathbb{Z}_k$ to IIA supergravity: see section 5.6.1.

In the main text we will need an expression for the $B$ field of the AdS$_5$ solution. We give it directly in terms of $x_7$, which is related to (6.151) via (6.153):

$$B = \frac{R^3}{48k}x_7\frac{(5-x_7^2)}{1+\frac{1}{3}x_7^2}\mathrm{vol}_{S^2} + \frac{1}{\sqrt{3}}\frac{x_7}{\sqrt{1-x_7^2}}\cos\theta\mathrm{vol}_{\Sigma_g}\ . \qquad (6.124)$$

This is similar to the one given for the AdS$_7$ solution in [I, Eq.(5.8)].

### 6.7.4. The Bah–Beem–Bobev–Wecht solution

The metric of the solution in [119] reads

$$ds^2_{11} = e^{2\lambda}\left[ds^2_{\mathrm{AdS}_5} + e^{2\nu+2\hat{A}(x_1,x_1)}(dx_1^2+dx_2^2)\right] + e^{-4\lambda}ds^2_{M_4}\ , \qquad (6.125)$$

where $ds^2_{\mathrm{AdS}_5}$ is the unit radius metric on AdS$_5$, and $\hat{A}(x_1,x_2)$ is the conformal factor of the constant curvature metric on the Riemann surface $\widehat{\Sigma}_g$ of genus $g$, obeying

$$(\partial^2_{x_1}+\partial^2_{x_2})\hat{A} + \kappa e^{2\hat{A}} = 0\ . \qquad (6.126)$$

The constant $\kappa$ is the Gaussian curvature of the Riemann surface which is set to 1, 0 or $-1$ for the sphere $S^2$, the torus $T^2$ or a hyperbolic surface respectively. $\nu$ is a real constant. The metric $ds^2_{M_4}$ is

$$ds^2_{M_4} = \left(1+\frac{4y^2}{qf}\right)dy^2 + \frac{qf}{k}\left(dq+\frac{12yk}{qf}dy\right)^2 + \\ + \frac{\hat{a}_1^2}{4}\frac{fk}{q}(d\chi+V)^2 + \frac{qf}{9}(d\psi+\hat{\rho})^2\ . \qquad (6.127)$$

The metric functions are

$$e^{6\lambda} = qf+4y^2\ , \qquad f(y)\equiv 1+6\frac{\hat{a}_2}{\hat{a}_1}y^2\ , \qquad k(q)\equiv\frac{\hat{a}_2}{\hat{a}_1}q^2+q-\frac{1}{36}\ , \qquad (6.128)$$

while the one-forms which determine the fibration of the $\psi$ and $\chi$ directions are given by

$$\hat{\rho} = (2-2g)V - \frac{1}{2}\left(\hat{a}_2 + \frac{\hat{a}_1}{2q}\right)(d\chi + V) \ , \qquad dV = \frac{\kappa}{2-2g} e^{2\hat{A}} dx_1 \wedge dx_2 \ . \tag{6.129}$$

The constants $\hat{a}_1$, $\hat{a}_2$ are fixed as

$$\hat{a}_1 \equiv \frac{2(2-2g)e^{2\nu}}{\kappa} \ , \qquad \hat{a}_2 \equiv 2(2-2g)\left(1 - \frac{6e^{2\nu}}{\kappa}\right) \ . \tag{6.130}$$

### Reduction to ten dimensions

We reduce the Bah–Beem–Bobev–Wecht solution to ten dimensions, along $\chi$. In order to do so, we rewrite the part of $ds^2_{M_4}$ involving $d\psi$ or $d\chi$ as

$$h_1^2(d\chi + A_1)^2 + h_2^2 D\psi^2 \ , \tag{6.131}$$

where

$$h_1^2(y,q) \equiv \frac{\hat{a}_1^2}{4}\frac{fk}{q} + \frac{1}{4}\frac{qf}{9}\left(\hat{a}_2 + \frac{\hat{a}_1}{2q}\right)^2 \ , \tag{6.132a}$$

$$h_2^2(y,q) \equiv \frac{qf}{9} - \frac{1}{4}\left(\frac{qf}{9}\right)^2\left(\hat{a}_2 + \frac{\hat{a}_1}{2q}\right)^2 \ , \tag{6.132b}$$

and

$$\rho = (2-2g)V \ , \qquad A_1 = V - \frac{qf}{9}\frac{1}{2}\left(\hat{a}_2 + \frac{\hat{a}_1}{2q}\right) h_1^{-1} D\psi \ . \tag{6.133}$$

Reducing along $d\chi$ yields then

$$ds^2_{10} = e^{2A}\left[ds^2_{\mathrm{AdS}_5} + e^{2\nu + 2\hat{A}}(dx_1^2 + dx_2^2)\right] + ds^2_{M_3} \ , \tag{6.134}$$

where

$$e^{4A-2\phi} ds^2_{M_3} = \left(1 + \frac{4y^2}{qf}\right) dy^2 + \frac{qf}{k}\left(dq + \frac{12yk}{qf}dy\right)^2 + h_2^2 D\psi^2 \ . \tag{6.135}$$

Furthermore,

$$\phi = \frac{3}{2}(\log h_1 - 2\lambda) \ , \qquad A = \frac{1}{2}\log h_1 \ . \tag{6.136}$$

The reduced solution fits into the generic branch of the first case of subsection 6.6.2 for $c_1 = \frac{9\hat{a}_1+\hat{a}_2}{108}$, $c_2 = \frac{(9\hat{a}_1+\hat{a}_2)\hat{a}_2}{18\hat{a}_1}$ and $c = \frac{\hat{a}_2}{3\hat{a}_1}$. The coordinate transformation relating $x$ to $y, q$ is:

$$x = -\frac{\hat{a}_1(18\hat{a}_1 + \hat{a}_2 + 18\hat{a}_2 q)}{36\hat{a}_2}\left(1 + 6\frac{\hat{a}_2}{\hat{a}_1}y^2\right) \ . \tag{6.137}$$

Certain generalizations of the Bah–Beem–Bobev–Wecht class of solutions have also appeared [130, 131]. It would be interesting to reduce these to solutions of IIA supergravity and verify that they fit in our classification of section 6.4.

# 6.8. Compactification solutions

We will now analyze further the case we started considering in section 6.6.3. We will see that it corresponds to a compactification of the AdS$_7$ solutions considered in chapter 5. Moreover, we will be able to find the most general explicit solution, thus providing a new infinite class of AdS$_5$ solutions.

## 6.8.1. Metric and fluxes

In section 6.6.3, we found that there are AdS$_5$ solutions associated with solutions of the ODE (6.103). Replacing the expression of $A$ given there, we have

$$\beta\,(y\beta'' - \beta') = \frac{1}{2}y(\beta')^2 - \frac{F_0}{144y}(\beta')^3 \ . \tag{6.138}$$

This equation is non-linear; however, it can be rewritten as

$$(q_5^2)' = \frac{2}{9}F_0 \ , \qquad q_5 \equiv -\frac{4y\sqrt{\beta}}{\beta'} \ . \tag{6.139}$$

We will see later that $q_5$ has actually a useful physical interpretation (similar to the $q$ introduced in section 5.5.8 of chapter 5): it will turn out to be related to D8-brane positions. In any case, the trick (6.139) allows us to solve the ODE (6.138): indeed we can write $16y^2\frac{\beta}{(\beta')^2} = \frac{2}{9}F_0(y - \hat{y}_0)$, which can now be integrated by quadrature.

We will postpone the detailed analysis of the solutions of (6.138) to sections 6.8.5 and 6.8.6. For the time being, in this subsection we will collect various features of the resulting AdS$_5$ solutions.

The internal metric for the class we are considering can be extracted from the general expression (6.40). However, at first its global meaning is not transparent.

It proves useful to trade the coordinate $x$ for a new coordinate $\theta$, defined by

$$\cos\theta = \frac{-3x+\epsilon}{\sqrt{\beta}} \ . \tag{6.140}$$

The metric then becomes

$$ds^2_{M_5} = e^{2A} ds^2_{\Sigma_g} + ds^2_{M_3} \ , \qquad ds^2_{M_3} = dr^2 + \frac{1}{9} e^{2A}(1-a_1^2) ds^2_{S^2} \ . \tag{6.141}$$

Here

$$ds^2_{S^2} = d\theta^2 + \sin^2\theta D\psi^2 \tag{6.142}$$

is the metric of the round $S^2$, fibered over $\Sigma_g$, which is a Riemann surface of Gaussian curvature $-3$ (recalling (6.82), and $c=-3$) and hence $g \geq 2$; The new coordinate $r$ is defined by

$$dr = \frac{e^{3A}}{\sqrt{\beta}} dy \ . \tag{6.143}$$

Moreover, from (6.37) and (6.84) we have

$$1 - a_1^2 = \frac{3\beta}{3\beta - y\beta'} \ , \qquad e^{4A} = -\frac{\beta'}{12y} \ , \qquad e^{\phi} = \frac{\sqrt{3} e^{5A}}{\sqrt{3\beta - y\beta'}} \ . \tag{6.144}$$

We can now remark that the $q_5$ defined in (6.139) is

$$q_5 \equiv e^{-\phi} R_{S^2} \equiv \frac{1}{3} e^{A-\phi} \sqrt{1-a_1^2} = -\frac{4y\sqrt{\beta}}{\beta'} \ . \tag{6.145}$$

$R_{S^2} = \frac{1}{3} e^A \sqrt{1-a_1^2}$ is the radius of the round $S^2$, as inferred from (6.141). The role of this particular combination of the radius and dilaton will become clearer in section 6.8.4.

From (6.145) and (6.144) we see that for the solution to make sense we must require

$$\beta \geq 0 \ , \qquad -\frac{\beta'}{y} \geq 0 \ . \tag{6.146}$$

We can now also obtain the fluxes, from the formulas in section 6.4.2. We have

$$F_2 = q_5 \left[ -(\mathrm{vol}_{S^2} + 3\cos\theta \mathrm{vol}_{\Sigma_g}) + \frac{1}{3} F_0 a_1 e^{A+\phi} \mathrm{vol}_{S^2} \right] \ , \tag{6.147}$$

where $\mathrm{vol}_{S^2} \equiv \sin\theta d\theta \wedge D\psi$. The four-form flux reads

$$F_4 = \frac{1}{3} \mathrm{vol}_{\Sigma_g} \wedge \left[ \frac{2y\beta}{3\beta - y\beta'} \cos\theta \mathrm{vol}_{S^2} + \sin^2\theta D\psi \wedge dy \right] \ . \tag{6.148}$$

When $F_0 \neq 0$, we need not give an expression for $H$: as usual for massive IIA, it can be written as $H = dB$, where

$$B = \frac{F_2}{F_0} + b \ , \qquad (6.149)$$

where $b$ is a closed two-form. When $F_0 = 0$, the only solution in the class we are considering in this section is the Maldacena–Núñez solution; an expression for $B$ is presented for that case in (6.124).

We can observe already now that the metric (6.141) and the flux (6.147) look related to those for the $AdS_7$ solutions of chapter 5; see (4.16) and (4.9) there. The expressions are very similar; one obvious difference is that the three-dimensional metric in (6.141) is fibered over $\Sigma_g$, and that the flux (6.147) has extra legs along $\Sigma_g$. Except for a few numerical factors, everything seems to correspond nicely; the role of $x$ introduced in section 5.5.1 of chapter 5 seems to be played here by $a_1$:

$$x \text{ in } (5.40) \quad \to \quad a_1 \text{ here.} \qquad (6.150)$$

Actually this correspondence can be justified a little better. In (5.40), $x$ is the zero-form part of $\mathrm{Im}\psi^1_+$ (see (5.36) and (5.55)), which is the calibration for a D6-brane extended along $AdS_7$. The analogue of this in the present case would be a D6-brane extended along $AdS_5 \times \Sigma_g$; the relevant calibration is the part along $u \wedge \bar{u}$ of $\mathrm{Im}\psi^1_+$ in (6.29): we see that the latter is indeed $\mathrm{Re}a = a_1$.

Motivated by this, in this section we will also use the name

$$x_5 \equiv a_1 \ . \qquad (6.151)$$

This $x_5$ is meant to evoke the $x$ in (5.40), and is not to be confused with the coordinate $x$ we temporarily used in sections 6.4 and 6.6.

## 6.8.2. Correspondence with $AdS_7$

We will now show that solutions of the type considered in section 6.6.3 are in one-to-one correspondence with the $AdS_7$ solutions of chapter 5. The map we will find is directly inspired from a similar map from $AdS_4$ to $AdS_7$ found in [120]. It would be possible to present our new $AdS_5$ solutions perfectly independently from the map to $AdS_7$; in fact, in finding the analytic solutions the map does not help at all. However, the existence of the map tells us right away that infinitely many regular solutions do exist, and what data they depend on.

Let us start from (6.138). Using the definition (6.143), the expressions (6.144) and the expression $x_5 = a_1 = -2ye^{-3A+\phi}$ from (6.151), (6.37), we can see that

$$\begin{aligned}
\partial_r \phi &= \frac{1}{4}\frac{e^{-A}}{\sqrt{1-x_5^2}}(11x_5 - 2x_5^3 + (2x_5^2-5)F_0 e^{A+\phi}) \ , \\
\partial_r x_5 &= -\frac{1}{2}e^{-A}\sqrt{1-x_5^2}(4 - x_5^2 + x_5 F_0 e^{A+\phi}) \ , \\
\partial_r A &= \frac{1}{4}\frac{e^{-A}}{\sqrt{1-x_5^2}}(3x_5 - F_0 e^{A+\phi}) \ .
\end{aligned} \tag{6.152}$$

Conversely, given a solution to this system, one may define $\beta = e^{10A-2\phi}(1-x_5^2)$, $y = -\frac{1}{2}x_5 e^{3A-\phi}$ (with an eye to (6.84), (6.37), which correspond to (6.144)); if one then eliminates $r$ from (6.152), the resulting equations imply $\beta' = -12ye^{4A}$ (the second in (6.144)), and (6.138). So the system (6.152) is in fact an equivalent way to characterize our solutions. It looks much more complicated than the original ODE (6.138). We write it because it bears an uncanny resemblance with the system (5.53): a few numerical factors have changed, and two new terms have appeared. This suggests that there might be a close relationship between solutions of one system and solutions of the other. This is in fact the case: to any solution $(\phi_5, x_5, A_5)$ of (6.152) one can associate a solution $(\phi_7, x_7, A_7)$ of (5.53) given by

$$\begin{aligned}
e^{\phi_7} &= \left(\frac{3}{4}\right)^{1/4}\frac{e^{\phi_5}}{\sqrt{1-\frac{1}{4}x_5^2}} \ , \qquad & e^{A_7} &= \left(\frac{4}{3}\right)^{3/4} e^{A_5} \ , \\
x_7 &= \left(\frac{3}{4}\right)^{1/2}\frac{x_5}{\sqrt{1-\frac{1}{4}x_5^2}} \ , \qquad & r_7 &= \left(\frac{4}{3}\right)^{1/4} r_5 \ .
\end{aligned} \tag{6.153}$$

Comparing (6.141) with (5.52), we find that the map acts on the metrics as

$$\begin{aligned}
e^{2A_5}(ds^2_{\mathrm{AdS}_5} + ds^2_{\Sigma_g}) + dr_5^2 + \frac{1-x_5^2}{9}e^{2A_5}ds^2_{S^2} \to \\
\sqrt{\frac{4}{3}}\left(\frac{4}{3}e^{2A_5}ds^2_{\mathrm{AdS}_7} + dr_5^2 + \frac{e^{2A_5}}{12}\frac{1-x_5^2}{1-\frac{1}{4}x_5^2}ds^2_{S^2}\right) \ .
\end{aligned} \tag{6.154}$$

Conversely, to any solution $(\phi_7, x_7, A_7)$ of (5.53), one can associate a solution

$(\phi_5, x_5, A_5)$ of (6.152) given by

$$e^{\phi_5} = \left(\frac{4}{3}\right)^{1/4} \frac{e^{\phi_7}}{\sqrt{1+\frac{1}{3}x_7^2}} , \qquad e^{A_5} = \left(\frac{3}{4}\right)^{3/4} e^{A_7} ,$$
$$x_5 = \left(\frac{4}{3}\right)^{1/2} \frac{x_7}{\sqrt{1+\frac{1}{3}x_7^2}} , \qquad r_5 = \left(\frac{3}{4}\right)^{1/4} r_7 . \tag{6.155}$$

This inverse map now acts on the metrics as

$$e^{2A_7} ds^2_{\mathrm{AdS}_7} + dr_7^2 + \frac{1-x_7^2}{16} e^{2A_7} ds^2_{S^2} \rightarrow \sqrt{\frac{3}{4}} \left( \frac{3}{4} e^{2A_7} (ds^2_{\mathrm{AdS}_5} + ds^2_{\Sigma_g}) + dr_7^2 + \frac{1}{12} \frac{1-x_7^2}{1+\frac{1}{3}x_7^2} e^{2A_7} ds^2_{S^2} \right) . \tag{6.156}$$

The simplicity of this map is basically a generalization of the simple Maldacena–Núñez solution [117], with the $1+\frac{1}{3}x_7^2$ factor ultimately playing the role of the $\Delta = 1 + \sin^2\theta$ factor in [117].

One can also apply (6.153) directly to (6.144), and infer the expressions for the variables of the seven-dimensional solution:

$$e^{A_7} = \frac{2}{3}\left(-\frac{\beta'}{y}\right)^{1/4} , \qquad x_7 = \sqrt{\frac{-y\beta'}{4\beta - y\beta'}} , \qquad e^{\phi_7} = \frac{(-\beta'/y)^{5/4}}{12\sqrt{4\beta - y\beta'}} . \tag{6.157}$$

Moreover, $dr_7 = \left(\frac{3}{4}\right)^2 \frac{e^{3A_7}}{\sqrt{\beta}} dy$.

In chapter 5, solving the system of ODEs (5.53) was only part of the problem. First, one had to take care of flux quantization; second, most solutions include D8's, and one must take care that supersymmetry be preserved also on top of them. We will see in section 6.8.4 that the relevant conditions also map nicely under (6.153); that will lead us to conclude that there are infinitely many AdS$_5$ solutions, each one of them corresponding to the AdS$_7$ solutions of chapter 5 (further generalized in [24]). Moreover, the map is quite simple: for example, it acts on the metrics as in (6.156).

## 6.8.3. Regularity analysis

We showed that solutions of (6.138) are in one-to-one correspondence with solutions of the system of ODEs relevant for AdS$_7$ solutions. However, (6.138) looks much simpler than that system; hence one may hope to learn more about both the AdS$_5$ and the AdS$_7$ solutions by studying it.

In this subsection we will see what boundary conditions on (6.138) have to be imposed in order to obtain compact and regular solutions.

We saw in (6.141) that the internal metric consists of an $M_3$ fibered over a Riemann surface $\Sigma_g$; $M_3$ is itself a fibration of $S^2$ over a one-dimensional space with coordinate $r$.

To make $M_3$ compact, we can use the same logic as for the AdS$_7$ solutions of chapter 5. One might think of making it compact by periodically identifying $r$, but this doesn't work for the same reason as in (5.60): the quantity $y = -\frac{1}{2}e^{3A-\phi}x_5$ is monotonic — from (6.152) we see $\partial_r y = e^{2A-\phi}\sqrt{1-x_5^2}$, which is always positive; or also, directly from (6.143) we see $\frac{\partial y}{\partial r} = e^{-3A}\sqrt{\beta}$. So periodically identifying $r$ is not an option. The other way to make $M_3$ compact is to make the $S^2$ shrink for two values of $r$, just like in chapter 5. This is what we will now devote ourselves to.

To make the $S^2$ shrink, we should make the coefficient $(1-a_1^2)$ in (6.141) go to zero, which, recalling (6.144), can be accomplished by making $\beta$ vanish. If $\beta$ has a single zero,

$$\beta = \beta_1(y-y_0) + O(y-y_0)^2 \ , \tag{6.158}$$

the metric (6.141) near $y_0$ is proportional to

$$\frac{dy^2}{4(y-y_0)} + (y-y_0)ds^2_{S^2} \ , \tag{6.159}$$

which in fact upon defining $r = \sqrt{y-y_0}$ turns into

$$dr^2 + r^2 ds^2_{S^2} \ , \tag{6.160}$$

which is the flat metric on $\mathbb{R}^3$. Hence if $\beta$ has a single zero at $y_0 \neq 0$ the metric is regular.

One might wonder what happens if $\beta$ has a double zero:

$$\beta = \beta_2(y-y_0)^2 + O(y-y_0)^3 \ . \tag{6.161}$$

In this case, (6.141) is proportional to $\frac{dy^2}{\sqrt{y-y_0}} + (y-y_0)^{3/2}ds^2_{S^2}$, which upon defining $\rho = y - y_0$ turns into

$$\frac{1}{\sqrt{\rho}}(d\rho^2 + \rho^2 ds^2_{S^2}) \ ; \tag{6.162}$$

we also have $e^A \sim \rho^{1/4}$, $e^\phi \sim \rho^{3/4}$. This is obviously not a regular point, but it is the local behavior appropriate for a D6 stack whose transverse directions are $\rho$ and the $S^2$.

Higher-order zeros do not lead to anything of physical relevance, and in fact they would not lead to solutions, as we will see later. However, given that we have obtained boundary conditions corresponding to a regular point and to presence of a D6 stack, it is natural to wonder whether we can find boundary conditions corresponding to presence of an O6. This is realized when

$$\beta = \beta_0 + \beta_{1/2}\sqrt{y-y_0} + O(y-y_0) \ ; \tag{6.163}$$

in this case the metric is proportional to $(y-y_0)^{1/4}\left(\frac{dy^2}{y-y_0} + 16\alpha_0^2 ds^2_{S^2}\right)$, with $\alpha_0 \equiv \frac{\beta_{1/2}}{\beta_0}$. With the definition $\rho = \sqrt{y-y_0}$, this turns into

$$\sqrt{\rho}\left(d\rho^2 + 4\alpha_0^2 ds^2_{S^2}\right) \ ; \tag{6.164}$$

moreover, $e^A \sim \rho^{-1/4}$, $e^\phi \sim \rho^{-3/4}$. These are the appropriate behaviors for fields near the beginning of an O6 hole: to see this, one can start from the flat space O6 metric, given by $ds^2_\perp = H^{1/2}(d\rho^2 + \rho^2 ds^2_{S^2})$, $e^A = H^{1/4}$, $e^\phi \propto H^{3/4}$, $H = 1 - \frac{\rho_0}{\rho}$, and expand around $\rho = \rho_0$, which is indeed the boundary of the O6 hole.

This concludes our study of the physically relevant boundary conditions for the ODE (6.138); as it will turn out, these are the only ones which are actually realized in its solutions. Later in this section we will turn to the task of finding such solutions.

### 6.8.4. Flux quantization and D8-branes

Before we look at explicit solutions, we will discuss flux quantization. We will also introduce D8-branes in our construction, as was done in chapter 5. This subsection is in many ways similar to 5.5.8, which the reader may want to consult for more details.

We will start with some preliminary comments about the $B$ field. In (6.149) we expressed it in terms of a closed two-form $b$. We will need this second term because the term $\frac{F_2}{F_0}$ in (6.149) will jump as we cross a D8 (since $F_0$ will jump there, by definition). More precisely, looking at $F_2$ we see that only the term proportional to $\mathrm{vol}_{S^2} + 3\cos\theta \mathrm{vol}_{\Sigma_g}$ jumps (since in the other term an $F_0$ cancels out). Thus we can limit ourselves to considering $b$ of the form

$$b = b_0(\mathrm{vol}_{S^2} + 3\cos\theta\, \mathrm{vol}_{\Sigma_g}) \ , \tag{6.165}$$

which is indeed closed (while $\mathrm{vol}_{S^2} = \sin\theta d\theta \wedge D\psi$ would not be, because of the presence of $\rho$). (6.149) now becomes

$$B = \left(b_0 - \frac{q_5}{F_0}\right)(\mathrm{vol}_{S^2} + 3\cos\theta \mathrm{vol}_{\Sigma_g}) + \frac{q_5}{3} x_5 e^{A+\phi} \mathrm{vol}_{S^2} \ . \tag{6.166}$$

At the poles, for regularity we should have that what multiplies $\mathrm{vol}_{S^2}$ should go to zero.

However, more precisely $B$ should be understood as a "connection on a gerbe". Concretely, this means that it is not necessarily a globally well-defined two-form. On a chart intersection $U \cap U'$, $B_U - B_{U'}$ can be any closed two-form whose periods are integer multiples of $4\pi^2$ (known as a "large gauge transformation"). This translates into the requirement that the coefficient of $\mathrm{vol}_{S^2}$ in (6.166) should wind $\pi\times$ an integer number of times in going from the north to the south pole. Alternatively, using Stokes' theorem, we see that the integral of $H$ between $r_{\mathrm{N}}$ and $r_{\mathrm{S}}$ (the positions of the two poles) is

$$\int_{M_3} H = \int_{S^2}\int_{r_{\mathrm{N}}}^{r_{\mathrm{S}}} dr H = \int_{S^2} (B(r_{\mathrm{N}}) - B(r_{\mathrm{S}})) \ ; \tag{6.167}$$

thus $\int_{M_3} H$ will be an integer multiple of $4\pi^2$, in agreement with flux quantization.

After these comments on the NSNS flux $H$, let us now consider the RR fluxes. First of all, the zero-form should satisfy $F_0 = \frac{n_0}{2\pi}$, $n_0 \in \mathbb{Z}$. For the higher forms, we should consider

$$\tilde{F}_2 \equiv F_2 - BF_0 \ , \qquad \tilde{F}_4 \equiv F_4 - B\wedge F_2 + \frac{1}{2} B\wedge BF_0 \ , \tag{6.168}$$

which are $d$-closed (unlike the original $F_2$ and $F_4$, which in our notation are $(d - H\wedge)$-closed). Flux quantization imposes that those should have integer periods. For the two-form we simply have

$$\tilde{F}_2 = -bF_0 = -b_0 F_0(\mathrm{vol}_{S^2} + 3\cos\theta \mathrm{vol}_{\Sigma_g}) \ . \tag{6.169}$$

Integrating this on the fiber $S^2$ and imposing that it is of the form $2\pi n_2$, $n_2 \in \mathbb{Z}$, we find

$$b_0 = -\frac{n_2}{2F_0} = -\pi\frac{n_2}{n_0} \ , \tag{6.170}$$

just like in chapter 5. A gauge transformation will change $b_0 \to b_0 + k\pi$, and simultaneously $n_2 \to n_2 - k$, so that (6.170) remains satisfied.

Near the north and south pole it is convenient to work in a gauge where $B$ is regular; then $\int \tilde{F}_2 \to \int F_2$, and $n_2$ is determined by setting to zero the limit near the pole of $\left(b_0 - \frac{q_5}{F_0} + \frac{q_5}{3} x_5 e^{A+\phi}\right)$, the coefficient of $\mathrm{vol}_{S^2}$ in (6.166). For a regular point, $n_2$ near the pole is zero, and both $q_5 \to 0$ and $q_5 x_5 e^{A+\phi} \to 0$. For a stack of $n_2$ D6-branes, $q_5 x_5 e^{A+\phi} \to 0$, and $q_5 \to -\frac{n_2}{2}$. In section 6.8.3, we saw that presence of a D6 corresponds to a double zero in $\beta$, (6.161). The condition we just saw will then discretize the parameter $\beta_2$, giving

$$\beta_2 = \left(\frac{4y_0}{n_2}\right)^2 . \tag{6.171}$$

An O6 point is different: $n_2 = \pm 1$ (depending on whether we are considering the north or south pole), $q_5 \to 0$, but $\frac{q_5}{3} x_5 e^{A+\phi}$ is non zero, and will have to tend to $-\frac{n_2}{2F_0}$. Again in section 6.8.3 we saw that an O6 corresponds in our class of solutions to the presence of a square root, (6.163). Flux quantization will then fix

$$\beta_0 = \left(\frac{18y_0}{F_0}\right)^2 . \tag{6.172}$$

The four-form $\tilde{F}_4$ can now be written, after some manipulations, as

$$\tilde{F}_4 = \left(\frac{3}{F_0}\left(-q_5^2 + \frac{n_2^2}{4}\right)\cos\theta \mathrm{vol}_{S^2} + \frac{1}{3}\sin^2\theta D\psi \wedge dy\right) \wedge \mathrm{vol}_{\Sigma_g} . \tag{6.173}$$

Using (6.139) we can also write $\tilde{F}_4 = d\tilde{C}_3$, where

$$\tilde{C}_3 = \frac{3}{2F_0}\left(-q_5^2 + \frac{n_2^2}{4}\right)\sin^2\theta D\psi \wedge \mathrm{vol}_{\Sigma_g} . \tag{6.174}$$

If both poles are regular points, $\tilde{C}_3$ is a regular form. Indeed, as we saw, at such a pole we should have $n_2 = 0$ and $q_5 \to 0$. So the coefficient $\left(-q_5^2 + \frac{n_2^2}{4}\right)$ will actually go to zero at the pole. Now, using the fact that $\beta$ has a single zero (6.158), from (6.143) and (6.145) we see that $q_5$ starts with a linear power in $r$. Hence we have

$$\tilde{C}_3 \sim r^2 \sin^2\theta D\psi \wedge \mathrm{vol}_{\Sigma_g} . \tag{6.175}$$

Now, $r^2\sin^2\theta d\psi$, going from spherical to cartesian coordinates $x^i$, $i = 1, 2, 3$, is proportional to $x^1 dx^2 - x^2 dx^1$, and hence is regular. All in all, we conclude that $\tilde{F}_4$ does not have any non-zero periods, since it is exact. In presence of a D6 or O6 point, it is best to go back to (6.173). The space is topologically an $S^3$ fibration over $\Sigma_g$; standard topological arguments tell us that its cohomology is just the product of that of $S^3$ and that of $\Sigma_g$. As such it would have no four-cycles. Thus so far flux quantization for $\tilde{F}_4$ is not an issue.

We will now introduce D8-branes. We will consider them to be extended along all directions except $r$. Their treatment is very similar to section 5.5.8 of chapter 5, and we will be brief. The defining feature of a D8 stack is that the Romans mass $F_0$ jumps as we go across them. Let us call $n_0$ and $n_0'$ the flux integers on the two sides. Moreover, we will allow the D8's to have non-zero worldsheet flux, which can also be thought of as a smeared D6 charge. This will make the flux integer for $\tilde{F}_2$ jump as well; we will call $n_2$ and $n'_2$ its value on the two sides. The "slope" $\mu \equiv \frac{\Delta n_2}{\Delta n_0} \equiv \frac{n'_2 - n_2}{n'_0 - n_0}$ needs to be an integer. With this notation, imposing that (6.166) be continuous we find the condition

$$q_5|_{\rm D8} = \frac{1}{2}\frac{n'_2 n_0 - n_2 n'_0}{n'_0 - n_0} = \frac{1}{2}(-n_2 + \mu n_0) = \frac{1}{2}(-n'_2 + \mu n'_0) \ . \tag{6.176}$$

This is to be read as a condition fixing the D8's position.

One might now also wonder whether the flux of $\tilde{F}_4$ along $\Sigma_g \times S^2$ might jump between D8's, as does the integral of $\tilde{F}_2$. But actually $\int_{S^2} \cos\theta {\rm vol}_{S^2} = 0$. So even in presence of D8's we need not worry about flux quantization for $\tilde{F}_4$.

Crucially, (6.176) is exactly the same condition that was found for D8-branes in (5.81). The function called $q$ in that chapter, which we will call $q_7$ here, is not exactly the same as the $q_5$ defined in (6.145): indeed $q_7 \equiv \frac{1}{4} e^{A_7 - \phi_7}\sqrt{1 - x_7^2}$. However, using the map (6.153), we see that the different overall factor is reabsorbed:[6]

$$q_5 = q_7 \ . \tag{6.177}$$

So (6.176) fixes the D8's at exactly the same position in an AdS$_5$ solution and in its AdS$_7$ solution.

Since (6.176) was found by imposing that $B$ should be continuous, it looks easy to impose the condition on flux quantization. As remarked earlier, by Stokes' theorem we can relate the integrality of $H$ to the periodicity of the coefficient of ${\rm vol}_{S^2}$ in $B$. (This periodicity was expressed visually as a dashed green line in figures 5.7(a) and 5.8(a) of chapter 5 and in [I, 24].) However, in presence of D8's one might encounter a region where $F_0 = 0$; generically such a region will exist (although there are also "limiting cases" where it does not exist; see [24, Sec. 4.2]). In such a region, (6.149) (and hence (6.166)) cannot be used; we have to resort to (6.124). This allows to write a general expression for the integral of $H$, as shown in [24, Eq. (4.7)].

Since we are going to simplify that formula for AdS$_7$ solutions, let us review it quickly here. To simplify things a bit, one derives first an expression for the integral in the "northern hemisphere", between $x_7 = 1$ and $x_7 = 0$; it can be shown that $x_7 = 0$ is in the massless region, where $F_0 = 0$. There might be many D8's; let D8$_n$ be the one right before the massless region, $\{n_{2,n}, n_{0,n} = 0\}$ the flux parameters right after it, and $\{n_{2,n-1}, n_{0,n-1}\}$ the ones right before it. Then we can divide the integral into a contribution from the massive region

[6] Actually, the condition that the system (6.152) be mapped to the similar system (5.53) for AdS$_7$ solutions only fixed the map (6.153) up to a constant. We fixed the constant so that (6.176) would look exactly equal to (5.81).

and one from the massless region:

$$\begin{aligned}
\int_{\text{north}} H &= \int_{r_{\rm N}}^{{\rm D8}_n} H + \int_{{\rm D8}_n}^{x=0} H \\
&= 4\pi \left[ q_7 \left( \frac{x_7}{4} e^{A_7+\phi_7} - \frac{1}{F_{0,n-1}} \right) - \frac{n_{2,n-1}}{2F_{0,n-1}} + \frac{3}{32} \frac{R^3}{n_{2,n}} \left( x_7 - \frac{x_7^3}{3} \right) \right]_{{\rm D8}_n} \\
&= 4\pi \left[ -\pi\mu_n + \frac{1}{4} q x_7 e^{A_7+\phi_7} - \frac{1}{4} q_7 x_7 e^{A_7+\phi_7} \frac{3-x_7^2}{1-x_7^2} \right]_{{\rm D8}_n} \\
&= 4\pi \left[ -\pi\mu_n + \frac{R^3}{16 n_{2,n}} x_7 \right]_{{\rm D8}_n} .
\end{aligned} \tag{6.178}$$

We have used that for the massless solution $-8q_7 \frac{e^{A_7+\phi_7}}{1-x_7^2} = -2\frac{e^{2A_7}}{\sqrt{1-x_7^2}} = \frac{R^3}{n_2}$, where $R$ is a constant. After this simplification, and putting together the contribution from $\int_{south} H$ from the "southern hemisphere", we can write

$$N \equiv -\frac{1}{4\pi^2} \int H = (|\mu_n| + |\mu_{n+1}|) + \frac{1}{4\pi} e^{2A(x=0)} (|x_n| + |x_{n+1}|) \ , \tag{6.179}$$

where $x_n$ and $x_{n+1}$ are the values of $x_7$ at the branes D8$_n$ and D8$_{n+1}$.[7]

To derive a similar expression for AdS$_5$ solutions, we follow a similar logic. It proves convenient to use from the very beginning $(A_7, x_7, \phi_7)$ variables, which are related to $(A_5, x_5, \phi_5)$ variables via (6.153). We can use (6.166) and (6.124), the latter of which is already expressed in terms of $x_7$. Some factors in the computation change, but remarkably the result turns out to be exactly the same as in (6.179). As a consequence, if the $H$ flux quantization is satisfied for an AdS$_7$ solution, it is also satisfied for an AdS$_5$ solution, and viceversa.

So the conclusion of this section is that the flux quantization conditions and the constraints fixing the D8-brane positions are all precisely mapped by (6.153), in such a way that if they are satisfied for an AdS$_7$ solution they are also automatically satisfied for an AdS$_5$ solution. This proves that the map (6.153) produces infinitely many AdS$_5$ solutions.

### 6.8.5. The simplest massive solution

We will now start studying solutions to (6.138), and their associated physics. We have already indicated in (6.139) how to solve it analytically. However, in this section we will warm up by a perturbative study, which we find instructive and which will allow us to isolate a particularly nice and useful solution.

[7] The $\mu_i$ and $x_i$ before the massless region are positive, while those after the massless region are negative.

In section 6.8.3 we studied the boundary conditions for the ODE (6.138). We can now proceed to study it in the neighborhood of such a solution. We will do so by assuming analytic behavior around $y_0$: $\beta = \sum_{k=1}^{\infty} \beta_k (y - y_0)^k$, by plugging this Taylor expansion in (6.138), and solving order by order.

Already at order zero we find

$$\left(\beta_1 - \frac{72 y_0^2}{F_0}\right) \beta_1^2 = 0 \; . \tag{6.180}$$

The first branch, $\beta_1 = \frac{72 y_0^2}{F_0}$, lets $\beta$ have a single zero, which as we saw after (6.158) corresponds to a regular point. The second branch, $\beta_1 = 0$, makes $\beta$ have a double zero, which as we saw after (6.161) corresponds to a D6. In this section we will use the first branch, leaving the second for section 6.8.6.

Continuing to solve (6.138) perturbatively after having set $\beta_1 = \frac{72 y_0^2}{F_0}$, we find a nice surprise: the perturbative expansion stops after three iterations. This leads to a very simple solution to (6.138):

$$\beta = \frac{8}{F_0}(y - y_0)(y + 2y_0)^2 \; . \tag{6.181}$$

This has the desired single zero at $y = y_0$, and it also has a double zero at $y = -2y_0$, signaling that $M_3$ has a D6 stack there. These are the qualitative features one expects from the solution in section 5.6.2; there, that solution was argued to exist (along with many others, which we shall discuss in due course) on numerical grounds — see in particular 5.6. It would also be possible to find (6.181) by finding the general solution, and imposing the presence of a simple zero; we will see this in section 6.8.6.

(6.181) looks superficially very similar to (6.96). Taking $c_1 = -y_0^2/4$, we see that (6.96) has *two* double zeros, at $y = \pm y_0$, corresponding to two D6 stacks. This is indeed correct for that massless solution: the two D6 stacks are generated by the reduction from eleven dimensions, in a similar way as in 5.6.1. Notice also that the massless limit of (6.181), on the other hand, does not exist, since $F_0$ appears there in the denominator.

Now that we have obtained one solution of (6.181), we can pause to explore what the resulting AdS$_5$ solution looks like; moreover, using the map (6.153), we can also produce an AdS$_7$ solution which will indeed be the one found numerically in section 5.6.2.

The conditions (6.146) give us two possibilities:

$$\{y_0 < 0, \; F_0 > 0, \; y \in [y_0, -2y_0]\} \;\; or \;\; \{y_0 > 0, \; F_0 < 0, \; y \in [-2y_0, y_0]\} \; . \tag{6.182}$$

We will assume the first possibility. One can then write the metric and fields most conveniently in terms of

$$\tilde{y} \equiv \frac{y}{y_0} \ , \tag{6.183}$$

which then has to belong to $[-2, 1]$. We have

$$ds^2_{M_5} = e^{2A} ds^2_{\Sigma_g} + \sqrt{-\frac{y_0}{8F_0}} \left( \frac{d\tilde{y}^2}{(1-\tilde{y})\sqrt{\tilde{y}+2}} + \frac{4}{9} \frac{(1-\tilde{y})(\tilde{y}+2)^{3/2}}{2-\tilde{y}} ds^2_{S^2} \right) \ , \tag{6.184}$$

and

$$e^{4A} = -2\frac{y_0}{F_0}(2+\tilde{y}) \ , \qquad e^{2\phi} = \sqrt{-\frac{1}{2y_0 F_0^3}} \frac{(\tilde{y}+2)^{3/2}}{2-\tilde{y}} \ . \tag{6.185}$$

We also need to implement flux quantization, which in this case is the statement that the D6 stack at the $\tilde{y} = -2$ point has an integer number $n_2$ of D6-branes. This constraint was discussed right below (6.170). From (6.145) and (6.181) we find $q_5 = \frac{1}{3}\sqrt{2F_0(y-y_0)}$, which implies

$$y_0 = -\frac{3}{8}\frac{n_2^2}{F_0} \ . \tag{6.186}$$

We did not replace this constraint in (6.184), as we did in (6.2), because later we will glue pieces of it together with other metrics and with itself, and in that context the parameter $y_0$ will be fixed by flux quantization a bit differently.

The AdS$_7$ solutions can now be found easily by applying the map (6.153), and in particular its action on the metric, (6.154). The internal metric on $M_3$ is

$$ds^2_{M_3} = \sqrt{-\frac{y_0}{6F_0}} \left( \frac{d\tilde{y}^2}{(1-\tilde{y})\sqrt{\tilde{y}+2}} + \frac{4}{3} \frac{(1-\tilde{y})(\tilde{y}+2)^{3/2}}{8-4\tilde{y}-\tilde{y}^2} ds^2_{S^2} \right) \ , \tag{6.187}$$

and

$$e^{4A} = -\left(\frac{4}{3}\right)^3 2\frac{y_0}{F_0}(\tilde{y}+2) \ , \qquad e^{2\phi} = \sqrt{-\frac{6}{y_0 F_0^3}} \frac{(\tilde{y}+2)^{3/2}}{8-4\tilde{y}-\tilde{y}^2} \ . \tag{6.188}$$

(6.187) and (6.188) give analytically the solution found numerically in section 5.6.2. The flux $F_2$ can be read off from the expression $F_2 = q(\frac{x_7}{4}F_0 e^{A+\phi} - 1)\mathrm{vol}_{S^2}$ in (5.78):

$$F_2 = \frac{k}{\sqrt{3}} \frac{(1-\tilde{y})^{3/2}(\tilde{y}+4)}{8-4\tilde{y}-\tilde{y}^2} \mathrm{vol}_{S^2} \ . \tag{6.189}$$

For both the AdS$_5$ and AdS$_7$ solutions, from (6.186) we can see that, making $n_2$ large, curvature and string coupling become as small as one wishes. This guarantees that the supergravity approximation is applicable. Similar limits can be taken for the solutions that we will present later. (This was shown in general in [24, Sec. 4.1].)

### 6.8.6. General massive solution

Let us now go back to (6.180) and see what happens if we use the branch $\beta_1 = 0$. This means that $\beta$ has a double zero, which corresponds to presence of a D6 stack at $y = -2y_0$.

The perturbative expansion for (6.138) now does not truncate anymore. It is possible to go to higher order, guess an expression for the $k$-th term $\beta_k$ in the Taylor expansion $\beta = \sum_{k=1}^{\infty} \beta_k (y - y_0)^k$, and resum this guess. (This is in fact the way we originally proceeded.) At this point it is of course much easier to use the trick explained below (6.139), and find the general solution directly. Assuming $y_0 > 0$, it reads

$$\beta = \frac{y_0^3}{b_2^3 F_0} \left(\sqrt{\hat{y}} - 6\right)^2 \left(\hat{y} + 6\sqrt{\hat{y}} + 6b_2 - 72\right)^2 , \tag{6.190}$$

where

$$\hat{y} \equiv 2b_2 \left(\frac{y}{y_0} - 1\right) + 36 , \qquad b_2 \equiv \frac{F_0}{y_0}\beta_2 . \tag{6.191}$$

An alternative expression for (6.190) is

$$\sqrt{\beta} = \sqrt{\frac{8}{F_0}} \sqrt{y - \tilde{y}_0}(y + 2\tilde{y}_0) - 36\sqrt{\frac{y_0^3}{b_2^3 F_0}}(b_2 - 12) , \tag{6.192}$$

where $\tilde{y}_0 = \left(1 - \frac{18}{b_2}\right) y_0$. Notice the similarity with (6.181).

This solution now depends on the two parameters $y_0$ and $b_2$, rather than just one as (6.181), and we expect it to be the most general solution to (6.138). To see whether this is true, let us analyze its features and compare them to what we expect from the qualitative study in section 5.6.2; we will do so using the first expression (6.190).

(6.190) has zeros at $\hat{y} = 36$ (which corresponds to $y = y_0$) and for $b_2 < 12$ also at $\hat{y} = (-3 + \sqrt{81 - 6b_2})^2$. Also, at $\hat{y} = 0$ it has a point where it behaves as $\beta \sim \beta_0 + \sqrt{\hat{y}} + O(\hat{y})$, which up to translation is the same as in (6.163), which corresponds to an O6 point. Taking also into account the constraints in (6.146), we find two possibilities, and one special case between them.

- If $b_2 < 12$, the solution is defined in the interval $\hat{y} \in [(-3+\sqrt{81 - 6b_2})^2, 36]$; there are two double zeros at both endpoints. This represents a solution with two D6 stacks at both ends, but where the numbers of D6s are not the same on the two sides (unlike for (6.96)). Under the map (6.153) to AdS$_7$, it becomes a solution that was briefly mentioned at the end of section 5.6.2; in terms of the graph in figure 5.6(b), its path would come from below and miss the green dot on the top side from the left, so as to end up in a D6 asymptotics on the top side as well.

- If $b_2 > 12$, the solution is defined for $\hat{y} \in [0, 36]$; there is a double zero at $\hat{y} = 36$, and an O6 singularity (see (6.163)) at $\hat{y} = 0$. This represents a solution with one D6 stack at one end, and one O6 at the other extremum. Under the map to AdS$_7$, it becomes another solution that was briefly mentioned in section 5.6.2; in terms of the graph in figure 5.6(b), its path would come from below and miss the green dot on the top side from the right, so as to end up in an O6 asymptotics on the top side.

- In the limiting case, $b_2 = 12$, the solution is again defined for $\hat{y} \in [0, 36]$; under the map to AdS$_7$ we expect to find the case where (again referring to 5.6(b)) we hit the green dot at the top, which should correspond to having a regular point. Indeed in this case (6.190) reduces to

$$\beta = \frac{y_0^3}{1728 F_0} \hat{y} (\hat{y} - 36)^2 \ , \tag{6.193}$$

which has a double zero in $\hat{y} = 36$ and a single zero in $\hat{y} = 0$; it is essentially (6.181). It would have been possible to obtain (6.181) this way, but we chose to highlight it in a subsection by itself because of its simplicity.

So the solution (6.190) has the features we expected from the qualitative analysis in section 5.6.2.

We record also here some data of the corresponding solutions. For the AdS$_5$ solution, the metric, warping and dilaton read

$$ds^2_{M_5} = e^{2A} ds^2_{\Sigma_g} + \frac{y_0^{5/4} d\hat{y}^2}{4(b_2^5 F_0^3 \hat{y}^3 \beta)^{1/4}} + \\ + \frac{(b_2^7 F_0 \hat{y})^{1/4}}{18 y_0^{7/4}} \frac{\beta^{3/4} ds^2_{S^2}}{2(b_2 - 18)^2 + 18(b_2 - 12)\sqrt{\hat{y}} - (b_2 - 18)\hat{y}} \ , \tag{6.194a}$$

$$e^{8A} = \frac{b_2 \beta}{F_0 y_0 \hat{y}} \ , \tag{6.194b}$$

$$e^{8\phi} = \frac{b_2^{11} \beta^3}{16 F_0^3 y_0^{11} \hat{y}^3 \left(2(b_2 - 18)^2 + 18(b_2 - 12)\sqrt{\hat{y}} - (b_2 - 18)\hat{y}\right)^4} \ . \tag{6.194c}$$

The AdS$_7$ solution reads

$$ds^2_{M_3} = \frac{y_0^{5/4} d\hat{y}^2}{4(b_2^5 F_0^3 \hat{y}^3 \beta)^{1/4}} + \\ + \frac{(b_2^7 F_0 \hat{y})^{1/4}}{3 y_0^{7/4}} \frac{\beta^{3/4} ds^2_{S^2}}{12(b_2-18)^2 + 144(b_2-12)\sqrt{\hat{y}} - 12(b_2-18)\hat{y} - \hat{y}^2} , \tag{6.195a}$$

$$e^{8A} = \frac{2^{12} b_2 \beta}{3^6 F_0 y_0 \hat{y}} , \tag{6.195b}$$

$$e^{8\phi} = \frac{144 b_2^{11} \beta^3}{F_0^3 y_0^{11} \hat{y}^3 \left(-12(b_2-18)^2 - 144(b_2-12)\sqrt{\hat{y}} + 12(b_2-18)\hat{y} + \hat{y}^2\right)^4} . \tag{6.195c}$$

Finally, flux quantization can be taken into account by using (6.170), (6.171) and the expansion of $\beta$ around its zeros (or around its zero and its square root point, for the O6–D6 case). We obtain two equations, which discretize the two parameters $b_2$ and $y_0$. The expressions are not particularly inspiring (especially in the D6–D6 case) and we will not give them here.

### 6.8.7. Some solutions with D8-branes

We will now show two simple examples of solutions with D8-branes. These will be the ones studied numerically in section 5.6.3; here we will give their analytic expressions. We will simply have to piece together solutions we have already studied; all we will have to work out is the position of the D8's.

The first example is a solution with only one D8 stack. This can be obtained by gluing two metrics of the type (6.184). We will assume

$$y_0 < 0 , \qquad F_0 > 0 ; \qquad y_0' > 0 , \qquad F_0' < 0 . \tag{6.196}$$

Following the logic in section 5.6.3, the flux quantization conditions can be satisfied by taking for example the two-form flux integer after the D8 stack to vanish, $n_2' = 0$, $n_2 = \mu(n_0' - n_0)$, $\mu \in \mathbb{Z}$, and

$$n_0' = n_0 \left(1 - \frac{N}{\mu}\right) , \tag{6.197}$$

where $N = \frac{1}{4\pi^2}\int H$ is the NSNS flux integer. (Recall that $F_0 = \frac{n_0}{2\pi}$, and similarly for $F_0'$.) As usual the metric can be written as $ds^2_{M_5} = e^{2A} ds^2_{\Sigma_g} + ds^2_{M_3}$, and putting together two copies of (6.184) we can write[8]

$$ds^2_{M_3} = \begin{cases} \dfrac{1}{\sqrt{8F_0}}\left(\dfrac{dy^2}{(y-y_0)\sqrt{-2y_0-y}} + \dfrac{4}{9}\dfrac{(y-y_0)(-2y_0-y)^{3/2}}{-y_0(y-2y_0)}ds^2_{S^2}\right); \\ \dfrac{1}{\sqrt{-8F_0'}}\left(\dfrac{dy^2}{(y_0'-y)\sqrt{2y_0'+y}} + \dfrac{4}{9}\dfrac{(y_0'-y)(2y_0'+y)^{3/2}}{y_0'(2y_0'-y)}ds^2_{S^2}\right). \end{cases} \tag{6.198}$$

The upper line should be used in the interval $y_0 < y < y_{\rm D8}$; we should instead use the bottom one in the interval $y_{\rm D8} < y < y_0'$. We reverted to using $y$ rather than $\hat{y}$, so as to be able to use the same coordinate before and after the D8 stack. Imposing that $A$ and $\phi$ (or, equivalently, that $\beta$ and $\beta'$) be continuous across the D8 stack, we get

$$y_0 = \frac{1}{2}\frac{2F_0 - F_0'}{F_0 + F_0'}y_{\rm D8}\ , \qquad y_0' = \frac{1}{2}\frac{2F_0' - F_0}{F_0 + F_0'}y_{\rm D8}\ . \tag{6.199}$$

We also have to impose (6.176), which fixes

$$y_{\rm D8} = y_0 + \frac{9(F_0')^2 n_2^2}{8F_0(F_0 - F_0')^2}\ , \tag{6.200}$$

which together with (6.199) and (6.197) gives

$$y_0 = -\frac{3}{2}F_0\pi^2(N^2 - \mu^2)\ , \tag{6.201}$$

$$y_0' = \frac{3}{2}F_0\pi^2(N-\mu)(2N-\mu)\ , \tag{6.202}$$

$$y_{\rm D8} = 3F_0\pi^2(N-2\mu)(N-\mu)\ . \tag{6.203}$$

One can also obtain the corresponding AdS$_7$ solution. This can be done using the map (6.156) on (6.198). Alternatively, we can just write one copy of (6.187) for $y_0 < y < y_{\rm D8}$, and a second copy of (6.187), formally obtained by $y \to -y$, $y_0 \to -y_0'$, $F_0 \to -F_0'$. This provides the analytic expression of the solution in figure 5.7.

We can also consider a configuration with two D8 stacks. We will take it to by symmetric, in the sense that the flux integers before the first D8 stack will be $(n_0, 0)$, between the two stacks $(0, n_2 = -k < 0)$, and after the second stack $(-n_0, 0)$. This corresponds to figure 5.8. Again we will assume $y_0 < 0$;

[8] The sign differences between the expression before and after the D8 have to do with the simplification of factors involving $\sqrt{F_0^2} = |F_0|$ from applying (6.144) to (6.181).

the positions of the two D8 stacks will be $y_{\rm D8} < 0$ and $y_{\rm D8'} = -y_{\rm D8} > 0$. We will give only the AdS$_7$ internal metric:

$$ds^2_{M_3} = \begin{cases} \frac{1}{\sqrt{6F_0}}\left(\frac{dy^2}{(y-y_0)\sqrt{-2y_0-y}} + \frac{4}{3}\frac{(y-y_0)(-2y_0-y)^{3/2}}{8y_0^2-4yy_0-y^2}ds^2_{S^2}\right) , \\ \frac{24^4R^6dy^2+(9^2R^6-32^2y^2)^2ds^2_{S^2}}{3\cdot 6^5(9^2R^6-32^2y^2)^{1/2}} , \\ \frac{1}{\sqrt{6F_0}}\left(\frac{dy^2}{(-y_0-y)\sqrt{-2y_0+y}} + \frac{4}{3}\frac{(-y_0-y)(-2y_0+y)^{3/2}}{8y_0^2+4yy_0-y^2}ds^2_{S^2}\right) . \end{cases} \tag{6.204}$$

The upper line should be used in the interval $y_0 < y < y_{\rm D8}$; the middle one in the interval $y_{\rm D8} < y < -y_{\rm D8}$, and finally the bottom one in the interval $-y_{\rm D8} < y < -y_0$. The metric in the middle region is the known massless metric in (5.91), with the change of coordinate (6.123).

We now have three unknowns: $R$, $y_0$, $y_{\rm D8}$. Continuity of $\beta$ and $\beta'$ this time only imposes one condition; we then have (6.176) and the condition (6.179). We get

$$y_0 = -\frac{9}{4}k\pi(N-\mu) , \qquad y_{\rm D8} = -\frac{9}{4}k\pi(N-2\mu) , \\ R^6 = \frac{64}{3}k^2\pi^2(3N^2-4\mu^2) , \tag{6.205}$$

where in this case $\mu = \frac{k}{n_0}$. Notice that the in this case the bound in [24, Eq.(4.10)] (which can also be found by (6.179)) implies $N > 2\mu$.

It would now be possible to produce solutions with a larger number of D8's. It is in fact possible to introduce an arbitrary number of them, although there are certain constraints on their numbers and their D6 charges [24, Sec. 4]. The most general solution can be labeled by the choice of two Young diagrams; there is also a one-to-one correspondence with the brane configurations in [19, 20]. One can in fact think of the AdS$_7$ solutions as a particular near-horizon limit of the brane configurations. For more details, see [24]. For these more general solutions, we expect to have to glue together not only pieces of the solution in subsection 6.8.5 and of the massless solution, but also pieces of the more complicated solution in 6.8.6.

### 6.8.8. Field theory interpretation

In this section we have found infinitely many new AdS$_5$ solutions in massive IIA, and we have established that they are in one-to-one correspondence with the $AdS_7$ solutions of [I, 24].

It is easy to guess the field theory interpretation of this correspondence. Recall first the Maldacena–Núñez $\mathcal{N} = 2$ solutions [117]. The original $AdS_7 \times S^4$ solution of M-theory has an SO(5) R-symmetry; when one compactifies on a Riemann surface $\Sigma_g$, one "mixes" the SO(2) of local transformations on $\Sigma_g$ with an $SO(2) \subset SO(5)$ subgroup; the commutant $SO(2)\times SO(3) \cong U(2)$ remains as the R-symmetry of the resulting $\mathcal{N} = 2$ $CFT_4$. This is reflected in the form of the metric of the $S^4$, that gets distorted (except for the directions protected by the R-symmetry).

In similar $\mathcal{N} = 1$ solutions [117, 119], the SO(2) is embedded in SO(5) in a more intricate way, so that its commutant is a U(1), which is indeed the R-symmetry of an $\mathcal{N} = 1$ theory.

For us, the $CFT_6$ has only $(1, 0)$ supersymmetry, and thus its R-symmetry is already only SU(2). The twisting is very similar to the usual one in [117]: it is signaled by the fact that the $\psi$ coordinate is fibered over the Riemann surface $\Sigma_g$.

When we mix this with the SO(2) of local transformations on $\Sigma_g$, the commutant is only a U(1). So in principle there is no symmetry protecting the shape of the internal $S^2$ in the $AdS_7$ solutions; indeed the metric (6.141) does not have SO(3) isometry, because the $\psi$ direction is fibered over $\Sigma_g$. What is a bit surprising is that the breaking is not more severe: (6.142) might have become considerably more complicated, with $\sin\theta$ for example being replaced by a different function. Likewise, in the fluxes, one can see that there is no SO(3) symmetry: the $\cos\theta$ in front of $\mathrm{vol}_{\Sigma_g}$, for example, breaks it. Still, there are various nice $\mathrm{vol}_{S^2}$ terms which were not guaranteed to appear.

In any case, we interpret our solutions as the twisted compactification of the $CFT_6$ dual to the $AdS_7$ solutions in [I, 24]. Recently, there has been a lot of progress in understanding such compactifications for the $(2, 0)$ theories [112, 118, 119], and it would be very interesting to extend those results to our $AdS_5$ solutions. Here, we will limit ourselves to pointing out a couple of preliminary results about the number of degrees of freedom.

A common way of estimating the number of degrees of freedom using holography in any dimension is to introduce a cut-off in AdS, and estimate the Bekenstein–Hawking entropy (see for example [28, Sec. 3.1.3]). This leads to $\frac{R^5_{AdS_7}}{G_{N,7}}$ in $AdS_7$, and to $\frac{R^3_{AdS_5}}{G_{N,5}}$ in $AdS_5$, where $G_{N,d}$ is Newton's constant in $d$ dimensions (see (4.13)). The latter can be computed as $\frac{1}{g_s^2}\mathrm{vol}_{10-d}$. In a warped compactification with non-constant dilaton, both $R_{AdS}$ and $g_s$ are non-constant, and should be integrated over the internal space. In our case, for $AdS_7$ this

leads to

$$\mathcal{F}_{0,6} \equiv \int_{M_3} e^{5A_7 - 2\phi_7} \mathrm{vol}_3 \tag{6.206}$$

and for $\mathrm{AdS}_5$ to $\mathcal{F}_{0,4} \equiv \int_{M_5} e^{3A_5 - 2\phi_5} \mathrm{vol}_5$. These can be thought of as the coefficient in the thermal partition function, $\mathcal{F} = \mathcal{F}_{0,d} V T^d$, where $V$ is the volume of space and $T$ is temperature. These computations however are basically the same for the coefficients of the conformal anomaly (as explained in section 4.2.1), at least at leading order (i.e. in the supergravity approximation).

As a consequence of our map (6.153), $\mathcal{F}_{0,6}$ and $\mathcal{F}_{0,4}$ are related. Taking into account the transformation of the volume form according to (6.154), we find

$$\mathcal{F}_{0,4} = \left(\frac{3}{4}\right)^4 \mathcal{F}_{0,6}\, \mathrm{vol}(\Sigma_g) \ . \tag{6.207}$$

The volume of $\Sigma_g$ can be easily computed using the Gauss–Bonnet theorem and the fact that its scalar curvature equals $-6$; we get:

$$\mathrm{vol}(\Sigma_g) = \frac{4}{3}\pi(g-1) \ . \tag{6.208}$$

Therefore the ratio of the degrees of freedom in four and six dimensions is universal, in that it depends only on $g$ and not on the precise $(1,0)$ theory we are considering in our class. This is reminiscent of what happens for compactifications of the $(2,0)$ theory; see e.g. [118, Eq. (2.8)], or [119, Eq. (2.22)].

We have not computed $\mathcal{F}_{0,6}$ in full generality for the $(1,0)$ theories. This would now be possible in principle, since the analytic expressions are now known. One first example is the solution in section 6.8.5. The corresponding brane configuration according to the identification in [24] consists in $k$ D6's ending on $N = \frac{k}{n_0}$ NS5-branes; see figure 6.1(a). We get

$$\mathcal{F}_{0,6} = \frac{512}{45} k^2 \pi^4 N^3 \ , \tag{6.209}$$

which reassuringly goes like $N^3$. (By way of comparison, for the massless case one gets $\mathcal{F}_{0,6} = \frac{128}{3} k^2 \pi^4 N^3$.)

We also computed $\mathcal{F}_{0,6}$ for the solution (6.204), which has two D8's and a massless region between them. The corresponding brane configuration would be $N$ NS5-branes in the middle with $k = \mu n_0$ D6's sticking out of them, ending on $n_0$ D8-branes both on the left and on the right; see figure 6.1(b). This case was considered in [24, Sec. 5], where approximate expressions for $\mathcal{F}_{0,6}$ were computed, using perturbation theory around the massless limit. Using (6.204) we can now obtain the exact result:

$$\mathcal{F}_{0,6} = \frac{128}{3} k^2 \pi^4 \left(N^3 - 4N\mu^2 + \frac{16}{5}\mu^3\right) \ . \tag{6.210}$$

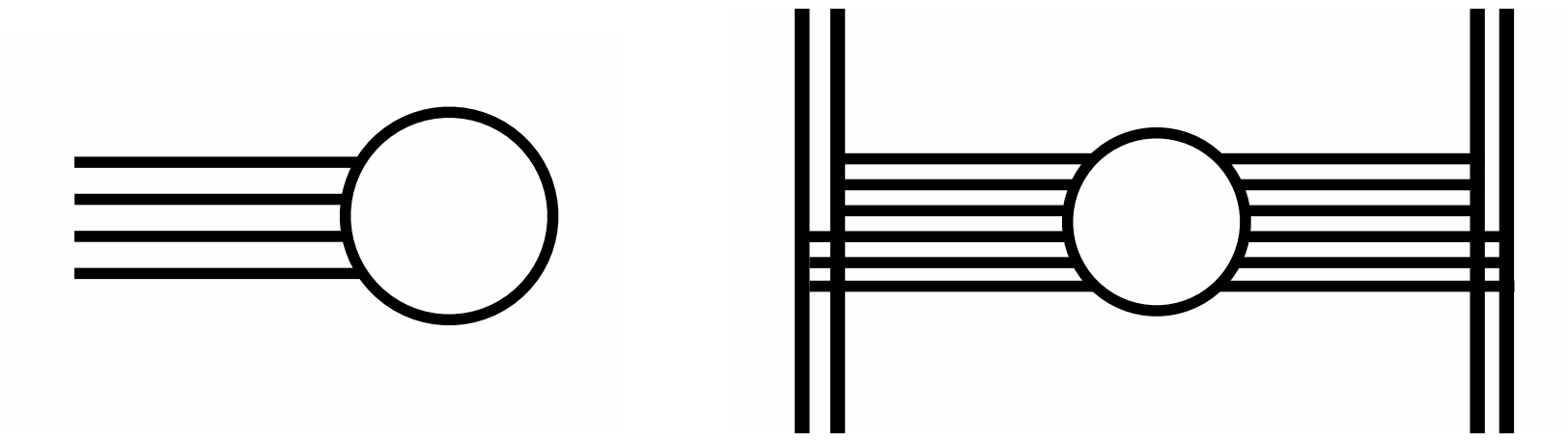

**Figure 6.1.:** Brane configurations for two sample theories. The circles represent stacks of $N$ NS5-branes; the horizontal lines represent D6-branes; the vertical lines represent D8-branes. In the second case, on each side we have $n_0 = 2$ D8-branes; $|\mu| = 3$ D6-branes end on each, for a total of $k = n_0|\mu| = 6$.

This agrees with [24, Sec. 5], but is now exact. Recall that $\mu = \frac{k}{n_0}$; since this number can be large, the second and third term are also large, and are not competing with stringy corrections. Using (6.207), and comparing with the $(2, 0)$ theory to fix the proportionality factors, we get that for the CFT$_4$ theory $a = c = \frac{1}{3}(g-1)\left(N^3 - 4N\mu^2 + \frac{16}{5}\mu^3\right)$. Stringy corrections will modify this result with terms linear in $N$ and probably in $\mu$.

## 6.9. Results and conclusions

We have classified supersymmetric AdS$_5 \times M_5$ solutions of massive type IIA supergravity, and we have found an infinite class of new analytic solutions.

The general classification, obtained in section 6.4, is summarized in section 6.4.4. We reduced the supersymmetry equations to six PDE's. A solution to this system completely determines the bosonic fields — metric, dilaton, and fluxes. The geometry of $M_5$ is given by a fibration of a three-dimensional manifold $M_3$ over a two-dimensional space $\mathcal{C}$.

We found an Ansatz that makes the PDE system solvable. As described in section 6.6, it consists in relating the metric on $\mathcal{C}$ to the warping function $A$. We recovered in this way several known massless solutions: the Bah–Beem–Bobev–Wecht [123, 119], Maldacena–Núñez [117], and Itsios–Núñez–Sfetsos–Thompson [124] solutions. More interestingly, we found new analytic ones when $F_0 \neq 0$.

This new class, analyzed in section 6.8, consists of infinitely many new solutions which preserve eight supercharges in five dimensions and are in one-to-one correspondence with the AdS$_7 \times M_3$ type IIA backgrounds classified in chapter 5. We have explicitly described the map between the former and the latter. The geometry of the $M_3$ fiber inside $M_5$ is a certain modification of the "distorted" $M_3 \cong S^3$ of the AdS$_7$ compactifications, whereas the base $\mathcal{C}$ is a Riemann surface with constant negative curvature and genus $g > 1$. An $S^2$ inside $M_3$ is twisted over $\mathcal{C}$, breaking the SU(2) isometry of $M_3 \cong S^3$ to U(1); this bears out the field-theoretic expectation of having a U(1) R-symmetry for the dual four-dimensional $\mathcal{N} = 1$ SCFT.

Importantly, we have been able to find analytic expressions for all these $\mathrm{AdS}_5$ solutions. Then, by means of the aforementioned one-to-one correspondence, we have obtained analytic versions of all the $\mathrm{AdS}_7$ solutions of chapter 5 (which were studied there only numerically). Thanks to the analytic expressions for the fields on the gravity side, we have computed explicitly the free energy for some examples of four-dimensional $\mathcal{N} = 1$, and six-dimensional $\mathcal{N} = (1, 0)$ SCFT's at large $N$ (using the AdS/CFT dictionary). It would be very interesting to find a field theory description of these theories, perhaps along the lines of [112].

Chapter 7

# AdS$_6$ solutions of type II supergravity

Sections 7.2 trough 7.6 of this chapter are based on the published article [II].

## 7.1. Five-dimensional field theories

Minimal $\mathcal{N} = 1$ supersymmetry in five dimensions contains eight $Q$ supercharges, which are rotated by the global SU(2) R-symmetry. This algebra can be enhanced to a superconformal one by adding eight $S$ supercharges. It is related by dimensional reduction to $\mathcal{N} = 2$ in four dimensions. The massless representations are a vector multiplet, containing a gauge field, a real scalar and a spinor, and a hypermultiplet, containing four real scalars and a spinor.

The Higgs branch of the theory is parameterized by the scalars in the hypermultiplet(s), and is a hyper-Kähler manifold. The Coulomb branch of the moduli space is as usual parameterized by the scalar(s) in the vector multiplet(s). Reduction to $d = 4$ $\mathcal{N} = 2$ fixes the Lagrangian of the theory on its Coulomb branch to be at most cubic in the prepotential $\mathcal{F}(\mathcal{A}^i)$, where $\mathcal{A}^i$ are the vector superfields [8]:

$$\mathcal{F} = c_0 + c_i\mathcal{A}^i + c_{ij}\mathcal{A}^i\mathcal{A}^j + c_{ijk}\mathcal{A}^i\mathcal{A}^j\mathcal{A}^k \ . \tag{7.1}$$

The coefficients $c_{ij}$, $c_{ijk}$ must be real, while $c_0$ and $c_i$ can be set to zero.

In the case of a single vector multiplet ($i = 1$), the term proportional to $c_{11}$ is the kinetic term for the fields contained in $\mathcal{A}^1$, hence $c_{11} \sim \frac{1}{g^2}$, with $g$ the gauge coupling. Notice that, according to our formula in 1, the inverse squared gauge coupling has mass dimension one in five dimensions: $\left[g^{-2}\right] = \left[M^1\right]$ .

$c_{ijk} \equiv c$ is instead a kind of anomaly, since the cubic term is absent in the classical theory but can be generated quantum-mechanically [8] (at one loop, being independent of $g \sim \frac{1}{\sqrt{c_{11}}}$ [132]).

The expression (7.1) is valid only locally. In particular, the theory can have singularities on its moduli space of vacua: The (effective) gauge coupling is given by $\frac{1}{g^2_{\text{eff}}} = \frac{1}{g^2} + c|\phi|$ ($\phi$ is the scalar in the vector multiplet), and it can clearly diverge at finite distance from its origin. This reflects the fact that a five-dimensional $\mathcal{N} = 1$ supersymmetric field theory is nonrenormalizable (see discussion in the introduction to this thesis, 1). (This actually happens only when the number of flavor hypermultiplets is (strictly) greater than seven [8].)

### Gauge theory on a probe D4

This kind of field theory can be realized on the worldvolume of a probe D4 localized at a point along $S^1/\mathbb{Z}_2$, on which we compactify type IIA string theory.[1] The D4-brane probes the background generated by $N_{\text{f}} = 16$ D8-branes. When the latter coincide we get a five-dimensional U(1) gauge theory with $N_{\text{f}} = 16$ hypermultiplets. If we also add a single O8$^-$ (see section 3.3.4), and put all sixteen D8's on top of it, the probe theory becomes an SU(2) gauge theory, and the flavor symmetry gets enhanced from SU($N_{\text{f}}$) to SO($2N_{\text{f}}$). It can be shown that the D8 charge of the combined D8-O8 system matches with the $c$ anomaly in both the U(1) and SU(2) case.

### Interacting fixed points in five dimensions

Consider now a more general situation with two O8-planes and many D8's along the segment $S^1/\mathbb{Z}_2$. The system is probed by a D4-brane. The D8's will contribute flavor hypermultiplets to the gauge theory on the D4, and their masses give the positions $\phi_i$ of the former along the segment. In fact, a real mass term for a five-dimensional hypermultiplet can be introduced in the theory, and is interpreted as the scalar in a *background vector multiplet* [133].

Let us place one O8 at $\phi = 0$ and one at $\phi = \frac{1}{R}$, with $R$ the radius of $S^1$. We then put $N_{\text{L}}$ D8's on top of the first O8-plane, $N_{\text{R}} \equiv 16 - p - N_{\text{L}}$ on top of the second, and $p$ D8's in between, at $0 < \phi_i < \frac{1}{R}$ $\forall i$. The probe theory is now SU(2) with $N_{\text{L}} + p$ flavors ($p$ have masses $\phi_i$). The novelty is that the effective gauge coupling $g^{-2}_{\text{eff}}$ of the theory depends on the generic point $\phi$ of its moduli space [8, (4.2)]. For big $R$, it stays finite. However, at some $R_0$ (depending on $N_{\text{L}}$, on the bare coupling $g$ at $\phi = 0$, and on the $\phi_i$) – i.e. at a certain point $\phi_0$ on the Coulomb branch – it can blow up. Schematically:

$$\frac{1}{g^2_{\text{eff}}(\phi)} \operatorname{Tr} F^2 \sim (\phi - \phi_0) \operatorname{Tr} F^2 \; ; \quad g_{\text{eff}} \to \infty \quad \text{as} \quad \phi \to \phi_0 \, . \tag{7.2}$$

[1] This is actually a type I' background.

Notice that a scalar field in five dimensions would have canonical mass dimension $[M^{3/2}]$. However, according to our current discussion, $g_{\text{eff}}^{-2}(\phi)$ is a *dynamical scalar field* with mass dimension one. On the other hand $F^2$ has always mass dimension four, thus the noncanonically normalized Yang–Mills interaction $\frac{1}{g^2}\,\text{Tr}\,F^2$ has dimension five in five dimensions.

(7.2) is very similar to (3.24) and (3.25). There too the gauge coupling was promoted to a dynamical scalar field; this in turn forced the whole Yang–Mills interaction to be irrelevant in six dimensions, and to become inadequate at the *strongly-coupled fixed point* (here $\phi \to \phi_0$).

The only difference is that in six dimensions the scalar belongs to a tensor multiplet, in five to a vector multiplet. This is perhaps not too surprising if one considers that, upon dimensional reduction, the $(1,0)$ tensor multiplet turns into the five-dimensional vector multiplet.

It is a famous result by Seiberg [8] that such fixed point theories exhibit an enhanced global symmetry given by $E_{N_\text{f}+1}$.[2]

### $(p,q)$-fivebrane webs and five-dimensional field theories

More general strongly coupled fixed points can be engineered by a five-dimensional version of the NS5-D6-D8 Hanany–Witten brane setup we considered in chapter 3. It involves $(p,q)$*-fivebranes* – which are the magnetic duals to the $(p,q)$-strings of type IIB (introduced in figure 3.3) – possibly ending on $(p,q)$-sevenbranes (with the same $(p,q)$ charges) – which are the nonperturbative generalizations of D7-branes (giving rise to *F-theory*). This construction exploits the fact that type IIB string theory exhibits a (quantum) $\text{SL}(2,\mathbb{Z})$ invariance, and contains general "doublets" of F1-string and D1-string (a $(1,0)$ and a $(0,1)$-string, respectively).

Consider the fivebrane web on the left of figure 7.1. The $x^5$ coordinate parameterizing the worldvolume of the D5, as it extends towards the NS5, will be affected by the latter. More precisely, we have to solve a Laplace equation with a source at the location of the D5 along the $x^6$ axis, where it meets the NS5 [134]:

$$\triangle\, x^5 = \delta(x^6) \quad \Leftrightarrow \quad x^5 = \frac{1}{2}(|x^6| + x^6)\ . \tag{7.3}$$

This means that the worldvolume of the D5 will bend beyond the NS5 by $\frac{\pi}{4}$: $x^5(x^6) = x^6$ for $x^6 > 0$.[4] However, such a configuration would break some more supersymmetry, and cannot be produced by one which preserves exactly one quarter of the maximum. We conclude that this result is not correct; in fact, we have to enforce $(p,q)$ *charge conservation* at each vertex in a fivebrane web. At each intersection point between various $(p,q)$-fivebranes we impose the following:

$$\sum_{\substack{\text{5-brane}_i\\ \text{at vertex}}} p_i = \sum_{\substack{\text{5-brane}_i\\ \text{at vertex}}} q_i = 0 \ . \tag{7.4}$$

[2] For $N_\text{f} = 0, \dots, 4$, by $E_{N_\text{f}+1}$ we mean, respectively: $\text{SU}(2)$, $\text{SU}(2) \times \text{U}(1)$, $\text{SU}(3) \times \text{SU}(2)$, $\text{SU}(5)$, $\text{Spin}(10)$.

[4] This is somewhat similar to the NS5 logarithmic bending exerted by a D$p$-brane ending on it (for $p < 6$) discussed in [42] and briefly touched upon in chapter 3.

| | $x^0$ | $x^1$ | $x^2$ | $x^3$ | $x^4$ | $x^5$ | $x^6$ | $x^7$ | $x^8$ | $x^9$ |
|---|---|---|---|---|---|---|---|---|---|---|
| D3 | × | × | · | · | · | × | × | · | · | · |
| D5 | × | × | × | × | × | × | 0 | · | · | · |
| NS5 | × | × | × | × | × | 0 | × | · | · | · |
| $(p,q)$-fivebrane | × | × | × | × | × | $px^5 = qx^6$ | | · | · | · |
| D7 | × | × | × | × | × | · | · | × | × | × |

**Table 7.1.:** Directions spanned by NS5, D5, and generic $(p,q)$-fivebranes. As in [134],[3] an NS5-brane is a $(1,0)$-fivebrane, hence it is stretched along the $x^6$ axis, while a D5-brane is a $(0,1)$-fivebrane, and it is stretched along the $x^5$ axis. A generic $(p,q)$-fivebrane spans the line $px^5 = qx^6$ in the $5,6$ plane. This condition fully preserves the remaining supersymmetry: D5's and NS5's break it to a quarter of the maximum thirty-two supercharges of type IIB, i.e. $\mathcal{N} = 1$ in $d = 5$. The rotational symmetry in the $7,8,9$ plane realizes the corresponding $\mathrm{SO}(3) \cong \mathrm{SU}(2)_R$ R-symmetry group. D7-branes (more generally $(p,q)$-sevenbranes) can be added to the web, and they contribute flavor hypermultiplets via strings connecting them to D5 and NS5-branes (in general $(p,q)$-fivebrane with same $(p,q)$ charges). However, their presence enforces a nontrivial monodromy on the type IIB axiodilaton field $\tau = C_0 + \frac{i}{g}$ (since they couple magnetically to the bulk RR potential $C_0$). Nonperturbative $(p,q)$-sevenbranes will enforce more general monodromies. Therefore, a $(p,q)$-fivebrane spans a line $px^5 = qx^6$ only if $C_0 = 0$ (i.e. when there are no sevenbranes around). Otherwise, they span a line whose slope in the $5,6$ plane is given by $\Delta x^5 + i\Delta x^6 = q + \tau p$. Finally, D3-branes spanning finite areas in the $5,6$ plane are strings in five dimensions (directions $0,\ldots,4$).

This condition, applied to the left configuration of figure 7.1, tells us that the D5 and NS5 have to merge into a $(1,1)$-fivebrane at their intersection; see the right side of figure 7.1.

In a generic fivebrane web, some of the branes will span finite lines in the $5,6$ planes (for instance, they are produced at a vertex by the merger of two other branes and end at another vertex). A five-dimensional $\mathcal{N} = 1$ field theory can then be obtained via Kaluza–Klein reduction along that finite line, and will live in flat Mink$_5$ (directions $0,\ldots,4$ in table 7.1).

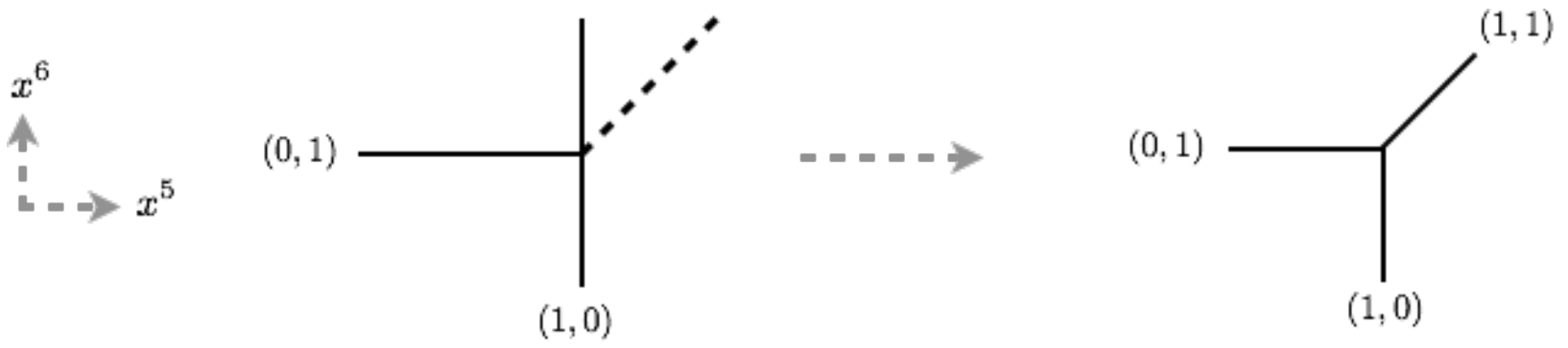

**Figure 7.1.:** We see the simplest $(p, q)$-fivebrane web. Instead of talking about Hanany–Witten setups, in five dimensions we will refer to them as *brane webs*, as the branes have to meet, merge and split at angles on the 5, 6 plane. On the left, we see the naive picture of a D5 ending from the left on an NS5, with the linear bending. On the right the correct picture, whereby a D5 and an NS5 meet and merge into a $(1, 1)$-fivebrane.

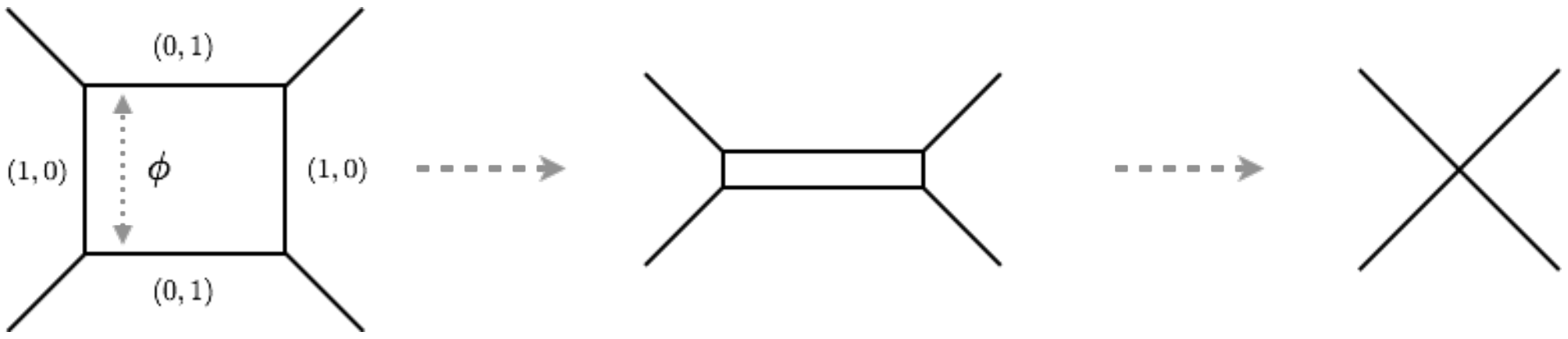

**Figure 7.2.:** On the left, a generic point along the Coulomb branch of a five-dimensional $\mathcal{N} = 1$ SU(2) gauge theory. The real scalar $\phi$ governs the distance between the D5's along direction 6. In the middle, a possible deformation. On the right, the superconformal fixed point.

Consider for instance the configuration on the left in figure 7.2. We have two *color* D5-branes sitting at $x_i^6$ along the 6 direction, two NS5-branes and many $(1, 1)$-fivebranes.[5] We expect to engineer an SU(2) gauge theory in five dimensions with $\mathcal{N} = 1$ supersymmetry when the finite segment they span along direction 5 has very small length $L$ ($L \to 0$ according to our general philosophy of section 3.2.4). Moreover, it can be argued [134] that the NS5-branes do not contribute to the low-energy physics (hence their degrees of freedom are decoupled from the massless spectrum) and the only deformation at our disposal is the distance between the D5-branes:

$$x_1^6 - x_2^6 \equiv \phi \ . \tag{7.5}$$

By supersymmetry, this (massless) real scalar must belong to a five-dimensional $\mathcal{N} = 1$ vector multiplet (there are no hypermultiplets since no other deformation scalars are available). When $\phi = 0$ the two D5's are on top of each other; hence the left picture in figure 7.2 corresponds to a generic point along the Coulomb branch of the gauge theory ($\phi$ acquires a nonzero vev). The middle picture is a possible deformation.

[5] We are neglecting signs in $p$ and $q$ since this does not affect our discussion. To be precise we should add them to the charge doublet $(1, 1)$ so that (7.4) is solved at each vertex.

Clearly, we expect something interesting happens in the situation depicted on the right of figure 7.2. To corroborate this hypothesis, remember that we can add D3-branes to the web extending along directions $0, 1, 5, 6$ (see table 7.1). In particular, they will end on both D5 and NS5-branes, acting like magnetic monopoles on their worldvolume, and will span the finite-area rectangle in figure 7.2. Hence, from the perspective of the five-dimensional spacetime (directions $0, \ldots, 4$), the D3's are just strings, with a tension proportional to the area of finite rectangles (more generally, polygons) in the $5, 6$ plane. This area will be piecewise quadratic in the scalars $\phi_i$ parameterizing the distances between consecutive D5-branes along the 6 direction (for an $\mathrm{SU}(N_\mathrm{c})$ gauge theory). One can match the area against the tension of a magnetic monopole as being computed from the prepotential $\mathcal{F}$ of a five-dimensional gauge theory [134]. For example, for the SU(3) gauge theory (depicted on the left of figure 7.3) the prepotential has been computed in [135].

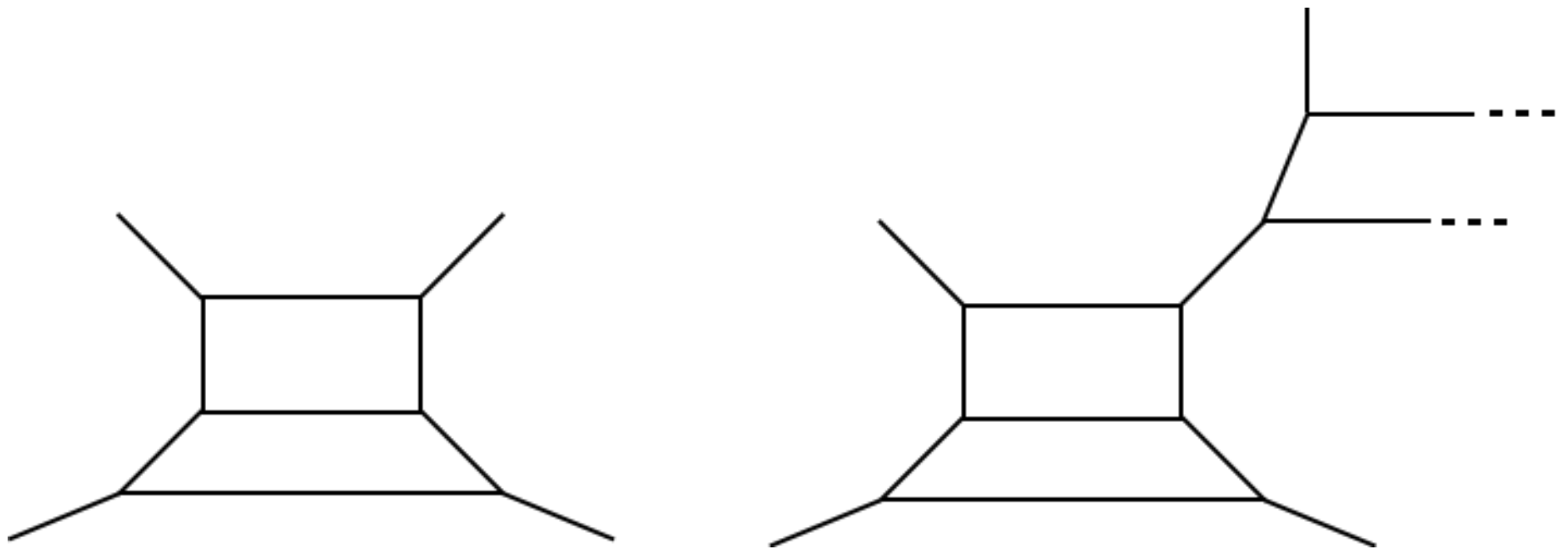

**Figure 7.3.:** On the left, a generic point along the Coulomb branch of a five-dimensional $\mathcal{N} = 1$ SU(3) gauge theory. On the right, the addition of $N_\mathrm{f} = 2$ flavor hypermultiplets introduced in the web by semi-infinite D5-branes extending towards $x^5 \to +\infty$.

When we shrink the finite rectangle completely (see the right side of figure 7.2) the monopole tension goes to zero, and we get tensionless strings in five dimensions. As we have learned by now, this signals that we are at an interacting fixed point (whereby the $\mathcal{N} = 1$ superalgebra gets enhanced to the superconformal one).

*In general, we reach a superconformal fixed point when all inner regions of a fivebrane web are shrunk to a point.*

We can also add flavors to the gauge theories as in six-dimensional Hanany–Witten brane setups, via semi-infinite color branes (see the right side of figure 7.3) or via $(p, q)$-sevenbranes extending along all directions but 5 and 6, see table 7.1. One should also keep in mind that, as in six dimensions, the number of flavors in a five-dimensional gauge theory is constrained: we must have $N_\mathrm{f} \leq 2N_\mathrm{c}$ [8]. However, there is no such restriction on the ranks of flavor groups coming from the geometric engineering of five-dimensional gauge theories. It has been argued [134] that theories with $N_\mathrm{f} > 2N_\mathrm{c}$ can be constructed from a brane web, and that they might correspond to the fixed points with enhanced $E_{N_\mathrm{f}+1}$ exceptional flavor symmetry of [8]. (They just cannot be deformed back to a gauge theory.) Another possible modification is the introduction of O5$^\pm$-planes along the D5's, projecting the SU gauge group to USp or SO.

Finally, a promising way to classify five-dimensional $\mathcal{N} = 1$ superconformal field theories is once again the use of holography. One could hope that, by classifying all AdS$_6$ vacua of perturbative type IIB string theory (that is, without nonperturbative sevenbranes), we could holographically compute interesting quantities on the field theory side for a vast class of theories, most notably the scaling behavior of their free energy. With this goal in mind, we now present the classification effort of [II].

## 7.2. Introduction

One of the interesting theoretical results of string theory is that it helps defining several nontrivial quantum field theories in dimensions higher than four, which are hard to study with traditional methods. For example, several five-dimensional superconformal field theories (SCFT$_5$'s) have been defined, using D4-branes in type I' [8, 136], M-theory on Calabi–Yau manifolds with shrinking cycles [136, 135], $(p, q)$-fivebrane webs [134] (sometimes also including $(p, q)$-sevenbranes [137]). These various realizations are dual to each other [138, 137]; some of these theories are also related by compactification [139] to the four-dimensional class $\mathcal{S}$ theories of [112].

However, not too many AdS$_6$ duals are known to these SCFT$_5$'s. Essentially the reason is that there is no D-brane stack whose near-horizon limit gives AdS$_6$. Indeed the string realizations quoted above originate from intersecting branes, whose localized metrics are notoriously difficult to find, as illustrated for example in [140]; even were they known, the relevant near-horizon limit would probably be far from obvious. One exception is when one of the branes is completely inside the other; in such cases some partially delocalized solutions [74] become actually localized. This was used by Brandhuber and Oz [104] to obtain the first AdS$_6$ solution in string theory. (It was also anticipated to exist [141] as a lift of a vacuum in the six-dimensional supergravity of [142].) It is in massive IIA, and it represents the near-horizon limit of a stack of D4's near an O8-D8 wall; thus it is dual to the theories in [8]. The internal space is half an $S^4$; the warping function $A$ and the dilaton $\phi$ go to infinity at its boundary. This is just a consequence of the presence of the O8-D8 system there, and it is a reflection of the peculiar physics of the corresponding SCFT$_5$'s. The fact that the dilaton diverges at the wall roughly corresponds to a Yang–Mills kinetic term of the type $\phi F_{\mu\nu}F^{\mu\nu}$; the scalar $\phi$ plays the role of $\frac{1}{g^2_{\mathrm{YM}}}$, and at the origin $\phi \to 0$ one finds a strongly coupled fixed point.

One can also study a few variations on the Brandhuber–Oz solution, such as orbifolding it [143] and performing T-duality [144, 106] or even the more recently developed [127, 145] nonabelian T-duality [106, 146]. The latter is not thought to be an actual duality, but rather a solution-generating duality; thus the solution should represent some new physics, although its global features are puzzling [146].

In this chapter, we attack the problem systematically, using the "pure spinor" techniques, emboldened by the success of this method for AdS$_7$ solutions of type II supergravity classified in chapter 5. In general, the procedure reformulates the equations for preserved supersymmetry in terms of certain differential forms defining $G$ structures on the "generalized tangent bundle" $T \oplus T^*$. It originates from generalized complex geometry [90, 91] and its first application was to Minkowski$_4$ or AdS$_4 \times M_6$ solutions of type II supergravity [94], in which case the relevant $G$ was SU(3)×SU(3). In [89] the method was extended (still in type II supergravity) to any ten-dimensional geometry; in this chapter we apply to AdS$_6 \times M_4$ the general system obtained there. We work in IIB, since in massive IIA the Brandhuber–Oz solution is unique [105], and in eleven-dimensional supergravity there are no solutions, as we show in appendix B.2.

As in chapter 5, the relevant structure on $T \oplus T^*$ is an "identity" structure (in other words, $G$ is the trivial group). Such a structure is defined by a choice of two Vielbeine $e^a_\pm$ (roughly associated with left- and right-movers in string theory). Just as in chapter 5, we actually prefer working with a single "average" Vielbein $e^a$ and with some functions on $M_4$ encoding the map between the two Vielbeine $e^a_\pm$. We then use these data to parameterize the forms appearing in the supersymmetry system. The supersymmetry equations then determine $e^a$ in terms of the functions on $M_4$, thus also determining completely the local form of the metric. As usual for this kind of formalism, the fluxes also come out as an output; less commonly, but again just as in chapter 5, the Bianchi identities are automatically satisfied.

When the dust settles, it turns out that we have completely reduced the problem to a system of two PDEs (see (7.45b), (7.46) below) on a two-dimensional space $\Sigma$. The metric is that of an $S^2$ fibration over $\Sigma$. This should not come as a surprise: an SCFT$_5$ has an SU(2) R-symmetry, which manifests itself in the gravity dual as the isometry group of the $S^2$. In chapter 5, for similar reasons the internal space $M_3$ was an $S^2$ fibration over an interval.

In AdS$_7$ the problem was reduced in chapter 5 to a system of first-order ODEs, which was then easy to study numerically; in our present case of supersymmetric AdS$_6$ solutions, we have PDEs, which are harder to study even numerically. Using EDS techniques (see for example [147, Chap. III] or [148, Sec. 10.4.1]) we have checked that the system is "well-formed": the general solution is expected to depend on two functions of one variable, which can be thought of as the values of the warping function $A$ and the dilaton $\phi$ at the boundary of $\Sigma$. (We expect regularity of the metric to fix those degrees of freedom as well, up to discrete choices.) We do recover two explicit solutions to the PDEs, corresponding to the abelian and nonabelian T-duals of the Brandhuber–Oz solution mentioned above.

Even though we do not present any new solutions in this chapter, it seems likely that our PDEs will describe $(p, q)$-fivebrane webs. For the AdS$_7$ case, it was conjectured [24] that the new solutions of chapter 5 arise as near-horizon limits of NS5-D6-D8 configurations previously studied in [19, 20]. The fact that those solutions have cohomogeneity one (namely, that all fields only depend on the coordinate on the base interval) matches with the details of the configuration. The coordinates $x^0, \ldots, x^5$ are common to all branes; the NS5's are located at $x^7 = x^8 = x^9 = 0$, while their positions in $x^6$ parameterize the tensor branch of the SCFT$_6$; the D6's are located at $x^7 = x^8 = x^9 = 0$, and extended along $x^6$; the D8's are extended along $x^7$, $x^8$, $x^9$, and located at various $x^6 = x^6_{\mathrm{D8}_i}$.

For AdS$_6$, the natural analogue of this story would involve $(p, q)$-fivebranes whose common directions would be $x^0, \ldots, x^4$, and which would be stretched along a line in the $5, 6$ plane (such that $\frac{x^5}{x^6} = \frac{p}{q}$). It is natural to conjecture that the solutions to our PDEs would correspond to near-horizon limits of such configurations, with the $5, 6$ plane somehow corresponding to our $\Sigma$; the remaining directions $x^7$, $x^8$, $x^9$ would provide our $S^2$ (as well as the radial direction of AdS$_6$). For such cases we would expect $\Sigma$ to have a boundary, at which the $S^2$ shrinks; the $(p, q)$-fivebranes would then be pointlike sources at this boundary. We hope to come back on this in the near future.

The chapter is organized as follows. In section 7.3 we present the system (7.13) of differential equations for supersymmetry, expressed in terms of differential forms $\Phi$ and $\Psi$ describing an identity structure on $M_4$; the derivation from (4.30), (4.31) is given in appendix B.1. In section 7.4 we parameterize the differential forms in terms of a Vielbein on $M_4$ and of four functions. We then plug this parameterization in the system, and obtain in section 7.5 our results on the metric and fluxes, and the two PDEs (7.45a), (7.46) that one needs to satisfy. Finally, in section 7.6, we make some general remarks about the PDEs, and recover the known examples.

## 7.3. Supersymmetry and pure spinor equations for AdS$_6$

We will start by presenting the system of pure spinor equations that we need to solve. Although this is similar to systems in other dimensions, there are some crucial differences, which we will try to highlight.

The original example of the pure spinor approach to supersymmetry was found for Mink$_4 \times M_6$ or AdS$_4 \times M_6$ solutions in type II supergravity [94], where the BPS conditions were reformulated in terms of certain differential equations on an SU(3) × SU(3) structure on the "generalized tangent bundle" $TM_6 \oplus T^*M_6$. Other examples followed over the years; for instance, [95] applied the strategy to Mink$_d \times M_{10-d}$ for even $d$ (for $d = 2$ the situation was improved in [149, 150, 151]); the case $\mathbb{R} \times M_9$ was considered in [152].

Partially motivated by the need of generating quickly pure-spinor-like equations for different setups, [89] formulated a system directly in ten dimensions, using the geometry of the generalized tangent bundle of $M_{10}$. This could have also been used in chapter 5 to generate a system for AdS$_7 \times M_3$ solutions in type II; in that case, however, it was more convenient to derive the system from the one of [95] for Mink$_6 \times M_4$, via a cone construction. This approach is not as readily available for our current case AdS$_6 \times M_4$; hence, we will attack it directly from [89].

We describe the derivation of our system from the ten-dimensional one of [89] in appendix B.1. The system in [89] contains two "symmetry" equations, (4.30b), that usually simply fix the normalizations of the pure spinors; two "pairing" equations (4.31) that often end up being redundant (although not always, see [153, 150]); and one "exterior" equation, (4.30a), that usually generates the pure spinor equations one is most interested in. This pattern is repeated for our case. One important difference is that the spinor decomposition we have to start with is clumsier than the one in other dimensions. Usually, the ten-dimensional spinors $\epsilon_a$ are the sum of two (or sometimes even one) tensor products. For AdS$_4 \times M_6$ in IIB, for example, we simply have $\epsilon_a = \zeta_{4+} \otimes \eta^a_{6+} + \text{c.c.}$. The analogue of this for Mink$_6 \times M_4$ in IIB would be

$$\begin{aligned} \epsilon_1 &= \zeta_{6+} \otimes \eta^1_{4+} + \zeta^c_{6+} \otimes \eta^{1\,c}_{4+} \\ \epsilon_2 &= \zeta_{6+} \otimes \eta^2_{4\mp} + \zeta^c_{6+} \otimes \eta^{2\,c}_{4\pm} \end{aligned} \qquad (\text{Mink}_6 \times M_4;\ \text{IIA/IIB})\ , \tag{7.6}$$

where $(\ )^c \equiv C(\ )^*$ denotes Majorana conjugation. For AdS$_6 \times M_4$, however, such an Ansatz cannot work: compatibility with the negative cosmological constant of AdS$_6$ demands that the $\zeta_6$ obey the Killing spinor equation on AdS$_6$,

$$\nabla_\mu \zeta_6 = \frac{1}{2}\gamma_\mu^{(6)}\zeta_6 \ , \tag{7.7}$$

and solutions to this equation cannot be chiral, while the $\zeta_{6\,+}$ in (7.6) are chiral. This issue does not arise in AdS$_4$ because in that case $(\zeta_{4\,+})^c$ has negative chirality; here $(\zeta_{6\,+})^c$ has positive chirality. This forces us to add "by hand" to (7.6) a second set of spinors with negative chirality, ending up with the unpromising-looking

$$\begin{aligned} \epsilon_1 &= \zeta_+\eta^1_+ + \zeta^c_+\eta^{1\,c}_+ + \zeta_-\eta^1_- + \zeta^c_-\eta^{1\,c}_- \\ \epsilon_2 &= \zeta_+\eta^2_\mp + \zeta^c_+\eta^{2\,c}_\mp + \zeta_-\eta^2_\pm + \zeta^c_-\eta^{2\,c}_\pm \end{aligned} \qquad (\text{AdS}_6\times M_4;\ \text{IIA/IIB}) \ , \tag{7.8}$$

where we have dropped the $_6$ and $_4$ labels (and the $\otimes$ sign), as we will do elsewhere. Attractive or not, (7.8) will turn out to be the correct one for our classification.

In the main text from now on we will consider the IIB case (unless otherwise stated). This is because AdS$_6\times M_4$ solutions in massive IIA were already analyzed in [105], where it was found that the only solution is the one in [104]. We did find it useful to check our methods on that solution as well; we sketch how that works in appendix B.3. As for the massless case, we found it more easily attacked by direct analysis in eleven-dimensional supergravity, which we present in appendix B.2, given that it is methodologically a bit outside the stream of our pure spinor analysis in IIB.

With the spinor Ansatz (7.8) in hand, we can apply the system (4.30), (4.31); the details of the derivation are described in appendix B.1. We first describe the forms appearing in the system. If we were interested in the Minkowski case, the system would only contain the bispinors $\eta^1_+\otimes\eta^{2\,\dagger}_+$ and $\eta^1_+\otimes(\eta^{2\,c}_+)^\dagger$.[6] (As usual in the pure spinor approach, we need not consider spinors of the type e.g. $\eta^1_+\otimes\eta^{1\,\dagger}_+$ to formulate a system which is necessary and sufficient.) Mathematically, this would describe an SU(2) $\times$ SU(2) structure on $TM_4\oplus T^*M_4$. Since in (7.8) we also have the negative chirality spinors $\eta^1_-$ and $\eta^{1\,c}_-$, there are many more forms we can build. We have the even forms:[7]

$$\phi^1_\pm = e^{-A_4}\eta^1_\pm\otimes\eta^{2\,\dagger}_\pm \ , \qquad \phi^2_\pm = e^{-A_4}\eta^1_\pm\otimes(\eta^{2\,c}_\pm)^\dagger \equiv e^{-A_4}\eta^1_\pm\otimes\overline{\eta^2_\pm} \ ; \tag{7.9a}$$

[6] As usual, we will identify forms with bispinors via the Clifford map $dx^{m_1}\wedge\ldots\wedge dx^{m_k}\mapsto\gamma^{m_1\ldots m_k}$.

[7] Notice that the $^1$ or $^2$ on $\phi$ has nothing to do with the $^1$ or $^2$ on the $\eta$'s; rather, it has to do with whether the second spinor is Majorana conjugated ($^2$) or not ($^1$). Another caveat is that the $_\pm$ does not indicate the degree of the form, as it is often the case in similar contexts; all the $\phi$'s in (7.9a) are even forms. One can think of the $_\pm$ as indicating whether these forms are self-dual or anti-self-dual.

and the odd forms:

$$\psi^1_\pm = e^{-A_4}\eta^1_\pm \otimes \eta^{2\,\dagger}_\mp \ , \qquad \psi^2_\pm = e^{-A_4}\eta^1_\pm \otimes (\eta^{2\,c}_\mp)^\dagger \equiv e^{-A_4}\eta^1_\pm \otimes \overline{\eta^2_\mp} \ . \tag{7.9b}$$

The factors $e^{-A_4}$ are inserted so that the bispinors have unit norm, in a sense to be clarified shortly; $A_4$ is the warping function, defined as usual by

$$ds^2_{10} = e^{2A_4} ds^2_{\mathrm{AdS}_6} + ds^2_{M_4} \ . \tag{7.10}$$

(From now on we call $A_4 \equiv A$.) Already by looking at (7.9a), we see that we have *two* $\mathrm{SU}(2)\times\mathrm{SU}(2)$ structures on $TM_4 \oplus T^*M_4$. If both of these structures come for example from SU(2) structures on $TM_4$, we see that we get an identity structure on $TM_4$, i.e. a Vielbein. In fact, this is true in general: (7.9a) always defines a Vielbein on $M_4$. We will see in section 7.4 how to parameterize both (7.9a) and (7.9b) in terms of the Vielbein they define.

In the meantime, we can already now notice that the (7.9a) and (7.9b) can be assembled more conveniently using the SU(2) R-symmetry. This is the group that rotates $\binom{\zeta}{\zeta^c}$ and each of $\binom{\eta^a_\pm}{\eta^{a\,c}_\pm}$ as a doublet. One can check that (7.8) is then left invariant, so it is a symmetry; since it acts on the external spinors, we call it an R-symmetry. It is the manifestation of the R-symmetry of a five-dimensional SCFT. Something very similar was noticed in chapter 5 for $\mathrm{AdS}_7$: the pure spinor system (5.22) naturally assembled into singlets and one triplet of SU(2). (Recall that a six-dimensional SCFT also has an SU(2) R-symmetry.) While in that chapter the SU(2) formalism was only stressed at the end of the computations, here the analysis is considerably more complicated, and SU(2) will be used from the very beginning to yield more manageable results. Let us define

$$\begin{aligned}\Phi_\pm &\equiv \begin{pmatrix}\eta^1_\pm \\ \eta^{1\,c}_\pm\end{pmatrix} \otimes \begin{pmatrix}\eta^{2\,\dagger}_\pm & \overline{\eta^2_\pm}\end{pmatrix} = \begin{pmatrix}\phi^1_\pm & \phi^2_\pm \\ -(\phi^2_\pm)^* & (\phi^1_\pm)^*\end{pmatrix} \\ &= \mathrm{Re}\phi^1_\pm \mathrm{Id}_2 + i(\mathrm{Im}\phi^2_\pm\sigma_1 + \mathrm{Re}\phi^2_\pm\sigma_2 + \mathrm{Im}\phi^1_\pm\sigma_3) \equiv \Phi^0_\pm \mathrm{Id}_2 + i\Phi^\alpha_\pm\sigma_\alpha \ ,\end{aligned} \tag{7.11a}$$

$$\begin{aligned}\Psi_\pm &\equiv \begin{pmatrix}\eta^1_\pm \\ \eta^{1\,c}_\pm\end{pmatrix} \otimes \begin{pmatrix}\eta^{2\,\dagger}_\mp & \overline{\eta^2_\mp}\end{pmatrix} = \begin{pmatrix}\psi^1_\pm & \psi^2_\pm \\ -(\psi^2_\pm)^* & (\psi^1_\pm)^*\end{pmatrix} \\ &= \mathrm{Re}\psi^1_\pm \mathrm{Id}_2 + i(\mathrm{Im}\psi^2_\pm\sigma_1 + \mathrm{Re}\psi^2_\pm\sigma_2 + \mathrm{Im}\psi^1_\pm\sigma_3) \equiv \Psi^0_\pm \mathrm{Id}_2 + i\Psi^\alpha_\pm\sigma_\alpha \ .\end{aligned} \tag{7.11b}$$

$\sigma_\alpha$, $\alpha = 1,2,3$, are the Pauli matrices. Here and in what follows, the superscript $^0$ denotes an SU(2) singlet, and not the zero-form part; the superscript $^\alpha$ denotes an SU(2) triplet, not a one-form. We hope this will not create confusion.

As we already mentioned, the forms $\Phi_\pm$, $\Psi_\pm$ will define an identity structure on $M_4$. However, not any random forms $\Phi_\pm$, $\Psi_\pm$ may be written as bispinors as in (7.11). In other cases, such as for SU(3) × SU(3) structures in six dimensions [94], it is useful to formulate a set of constraints on the forms that guarantee that they come from spinors; this allows to completely forget about the original spinors, and formulate supersymmetry completely in terms of some forms satisfying some constraints. In the present case, it would be possible to set up such a fancy approach, by saying that $\Phi_\pm$ and $\Psi_\pm$ should satisfy a condition on their inner products. For example we could impose that the $\Phi$'s and $\Psi$'s be pure spinors on $M_4$ obeying the compatibility conditions[8]

$$(\Phi^\alpha_\pm, \Phi^\beta_\pm) = (\Psi^\alpha_\pm, \Psi^\beta_\pm) = \delta^{\alpha\beta}(\Phi^0_\pm, \Phi^0_\pm) = \delta^{\alpha\beta}(\Psi^0_\pm, \Psi^0_\pm) \ . \tag{7.12}$$

As in chapter 5, this would however be an overkill, since in section 7.4 we will directly parameterize $\Phi_\pm$ and $\Psi_\pm$ in terms of a Vielbein and some functions on $M_4$. This will achieve the end of forgetting about the spinors $\eta^a_\pm$ by different means.

We can finally give the system of equations equivalent to preserved supersymmetry:

$$d_H\left[e^{3A-\phi}(\Psi_- - \Psi_+)^0\right] - 2e^{2A-\phi}(\Phi_- + \Phi_+)^0 = 0 \ , \tag{7.13a}$$
$$d_H\left[e^{4A-\phi}(\Phi_- - \Phi_+)^\alpha\right] - 3e^{3A-\phi}(\Psi_- + \Psi_+)^\alpha = 0 \ , \tag{7.13b}$$
$$d_H\left[e^{5A-\phi}(\Psi_- - \Psi_+)^\alpha\right] - 4e^{4A-\phi}(\Phi_- + \Phi_+)^\alpha = 0 \ , \tag{7.13c}$$
$$d_H\left[e^{6A-\phi}(\Phi_- - \Phi_+)^0\right] - 5e^{5A-\phi}(\Psi_- + \Psi_+)^0 = -\frac{1}{4}e^{6A} *_4 \lambda F \ , \tag{7.13d}$$
$$d_H\left[e^{5A-\phi}(\Psi_- + \Psi_+)^0\right] = 0 \ ; \tag{7.13e}$$
$$||\eta^1||^2 = ||\eta^2||^2 = e^A \ . \tag{7.13f}$$

As usual, $\phi$ here is the dilaton; $d_H = d - H\wedge$; $A$ was defined in (7.10); $\lambda$ is a sign operator defined in footnote 8; $F = F_1 + F_3$ is the "total" allowed internal RR flux, which also determines the external flux via

$$F_{(10)} = F + e^{6A}\text{vol}_6 \wedge *_4\lambda F \ . \tag{7.14}$$

Again, we remind the reader that the superscript $^0$ denotes a singlet part, and $^\alpha$ a triplet part, as in (7.11).

The last equation, (7.13f), can be reformulated in terms of $\Phi$ and $\Psi$. Since $\|\eta^a\|^2 \equiv \|\eta^a_+\|^2 + \|\eta^a_-\|^2$, we can define $||\eta^1_+|| = e^{A/2}\cos(\alpha/2)$, $||\eta^1_-|| = e^{A/2}\sin(\alpha/2)$,

[8] The Chevalley–Mukai pairing is defined as $(\alpha, \beta) = (\alpha \wedge \lambda(\beta))_4$, where on a $k$-form $\lambda\omega_k = (-)^{\lfloor \frac{k}{2} \rfloor}\omega_k$.

$||\eta^2_+|| = e^{A/2}\cos(\tilde\alpha/2)$, $||\eta^2_-|| = e^{A/2}\sin(\tilde\alpha/2)$, where $\alpha, \tilde\alpha \in [0,\pi]$; we then get

$$\begin{aligned}
(\Phi^0_+, \Phi^0_+) &= \frac{1}{8}\cos^2(\alpha/2)\cos^2(\tilde\alpha/2) \ , \\
(\Phi^0_-, \Phi^0_-) &= -\frac{1}{8}\sin^2(\alpha/2)\sin^2(\tilde\alpha/2) \ ; \\
(\Psi^0_+, \Psi^0_-) &= \frac{1}{8}\cos^2(\alpha/2)\sin^2(\tilde\alpha/2) \ , \\
(\Psi^0_-, \Psi^0_+) &= -\frac{1}{8}\sin^2(\alpha/2)\cos^2(\tilde\alpha/2) \ .
\end{aligned} \tag{7.15}$$

Just as (7.12), however, such a fancy formulation will be ultimately made redundant by our parameterization of $\Phi$ and $\Psi$ in section 7.4, which will satisfy (7.12) automatically, and where we will take care to implement (7.13f), so that (7.15) will be satisfied too.

We can check immediately that (7.13) imply the equations of motion for the flux, by acting on (7.13d) with $d_H$ and using (7.13e). The equations of motion for the metric and dilaton are then satisfied (as shown in general in [87] for IIA, and in [88] for IIB); the equations of motion for $H$ are also implied, since they are [102] for Minkowski$_4$ compactifications (which include Minkowski$_5$ as a particular case, and hence also AdS$_6$ by a conical construction). We will see later that the Bianchi identities for $F$ and $H$ are also automatically satisfied for this case, as was the case for the solutions of chapter 5.

It is also interesting to compare the system (7.13) with the above-mentioned system for Mink$_6$ in [95]. First of all the second summands in the left-hand side of (7.13a), (7.13d) implicitly come with a factor proportional to $\sqrt{-\Lambda}$ that we have set to one (since it can be reabsorbed in the warping factor $A$). To take the Mink$_6$ limit, we can imagine to restore those factors, and then take $\Lambda \to 0$. Hence all the second summands in the left-hand side of (7.13a), (7.13d) will be set to zero. This is not completely correct, actually, because implicit in (7.13a), (7.13c) there are more equations, that one can get by acting on them with $d_H$ (before taking the $\Lambda \to 0$ limit); we have to keep these equations as well. So far the limit works in the same way as for taking the $\Lambda \to 0$ limit from AdS$_4$ to Minkowski$_4$ in [94]. In the present case, however, there is one more thing to take into account. As we have seen, in the Minkowski$_6$ case the spinor Ansatz can be taken to be (7.6) rather than the more complicated (7.8) we had to use for AdS$_6$. To go from (7.8) to (7.6), we can simply set $\eta^1_- = 0$ and $\eta^2_\pm = 0$. This sets to zero some of our bispinors; for the IIB case on which we are focusing, it sets to zero everything but $\Phi_+$. This makes some of the equations disappear;

some others become redundant. All in all, we are left with

$$d_H(e^{2A-\phi}\Phi^0_+) = 0 \ , \quad d_H(e^{4A-\phi}\Phi^\alpha_+) = 0 \ , \quad d_H(e^{6A-\phi}\Phi^0_+) = -\frac{1}{4}e^{6A} *_4 \lambda F \ , \tag{7.16}$$

which is [95, Eq. (4.11)] in our SU(2) covariant language. (In section 5.3.1, this system was quoted in a slightly different way: the last equation was mixed with the first, to yield $e^\phi F = 16 *_4 \lambda(dA \wedge \Phi^0_+)$.)

In summary, in this section we have presented the system (7.13), which is equivalent to preserved supersymmetry for backgrounds of the form AdS$_6 \times M_4$. The forms $\Phi$ and $\Psi$ are not arbitrary: they obey certain algebraic constraints expressing their origin as spinor bilinears in (7.11), (7.9). We will now give the general solution to those constraints, and then proceed in section 7.5 to analyze the system.

## 7.4. Parameterization of the pure spinors

We have introduced in section 7.3 the even forms $\Phi_\pm$ and the odd forms $\Psi_\pm$ (see (7.11), (7.9a), (7.9b)). These are the main characters in the system (7.13), which is equivalent to preserved supersymmetry. Before we start using the system, however, we need to characterize what sorts of forms $\Phi_\pm$ and $\Psi_\pm$ can be: this is what we will do in this section.

### 7.4.1. Even forms

We will first deal with $\Phi_\pm$. We will actually first focus on $\Phi_+$, and then quote the results for $\Phi_-$. The computations in this subsection are actually pretty standard, and we will be brief.

Let us start with the case $\eta^1_+ = \eta^2_+ \equiv \eta_+$. Assume also for simplicity that $||\eta_+||^2 = 1$. In this case the bilinears define an SU(2) structure:

$$\eta_+\eta_+^\dagger = \frac{1}{4}e^{-ij_+} \ , \qquad \eta_+\overline{\eta_+} = \frac{1}{4}\omega_+ \ , \tag{7.17}$$

where the two-forms $j_+$, $\omega_+$ satisfy

$$j_+ \wedge \omega_+ = 0 \ , \qquad \omega_+^2 = 0 \ , \qquad \omega_+ \wedge \overline{\omega_+} = 2j_+^2 = -\mathrm{vol}_4 \ . \tag{7.18}$$

We can also compute

$$\eta^c_+\eta^{c\,\dagger}_+ = \frac{1}{4}e^{ij_+} \ , \qquad \eta^c_+\eta_+^\dagger = -\frac{1}{4}\overline{\omega_+} \ . \tag{7.19}$$

Let us now consider the case with two different spinors, $\eta^1_+ \neq \eta^2_+$; let us again assume that they have unit norm. We can define (in a similar way as in [100])

$$\eta_{0+} = \frac{1}{2}(\eta^1_+ - i\eta^2_+) \ , \qquad \tilde{\eta}_{0+} = \frac{1}{2}(\eta^1_+ + i\eta^2_+) \ . \tag{7.20}$$

Consider now $a_+ = \eta^{2\,\dagger}_+ \eta^1_+$, $b_+ = \overline{\eta^2_+}\eta^1_+$. $\{\eta^2_+, \eta^{2\,c}_+\}$ is a basis for spinors on $M_4$; $a_+$, $b_+$ are then the coefficients of $\eta^1_+$ along this basis. Since $\eta^a_+$ have both unit norm, we have $|a_+|^2 + |b_+|^2 = 1$. By multiplying $\eta^a_+$ by phases, we can assume that $a_+$ and $b_+$ are for example purely imaginary, and we can then parameterize them as $a_+ = -i\cos(\theta_+)$, $b_+ = i\sin(\theta_+)$. Going back to (7.20), we can now compute their inner products:

$$\eta^\dagger_{0+}\eta_{0+} = \cos^2\left(\frac{\theta_+}{2}\right) \ , \qquad \eta^\dagger_{0+}\tilde{\eta}_{0+} = 0 \ , \qquad \overline{\eta_{0+}}\tilde{\eta}_{0+} = \frac{1}{2}\sin(\theta_+) \ . \tag{7.21}$$

From this we can in particular read off the coefficients of the expansion of $\tilde{\eta}_{0+}$ along the basis $\{\eta_{0+}, \eta^c_{0+}\}$. This gives:

$$\tilde{\eta}_{0+} = \frac{1}{||\eta_{0+}||^2}(\eta^\dagger_{0+}\tilde{\eta}_{0+}\eta_{0+} + \overline{\eta_{0+}}\tilde{\eta}_{0+}\eta^c_{0+}) = \tan\left(\frac{\theta_+}{2}\right)\eta^c_{0+} \ . \tag{7.22}$$

Recalling (7.20), and defining now $\eta_{0+} = \cos\left(\frac{\theta_+}{2}\right)\eta_+$, we get

$$\begin{aligned} \eta^1_+ &= \cos\left(\frac{\theta_+}{2}\right)\eta_+ + \sin\left(\frac{\theta_+}{2}\right)\eta^c_+ \ , \\ \eta^2_+ &= i\left(\cos\left(\frac{\theta_+}{2}\right)\eta_+ - \sin\left(\frac{\theta_+}{2}\right)\eta^c_+\right) \ . \end{aligned} \tag{7.23}$$

From this it is now easy to compute $\eta^1_+\eta^{2\,\dagger}_+$ and $\eta^1_+\overline{\eta^2_+}$. Recall, however, that in the course of our computation we have first fixed the norms and then the phases of $\eta^a_+$. The norms of the spinors we need in this chapter are not one; they were actually already parameterized before (7.15), so as to satisfy (7.13f). The factor $e^A$, however, simplifies with the $e^{-A}$ in the definition (7.9a). Let us also restore the phases we earlier fixed, by rescaling $\eta^1_\pm \to e^{iu_\pm}\eta^1_\pm$, $\eta^2_\pm \to e^{it_\pm}\eta^2_\pm$. All in all

we get

$$\phi^1_+ = \frac{1}{4}\cos(\alpha/2)\cos(\tilde\alpha/2)e^{i(u_+-t_+)}\cos(\theta_+)\,\cdot$$
$$\cdot\exp\left[-\frac{1}{\cos(\theta_+)}(ij_+ + \sin(\theta_+)\text{Re}\omega_+)\right]\,, \tag{7.24a}$$
$$\phi^2_+ = \frac{1}{4}\cos(\alpha/2)\cos(\tilde\alpha/2)e^{i(u_++t_+)}\sin(\theta_+)\,\cdot$$
$$\cdot\exp\left[\frac{1}{\sin(\theta_+)}(\cos(\theta_+)\text{Re}\omega_+ + i\text{Im}\omega_+)\right]\,. \tag{7.24b}$$

The formulas for $\phi^{1,2}_-$ can be simply obtained by changing $\cos(\alpha/2) \to \sin(\alpha/2)$, $\cos(\tilde\alpha/2) \to \sin(\tilde\alpha/2)$, and $_+ \to _-$ everywhere. The only difference to keep in mind is that the last equation in (7.18) is now replaced with $\omega_- \wedge \overline{\omega_-} = 2j_-^2 = \text{vol}_4$.

### 7.4.2. Odd forms

We now turn to the bilinears of "mixed type", i.e. the $\psi^{1,2}_\pm$ we defined in (7.9b), which result in odd forms. We will again start from the case in which $\eta^1_\pm = \eta^2_\pm \equiv \eta_\pm$.

There are two vectors we can define:

$$v_m = \eta^{2\,\dagger}_- \gamma_m \eta^1_+\,, \quad w_m = \overline{\eta^2_-}\gamma_m\eta^1_+\,. \tag{7.25}$$

In bispinor language, we can compute

$$\eta_+\eta_-^\dagger = \frac{1}{4}(1+\gamma)v\,, \quad \eta^c_+\eta^{c\,\dagger}_- = \frac{1}{4}(1+\gamma)\overline{v}\,, \tag{7.26a}$$
$$\eta_-\eta_+^\dagger = \frac{1}{4}(1-\gamma)\overline{v}\,, \quad \eta^c_-\eta^{c\,\dagger}_+ = \frac{1}{4}(1-\gamma)v\,, \tag{7.26b}$$

and

$$\eta_+\eta^{c\,\dagger}_- = \frac{1}{4}(1+\gamma)w\,, \quad \eta^c_+\eta^\dagger_- = -\frac{1}{4}(1+\gamma)\overline{w}\,, \tag{7.26c}$$
$$\eta_-\eta^{c\,\dagger}_+ = -\frac{1}{4}(1-\gamma)w\,, \quad \eta^c_-\eta^\dagger_+ = \frac{1}{4}(1-\gamma)\overline{w}\,. \tag{7.26d}$$

(In four Euclidean dimensions, the chiral $\gamma = *_4\lambda$, so that $(1+\gamma)v = v + *_4 v$, and so on. See [89, App. A] for more details.) For the more general case where

$\eta^1_\pm \neq \eta^2_\pm$, we can simply refer back to (7.23). For example we get

$$\psi^1_+ = \frac{e^{i(u_+ - t_-)}}{4} \cos(\alpha/2)\sin(\tilde{\alpha}/2)(1+\gamma)\left[\cos\left(\frac{\theta_+ + \theta_-}{2}\right)\mathrm{Re}v + \right. \\ \left. + i\cos\left(\frac{\theta_+ - \theta_-}{2}\right)\mathrm{Im}v - \sin\left(\frac{\theta_+ + \theta_-}{2}\right)\mathrm{Re}w + i\sin\left(\frac{\theta_+ - \theta_-}{2}\right)\mathrm{Im}w\right] . \tag{7.27}$$

For the time being we do not show the lengthy expressions for the other odd bispinors $\psi^2_+$ and $\psi^{1,2}_-$, because they will all turn out to simplify quite a bit as soon as we impose the zero-form equations in (7.13).

The $v$ and $w$ we just introduced are a complex Vielbein; let us see why. First, a standard Fierz computation gives

$$v \cdot \eta_+ = 0 \ , \qquad \overline{v} \cdot \eta_+ = 2\eta_- \ , \tag{7.28}$$

where $\cdot$ denotes Clifford product. Multiplying from the left by $\eta_-^\dagger$, we obtain

$$v^2 = 0 \ , \qquad v \llcorner \overline{v} = v^m \overline{v}_m = 2 \ . \tag{7.29}$$

Similarly to (7.28), we can compute the action of $w$:

$$w \cdot \eta_\pm = 0 \ , \qquad \overline{w} \cdot \eta_\pm = \pm 2\eta^c_\mp \ . \tag{7.30}$$

Multiplying by $\overline{\eta_\mp}$, we get

$$w^2 = 0 \ , \qquad w \llcorner \overline{w} = 2 \ . \tag{7.31}$$

From (7.28) we can also get $v \cdot \eta_+ \overline{\eta_-} = 0$, $\overline{v} \cdot \eta_+ \overline{\eta_-} = 2\eta_- \overline{\eta_-}$, whose zero-form parts read

$$v \llcorner w = 0 = \overline{v} \llcorner w \ . \tag{7.32}$$

Together, (7.29), (7.31), (7.32) say that

$$\{\mathrm{Re}v,\ \mathrm{Re}w,\ \mathrm{Im}v,\ \mathrm{Im}w\} \tag{7.33}$$

are a Vielbein.

We can also now try to relate the even forms of section 7.4.1 to this Vielbein. From (7.28) we also see $v \cdot \eta_+ \overline{\eta_+} = 0$, which says $v \wedge \omega_+ = 0$; similarly one gets $\overline{v} \wedge \omega_- = 0$. Also, (7.30) implies that $w \cdot \eta_+ \overline{\eta_+} = w \cdot \omega_+ = 0$, and thus that $w \wedge \omega_\pm = 0$. So we have $\omega_+ \propto v \wedge w$, $\omega_- \propto \overline{v} \wedge w$. One can fix the proportionality constant by a little more work:

$$\omega_+ = -v \wedge w \ , \qquad \omega_- = \overline{v} \wedge w \ . \tag{7.34a}$$

Similar considerations also determine the real two-forms:

$$j_\pm = \pm\frac{i}{2}(v\wedge\overline{v} \pm w\wedge\overline{w}) \ . \tag{7.34b}$$

So far we have managed to parameterize all the pure spinors $\Phi_\pm$, $\Psi_\pm$ in terms of a Vielbein given by (7.33). The expressions for $\Phi_+$ are given in (7.24); $\Phi_-$ is given by changing $(\cos(\alpha/2), \cos(\tilde\alpha/2)) \to (\sin(\alpha/2), \sin(\tilde\alpha/2))$, and $_+ \to _-$ everywhere. The forms $j_\pm$, $\omega_\pm$ are given in (7.34) in terms of the Vielbein. Among the odd forms of $\Psi_\pm$, we have only quoted one example, (7.27); similar expressions exist for $\psi^2_+$ and for $\psi^{1,2}_-$. We will summarize all this again after the simplest supersymmetry equations will allow us to simplify the parameterization quite a bit.

## 7.5. General analysis

We will now use the parameterization obtained for $\Phi$ and $\Psi$ in section 7.4 in the system (7.13). As anticipated in the introduction, we will reduce the system to the two PDEs (7.45a), (7.46), and we will determine the local form of the metric and of the fluxes in terms of a solution to those equations.

### 7.5.1. Zero-form equations

The only equations in (7.13) that have a zero-form part are (7.13a) and (7.13c):

$$(\Phi_+ + \Phi_-)^0_0 = 0 \ , \qquad (\Phi_+ + \Phi_-)^\alpha_0 = 0 \ . \tag{7.35}$$

The subscript $_0$ here denotes the zero-form part. (Recall that the superscripts $^0$ and $^\alpha$ denote SU(2) singlets and triplets respectively.) To simplify the analysis, it is useful to change variables so as to make the SU(2) R-symmetry more manifest; this will lead us to definitions similar to those made in section 5.5.5.

In (7.24), apart for the overall factor $\cos(\alpha/2)\cos(\tilde\alpha/2)/4$, we have $\phi^1_{+\,0} \propto e^{i(u_+-t_+)}\cos(\theta_+)$, $\phi^2_{+\,0} \propto e^{i(u_++t_+)}\sin(\theta_+)$. The singlet is $\mathrm{Re}\phi^1_{+\,0}$, and is proportional to $\cos(\theta_+)\cos(u_+-t_+)$, and it is a good idea to give it a name, say $x_+$. On the other hand, the triplet is $\{\mathrm{Im}\phi^2_+, \mathrm{Re}\phi^2_+, \mathrm{Im}\phi^1_+\} \propto \{\sin(\theta_+)\sin(u_++t_+), \sin(\theta_+)\cos(u_++t_+), \cos(\theta_+)\sin(u_+-t_+)\}$. If we sum their squares, we obtain:

$$\sin^2(\theta_+) + \cos(\theta_+)^2\sin^2(u_+-t_+) = x_+^2\tan^2(u_+-t_+) + \sin^2(\theta_+) = 1 - x_+^2 \ . \tag{7.36}$$

This suggests that we parameterize the triplet using the combination $\sqrt{1-x_+^2}\,y^\alpha$, where $y^\alpha$ should obey $y_\alpha y^\alpha = 1$ and can be chosen to be the $\ell = 1$ spherical harmonics on $S^2$. What we are doing is essentially changing variables on an $S^3$, going from coordinates that exhibit it as an $S^1 \times S^1$ fibration over an interval to coordinates that exhibit it as an $S^2$ fibration over an interval:

$$\left\{\cos(\theta_+)e^{i(u_+-t_+)}, \sin(\theta_+)e^{i(u_++t_+)}\right\} \to \left\{x_+, \sqrt{1-x_+^2}y^\alpha\right\} \; . \tag{7.37}$$

An identical discussion can of course be given for $\phi_-^{1,2}$. Summing up, we are led to the following definitions:

$$x_\pm \equiv \cos(\theta_\pm)\cos(u_\pm - t_\pm) \; , \quad \sin\beta_\pm \equiv \frac{\sin(\theta_+)}{\sqrt{1-x_+^2}} \; , \quad \gamma_\pm \equiv \frac{\pi}{2} - u_\pm - t_\pm \; , \tag{7.38}$$

and

$$y_\pm^\alpha \equiv \Big(\sin(\beta_\pm)\cos(\gamma_\pm),\; \sin(\beta_\pm)\sin(\gamma_\pm),\; \cos(\beta_\pm)\Big) \; , \tag{7.39}$$

in terms of which

$$\begin{aligned} \Phi_{+\,0} &= \cos(\alpha/2)\cos(\tilde\alpha/2)\left(x_+ + i y_+^\alpha\sqrt{1-x_+^2}\sigma_\alpha\right) \; , \\ \Phi_{-\,0} &= \sin(\alpha/2)\sin(\tilde\alpha/2)\left(x_- + i y_-^\alpha\sqrt{1-x_-^2}\sigma_\alpha\right) \; . \end{aligned} \tag{7.40}$$

Going back to (7.35), summing the squares of all four equations we get $\cos^2(\alpha/2)\cos^2(\tilde\alpha/2) = \sin^2(\alpha/2)\sin^2(\tilde\alpha/2)$. Given that $\alpha$ and $\tilde\alpha \in [0,\pi]$, this is uniquely solved by

$$\tilde\alpha = \pi - \alpha \; . \tag{7.41}$$

Now (7.35) reduces to

$$-x_- = x_+ \equiv x \; , \qquad -y_-^\alpha = y_+^\alpha \equiv y^\alpha \; . \tag{7.42}$$

In terms of the original parameters, this means $\theta_+ = \theta_-$, $u_- = u_+$, $t_- = t_+ + \pi$.

The parameterization obtained in section 7.4 now simplifies considerably:

$$\phi_\pm^1 = \pm\frac{1}{8}\sin\alpha\cos\theta\, e^{i(u-t)} \exp\left[-\frac{1}{\cos\theta}(ij_\pm + \sin\theta \mathrm{Re}\omega_\pm)\right] \; , \tag{7.43a}$$

$$\phi_\pm^2 = \pm\frac{1}{8}\sin\alpha\sin\theta\, e^{i(u+t)} \exp\left[\frac{1}{\sin\theta}(\cos\theta \mathrm{Re}\omega_+ + i\mathrm{Im}\omega_+)\right] \; ; \tag{7.43b}$$

$$\psi_\pm^1 = \mp\frac{1}{8}(1\pm\cos\alpha)e^{i(u-t)}(1\pm\gamma)\left[\cos\theta\mathrm{Re}v \pm i\mathrm{Im}v \mp \sin\theta\mathrm{Re}w\right] \; , \tag{7.43c}$$

$$\psi_\pm^2 = \mp\frac{1}{8}(1\pm\cos\alpha)e^{i(u+t)}(1\pm\gamma)\left[\sin\theta\mathrm{Re}v \pm i\mathrm{Im}w \pm \cos\theta\mathrm{Re}w\right] \; . \tag{7.43d}$$

We temporarily reverted here to a formulation where $\mathrm{SU}(2)_\mathrm{R}$ is not manifest; however, in what follows we will almost always use the SU(2) covariant variables $x$ and $y^\alpha$ introduced above.

### 7.5.2. Geometry

We will now describe how we analyzed the higher-form parts of (7.13), although not in such detail as in section 7.5.1.

The only equations that have a one-form part are (7.13b). From (7.43c), (7.43d), we see that the second summand $(\Psi_+ + \Psi_-)^\alpha_1$ is a linear combination of the forms in the Vielbein (7.33). The first summand consists of derivatives of the parameters we have previously introduced. This gives three constraints on the four elements of the Vielbein. We used it to express Im$v$, Re$w$, Im$w$ in terms of Re$v$;[9] the resulting expressions are at this point still not particularly illuminating, and we will not give them here. These expressions are not even manifestly SU(2)-covariant at this point; however, once one uses them into $\Phi_\pm$ and $\Psi_\pm$, one does find SU(2)-covariant forms. Just by way of example, we have

$$\begin{aligned}(\Phi_+ + \Phi_-)^\alpha_2 =& -\frac{1}{3}e^{-3A+\phi}\sin\alpha\,\mathrm{Re}v \wedge d\left(y^\alpha \sin\alpha\, e^{4A-\phi}\sqrt{1-x^2}\right) \ ,\\ (\Psi_- - \Psi_+)^\alpha_1 =& y^\alpha\sqrt{1-x^2}\sin^2(\alpha)\mathrm{Re}v \, + \\ &+ \frac{1}{3}e^{-3A+\phi}\cos\alpha\, d\left(y^\alpha \sin\alpha\, e^{4A-\phi}\sqrt{1-x^2}\right) \ .\end{aligned}$$

We chose these particular 2-form and 1-form triplet combinations because they are involved in the 2-form part of (7.13c). The result is a triplet of equations of the form $y^\alpha E_2 + dy^\alpha \wedge E_1 = 0$, where $E_i$ are $i$-forms and SU(2)$_\mathrm{R}$ singlets. If we multiply this by $y_\alpha$, we obtain $E_2 = 0$ (since $y_\alpha dy^\alpha = 0$); then also $E_1 = 0$ necessarily. The latter gives a simple expression for Re$v$, the one-form among the Vielbein (7.33) that we had not determined yet:

$$\mathrm{Re}v = -\frac{e^{-A}}{\sin\alpha}d(e^{2A}\cos\alpha) \ . \tag{7.44}$$

Once this is used, the two-form equation $E_2 = 0$ is automatically satisfied.

There are some more two-form equations from (7.13). The easiest is (7.13e), which gives

$$d\left(\frac{e^{4A-\phi}}{x}\cot\alpha\, d(e^{2A}\cos\alpha) + \frac{1}{3x}e^{2A}\sqrt{1-x^2}d\left(e^{4A-\phi}\sqrt{1-x^2}\sin\alpha\right)\right) = 0 \ . \tag{7.45a}$$

Locally, this can be solved by saying

$$xdz = e^{4A-\phi}\cot\alpha\, d(e^{2A}\cos\alpha) + \frac{1}{3}e^{2A}\sqrt{1-x^2}d\left(e^{4A-\phi}\sqrt{1-x^2}\sin\alpha\right) \tag{7.45b}$$

[9] Doing so requires $x \neq 0$; the case $x = 0$ will be analyzed separately in section 7.5.4.

for some function $z$. The two-form part of (7.13a) reads, on the other hand,

$$e^{-8A}d(e^{6A}\cos\alpha)\wedge dz = d(xe^{2A-\phi}\sin\alpha)\wedge d(e^{2A}\cos\alpha)\ . \tag{7.46}$$

If one prefers, $dz$ can be eliminated, giving

$$3\sin(2\alpha)dA\wedge d\phi = d\alpha\wedge\big(6dA\ + \\ +\sin^2(\alpha)\left(-dx^2-2(x^2+5)dA+(1+2x^2)d\phi\right)\big)\ . \tag{7.47}$$

We will devote the whole section 7.6 to analyze the PDEs (7.45a), (7.46) and we will also exhibit two explicit solutions.

Taking the exterior derivative of (7.46) one sees that $d\alpha\wedge dA\wedge dz=0$. Wedging (7.45a) with an appropriate one-form, one also sees $d\alpha\wedge dA\wedge dx=0$. Taken together, these mean that only two among the remaining variables $(\alpha,x,A,\phi)$ are really independent. For example we can take $\alpha$ and $x$ to be independent, and

$$A=A(\alpha,x)\ ,\qquad \phi=\phi(\alpha,x)\ . \tag{7.48}$$

We are not done with the analysis of (7.13), but there will be no longer any purely geometrical equations: the remaining content of (7.13) determines the fluxes, as we will see in the next subsection. Let us then pause to notice that at this point we have already determined the metric: three of the elements of the Vielbein (7.33) were determined already at the beginning of this section in terms of Re$v$, and the latter was determined in (7.44). This gives the metric

$$ds^2=\frac{\cos\alpha}{\sin^2(\alpha)}\frac{dq^2}{q}+\frac{1}{9}q(1-x^2)\frac{\sin^2(\alpha)}{\cos\alpha}\left(\frac{1}{x^2}\left(\frac{dp}{p}+3\cot^2(\alpha)\frac{dq}{q}\right)^2+ds^2_{S^2}\right)\ , \tag{7.49}$$

where the $S^2$ is spanned by the functions $\beta$ and $\gamma$ introduced in (7.39) (namely, $ds^2_{S^2}=d\beta^2+\sin^2(\beta)d\gamma^2$), and we have eliminated $A$ and $\phi$ in favor of

$$q\equiv e^{2A}\cos\alpha\ ,\qquad p\equiv e^{4A-\phi}\sin\alpha\sqrt{1-x^2}\ . \tag{7.50}$$

These variables could also be used in the equations (7.45a), (7.46) above, with marginal simplification. Notice that positivity of (7.49) requires $|x|\leq 1$.

Thus we have found in this section that the internal space $M_4$ is an $S^2$ fibration over a two-dimensional space $\Sigma$, which we can think of as spanned by the coordinates $(\alpha,x)$.

### 7.5.3. Fluxes

We now turn to the three-form part of (7.13b). This is an SU(2)$_\mathrm{R}$ triplet. It can be written as $y^\alpha H=\epsilon^{\alpha\beta\gamma}y^\beta dy^\gamma\wedge\tilde{E}_2+y^\alpha\mathrm{vol}_{S^2}\wedge\tilde{E}_1$, where $\tilde{E}_i$ are $i$-forms and SU(2)$_\text{R}$ singlets. Actually, from (7.45a) and (7.46) it follows that $\tilde{E}_2 = 0$; we are then left with a single equation setting $H = \text{vol}_{S^2} \wedge \tilde{E}_1$:

$$H = -\frac{1}{9x} e^{2A} \sqrt{1-x^2} \sin\alpha \left[ -\frac{6dA}{\sin\alpha} + 2e^{-A}(1+x^2) d(e^A \sin\alpha) + \right. \\ \left. + \sin\alpha\, d(\phi + x^2) \right] \wedge \text{vol}_{S^2} \ . \quad (7.51)$$

As expected, $H$ is a singlet under SU(2)$_\text{R}$.

All the four-form equations in (7.13e), (7.13a), (7.13c) turn out to be automatically satisfied. We can then finally turn our attention to (7.13d), which we have ignored so far. It gives the following expressions for the fluxes:

$$F_1 = \frac{e^{-\phi}}{6x\cos\alpha} \left[ \frac{12dA}{\sin\alpha} + 4e^{-A}(x^2-1) d(e^A \sin\alpha) + \right. \\ \left. + e^{2\phi} \sin\alpha\, d(e^{-2\phi}(1+2x^2)) \right] ; \quad (7.52a)$$

$$F_3 = \frac{e^{2A-\phi}}{54} \sqrt{1-x^2} \frac{\sin^2(\alpha)}{\cos\alpha} \left[ \frac{36dA}{\sin\alpha} + 4e^{-A}(x^2-7) d(e^A \sin\alpha) + \right. \\ \left. + e^{2\phi} \sin\alpha\, d(e^{-2\phi}(1+2x^2)) \right] \wedge \text{vol}_{S^2} \ . \quad (7.52b)$$

The Bianchi identities

$$dH = 0 \ , \quad dF_1 = 0 \ , \quad dF_3 + H \wedge F_1 = 0 \ , \quad (7.53)$$

are all automatically satisfied, using of course the PDEs (7.45a), (7.46). As usual, this statement is actually true only if one assumes that the various functions appearing in those equations are smooth. As in chapter 5, one can introduce sources by relaxing this condition.

### 7.5.4. The case $x = 0$

In section 7.5.2, we used the three-form part of (7.13b) to express Im$v$, Re$w$, Im$w$ in terms of Re$v$. This actually can only be done for $x \neq 0$: the expressions we get contain $x$ in the denominator, as can be seen for example in (7.45a). This left out the case $x = 0$; we will analyze it in this section, showing that it leads to a single solution, discussed in [144, 106] — namely, to a T-dual of the AdS$_6$ solution found in [104] and reviewed in our language in appendix B.3.

Keeping in mind that $-x_- = x_+ = x$ (from (7.42)), from (7.38) we have $x = \cos(\theta)\cos(u-t)$. Imposing $x = 0$ then means either $\theta = \frac{\pi}{2}$ or $u - t = \frac{\pi}{2}$. Of these two possibilities, the first does not look promising, because on the $S^3$ parameterized by $(\cos(\theta)e^{i(u-t)}, \sin(\theta)e^{i(u+t)})$ it effectively restricts us to an $S^1$: only the function $u+t$ is left in the game, and indeed going further in the analysis one finds that the metric becomes degenerate.[10] The second possibility, $u-t=\frac{\pi}{2}$, restricts us instead to an $S^2 \subset S^3$; we will now see that this possibility survives. It gives

$$\beta = \theta \ , \qquad t = -\frac{1}{2}\gamma \ , \qquad u = \frac{\pi}{2} - \frac{1}{2}\gamma \ . \tag{7.54}$$

This leads to a dramatic simplification in the whole system. The one-form equations from (7.13b) do not involve Im$v$ any more; we can now use them to solve for Re$v$, Re$w$, Im$w$ (rather than for Im$v$, Re$w$, Im$w$ as we did in previous subsections, for $x \neq 0$). This strategy would actually have been possible for $x \neq 0$ too, but it would have led to far more involved expressions; for this reason we decided to isolate the $x=0$ case and to treat it separately in this subsection. We get

$$\mathrm{Re}v = \frac{e^{-3A+\phi}}{3\cos\alpha} d(\sin\alpha e^{4A-\phi}) \ , \tag{7.55a}$$

$$\mathrm{Re}w = \frac{e^A}{3}\sin\alpha \, d\beta \ , \tag{7.55b}$$

$$\mathrm{Im}w = -\frac{e^A}{3}\sin\alpha \, \sin\beta d\gamma \ . \tag{7.55c}$$

We now turn to the two-form equation in (7.13c). As in the previous subsections of this section, this can be separated into a 2-form multiplying $y^\alpha$ and a one-form multiplying $dy^\alpha$, which have to vanish separately:

$$d(e^{5A-\phi}\mathrm{Re}v) = 0 \ , \quad e^{5A-\phi}(3-4\sin^2(\alpha))\mathrm{Re}v = d(e^{6A-\phi}\sin\alpha\cos\alpha) \ . \tag{7.56}$$

Hitting the second equation with $d$ and using the first, we find $\sin\alpha\cos\alpha \, d\alpha \wedge \mathrm{Re}v = 0$, and hence, recalling (7.55), to $\sin\alpha d\alpha \wedge d(4A-\phi) = 0$. Now, $\sin\alpha$ is not allowed to vanish because of (7.55) (recall that Re$v$, Re$w$, Im$w$ are part of a Vielbein); hence $d\alpha \wedge d(4A-\phi) = 0$. This can be interpreted as saying that $4A-\phi$ is a function of $\alpha$. On the other hand, using (7.55) in the first in (7.56), we get $d(\frac{e^{2A}}{\cos\alpha}) \wedge d(\sin\alpha e^{4A-\phi}) = 0$, which shows that $A = A(\alpha)$, and hence also that $\phi = \phi(\alpha)$. Going back to the second in (7.56), it now reads

$$2(\cos^2(\alpha)+2)\partial_\alpha A + \sin^2(\alpha)\partial_\alpha\phi = \sin(2\alpha) \ . \tag{7.57}$$

Turning to (7.13e), its two-form part reads

$$d(e^{5A-\phi}\mathrm{Im}v) = 0 \quad \Rightarrow \quad \mathrm{Im}v = e^{-(5A-\phi)}dz \tag{7.58}$$

[10] At the stage of (7.55) below, one would find Re$w \propto$ Im$v$.

for some function $z$. This completes (7.55).

Finally, (7.13a) gives

$$\left(d(e^{-2A}\cos\alpha) + 2e^{-3A}\sin\alpha\,\mathrm{Re}v\right)\wedge \mathrm{Im}v = 0\ . \tag{7.59}$$

In view of (7.58), the parenthesis has to vanish by itself; this leads to

$$4(7\cos^2(\alpha) - 4)\partial_\alpha A + 4\sin^2(\alpha)\partial_\alpha \phi = -\sin(2\alpha)\ . \tag{7.60}$$

Notice that now (7.57) and (7.60) are two *ordinary* (as opposed to partial) differential equations, which can be solved explicitly:

$$e^A = \frac{c_1}{\cos^{1/6}(\alpha)}\ ,\qquad e^\phi = \frac{c_2}{\sin\alpha\cos^{2/3}(\alpha)}\ , \tag{7.61}$$

where $c_i$ are two integration constants. These are exactly the warping and dilaton presented in [146, (A.1)], for $c_1 = \frac{3}{2}Lm^{-1/6}$, $c_2 = 4/(3L^2m^{2/3})$. It is now possible to derive the fluxes, as we did in subsection 7.5.3 for $x \neq 0$, and check that they coincide with those in [146, (A.1)].

The metric can now be computed too, using the Vielbein (7.56), (7.58); it also agrees with the one given in [146, (A.1)]. It inherits the singularity at $\alpha = \frac{\pi}{2}$ from the Brandhuber–Oz solution [104]; moreover, it now has a singularity at $\alpha = 0$. The latter is actually the singularity one always gets when one T-dualizes along a Hopf direction in a $S^3$ that shrinks somewhere. It represents an NS5 smeared along the T-dual $S^1$; one expects worldsheet instantons to modify the metric so that the NS5 singularity gets localized along that direction, as in [154]. As for the singularity at $\alpha = \frac{\pi}{2}$, it now cannot be associated with an O8-D8 system as it was in IIA, since we are in IIB. It probably now represents a smeared O7-D7 system; it is indeed always the case that T-dualizing a brane along a parallel direction in supergravity gives a smeared version of the correct D-brane solution on the other side, as we just saw for the NS5-brane. It is possible that again instanton effects localize the singularity, this time to an O7-D7 system. (Even more correctly, we should expect the O7 to split into an $(1,1)$-sevenbrane and an $(1,-1)$-sevenbranes, as pointed out in [137] following [155].)

Notice finally that, although we have found it convenient to treat the $x = 0$ case separately from the rest, it is in fact a particular case of the general treatment (although a slightly degenerate one). Indeed one can check that (7.45b) is satisfied by (7.61); in contrast to the general case, this does not determine a function $z$, but we can use (7.47), where $z$ has been eliminated, instead of (7.46), which contains $z$. Thus the solution presented in this subsection is already an example of our general formalism. In section 7.6.2 we will see another, more elaborate example.

## 7.6. The PDEs

In section 7.5, we reduced the problem of finding $\mathrm{AdS}_6 \times M_4$ solutions to the two PDEs (7.45a), (7.46). As anticipated in the introduction, we will not try to find the most general solution to these equations in this chapter. In this section we will make some general remarks about the PDEs, and we will recover via a simple Ansatz the known solution [106], originally obtained via nonabelian T-duality. (As we mentioned in that section, one can also see the $x = 0$ case as a particular solution to the PDEs.)

### 7.6.1. General considerations

We derived in section 7.5.2 the two equations (7.45a), (7.46). Recall that $z$ is an auxiliary variable, defined by (7.45b). As we already remarked, among the four remaining variables $(\alpha, x, A, \phi)$, only two (for example $\alpha$ and $x$) are independent. The other two, $A$ and $\phi$, can be taken to be dependent as in (7.48). The equations (7.45a) and (7.46) can then be reexpressed as two scalar PDEs in the two dimensions spanned by $\alpha$ and $x$:

$$\begin{aligned} 3\sin(2\alpha)(A_\alpha\phi_x - A_x\phi_\alpha) = 6A_x + \sin^2\alpha\big(&-2x - 2(x^2+5)A_x + \\ &+ (1+2x^2)\phi_x\big) \ , \end{aligned} \tag{7.62a}$$

$$\begin{aligned} \cos\alpha(2+3x\phi_x) + \sin\alpha\phi_\alpha = {}& 2x\left(\frac{3}{\sin\alpha} + (x^2-4)\sin\alpha\right)(A_\alpha\phi_x - A_x\phi_\alpha) + \\ & - 2x\cos\alpha\left(\frac{3}{\sin^2\alpha} - (5+x^2)\right)A_x + \\ & + 2\left(\frac{3}{\sin\alpha} - (1+x^2)\sin\alpha\right)A_\alpha \ , \end{aligned} \tag{7.62b}$$

where $A_\alpha \equiv \partial_\alpha A$ etc. As we will see, they are actually easier to study in their original form manifestations (7.45a) and (7.46).

These equations are nonlinear, and as such they are rather hard to study. Even so, there are quite a few techniques that have been developed over the years to tackle such systems. Perhaps the first natural question is how many solutions one should expect. For a first-order system of ODEs, it is roughly enough to compare the number of equations to the number of functions. If there are $n$ equations and $n$ functions, the system is neither over- nor under-constrained: geometrically, the system gives a vector field in an open set in $\mathbb{R}^{n+1}$ (including time), and solving the system means finding integral curves to this vector field. (When the system is "autonomous", i.e. it does not depend explicitly on time, one can more simply consider a vector field on $\mathbb{R}^n$).

The picture is more complicated for a system of PDEs. In general, if we have $k$ "times" and $m$ functions, the system will define a distribution of dimension $k$ (namely, a choice of subspaces $V_x \subset T_x\mathbb{R}^{k+m}$ of dimension $k$ for every point $x \in \mathbb{R}^{k+m}$); solving the system then means finding "integral submanifolds" for the distribution, namely submanifolds $S \subset \mathbb{R}^{k+m}$ such that $V_x$ is tangent to $S$ for every $x \in S$. This distribution is in general however not guaranteed to admit integral submanifolds. (A famous example is given by Frobenius theorem: a distribution defined by the span of vector fields $v_i$ will only be integrable if all the Lie brackets $[v_i, v_j]$ are linear combinations of the $v_i$ themselves.) Fortunately, the machinery of "exterior differential systems" (EDS) has been developed to deal with these issues, culminating in the Cartan–Kähler theorem (see for example [147, Chap.III], or [148, Sec. 10.4.1] in slightly more informal language).

Describing and applying such methods in detail is beyond our scope, but here is a sketch. First one defines a "differential ideal", namely a vector space of the equations in the system and their exterior derivatives. In our case, denote by $E_i$ the two two-forms that have to vanish in (7.45a) and (7.46); the ideal is then the linear span $I = \langle E_1, E_2, dE_2 \rangle$ (since $dE_1 = 0$ automatically). We then want to construct the distribution $V$ on which the forms in $I$ vanish, in the sense that each multi-vector built from vectors in the distribution has zero pairing with the forms in $I$. One proceeds iteratively. We first consider a single vector field $e_1$ on which the forms vanish (in our case this is trivial, since there are no one-forms in $I$; we can take for example $e_1 = \partial_\alpha$). We then add a second vector: this is done by solving the "polar equations" $H(E_1) \equiv \{v \llcorner e_1 \llcorner E_i = 0\}$. The rank of this system is denoted by $c_1$. (In general there might be a $c_0$ too, but in our case the first choice of a vector was free because there are no one-forms in $I$; $c_0$ is then considered to be 0.) For us it turns out that $c_1 = 2$. In general one would go on by choosing a solution $e_2$ to the polar equations above, and would consider new polar equations $H(E_2) \equiv \{v \llcorner e_1 \llcorner E_i = v \llcorner e_2 \llcorner E_i = 0, v \llcorner e_1 \llcorner e_2 \llcorner dE_2 = 0\}$; the rank of this new system would be denoted by $c_2$, which in our case also happens to be 2. However, solving our PDEs means finding a two-dimensional integral manifold, and hence we can stop at the second step and disregard the higher polar equations $H(E_2)$. (The general theory would also show that for our system there is actually no three-dimensional integral manifold.) We can then apply the so-called "Cartan test" and a corollary to the Cartan–Kähler theorem (respectively Thm. 1.11 and Cor. 2.3 in [147]) to infer that an integral submanifold of dimension 2 actually does exist. The proof of the theorem also says that the general solution depends on $s_1 = c_1 - c_0 = 2$ functions of one variable. ($s_i = c_i - c_{i-1}$ are called "Cartan characters".) These two functions can be thought of as functions at the boundary of the two-dimensional domain in $\alpha$ and $x$ on which the solution exists.

Having determined the structure of the solutions, it would be nice to find as many as possible of them. A strategy which is common in this context is to impose some extra symmetry. This is less obvious than usual to implement. We cannot for example just assume that $A$ and $\phi$ do not depend on one of the coordinates $\alpha$ and $x$: the metric (7.49) would become degenerate. Another perhaps more promising idea is to use the so-called "method of characteristics" to reduce the problem to a system of ODEs. We plan to return on this in the future.

Finally, let us point out that two solutions to our PDEs are already known. One is the case $x = 0$, which we studied in section 7.5.4. Although we had to treat it separately, we also mentioned that it is a solution of the general system of PDEs (once we eliminate $dz$ from (7.46), obtaining (7.47)).

We will now see another particular solution. Although the global properties of the resulting $M_4$ are even more puzzling than those of the solution in section 7.5.4, it might be possible to generalize it to new solutions which are better-behaved; for example, one might start by studying perturbations around it.

### 7.6.2. A local solution: nonabelian T-duality

Many PDEs are reduced to ODEs by a separation of variables Ansatz. For our nonlinear PDEs, this does not work. However, we will now see that a particular case does lead to a solution, namely:

$$\phi = f(\alpha) + \log(x) \ , \qquad A = A(\alpha) \ . \tag{7.63}$$

Notice that this Ansatz restricts $x$ to be in $(0, 1]$. (We already observed after (7.49) that $|x| \leq 1$ in general.)

We begin by considering (7.45b). With (7.63), after a few manipulations it reduces to

$$dz = d\left(e^{6A-f}\frac{\sin\alpha}{6x^2}\right) - \frac{1}{3}e^{2A}d(e^{4A-f}\sin\alpha) + \\ + \frac{1}{x^2}\left[-\frac{1}{6}e^{4A}d(e^{2A-f}\sin\alpha) + e^{4A-f}\cot\alpha\, d(e^{2A}\cos\alpha)\right] \ . \tag{7.64}$$

The first line in (7.64) is manifestly exact, since everything is a function of $\alpha$ alone. The second line is of the form $\frac{1}{x^2}d(\text{function}(\alpha))$, and cannot be exact unless it vanishes, which leads to

$$d(e^{2A-f}\sin\alpha) = 6e^{-f}\cot\alpha\, d(e^{2A}\cos\alpha) \ . \tag{7.65}$$

The first line in (7.64) then determines $dz$ (and can be integrated to produce $z$). We can now use this expression for $dz$ in (7.46). Most terms in (7.46) actually vanish because they involve wedges of forms proportional to $d\alpha$; the only one surviving is of the form $d(e^{6A}\cos\alpha)\wedge dx$. In other words, we are forced to take

$$e^A = c_1(\cos\alpha)^{-1/6} \ , \tag{7.66}$$

with $c_1$ an integration constant. Plugging this back into (7.65) we get

$$e^f = c_2 \frac{(\cos\alpha)^{-1/3}}{\sin^3\alpha} \tag{7.67}$$

for $c_2$ another integration constant.

This is actually the solution found in [106]. To see this, one needs to identify

$$\alpha = \theta \ , \qquad x = \frac{e^{2\hat{A}}}{\sqrt{r^2 + e^{4\hat{A}}}} \ , \tag{7.68}$$

where $\hat{A}$ is the function denoted by $A$ in [106]. One can check that indeed the fluxes (7.51), (7.52) and metric (7.49) give the expressions in [106]. The metric one gets has a singularity at $\alpha = \pi/2$, just like the solution [104], and a new singularity at $\alpha = 0$ [146]. More worryingly, it is noncompact; it might be possible to find a suitable analytic continuation, with the help of the PDEs (7.45a), (7.46) found in this chapter.

# Appendices

# Appendix A

# More on $\mathrm{AdS}_7$

## A.1. Supercharges

At the beginning of section 5.3.2 we reviewed an old argument that shows how a solution of the form $\mathrm{AdS}_7 \times M_3$ can also be viewed as a solution of the type $\mathrm{Mink}_6 \times M_4$. In this appendix we show how the $\mathrm{AdS}_7 \times M_3$ supercharges get translated in the $\mathrm{Mink}_6 \times M_4$ framework.

A decomposition of gamma matrices appropriate to six-dimensional compactifications reads

$$\Gamma^{(6+4)}_{\mu} = e^{A_4}\gamma^{(6)}_{\mu} \otimes 1 \ , \qquad \Gamma^{(6+4)}_{m+5} = \gamma^{(6)} \otimes \gamma^{(4)}_{m} \ . \tag{A.1}$$

Here $\gamma^{(6)}_{\mu}$, $\mu = 0, \ldots, 5$, are a basis of six-dimensional gamma matrices, while $\gamma^{(4)}_{m}$, $m = 1, \ldots, 4$ are a basis of four-dimensional gamma matrices. For a supersymmetric $\mathrm{Mink}_6 \times M_4$ solution, the supersymmetry parameters can be taken to be

$$\begin{aligned} \epsilon^{(6+4)}_1 &= \zeta^0_+ \otimes \eta^1_+ + \zeta^{0\,c}_+ \otimes \eta^{1\,c}_+ \ , \\ \epsilon^{(6+4)}_2 &= \zeta^0_+ \otimes \eta^2_\mp + \zeta^{0\,c}_+ \otimes \eta^{2\,c}_\mp \ , \end{aligned} \tag{A.2}$$

where $\zeta_+$ is a constant spinor; $_\mp$ denotes the chirality, and $^c$ Majorana conjugation both in six and four dimensions. Supersymmetry implies that the norms of the internal spinors satisfy $||\eta^1||^2 \pm ||\eta^2||^2 = c_\pm e^{\pm A_4}$, where $c_\pm$ are constant.

On the other hand, for seven-dimensional compactifications a possible gamma matrix decomposition reads

$$\begin{aligned} \gamma^{(7+3)}_{\mu} &= e^{A_3}\gamma^{(7)}_{\mu} \otimes 1 \otimes \sigma_2 \ , \\ \gamma^{(7+3)}_{i+6} &= 1 \otimes \sigma_i \otimes \sigma_1 \ . \end{aligned} \tag{A.3}$$

This time $\gamma^{(7)}_{\mu}$, $\mu = 0, \ldots, 6$, are a basis of seven-dimensional gamma matrices, and $\sigma_i$, $i = 1, 2, 3$, are a basis of gamma matrices in three dimensions (which in flat indices can be taken to be the Pauli matrices). For a supersymmetric solution of the form AdS$_7 \times M_3$, the supersymmetry parameters are now of the form

$$\begin{aligned} \epsilon_1^{(7+3)} &= (\zeta \otimes \chi_1 + \zeta^c \otimes \chi_1^c) \otimes v_+ \ , \\ \epsilon_2^{(7+3)} &= (\zeta \otimes \chi_2 \mp \zeta^c \otimes \chi_2^c) \otimes v_\mp \ . \end{aligned} \tag{A.4}$$

Here, $\chi_{1,2}$ are spinors on $M_3$, with $\chi_{1,2}^c \equiv B_3 \chi_{1,2}^*$ their Majorana conjugates; a possible choice of $B_3$ is $B_3 = \sigma_2$. $\zeta$ is a spinor on AdS$_7$, and $\zeta^c \equiv B_7 \zeta^*$ is its Majorana conjugate; there exists a choice of $B_7$ which is real and satisfies $B_7 \gamma_\mu = \gamma_\mu^* B_7$. (It also obeys $B_7 B_7^* = -1$, which is the famous statement that one cannot impose the Majorana condition in seven Lorentzian dimensions.) The ten-dimensional conjugation matrix can then be taken to be $B_{10} = B_7 \otimes B_3 \otimes \sigma_3$; the last factor in (A.4), $v_\pm$, are then spinors chosen in such a way as to give the $\epsilon_i^{(7+3)}$ the correct chirality, and to make them Majorana; with the above choice of $B_{10}$, $v_+ = \frac{1}{\sqrt{2}}\begin{pmatrix} 1 \\ -1 \end{pmatrix}$, $v_- = \frac{1}{\sqrt{2}}\begin{pmatrix} 1 \\ 1 \end{pmatrix}$. The minus sign (for the IIA case) in front of the term $\zeta^c \otimes \chi_2^c$ in (A.4) is due to the fact that, both in seven Lorentzian and three Euclidean dimensions, conjugation does not square to one: $(\zeta^c)^c = -\zeta$, $(\chi^c)^c = -\chi$.

The presence of the cosmological constant in seven dimensions means that $\zeta$ is not constant, but rather that it satisfies the so-called Killing spinor equation, which for $R_{\rm AdS} = 1$ reads

$$\nabla_\mu \zeta = \frac{1}{2} \gamma_\mu^{(7)} \zeta \ . \tag{A.5}$$

One class of solutions to this equation [156, 157] is simply of the form

$$\zeta_+ = \rho^{1/2} \zeta_+^0 \ . \tag{A.6}$$

The coordinate $\rho$ appears in (5.18), which expresses AdS$_7$ as a warped product of Mink$_6$ and $\mathbb{R}$. $\zeta_+^0$ is a spinor constant along Mink$_6$ and such that $\gamma_{\hat\rho} \zeta_+^0 = \zeta_+^0$ (the hat denoting a flat index).

Just like for Mink$_6 \times M_4$, supersymmetry again implies that the norms of the internal spinors $\chi^{1,2}$ should be related to the warping function: $||\chi_1||^2 \pm ||\chi_2||^2 = c_\pm e^{\pm A_3}$, where $c_\pm$ are constant. We will now see, however, that for AdS$_7 \times M_3$ actually $c_- = 0$. We use the ten-dimensional system (4.30), (4.31). As we mentioned in section 5.3, it can be used to derive quickly the system 5.3.1, while applying it directly to AdS$_7 \times M_3$ to derive (5.22) is more lengthy. For our purposes, however, it will be enough to apply one equation of that system to the AdS$_7 \times M_3$ setup, namely

$$d\tilde{K} = \iota_K H \tag{A.7}$$

This is the second equation in (4.30b) ((3.1b) in [89], although it had already appeared in [158, 159, 152]. $K$ and $\tilde{K}$ are the ten-dimensional vector and one-form defined by $K = \frac{1}{64}(\bar{\epsilon}_1 \Gamma_M \epsilon_1 + \bar{\epsilon}_2 \Gamma_M \epsilon_2) dx^M$ and $\tilde{K} = \frac{1}{64}(\bar{\epsilon}_1 \Gamma_M \epsilon_1 - \bar{\epsilon}_2 \Gamma_M \epsilon_2) dx^M$. Plugging the decomposed spinors (A.4) in these definitions and calling $\beta_1 = e^{A_3}(\frac{1}{8}\bar{\zeta}\gamma^{(7)}_\mu \zeta) dx^\mu$, the part of (A.7) along AdS$_7$ leads to $e^{A_3} d_7 \beta_1 (||\chi_1||^2 - ||\chi_2||^2) = (d_7 \beta_1) c_- = 0$, where $d_7$ is the exterior derivative along AdS$_7$. (The right hand side does not contribute, because $H$ has only internal components.) On the other hand, using the Killing spinor equation (A.5) in AdS$_7$, we have that $d_7 \beta_1 = e^{2A_3}(\bar{\zeta}\gamma^{(7)}_{\mu\nu}\zeta) dx^{\mu\nu} \equiv \beta_2$. A spinor in seven dimensions can be in different orbits (defining an SU(3) or an SU(2)$\ltimes\mathbb{R}^5$ structure [160, 161]), but for none of them the bilinear $\beta_2$ is identically zero. Consequently, the norms of the two Killing spinors have to be equal, namely $c_- = 0$.

Let us now see how to translate the spinors $\epsilon_i$ for an AdS$_7 \times M_3$ solution into a language relevant for Mink$_6 \times M_4$. First, we split the seven-dimensional gamma matrices $\gamma^{(7)}_\mu$; the first six give a basis of gamma matrices in six dimensions, $\tilde{\gamma}^{(6)}_\mu = \rho \gamma^{(7)}_\mu$, $\mu = 0, \ldots, 5$, while the radial direction, $\gamma^{(7)}_{\hat{\rho}} = \gamma^{(6)}$ becomes the chiral gamma in six dimensions. (The hat denotes a flat index.) This split is by itself not enough to turn (A.3) into (A.1), because the three-dimensional gamma's in (A.3) have no $\gamma^{(6)}$ in front. This can be cured by applying a change of basis:

$$\Gamma^{(6+4)}_M = O \Gamma^{(7+3)}_M O^{-1} \ , \quad O = \frac{1}{\sqrt{2}}(1 - i\Gamma^{(7+3)}_{\hat{\rho}}) \ , \tag{A.8}$$

with, however, a change of basis in six dimensions: $\gamma^{(6)}_\mu \to -i\gamma^{(6)}\gamma^{(6)}_\mu$. Likewise, the spinors (A.4) are related to (A.2) by

$$\epsilon^{(6+4)}_i = O \epsilon^{(7+3)}_i \ , \tag{A.9}$$

if we take

$$\eta_1 = \rho^{1/2} \chi_1 \otimes v_+ = \frac{1}{\sqrt{2}} \rho^{1/2} \chi_1 \otimes \begin{pmatrix} 1 \\ -1 \end{pmatrix} \ , \tag{A.10a}$$

$$\eta_2 = \rho^{1/2} \chi_2 \otimes v_\mp = \frac{1}{\sqrt{2}} \rho^{1/2} \chi_2 \otimes \begin{pmatrix} 1 \\ \pm 1 \end{pmatrix} \ . \tag{A.10b}$$

Notice that the two $\eta^i$ have equal norm, because the $\chi^i$ have equal norm, as shown earlier. Moreover, since the norm of the $\chi^i$ is $e^{A_3/2}$, and because of the factor $\rho^{1/2}$ in (A.10), the $\eta^i$ have norm equal to $\rho^{1/2} e^{A_3/2}$; recalling (5.19), this is equal to $e^{A_4/2}$, as it should.

Besides (A.6), there is also a second class of solution to the Killing spinor equation $\nabla_\mu \zeta = \frac{1}{2}\gamma^{(7)}_\mu \zeta$ on AdS$_7$: it reads $\zeta = (\rho^{-1/2} + \rho^{1/2} x^\mu \gamma^{(7)}_\mu)\zeta^0_-$, where now $\gamma_{\hat\rho}\zeta^0_- = -\zeta^0_-$. If we plug this into (A.4) and use the above procedure (A.9) to translate it in the Mink$_6 \times M_4$ language, we find a generalization of (A.2) where both a positive and negative chirality six-dimensional spinor appear (namely, $x^\mu\gamma_\mu\zeta^0_-$ and $\zeta^0_-$) instead of just a positive chirality spinor $\zeta^0_+$. Because of the $x^\mu\gamma_\mu$ factor, this spinor Ansatz would break Poincaré invariance if used by itself; if four supercharges of the form (A.2) are preserved, Poincaré invariance is present, and these additional supercharges simply signal that an AdS$_7 \times M_3$ solution is $\mathcal{N} = 2$ in terms of Mink$_6 \times M_4$.

## A.2. Killing spinors on $S^4$

The AdS$_7 \times S^4$ is a familiar Freund–Rubin solution; the flux is taken to be proportional to the internal volume form, $G_4 = g\,\mathrm{vol}_{S^4}$. The eleven-dimensional supersymmetry transformation reads

$$\left(\nabla_M + \frac{1}{144}G_{NPQR}(\gamma^{NPQR}{}_M - 8\gamma^{NPQ}\delta^R_M)\right)\epsilon_{11} = 0\ . \tag{A.11}$$

Decomposing $\epsilon_{11} = \sum_{a=1}^4 \zeta_a \otimes \eta_a + \text{c.c.}$, and using (A.5), one reduces the requirement of supersymmetry (for $R_{\text{AdS}} = 1$) to taking $g = 3/4$, and to the equation

$$(\nabla_m - \frac{1}{2}\gamma\gamma^m)\eta = 0 \tag{A.12}$$

on $S^4$. This is an alternative form of the Killing spinor equation; it was solved in [162] in any dimension. However, we are using different coordinates, adapted to the $S^1$ reduction used in section 5.6.1; we will here solve (A.12) again, using more or less the same method.

The idea is to start from the easiest components of the equation, and to work one's way to the more complicated ones. Our coordinates in section 5.6.1 are $\alpha$, $\beta$, $\gamma$, $y$, the latter being the reduction coordinate. Our vielbein reads $e^1 = d\alpha$, $e^2 = \frac{1}{2}\sin(\alpha)d\beta$, $e^3 = \frac{1}{2}\sin(\alpha)\sin(\beta)d\gamma$, $e^4 = \frac{1}{2}\sin(\alpha)(dy + \cos(\beta)d\gamma)$. We begin with the $\alpha$ component of (A.12):

$$\partial_\alpha\eta = \frac{1}{2}\gamma\gamma_1\eta \quad\Rightarrow\quad \eta = e^{\frac{1}{2}\alpha\gamma\gamma_1}\eta_1\ . \tag{A.13}$$

The next component we use is

$$\left(\partial_\beta - \frac{1}{4}\cos(\alpha)\right)\eta = \frac{1}{4}\sin(\alpha)\gamma\gamma_2\eta\ . \tag{A.14}$$

This can be manipulated as follows:

$$0 = \left(\partial_\beta - \frac{1}{4}e^{\alpha\gamma\gamma_1}\gamma_{12}\right)\eta = e^{\frac{1}{2}\alpha\gamma\gamma_1}\left(\partial_\beta - \frac{1}{4}\gamma_{12}\right)\eta_1 \;\Rightarrow\; \eta_1 = e^{\frac{1}{4}\beta\gamma_{12}}\eta_2 \ . \tag{A.15}$$

We proceed in a similar way for the two remaining coordinates; the details are complicated, and we omit them here. The final result is

$$\eta = \exp\left[\frac{\alpha}{2}\gamma\gamma_1\right]\exp\left[\frac{\beta}{4}\gamma_{12} + \frac{\beta-\pi}{4}\gamma_{34}\right]\exp\left[\frac{y+\gamma}{4}\gamma_{13} + \frac{y-\gamma}{4}\gamma_{24}\right]\eta_0 \tag{A.16}$$

where $\eta_0$ is a constant spinor. When we reduce, we demand that $\partial_y\eta = 0$, which becomes $(\gamma_{13}+\gamma_{24})\eta_0 = 0$; this condition indeed keeps two out of four spinors, as anticipated in our discussion in section 5.6.1.

## A.3. Sufficiency of the system (5.22)

In section 5.3.2 we obtained the system of equations (5.22) starting from (5.14) and using the fact that AdS$_7$ can be considered as a warped product of Mink$_6$ and $\mathbb{R}$. In this section we will explain how one can show that (5.22) is completely equivalent to supersymmetry for AdS$_7 \times M_3$ with a direct computation. Our strategy will be very similar to the one in [98, Sec. A.4], with some relevant differences that we will promptly point out.

To begin with, we write the system of equations resulting from setting to zero the type II supersymmetry variations (of gravitinos and dilatinos) using the spinorial decomposition (A.4):[1]

$$\left(D_m - \frac{1}{4}H_m\right)\chi_1 - \frac{e^\phi}{8}F\sigma_m\chi_2 = 0\ , \tag{A.17a}$$

$$\left(D_m + \frac{1}{4}H_m\right)\chi_2 - \frac{e^\phi}{8}\lambda(F)\sigma_m\chi_1 = 0\ , \tag{A.17b}$$

$$\frac{1}{2}e^{-A}\chi_1 - \frac{i}{2}\partial A\chi_1 + i\frac{e^\phi}{8}F\chi_2 = 0\ , \tag{A.17c}$$

$$\frac{1}{2}e^{-A}\chi_2 + \frac{i}{2}\partial A\chi_2 - i\frac{e^\phi}{8}\lambda(F)\chi_1 = 0\ , \tag{A.17d}$$

$$\left(D - \frac{1}{4}H\right)\chi_1 + i\frac{7}{2}e^{-A}\chi_1 + \left(\frac{7}{2}\partial A - \partial\phi\right)\chi_1 = 0\ , \tag{A.17e}$$

$$\left(D + \frac{1}{4}H\right)\chi_2 - i\frac{7}{2}e^{-A}\chi_2 + \left(\frac{7}{2}\partial A - \partial\phi\right)\chi_2 = 0\ . \tag{A.17f}$$

As in [98, Sec. A.4], we introduce a set of intrinsic torsions $p^a_m$, $q^a_m$, and $T^a$, $\hat{T}^a$, with $a = 1, 2$:

$$\left(D_m - \frac{1}{4}H_m\right)\chi_1 \equiv p^1_m\chi_1 + q^1_m\chi^c_1\ , \quad \left(D_m + \frac{1}{4}H_m\right)\chi_2 \equiv p^2_m\chi_2 + q^2_m\chi^c_2\ , \tag{A.18a}$$

$$\left(D - \frac{1}{4}H\right)\chi_1 \equiv T^1\chi_1 + \hat{T}^1\chi^c_1\ , \quad \left(D + \frac{1}{4}H\right)\chi_2 \equiv T^2\chi_2 + \hat{T}^2\chi^c_2\ , \tag{A.18b}$$

where $D = \gamma^m_{(7)}D_m$, $H_m \equiv \frac{1}{2}H_{mnp}\gamma^{np}_{(7)}$, $H \equiv \frac{1}{6}H_{mnp}\gamma^{mnp}_{(7)}$ as usual. We used the fact that $\chi_1$ and $\chi^c_1$ (or $\chi_2$ and $\chi^c_2$) constitute a basis for the three-dimensional spinors. Taking tensor products of these two bases, we also obtain a basis for bispinors, on which we can now expand $F$:

$$F \equiv R_{00}\,\chi_1\otimes\chi_2^\dagger + R_{10}\,\chi^c_1\otimes\chi^\dagger_2 + R_{01}\,\chi_1\otimes\chi^{c\dagger}_2 + R_{11}\,\chi^c_1\otimes\chi^{c\dagger}_2\ . \tag{A.19}$$

Using (A.18) and (A.19) in (A.17), we can rewrite the conditions for unbroken supersymmetry as a set of equations relating the intrinsic torsions to the coefficients $R_{ij}$. Let us call this system of equations the “spinorial system”. Using instead (A.18) and (A.19) in (5.22), we obtain a second set of equations, again in terms of the intrinsic torsions and $R_{ij}$; let us call this system the “form system”. Our aim is to show the equivalence between the spinorial and the form systems.

[1] We choose to show the equivalence in the IIA case, hence we pick $\epsilon^{(7+3)}_1$ and $\epsilon^{(7+3)}_2$ with opposite chirality.

Although we are using the same technique appearing in [98, Sec. A.4] (there applied to four-dimensional vacua), proving this equivalence in the case at hand is more involved. Relying on a superficial counting, it would seem that the form system contains fewer equations than the spinorial one. To see why this happens, we first notice that the definitions (A.18) are redundant. Indeed the torsions $T^a$ and $\hat{T}^a$ can be rewritten in terms of the torsions $p^a$, $q^a$ and $H$; however, in three dimensions, $\gamma^{(7)}_{mnp}$, hence $H$, is proportional to the identity (use (5.24) with $\alpha = H$). Thus in (A.18b) four complex numbers ($T$'s and $\hat{T}$'s) are used to describe a single real number $H$. This suggests that some of the equations in the spinorial system are redundant and could be dropped. However, this redundancy is not manifest.

To make it manifest, we could use the following strategy. On the one hand (A.17a) and (A.17b) give a natural expansion of the torsions $p^a$ and $q^a$ in terms of the vielbein $e^b$, with $a \neq b$, defined by the spinor $\chi_b$ (see (5.25) and (5.27)); that is, they transform into equations for the components $q^1 \cdot e^2_3$, $q^1 \cdot e^2_1$ and so forth. On the other hand the intrinsic torsions $T^a$ and $\hat{T}^a$ give expressions like $q^1 \cdot e^1_3$, $q^1 \cdot e^1_1$. Therefore, we would need a formula relating the vielbein $e^1$ defined by $\chi_1$ to the vielbein $e^2$ defined by $\chi_2$.

Actually, there exists a simpler method. Indeed we can use the following equations,

$$\begin{aligned} & d_H(e^{2A_3-\phi}\mathrm{Re}\psi^1_-) = 0 \ , \\ & d_H(e^{4A_3-\phi}\mathrm{Im}\psi^1_-) = 0 \ , \\ & d_H(e^{4A_3-\phi}\psi^2_-) = 0 \ , \end{aligned} \tag{A.20}$$

obtained by simply applying $d_H$ to the equations (5.22a), (5.22b) and (5.22c) respectively (in other words, they are redundant with respect to the original system (5.22)). If we now express (A.20) in terms of (A.18), and add the resulting equations to the form system we obtained earlier, we obtain a new, equivalent expression for the form system. With some linear manipulations, it can now be shown that it is equivalent to the spinorial system. This concludes our alternative proof that (5.22) is completely equivalent to the requirement of unbroken supersymmetry.

# Appendix B

# More on $AdS_6$

## B.1. Derivation of (7.13)

The starting point to obtain (7.13) is the system of equations (4.30), (4.31), i.e. equation (3.11) in [89], which was shown in that reference to be equivalent to $\mathcal{N} = 1$ supersymmetry on any $M_{10}$; here we will specialize the ten-dimensional spacetime $M_{10}$ to $\mathrm{AdS}_6 \times M_4$. Actually, the equations appearing in (7.13) strictly derive from equations (4.30). In section B.1.1 we show such derivation.

Furthermore, to prove the equivalence between the system (7.13) and the conditions imposed by $\mathcal{N} = 1$ supersymmetry on $\mathrm{AdS}_6 \times M_4$, we need to show that the two remaining "pairing" equations (4.31) (equations (3.1c,d) in [89]) are completely redundant on such background: this is done in subsection B.1.2.

### B.1.1. Derivation of the system

Let us quote equations (4.30):

$$d_H(e^{-\phi}\Phi) = -(\tilde{K} \wedge + \iota_K)F_{(10)} \; ; \tag{B.1a}$$

$$L_K g = 0 \; , \quad d\tilde{K} = \iota_K H \; . \tag{B.1b}$$

$\Phi = \epsilon_1 \otimes \overline{\epsilon_2}$ is the key ten-dimensional polyform,[1] which is adapted to our background; $g$ is the ten-dimensional metric while $K$ and $\tilde{K}$ are ten-dimensional one-forms which will be defined momentarily.

The decomposition of the ten-dimensional spinors $\epsilon_a$ suggests we decompose accordingly the ten-dimensional gamma matrices:

$$\Gamma^{(6+4)}_{\mu} = e^A \gamma^{(6)}_{\mu} \otimes 1 \; , \qquad \Gamma^{(6+4)}_{m+5} = \gamma^{(6)} \otimes \gamma^{(4)}_{m} \; . \tag{B.2}$$

[1] It should not be confused with the SU(2) covariant internal even forms $\Phi_\pm$.

Here $\gamma^{(6)}_\mu$, $\mu = 0, \ldots, 5$, are a basis of six-dimensional gamma matrices ($\gamma^{(6)}$ is the chiral gamma), while $\gamma^{(4)}_m$, $m = 1, \ldots, 4$ are a basis of four-dimensional gamma matrices. We can now expand via Fierz identities (see formula (A.12) in [89]) the bilinear $\epsilon_1 \otimes \overline{\epsilon_2}$, by plugging in the decomposition (7.8) and (B.2). We get a sum of terms such as the following:

$$\sum_{k=0}^{6} \frac{1}{8k!} \left(\overline{\zeta_+} \gamma^j_{(6)} \gamma^{(6)}_{\mu_k \ldots \mu_1} \zeta_+\right) \gamma^{\mu_1 \ldots \mu_k}_{(6)} \sum_{j=0}^{4} \frac{1}{4j!} \left(\eta^{2\dagger}_{\mp} \gamma^{(4)}_{m_j \ldots m_1} \eta^1_+\right) \gamma^{m_1 \ldots m_j}_{(4)}$$
$$= \mp \zeta_+ \overline{\zeta_+} \wedge \eta^1_+ \eta^{2\dagger}_{\mp} \; . \tag{B.3}$$

What we mean by e.g. $\zeta_+ \overline{\zeta_+}$ is the six-dimensional polyform corresponding to this bilinear via the Clifford map (see footnote 6). All in all we get:

$$\begin{aligned} \Phi = \; & \mp \zeta_+ \overline{\zeta_+} \wedge \eta^1_+ \eta^{2\,\dagger}_{\mp} \mp \zeta_+ \overline{\zeta^c_+} \wedge \eta^1_+ \overline{\eta^2_{\mp}} + \zeta_- \overline{\zeta_-} \wedge \eta^1_- \eta^{2\,\dagger}_{\pm} + \zeta_- \overline{\zeta^c_-} \wedge \eta^1_- \overline{\eta^2_{\pm}} + \\ & + \zeta_+ \overline{\zeta_-} \wedge \eta^1_+ \eta^{2\,\dagger}_{\pm} + \zeta_+ \overline{\zeta^c_-} \wedge \eta^1_+ \overline{\eta^2_{\mp}} \pm \zeta_- \overline{\zeta_+} \wedge \eta^1_- \eta^{2\,\dagger}_{\mp} \pm \zeta_- \overline{\zeta^c_+} \wedge \eta^1_- \overline{\eta^2_{\mp}} \end{aligned} \tag{B.4}$$

plus the complex conjugates of all summands. Their is due to relations such as $\zeta^c_\pm \overline{\zeta_\pm} = -(\zeta_\pm \overline{\zeta^c_\pm})^*$ and $\eta^{1\,c}_\pm \eta^{2\,\dagger}_\pm = -(\eta^1_\pm \overline{\eta^2_\pm})^*$.

Since we already know from (7.9a) and (7.9b) the forms defined by the bispinors along the internal space $M_4$, we just need to compute the bispinors along AdS$_6$, as $\zeta_+ \overline{\zeta_+}$. The structure of these bispinors actually depends on how $\zeta_+$ is chosen. One way to see this is to notice that some of the algebraic relations depend on whether the bilinear $\overline{\zeta_+} \zeta_-$ vanishes or not. A more invariant way to describe the situation is to notice that a pair $\zeta_\pm$ of chiral spinors has the same properties as another pair $\zeta'_\pm$ if they can be related via a Lorentz transformation, $\zeta'_\pm = \Lambda \zeta_\pm$; or in other words if they lie in the same *orbit*. The orbits for SO(1, 5) have been studied in [163, Sec. 2.4.5.2]. Two orbits correspond to the case where either $\zeta_+$ or $\zeta_-$ is zero; these are not compatible with the Killing spinor equation (7.7), and are therefore not interesting to us. There is then a one-parameter family of orbits whose stabilizer (i.e. the little group under the SO(1, 5) action) is the abelian group $\mathbb{R}^4$; each of these orbits has dimension 11. Finally, there is a four-parameter family of orbits whose stabilizer is SU(2); each of these orbits has dimension 12.

The properties of the forms that one can define from spinor bilinears depend on whether we consider an orbit with stabilizer $\mathbb{R}^4$ or SU(2). The system (4.30) will give systems of equations which are superficially different for these two types of orbits. However, the original system for supersymmetry is linear in the supersymmetry parameters $\epsilon_a$. So its solution space should be a linear space, which must in fact have dimension 8 (since this is the smallest number of supercharges for a superalgebra in this dimension). Even if two choices of spinor pairs on this linear space might give superficially different systems of equations, eventually these two different systems must agree. So we can choose the spinor pair in such a way as to get the most convenient system of equations. It turns out that this is one of the orbits with $\mathbb{R}^4$ stabilizer.

To get more concrete, let us decompose the external spinors splitting the external index $\mu$ into a "lightcone" part, $a = +, -$, and a four-dimensional Euclidean part, $m = 1, \ldots, 4$:

$$\gamma^a_{(6)} = \sigma_a \otimes 1_{(4)} = \frac{1}{2}(\gamma^0_{(6)} \pm \gamma^1_{(6)}) \ , \quad \gamma^m_{(6)} = \sigma_3 \otimes \gamma^m_{(4)} \ , \tag{B.5}$$

with $\sigma_\pm = \frac{1}{2}(\pm\sigma_1 + i\sigma_2)$. The matrices $\gamma^\mu_{(6)}$ satisfy the algebra Cl$(1,5)$ with lightcone metric $\tilde{\eta}^{\mu\nu} = \begin{bmatrix} 0 & -\frac{1}{2} \\ -\frac{1}{2} & 0 \end{bmatrix} \oplus \delta^{mn}_{(4)}$, so that $\gamma^{(6)}_\pm = -2\gamma^\mp_{(6)}$ and $\gamma^{(6)}_m = \gamma^m_{(6)}$.

Using this decomposition, we choose now a spinor pair of the form

$$\zeta_\pm \equiv \begin{pmatrix} 1 \\ 0 \end{pmatrix} \otimes \chi_\pm \ , \tag{B.6}$$

with $\chi_\pm$ a chiral spinor in four dimensions. This corresponds to an orbit with $\mathbb{R}^4$ stabilizer. (Orbits with SU(2) stabilizer would correspond to taking $\zeta_+ = \binom{1}{0} \otimes \chi_+$, $\zeta_- = \binom{0}{1} \otimes \chi_-$.) One consequence of this (which would not be true for the SU(2) orbit) is that the one-form part of the bilinears $\zeta_+\overline{\zeta_+}$ and $\zeta_-\overline{\zeta_-}$ coincide; we will call it $z$. It is light-like, and it only has components in the two-dimensional part of the decomposition (B.5). As for the bilinears in the four dimensions $1, \ldots, 4$, they can be evaluated in the same way as those along $M_4$, in terms of two one-forms that we will call $V$ and $W$ and which satisfy exactly the same properties as the forms $v$ and $w$ introduced in (7.25).

$z$ and the real and imaginary parts of $V$ and $W$ are independent, and in fact orthogonal. They are not quite a vielbein: if we think of $z$ as of the element of a vielbein in the null direction $-$, we are missing another element in direction $+$. As stressed in [89], this cannot be obtained as a bilinear of the supersymmetry parameters; we will see in section B.1.2 that the remaining equations in the ten-dimensional system of [89] require picking such a null vector as an auxiliary piece of data. In conclusion,

$$\{z = e_-,\ e_+,\ \mathrm{Re}V,\ \mathrm{Im}V,\ \mathrm{Re}W,\ \mathrm{Im}W\} \tag{B.7}$$

is a vielbein in AdS$_6$.

We will also define $\Omega_+ = -V \wedge W$, $\Omega_- = \bar{V} \wedge W$, $J_\pm = \pm\frac{i}{2}(V \wedge \bar{V} \pm W \wedge \bar{W})$, just as in (7.34), (7.34b) for $M_4$. With all these definitions, we can evaluate

$$\zeta_\pm\overline{\zeta_\pm} = z \wedge e^{-iJ_\pm} \ , \tag{B.8a}$$
$$\zeta_+\overline{\zeta_-} = -z \wedge (V + *_4 V) \ , \tag{B.8b}$$
$$\zeta_-\overline{\zeta_+} = -z \wedge (\overline{V} - *_4 \overline{V}) \ , \tag{B.8c}$$
$$\zeta_\pm\overline{\zeta_\pm^c} = z \wedge \Omega_\pm \ , \tag{B.8d}$$
$$\zeta_\pm\overline{\zeta_\mp^c} = \mp z \wedge (W \pm *_4 W) \ . \tag{B.8e}$$

Specializing to IIB from now on, we can now plug (B.8) into (B.4); we have:

$$\begin{aligned}\Phi_{\rm IIB} = e^A \big[ &(z \wedge e^{-iJ_+}) \wedge \phi_+^1 + (z \wedge e^{-iJ_-}) \wedge \phi_-^1 \\ &+ z \wedge \Omega_+ \wedge \phi_+^2 + z \wedge \Omega_- \wedge \phi_-^2 \\ &- z \wedge (V + *_4 V) \wedge \psi_+^1 + z \wedge (\overline{V} - *_4 \overline{V}) \wedge \psi_-^1 \\ &- z \wedge (W + *_4 W) \wedge \psi_+^2 - z \wedge (W - *_4 W) \wedge \psi_-^2 + \text{c.c.} \big] \ .\end{aligned} \tag{B.9}$$

This is an odd form, as should be the case for IIB.

To evaluate (B.1a), we need to compute the ten-dimensional exterior derivative of $e^{-\phi}\Phi$; schematically, it takes the form:

$$\begin{aligned} d_H(e^{-\phi}\Phi) &= d_H \left( \sum \text{ext} \wedge e^{A-\phi}\, \text{int} \right) \\ &= \sum d_6 \text{ext} \wedge e^{A-\phi}\, \text{int} + (-)^{\deg(\text{ext})} \text{ext} \wedge d_H(e^{A-\phi}\, \text{int}) \ . \end{aligned} \tag{B.10}$$

$d_6$ is the differential along the AdS$_6$ coordinates, while $d_H = d_4 - H\wedge$ in the last identity is a combination of the exterior differential $d_4$ along $M_4$ and of the NS three-form $H$ (which only has components along $M_4$). Since we are looking for vacuum solutions to (B.1a) which are compatible with supersymmetry on AdS$_6$, we need to take the external spinors $\zeta_\pm$ to be the chiral components of a Killing spinor $\zeta$ on this spacetime, i.e. $\nabla_\mu \zeta = \frac{1}{2}\mu\gamma_\mu\zeta$. The norm of the complex constant $\mu$ (which is proportional to $\sqrt{-\Lambda}$) can be reabsorbed in the warping function $A$; its phase can be reabsorbed by multiplying $\eta_\pm^a$ by $e^{\pm i\theta}$. Hence in what follows we will set $\mu = 1$, resulting in the equation (7.7) that we already quoted in the main text.

Exploiting (7.7) we can now compute the derivatives of the external forms

(B.8):

$$d_6(\zeta_\pm\overline{\zeta_\pm}) = \; -2z\wedge(\mathrm{Re}V + 2i *_4 \mathrm{Im}V) \; , \tag{B.11a}$$
$$d_6(\zeta_\pm\overline{\zeta_\mp}) = \; \pm 3iz\wedge \mathrm{Re}V\wedge \mathrm{Im}V \; + \pm 5z\wedge \mathrm{Re}v\wedge\mathrm{Im}V\wedge\mathrm{Re}W\wedge\mathrm{Im}W \; , \tag{B.11b}$$
$$d_6(\zeta_\pm\overline{\zeta^c_\pm}) = \; -4z\wedge *_4 W \; , \tag{B.11c}$$
$$d_6(\zeta_\pm\overline{\zeta^c_\mp}) = \; \pm 3z\wedge\mathrm{Re}V\wedge W \; . \tag{B.11d}$$

As an illustration, (B.11a) is computed as follows:

$$\begin{aligned} d_6(\zeta_+\overline{\zeta_+}) &= \frac{1}{2}\left[\gamma^\mu_{(6)}, \nabla_\mu(\zeta_+\overline{\zeta_+})\right] \\ &= \frac{1}{4}(\gamma^\mu\gamma_\mu\zeta_-\overline{\zeta_+} - \gamma^\mu\zeta_+\overline{\zeta_-}\gamma_\mu - \gamma_\mu\zeta_-\overline{\zeta_+}\gamma^\mu + \\ &\quad + \zeta_+\overline{\zeta_-}\gamma_\mu\gamma^\mu) \\ &= \frac{1}{2}(-3z\wedge(\overline{V} - *_4\overline{V}) - 3z\wedge(V + *_4 V) + z\wedge(V - *_4 V) + \\ &\quad + z\wedge(\overline{V} + *_4\overline{V})) \\ &= \; -2z\wedge(\mathrm{Re}V + 2i *_4 \mathrm{Im}V) \; , \end{aligned} \tag{B.12}$$

having used the formula $\gamma^\mu\omega_k\gamma_\mu = (-)^k(d-2k)\omega_k$ for a $k$-form $\omega_k$ in $d$ dimensions.

The left-hand side $d_H(e^{-\phi}\Phi)$ of (B.1a) then contains only unknown derivatives of the internal forms, since those of the external forms have been traded for the right-hand sides of (B.11). Once we compute its right-hand side, the complete equation will only involve internal forms and will be valid for any of the sixteen independent components of $\zeta = \zeta_+ + \zeta_-$, as appropriate for an $\mathcal{N} = 1$ vacuum in six dimensions.

Before computing the right-hand side of (B.1a), namely $-(\tilde{K}\wedge + \iota_K)F$, we will look at the simpler (B.1b): as it happens in other dimensions, they imply that the norms of the internal spinors are related to the warping function $A$. Let us see how. First, recall the definitions of $K$ and $\tilde{K}$ [89]:

$$K = \frac{1}{64}(\overline{\epsilon}_1\Gamma_M\epsilon_1 + \overline{\epsilon}_2\Gamma_M\epsilon_2)\, dx^M \; , \quad \tilde{K} = \frac{1}{64}(\overline{\epsilon}_1\Gamma_M\epsilon_1 - \overline{\epsilon}_2\Gamma_M\epsilon_2)\, dx^M \; . \tag{B.13}$$

Plugging in these formulas the decomposition (7.8), we obtain:

$$K = \frac{e^{-A}}{4} z\,(\|\eta^1\|^2 + \|\eta^2\|^2) \; , \quad \tilde{K} = \frac{e^{-A}}{4} z\,(\|\eta^1\|^2 - \|\eta^2\|^2) \; . \tag{B.14}$$

The external part of the second equation in (B.1b) gives $e^{-A}d_6 z\,(\|\eta^1\|^2 - \|\eta^2\|^2) = 0$ (the right-hand side vanishes since $H$ is purely internal). One can explicitly compute $d_6 z$, recalling that $z$ is the one-form part of $\zeta_\pm\overline{\zeta_\pm}$; using (7.7), one can show that it is nonvanishing. Thus we get:

$$\|\eta^1\|^2 = \|\eta^2\|^2 \; . \tag{B.15}$$

Hence $K = \frac{e^{-A}}{2} z\,\|\eta^1\|^2$ and $\tilde{K} = 0$. On the other hand, the first equation in (B.1b) says that $K$ is a Killing vector with respect to the ten-dimensional metric $g$: its external part says that $z$ is Killing with respect to $g_{\mathrm{AdS}_6}$ (this is obvious, since $z$ is a bilinear constructed out of Killing spinors), while its internal part implies $\partial_m\left(\frac{e^{-A}}{2}\|\eta^1\|^2\right) = 0$, which upon integration gives

$$\|\eta^1\|^2 = e^A \; , \tag{B.16}$$

where without loss of generality we have set to one the integration constant. Putting (B.15) and (B.16) together we get (7.13f). Moreover $K = z/2$. Recalling (7.14) we now have:

$$\begin{aligned} -(\tilde{K}\wedge + \iota_K)F_{(10)} &= -\iota_K(e^{6A}\mathrm{vol}_6 \wedge *_4\lambda F) \\ &= -\frac{e^{6A}}{2} *_6 z \wedge *_4\lambda F \\ &= \frac{e^{6A}}{2}(z\wedge \mathrm{Re}V \wedge \mathrm{Im}V \wedge \mathrm{Re}W \wedge \mathrm{Im}W)\wedge *_4\lambda F \; . \end{aligned} \tag{B.17}$$

Putting everything together, we can now separate the various terms in (B.1a) that multiply different wedge products of the one-forms in (B.7); since those forms are a vielbein in AdS$_6$, they are linearly independent, and each term has to be set to zero separately. In particular, we see from (B.17) that the RR flux only contributes to one equation. This gives rise to many equations that can then be arranged in SU(2)$_R$ representations by recalling the definitions (7.11) of the SU(2) covariant forms $\Phi_\pm$ and $\Psi_\pm$. This finally results in the system (7.13).

### B.1.2. Redundancy of pairing equations

We will now show that equations (4.31),[2]

$$\left(e_{+_1}\cdot\Phi\cdot e_{+_2},\Gamma^{MN}\left[\pm d_H(e^{-\phi}\Phi\cdot e_{+_2})+\frac{1}{2}e^{\phi}d^\dagger(e^{-2\phi}e_{+_2})\Phi-F_{(10)}\right]\right)=0\ , \tag{B.18a}$$

$$\left(e_{+_1}\cdot\Phi\cdot e_{+_2},\left[d_H(e^{-\phi}e_{+_1}\cdot\Phi)-\frac{1}{2}e^{\phi}d^\dagger(e^{-2\phi}e_{+_2})\Phi-F_{(10)}\right]\Gamma^{MN}\right)=0\ , \tag{B.18b}$$

are completely redundant when specialized to AdS$_6\times M_4$ solutions in IIB, i.e. they are automatically satisfied by the expressions for bispinors and fluxes we found in section 7.5. Since the analysis of the case at hand is similar to the ones presented in [89] and [150] (for four- and two-dimensional Minkowski vacua respectively), we will briefly describe the main computations and point out the novelties arising for an AdS vacuum.

Firstly, we need to choose the vectors $e_{+_a}$. Intuitively, these auxiliary vectors are needed because the form $\Phi$ is not enough by itself to specify a vielbein; for more details, see [89]. The $e_{+_a}$ can be chosen quite freely, provided they satisfy the constraints

$$e_{+_a}^2=0\ ,\quad e_{+_a}\cdot K_a=\frac{1}{2}\ . \tag{B.19}$$

Since $K_1=K_2=K=\frac{1}{2}z$ has only external indices, we will set

$$e_{+_1}=e_{+_2}\equiv e_+\ , \tag{B.20}$$

and we will consider $e_+$ to be purely external as well. This is just the one-form that in (B.7) we had to leave undetermined; as we anticipated there, it is an auxiliary piece of data and cannot be determined as a bilinear of $\zeta_\pm$. For Minkowski vacua, $K$ is a constant vector, and one can then simply take $e_+$ to be constant too. In AdS, however, the requirement that $K$ be a Killing vector does not imply that it is constant, and hence there is no reason to have $e_+$ constant either. However, we will argue that $e_+$ can be chosen in such a way to at least make the $d_6^\dagger e_+$ terms in (B.18) vanish. To this end, let us first define the spinors $\tilde\zeta_\pm$ along the lines of (B.6):

$$\tilde\zeta_\pm\equiv\begin{pmatrix}0\\1\end{pmatrix}\otimes\chi_\pm\ , \tag{B.21}$$

[2] The Clifford action from the left (right) of a ten-dimensional gamma matrix on a $k$-form $\omega_k$ is given by [89]:

$$\Gamma^M\omega_k=(dx^M\wedge+g^{MN}\iota_N)\omega_k\ ,\quad \omega_k\Gamma^M=(-)^k(dx^M\wedge-g^{MN}\iota_N)\omega_k\ .$$

and the one-form

$$e_+ \equiv (\tilde{\zeta}_+ \overline{\tilde{\zeta}_+})_{\text{one-form}} \propto \overline{\tilde{\zeta}_+} \gamma^{(6)}_\mu \tilde{\zeta}_+ \, dx^\mu \ , \tag{B.22}$$

which satisfies $e_+^2 = 0$, $e_+ \cdot K \neq 0$; thus, by appropriate rescaling, taking (B.20) and (B.22) will indeed satisfy (B.19). Since (B.21) now also satisfies the Killing spinor equations (7.7), $d_6^\dagger e_+$ vanishes.

Another difference with respect to the Minkowski case comes from the term $d_H(e^{-\phi}\Phi \cdot e_+)$. Using the formula $\left\{d, \cdot e_+ (-)^{\text{deg}}\right\} = e^{-A}\partial_+ + dA \wedge e_+ \cdot$, we can write it as

$$d_H(e^{-\phi}\Phi \cdot e_+) = (d_H(e^{-\phi}\Phi)) \cdot e_+ - e^{-\phi} dA \wedge e_+ \cdot \Phi - e^{-(A+\phi)}\partial_+\Phi \ . \tag{B.23}$$

As usual, the first term on the right hand side vanishes inside a pairing,[3] while the last one does not (contrary to the Minkowski case), and we must evaluate it. Since $\partial_+\Phi = \delta^{+\mu}\nabla_\mu\Phi = \delta^{+\mu}\nabla_\mu(\epsilon_1\overline{\epsilon_2})$, we can use the decomposition (7.8) and the equations (7.7) to conclude that

$$\partial_+\Phi = \frac{1}{2} e_+ \cdot \hat{\Phi} + \dots \ , \tag{B.24}$$

where the dots denote terms that vanish in the pairing in (B.18a), and where we defined

$$\hat{\Phi} \equiv (\hat{\epsilon}_1\overline{\epsilon_2}) \ , \qquad \hat{\epsilon}_1 \equiv \zeta_-\eta_+^1 + \zeta_-^c \eta_+^{1\,c} + \zeta_+\eta_-^1 + \zeta_+^c \eta_-^{1\,c} \ . \tag{B.25}$$

To sum up, for type IIB AdS$_6 \times M_4$ vacua we can rewrite (B.18a) as

$$\left(e_+ \cdot \Phi \cdot e_+, \Gamma^{MN}\left[e^{-\phi} dA \wedge (e_+ \cdot \Phi) + \frac{e^{-(A+\phi)}}{2} e_+ \cdot \hat{\Phi} - 2F\right]\right) = 0 \ ; \tag{B.26}$$

to rewrite the flux term we have made use of the formula

$$\left(e_+ \cdot \Phi \cdot e_+, F_{(10)}\right) = 2\left(e_+ \cdot \Phi \cdot e_+, F\right) \ . \tag{B.27}$$

From now on the analysis parallels the one for Minkowski vacua, and we will not repeat it here. Specializing (B.18a), (B.18b) to the case $M = m$, $N = n$

[3] This is because $e_+^2 = 0$. Just replace $C$ with $(d_H(e^{-\phi}\Phi)) \cdot e_+$ in the following formula [89, Sec. B.4]:

$$(e_+ \cdot \Phi \cdot e_+, C) = -\frac{(-)^{\deg(\Phi)}}{32} \overline{\epsilon_1} e_+ C e_+ \epsilon_2 \ .$$

does not give any equations; specializing them to the cases $M = \mu$, $N = \nu$ and $M = m$, $N = \nu$ gives[4]

$$\left(\Psi_+^0 + \Psi_-^0, F\right) = e^{-\phi} \ , \tag{B.29a}$$

$$\left(\Psi_+^\alpha - \Psi_-^\alpha, F\right) = 0 \ , \tag{B.29b}$$

$$\left(dx_m \wedge (\Phi_+^0 - \Phi_-^0), F\right) = -e^{A-\phi}\partial_m A \ , \tag{B.29c}$$

$$\left(\iota_m(\Phi_+^0 - \Phi_-^0), F\right) = 0 \ . \tag{B.29d}$$

It can be shown that these equations transform into identities upon plugging in the expressions for the solutions to the system (7.13). This completes the proof of the redundancy of (B.18a) and (B.18b) for $\mathrm{AdS}_6 \times M_4$ vacua in type IIB.

## B.2. $\mathrm{AdS}_6$ solutions of eleven-dimensional supergravity

We will show here that there are no $\mathrm{AdS}_6 \times M_5$ solutions of eleven-dimensional supergravity.[5] This case is easy enough that we will deal with it by using the original fermionic form of the supersymmetry equations, without trying to reformulate them in terms of bilinears as we did in the main text for IIB.

The bosonic fields of eleven-dimensional supergravity consist of a metric $g_{11}$ and a three-form potential $C$ with four-form field strength $G = dC$. The action is

$$S = \frac{1}{(2\pi)^8 l_{\mathrm{Pl}}^9} \int R *_{11} 1 - \frac{1}{2} G \wedge *_{11} G - \frac{1}{6} C \wedge G \wedge G \ , \tag{B.30}$$

with $l_{\mathrm{Pl}}$ the eleven-dimensional Planck length.

We take the eleven-dimensional metric to have the warped product form

$$ds_{11}^2 = e^{2A} ds_{\mathrm{AdS}_6}^2 + ds_{M_5}^2 \ . \tag{B.31}$$

In order to preserve the $\mathrm{SO}(2,5)$ invariance of $\mathrm{AdS}_6$ we take the warping factor to be a function of $M_5$, and $G$ to be a four-form on $M_5$. Preserved supersymmetry is equivalent to the existence of a Majorana spinor $\epsilon$ satisfying the equation

$$\nabla_M \epsilon + \frac{1}{288} \left( \Gamma_M^{\ NPQR} - 8\delta_M^N \Gamma^{PQR} \right) G_{NPQR}\, \epsilon = 0 \ . \tag{B.32}$$

[4] As a curiosity, notice that (B.29c) can also be written as

$$\sqrt{g} * \left( (\Phi_+^0 - \Phi_-^0) \wedge \lambda(F) \right) = -e^{A-\phi} dA \ . \tag{B.28}$$

[5] This conclusion was also reached independently by F. Canoura and D. Martelli.

We may decompose the eleven-dimensional gamma matrices via

$$\Gamma^{(6+5)}_{\mu} = e^{A}\gamma^{(6)}_{\mu} \otimes 1 \ , \quad \Gamma^{(6+5)}_{m+5} = \gamma^{(6)} \otimes \gamma^{(5)}_{m} \ . \tag{B.33}$$

Here $\gamma^{(6)}_{\mu}$, $\mu = 0, \dots, 5$ are a basis of six-dimensional gamma matrices ($\gamma^{(6)}$ is the chiral gamma), while $\gamma^{(5)}_{m}$, $m = 1, \dots, 5$ are a basis of five-dimensional gamma matrices. The spinor Anzatz preserving $\mathcal{N} = 1$ supersymmetry in AdS$_6$ is

$$\epsilon = \zeta_{+}\eta_{+} + \zeta_{-}\eta_{-} + \text{c.c.} \tag{B.34}$$

where $\zeta_{\pm}$ are the chiral components of a Killing spinor on AdS$_6$ satisfying

$$\nabla_{\mu}\zeta_{\pm} = \frac{1}{2}\gamma^{(6)}_{\mu}\zeta_{\mp} \ , \tag{B.35}$$

while $\eta_{\pm}$ are Dirac spinors on $M_5$.

Substituting (B.34) in (B.32) leads to the following equations for the spinors $\eta_{\pm}$:

$$\frac{1}{2}e^{-A}\eta_{\mp} \pm \frac{1}{2}\gamma^{m}_{(5)}\partial_{m}A\,\eta_{\pm} + \frac{1}{12} *_5 G_{m}\gamma^{m}_{(5)}\eta_{\pm} = 0 \ , \tag{B.36a}$$

$$\nabla_{m}\eta_{\pm} \pm \frac{1}{4} *_5 G_{m}\eta_{\pm} \mp \frac{1}{6} *_5 G_{n}\gamma^{(5)}_{m}\gamma^{n}_{(5)}\eta_{\pm} = 0 \ . \tag{B.36b}$$

Using (B.36) it is possible to derive the following differential conditions on the norms $\eta^{\dagger}_{\pm}\eta_{\pm} \equiv e^{B_{\pm}}$ of the internal spinors:

$$*_5 G = \mp 6\, d_5 B_{\pm} \ , \tag{B.37}$$

$$B_{+} = -B_{-} + \text{const.} \ . \tag{B.38}$$

We can absorb the constant in a redefinition of $\eta_{-}$ so that $B_{+} = -B_{-} \equiv B$; thus

$$*_5\, G = -6\, d_5 B \ . \tag{B.39}$$

The equation of motion for $G$ is then automatically satisfied; in absence of sources, the Bianchi identity reads $d_5 G = 0$, resulting in $*_5 G$ being harmonic. This is in contradiction with $*_5 G$ being exact. This still leaves open the possibility of adding M5-branes extended along AdS$_6$, which would modify the Bianchi identity to $d_5 G = \delta_{\mathrm{M5}}$. However, we will now show that even that possibility is not realized.

Defining $\tilde{\eta}_{\pm} \equiv e^{-B/2}\eta_{\pm}$ we can rewrite (B.36b) as

$$\nabla_{m}\tilde{\eta}_{\pm} \pm \partial_{n}B\,\gamma^{n}_{m}\tilde{\eta}_{\pm} = 0 \ . \tag{B.40}$$

Upon rescaling the metric $ds^2_{M_5} \to e^{-4B} ds^2_{M_5'}$ the equation for $\tilde{\eta}_+$ becomes

$$\nabla'_m \tilde{\eta}_+ = 0 \ . \tag{B.41}$$

In five dimensions the only compact manifold admitting parallel spinors is the torus $T^5$, so we are forced to set $ds^2_{M_5'} = ds^2_{T^5}$. Similarly if we rescale the metric $ds^2_{M_5} \to e^{4B} ds^2_{M_5''}$ the equation for $\tilde{\eta}_-$ becomes

$$\nabla''_m \tilde{\eta}_- = 0 \ , \tag{B.42}$$

so that $ds^2_{M_5''} = ds^2_{T^5}$.[6] We are thus led to the relation

$$e^{-4B} ds^2_{M_5'} = e^{4B} ds^2_{M_5''} \ . \tag{B.43}$$

Since $ds^2_{M_5'} = ds^2_{M_5''} = ds^2_{T^5}$, this implies $B = 0$, and hence $G = 0$ (from (B.39)). This makes the whole system collapse to flat space.

## B.3. The massive IIA solution

We have shown in appendix B.2 that there are no AdS$_6$ solutions in eleven-dimensional supergravity — and hence in massless IIA. As for massive IIA, it was shown in [105] that the only solution is the one in [104]. In this section, we show how that solution fits in the IIA version of the formalism presented in the main text.

For the bispinors $\Phi$ and $\Psi$, we will keep using the definitions given in section 7.3 and the parameterizations given in section 7.4. The main difference is the system for supersymmetry, which in IIB was (7.13), and in IIA reads instead

$$d_H \left[ e^{3A-\phi} (\Phi_- + \Phi_+)^0 \right] + 2 e^{2A-\phi} (\Psi_- - \Psi_+)^0 = 0 \ , \tag{B.44a}$$
$$d_H \left[ e^{4A-\phi} (\Psi_- + \Psi_+)^\alpha \right] + 3 e^{3A-\phi} (\Phi_- - \Phi_+)^\alpha = 0 \ , \tag{B.44b}$$
$$d_H \left[ e^{5A-\phi} (\Phi_- + \Phi_+)^\alpha \right] + 4 e^{4A-\phi} (\Psi_- - \Psi_+)^\alpha = 0 \ , \tag{B.44c}$$
$$d_H \left[ e^{6A-\phi} (\Psi_- + \Psi_+)^0 \right] + 5 e^{5A-\phi} (\Phi_- - \Phi_+)^0 = -\frac{1}{4} e^{6A} *_4 \lambda F \ , \tag{B.44d}$$
$$d_H \left[ e^{5A-\phi} (\Phi_- - \Phi_+)^0 \right] = 0 \ ; \tag{B.44e}$$
$$||\eta^1||^2 = ||\eta^2||^2 = e^A \ . \tag{B.44f}$$

[6] One might try to avoid this conclusion by setting $\tilde{\eta}_-$ to zero. However, (B.36a) would then also set $\tilde{\eta}_+$ to zero.

The bispinors $\Phi$ and $\Psi$ can be easily extracted from the supersymmetry parameters: in terms of the vielbein $\{e^\alpha, e^4\}$,

$$e^\alpha = -w^{-1/6}\frac{1}{2}\sin\alpha\,\hat{e}^\alpha\ , \qquad e^4 = -w^{-1/6}d\alpha\ , \qquad w \equiv \frac{3}{2}F_0\cos\alpha\ , \quad \text{(B.45)}$$

where $\hat{e}^\alpha$ are the left-invariant one-forms on $S^3$, satisfying

$$d\hat{e}^\alpha = \frac{1}{2}\epsilon^\alpha{}_{\beta\gamma}\hat{e}^\beta \wedge \hat{e}^\gamma\ , \qquad \text{(B.46)}$$

we have

$$\Phi_\pm = \frac{1}{8}(\pm 1 - \cos\alpha)\left((1 \pm \mathrm{vol}_4)\mathrm{Id}_2 + i\left(\frac{1}{2}\epsilon^\alpha{}_{\beta\gamma}e^\beta \wedge e^\gamma \mp e^\alpha \wedge e^4\right)\sigma_\alpha\right)\ ; \qquad \text{(B.47a)}$$

$$\Psi_\pm = \frac{1}{8}\sin\alpha\,(1 \pm *_4)\Big(\mp e^4\mathrm{Id}_2 + ie^\alpha\sigma_\alpha\Big)\ , \qquad \text{(B.47b)}$$

being $\sigma_\alpha$ the Pauli matrices.

The physical fields then read:

$$e^\phi = w^{-5/6}\ , \quad e^A = \frac{3}{2}w^{-1/6}\ , \quad ds^2_{M_4} = e^\alpha e^\alpha + e^4 e^4\ , \quad F_4 = \frac{10}{3}w\,\mathrm{vol}_4\ . \qquad \text{(B.48)}$$

# Appendix C

# More on $\mathrm{AdS}_5$

## C.1. Supersymmetry variations and the Killing vector

Setting to zero the type IIA supersymmetry variations (of gravitinos and dilatinos) yields the following set of equations[1]

$$0 = \left(\nabla_M + \frac{1}{4} H_M\right)\epsilon_1 + \frac{e^\phi}{16}\lambda(F)\Gamma_M\epsilon_2\ , \tag{C.1a}$$

$$0 = \left(\nabla_M - \frac{1}{4} H_M\right)\epsilon_2 + \frac{e^\phi}{16} F\,\Gamma_M\epsilon_1\ , \tag{C.1b}$$

$$0 = \left(\nabla - \partial\phi + \frac{1}{4} H\right)\epsilon_1\ , \tag{C.1c}$$

$$0 = \left(\nabla - \partial\phi - \frac{1}{4} H\right)\epsilon_1\ , \tag{C.1d}$$

where suppressed indices are contracted with antisymmetric products of gamma matrices and $\epsilon_1$, $\epsilon_2$ are Spin$(9,1)$ Majorana–Weyl spinors of opposite chirality.

We wish to obtain a set of differential and algebraic equations for the Spin$(5)$ spinors $\eta_1$, $\eta_2$ and so we decompose the the generators of Cl$(9,1)$ as

$$\Gamma_\mu = e^A\gamma_\mu^{(4,1)}\otimes 1\otimes\sigma_3 \qquad \Gamma_i = 1\otimes\gamma_m^{(5)}\otimes\sigma_1\ , \tag{C.2}$$

where $\mu = 0,\ldots,4$, $m = 1,\ldots,5$ and $\sigma_1$ and $\sigma_3$ are the Pauli matrices; $\gamma_\mu^{(4,1)}$ generate Cl$(4,1)$ and $\gamma_m^{(5)}$ Cl$(5)$. Accordingly, the chirality matrix $\Gamma_{(10)}$ and the intertwiner $B_{10}$ relating $\Gamma_M$ and $\Gamma_M^*$, are decomposed as

$$\Gamma_{(10)} = 1\otimes 1\otimes\sigma_2\ , \qquad B_{10} = B_{4,1}\otimes B_5\otimes\sigma_1\ . \tag{C.3}$$

[1] The first two equations follow from setting the gravitino variation $\delta\psi_M$ to zero, while the last two equations follow from $\Gamma^M\delta\psi_M - \delta\lambda = 0$ where $\lambda$ is the dilatino.

Furthermore, the supersymmetry parameters $\epsilon_1$, $\epsilon_2$ split as

$$\epsilon_1 = (\zeta \otimes \eta_1 + \zeta^c \otimes \eta_1^c) \otimes \theta \ , \tag{C.4a}$$
$$\epsilon_2 = (\zeta \otimes \eta_2 + \zeta^c \otimes \eta_2^c) \otimes \theta^* \ , \tag{C.4b}$$

where $\eta^c_{1,2} = B_5 \eta^*_{1,2}$ and $\zeta^c = B_{4,1}\zeta^*$. $\zeta$ is a Spin$(4,1)$ spinor obeying the AdS$_5$ Killing spinor equation

$$\nabla_\mu \zeta = \frac{1}{2}\gamma_\mu \zeta \ , \tag{C.5}$$

while $\theta$ obeys $\sigma_2\theta = \theta$ and $\sigma_1\theta = \theta^*$.

Applying the above decomposition, the equations (C.1) become

$$0 = \left(\nabla_i + \frac{1}{4}H_i\right)\eta_1 + \frac{e^\phi}{16}\lambda(F)\gamma_m^{(5)}\eta_2 \ , \tag{C.6a}$$
$$0 = \left(\nabla_i - \frac{1}{4}H_i\right)\eta_2 + \frac{e^\phi}{16}F\,\gamma_m^{(5)}\eta_1 \ , \tag{C.6b}$$
$$0 = \left(\frac{i}{2}e^{-A} - \frac{1}{2}\partial A\right)\eta_1 - \frac{e^\phi}{16}\lambda(F)\eta_2 \ , \tag{C.6c}$$
$$0 = \left(\frac{i}{2}e^{-A} + \frac{1}{2}\partial A\right)\eta_2 + \frac{e^\phi}{16}F\eta_1 \ , \tag{C.6d}$$
$$0 = \left(\frac{5i}{2}e^{-A} - \nabla - \frac{5}{2}\partial A + \partial\phi - \frac{1}{4}H\right)\eta_1 \ , \tag{C.6e}$$
$$0 = \left(\frac{5i}{2}e^{-A} + \nabla + \frac{5}{2}\partial A - \partial\phi - \frac{1}{4}H\right)\eta_2 \ . \tag{C.6f}$$

Using equations (C.6a) and (C.6b) it is straightforward to show that $\xi \equiv \frac{1}{2}(\eta_1^\dagger\gamma^m\eta_2 - \eta_2^\dagger\gamma^m\eta_2)\partial_m$ satisfies

$$\nabla_{(m}\,\xi_{n)} = 0 \ , \tag{C.7}$$

i.e. that $\xi$ is a Killing vector, while equations (C.6c) and (C.6d) yield $L_\xi A = 0$. That $L_\xi\phi = 0$ follows from the algebraic equations obtained from (C.6e) and (C.6f) afer eliminating $\nabla$, using (C.6a) and (C.6b).[2]

[2]These conditions also follow directly from setting the dilatino variation $\delta\lambda$ to zero.